\documentclass{aa}  

\usepackage{graphicx}
\usepackage{natbib}
\bibpunct{(}{)}{;}{a}{}{,}
\usepackage{stfloats}
\usepackage{enumitem} %enumerate with a), b), c)

\usepackage{txfonts}
\usepackage{hyperref}

\begin{document} 

\title{Environments of a sample of AzTEC submillimetre galaxies in the COSMOS field}
   \titlerunning{Environments of a sample of AzTEC submillimetre galaxies in the COSMOS field}

   \author{N. \'Alvarez Crespo\inst{1,2},
           V. Smol\u ci\' c \inst{1},
           A. Finoguenov \inst{3}, 
            L. Barrufet \inst{4}
          and M. Aravena \inst{5}
          }
\authorrunning{N. \'Alvarez Crespo et al.}
   \institute{Department of Physics, University of Zagreb, 
              Bijeni\u cka cesta 32, 10000 Zagreb, Croatia
          \and
      European Space Agency (ESA), European Space Astronomy Centre (ESAC), Camino Bajo del Castillo s/n, 28692 Villanueva de la Ca\~{n}ada, Madrid, Spain\\
                   \email{nalvarez@sciops.esa.int}
                   \and
        Department of Physics, University of Helsinki, Gustaf H\"allstr\"omin
        katu 2, FI-00014 Helsinki, Finland
        \and
         Geneva Observatory, University of Geneva, Ch. des Mail-lettes 51, 1290 Versoix, Switzerland
        \and
        N\'ucleo de Astronom\'{\i}a de la Facultad de Ingenier\'{\i}a y Ciencias, Universidad Diego Portales, Av. Ej\'ercito Libertador 441, Santiago, Chile
        }

%   \date{Received September 15, 1996; accepted March 16, 1997}

\abstract
% 5 {} token are mandatory
{} 
{Submillimetre galaxies (SMGs) are bright sources at submillimetre wavelengths ($F_{850\mu m} >$ 2 - 5 mJy). Made up of mostly of high-z galaxies ($z >$ 1), SMGs are amongst the most luminous 
dusty galaxies in the Universe.
These galaxies are thought to be the progenitors of the massive elliptical galaxies in the local Universe and to reside in massive haloes at early epochs. Studying their environments and clustering strength is thus important to put these galaxies in a cosmological context. }
{We present an environmental study of a sample of 116 SMGs  in 96 ALMA observation fields, which were initially discovered with the AzTEC
camera on ASTE and identified with high-resolution
1.25 mm ALMA imaging within the COSMOS survey field, having either spectroscopic or unambiguous photometric redshift.
We analysed their environments making use of the latest release of the COSMOS photometric catalogue, COSMOS2015, a catalogue 
that contains precise photometric redshifts for more than half a million objects over the 2deg$^{2}$ COSMOS field. 
We searched for dense galaxy environments computing the so-called overdensity parameter as a function of distance within a radius of 5' from the SMG.
We validated this approach spectroscopically for those SMGs for which spectroscopic redshift is available.
 As an additional test, we searched for extended X-ray emission as a proxy for the  hot intracluster medium, performing an  X-ray stacking analysis in the 0.5 – 2 keV band 
with a 32" aperture and our SMG position
using all available {\it XMM–Newton} and {\it Chandra} X-ray observations of the COSMOS field.
}
{We find that 27\% (31 out of 116) of the SMGs in our sample are located in a galactic dense environment; this fraction that is similar to previous studies. The spectroscopic redshift is known for 15 of these 31 sources, thus
this photometric approach is tested using spectroscopy. We are able to confirm that 7 out of 15 SMGs lie in high-density peaks. 
However, the search for associated extended X-ray emission via an X-ray stacking analysis leads to a detection that is not statistically significant. }
{}


   \maketitle
%
%-------------------------------------------------------------------

\section{Introduction}
Submillimetre galaxies (SMGs) are amongst the most luminous dusty galaxies in the Universe \citep{2017MNRAS.464.1380W},
 which emit most of their energy at submillimetre (sub-mm) wavelengths, from 200 $\mu$m to about 1mm \citep{2017MNRAS.465.1789G}.
They were initially discovered in extragalactic submillimetre surveys using the Submillimetre Common-User Bolometer Array 
\citep[SCUBA; ][]{1999MNRAS.303..659H}  \citep{1998Natur.394..248B,1998Natur.394..241H}. 
These galaxies are highly star forming, reaching star formation rates (SFRs) up to thousands of $M_\sun$ yr$^{-1}$ \citep[see e.g.][]{2014PhR...541...45C,2018A&A...619A.169R}.
 The bulk of this population 
 has a 
redshift distribution that ranges between $z \sim$ 2 and 3 with a tail reaching up to $z \sim$ 5 \citep{2016ApJ...822...80S,2017A&A...608A..15B,2019MNRAS.487.4648S}.
At these high redshifts, SMGs are found to generate a significant fraction of
the energy release of all galaxies at such distances.
Most of their radiation comes from thermal continuum emission from dust grains from the interstellar medium (ISM) heated by a
 population of optically obscured young massive stars \citep[for a review see ][]{2002PhR...369..111B}. 

%Clusters and proto-clusters of galaxies
Galaxy clusters are the largest gravitationally bound systems in the Universe and are usually located at the knots of the filamentary large-scale
structures. These clusters are typically composed  of 30 to 100s of galaxies, with total masses up to $10^{15} M_{\odot}$ and sizes of around 1 Mpc 
\citep[see e.g.][]{1996astro.ph.11148B}.
Their progenitors are proto-clusters, which are structures found at redshifts 2 $< z <$ 7 that extend over tens of Mpc forming a structure that eventually coalesces into a galaxy cluster
\citep[see e.g.][]{2005Natur.435..629S,2009MNRAS.394..577O}. 
Proto-clusters are rare and difficult to observe, although some observational efforts have been made
to search, identify, and characterise them  \citep[see e.g.][]{2005ApJ...620L...1O,2010A&A...522A..58G,2012ApJ...750..137T}. 
The detection of proto-clusters is important to understand hierarchical
structure formation and stellar mass growth in galaxies
at early times.

%SMGs in clusters
Most likely, SMGs are the progenitors of massive elliptical galaxies observed in the local universe \citep[see e.g.][]{1999ApJ...518..641L,2013Natur.498..338F,2014ApJ...782...68T}.
Since early-type galaxies are predominantly found in clusters, it is important to address the question of whether SMGs are preferably found in regions with enhanced galaxy density.  
Studying the clustering properties of SMGs can also provide constraints on their nature in a cosmological context.
Some models depict SMGs as a long-lived episode of star formation in the most massive galaxies, driven by the early fast collapse
of the dark matter halo \citep{2012MNRAS.422.1324X}, yielding strong clustering for these sources.
On the other hand, other models in which SMGs are
short-lived bursts in less massive galaxies, predict 
weaker clustering \citep{2011MNRAS.417.2057A}. 
So far, most of the studies on the environments of SMGs have been done in individual sources or small samples 
 \citep{2000ApJ...542...27I,2002MNRAS.331..817S,2010ApJ...708L..36A,2018ApJ...856...72O,2020MNRAS.tmp.1394H},
which has not really allowed us to test the abundance of structures around these galaxies and has forbidden clustering measurements.

The Cosmic Evolution Survey \citep[COSMOS; ][]{2007ApJS..172....1S} includes multiwavelength imaging and spectroscopy from X-ray to radio wavelengths over an area of 1.4 $\times$ 1.4 deg
at a sufficient depth to provide a comprehensive view of galaxy formation and large-scale structure. 
In this work we aim to improve the clustering measurements of SMGs by extending the current statistical sample.
We evaluate overdensities of a sample of 116 SMGs in the COSMOS field
that were initially discovered with the AzTEC camera on the Atacama Submillimeter Telescope Experiment   \citep[ASTE; ][]{2011MNRAS.415.3831A}, and subsequently identified 
 with high-resolution 1.25 mm ALMA imaging  in 96 different fields   \citep{2017A&A...608A..15B}.
 
% Description sections
This paper is organised as follows. In Sect.~\ref{sec:data} we introduce our
SMG sample and the COSMOS catalogues used for the environmental study. In Sect. ~\ref{sec:method} we describe the methodology used to measure the overdensities
 and their significance and false detection rate. 
In Sect. ~\ref{sec:result} we present the results of our analysis.
Then in Sect. ~\ref{sec:spectr}  we verify the overdensities using spectroscopic redshift. 
Later in Sect.  ~\ref{sec:discussion}  we
discuss our results comparing them to what previously found in the literature, and finally in Sect. ~\ref{sec:conclusion} we present our conclusions. 

Unless otherwise stated, we assume a flat $\Lambda$CDM cosmology with a Hubble constant $H_0 =$ 73 km s$^{-1}$ Mpc $^{-1}$, total matter density $\Omega_m =$ 0.27, and dark energy density
$\Omega_\Lambda =$ 0.73 \citep{2007ApJS..170..377S,2011ApJS..192...16L}.
Magnitudes throughout this paper refer to the AB magnitude system \citep{1974ApJS...27...21O}.

%--------------------------------------------------------------------
\section{Data}
\label{sec:data}

\subsection{SMG sample}
 The sources studied  in this work are the so-called "strict" subsample from \cite{2017A&A...608A..15B}. These 116 SMGs have been detected using high spatial resolution ($\sim$ 1") targeted observations in cycle 2 
ALMA operations of 96 fields within the COSMOS survey, designed to include a sample
of AzTEC/ASTE sources with 1.1 mm flux densities $\geq$ 1 mJy \citep{2011MNRAS.415.3831A}. 
 For this work  we use those sources that are defined as the "strict" sample, for which either spectroscopic or
  unambiguous photometric redshift is determined. The median redshift of this sample is $\tilde{z}$ = 2.3 $\pm$ 0.6.

For 30 SMGs the spectroscopic redshift is available.  Seven of these values were calculated using CO measurements (AzTEC/C1a, C2a,
C3a, C5, C6a, C6b and C17), and the rest were taken from the COSMOS spectroscopic redshift catalogue (available internally for members of the COSMOS collaboration). 
For those sources with photometric redshift, they were measured by \cite{2017A&A...608A..15B} cross correlating the ALMA positions with the latest release of the COSMOS photometric catalogue, COSMOS2015.

 The complete list of SMGs is given in Table A.1 of \cite{2017A&A...608A..15B}. Throughout this paper we use the nomenclature given by these authors; for alternative names, see the second column of their Table A.1.  Of these 96 different ALMA observations, 19 are  multiple component observation fields, meaning there are several resolved SMGs inside each field. The 39 SMGs belonging to multi-component systems are described by an alphabetical tag in descendant order of brightness for each observing field. Some of these sources are physically related, while others belong to the same ALMA observation only as a result of chance alignment. This is explored in more detail in Sect.~\ref{subsec:multicomponent}.

\subsection{Catalogues}
To study the environments of SMGs, we
 made use of the latest version of the COSMOS\footnote{For more information we refer to the COSMOS webpage\\
 \url{http://cosmos.astro.caltech.edu} .}
 photometric catalogue \citep[COSMOS2015 hereafter; ][]{2016ApJS..224...24L}. This catalogue includes photometric measurements from the 
ultraviolet (UV) to the infrared (IR) wavelengths,
including 6 broad optical bands ({\it B, V, g, r, i, z$^{++}$}), 12 medium bands, and 2 narrow bands, as well as {\it Y, J H} and {\it K$_S$} data from the UltraVISTA Data Release 2 new HyperSuprime-Cam
Subaru {\it Y} band and new SPLASH 3.6 and 4.5 $\mu$m {\it Spitzer}/Infrared Array Camera (IRAC) data \citep{2007ApJS..172...86S,2007ApJS..172...99C,2012A&A...544A.156M,2013A&A...556A..55I}.

Furthermore, we used a catalogue of spectroscopic redshifts in the COSMOS field available internally for members of the COSMOS collaboration. 
It is composed of 36,274 spectroscopic redshifts, both available only internally to the COSMOS collaboration and from the following surveys:
\begin{enumerate}
\item The zCOSMOS-bright survey contributed 8,608 galaxies at 0.1 $\leq z \leq$ 1.2 performed with the VIsible Multi-Object Spectrograph (VIMOS) at the
Very Large Telescope (VLT) covering the entire COSMOS field  \citep{2007ApJS..172...70L,2009ApJS..184..218L}.
\item The zCOSMOS-deep survey for fainter sources, contributing 767 galaxies with secure redshifts at  1.4 $\leq z \leq$ 3.0 and covering the central 1 deg$^2$ of the COSMOS field (Lilly et al. in prep.).
\item The 6,617 galaxies with high- quality spectra from the DEIMOS 10K Spectroscopic Survey Catalog of the COSMOS Field, which is a survey
 that samples a broad redshift distribution in up to $z =$ 6 \citep{2018ApJ...858...77H}.
 \item The 2,022 galaxies  observed through multi-slit spectroscopy with the Deep Imaging Multi-Object Spectrograph (DEIMOS) on the Keck II telescope, having a broad
redshift distribution 0.02 $< z <$ 6 \citep{2010ApJ...721...98K}.
\item A number of 998 galaxies from VIMOS Ultra Deep Survey (VUDS), a spectroscopic redshift survey of very faint galaxies that covers the central 1 deg$^2$ of the COSMOS field at 2 $< z <$ 6 \citep{2015A&A...576A..79L,2017A&A...600A.110T}. 
\item A catalogue based on [OII] flux-calibrated spectroscopy of a total of 788 galaxies at 0.1 $< z<$ 1.65, obtained with the VLT/FORS-2 instrument  \citep{2015A&A...575A..40C}.
\item The spectroscopic survey of galaxies in the COSMOS field using the Fiber Multi-object
Spectrograph (FMOS), a near-IR instrument on the Subaru Telescope at 1.34 $\leq z \leq$ 1.73 contributed 178 galaxies \citep{2015ApJS..220...12S}.
\item The COSMOS field from the MOSFIRE Deep Evolution Field (MOSDEF) survey at 0.8$\leq z \leq$ 3.71 contributed 80 galaxies \citep{2015ApJS..218...15K}.
\item A sample of 26 galaxies at 0.82 $\leq z \leq$1.50 observed via near-IR spectroscopy with  Subaru-FMOS and selected from a sample of IR luminous galaxies \citep{2012MNRAS.426.1782R}.
\item Wide Field Camera 3 (WFC3) of the  Hubble Space Telescope grism spectroscopy and imaging for
a sample of 11 galaxies at 1.88 $\leq z \leq$ 2.54 \citep{2014ApJ...797...17K}.
\item The XSHOOTER spectrograph at the VLT at  1.98 $\leq z \leq$ 2.48 observed 14 galaxies \citep{2015MNRAS.451.2050Z}.
 \item Near-IR spectroscopic observations of 10 passive galaxies with Subaru-MOIRCS at 1.24 $\leq z \leq$ 2.09 \citep{2012ApJ...755...26O}.
 \end{enumerate}

The COSMOS2015 photometric redshift accuracy was estimated by \cite{2017A&A...597A...4S} to test whether photometric redshifts could
 be efficiently used to search for overdensities and especially at
high redshifts ($z >$ 3.5), where the photometric redshift uncertainty is higher than lower redshifts. 
These authors performed a comparison study between the COSMOS2015 and the COSMOS spectroscopic redshift catalogue
by measuring the distribution of $\Delta z / (1+z_{spec})$, where $\Delta z$ is the difference
between the spectroscopic and photometric redshifts. They find the standard deviation of the distribution to be  $\sigma_{\Delta z / (1+z_{spec})} =$ 0.0067 for $z_{phot} \leq$ 3.5
and  $\sigma_{\Delta z / (1+z_{spec})} =$ 0.0155 for $z_{phot} >$ 3.5, verifying a good photometric redshift accuracy.

\section{Methodology}
\label{sec:method}

\subsection{Overdensity  parameter}
\label{subsec:overdparameter}
To search for overdensities in the SMG fields, we use the galaxies in the COSMOS2015 photometric catalogue that lie 
within the redshift range $z_{phot} = z_{SMG} \pm 3 \sigma_{\Delta z / (1+z_{SMG})} (1+z_{SMG})$ from each SMG, being 
$\sigma_{\Delta z / (1+z_{spec})} =$ 0.0067 for $z_{phot} \leq$ 3.5
and  $\sigma_{\Delta z / (1+z_{spec})} =$ 0.0155 for $z_{phot} >$ 3.5. This interval is selected large enough to account for the uncertainties of the photometric sources.

First we compute the galaxy overdensity parameter to all those sources from the COSMOS2015 photometric catalogue
 within a distance of $r =$ 0'.5, 1', 2'.5, and 5' from the position of the central SMG and in the above-mentioned redshift range.
The galaxy overdensity parameter is defined as a function of radius as

\begin{equation}
\delta_g (r)= \frac{\Sigma_r (r)- \Sigma_{bg}}{\Sigma_{bg}}=\frac{\Sigma_r (r)}{\Sigma_{bg}}-1
,\end{equation}
where $\Sigma_r$ is the local galaxy surface density calculated as $\Sigma_r = N_r/ A_r$, where $N_r$ is the number of galaxies within the given radius and redshift bin in a
 search window area of  $A_r= \pi \times r^2$. 
Correspondingly, $\Sigma_{bg}$ is the background galaxy surface density and is defined as $\Sigma_{bg}= N_{bg}/ A_{bg}$. The quantity $N_{bg}$ is the number of galaxies satisfying the photometric
redshift interval within the entire area $A_{bg}$ of the COSMOS field, to take  masked areas due to saturated stars and/or corrupted data into account.
Conforming to this definition, $\delta_g (r) >$ 0 corresponds to an overdensity and  $\delta_g (r) <$ 0 means an underdense region.

\subsection{Significance and false detection rate}
\label{subsec:falsedetrate}
The probability of observing $\geq N_r$  objects when the expected number is
$n_r = \Sigma_{bg} \times A_r$ is analytically defined  by the Poisson distribution $p(\geq N_r,n_r) = 1 - \Sigma_{i=0}^{N_r} (e^{-n_r} n_{r}^{i}/i!)$. 
We calculate the significance of the overdensity parameter by computing the Poisson probability for
each different radius and the value $\delta_g (r)$ is considered as robust when  p $\leq$ 0.05.  

Furthermore, we estimate the false detection
rate by generating 10 mock catalogues for each SMG at a certain $z_{SMG}$  by randomly shifting their positions over the inner 1 deg$^2$ of the COSMOS field.
For each one of these mock catalogues that correspond to a certain SMG at $z_{SMG}$, we generate other 1,000 mock catalogues by randomly distributing in the sky 
the number of galaxies at $z_{phot} = z_{SMG} \pm 3 \sigma_{\Delta z / (1+z_{SMG})} (1+z_{SMG})$ in the COSMOS2015 photometric 
catalogue, leading to 10,000 mock catalogues for each SMG.
The false detection probability is given by the fraction of events in which $N_r$  are found within a radius $r$ over these 10,000 mock catalogues.

\section{Analysis and results}
\label{sec:result}

\subsection{Overdensity  parameter}
The results from our overdensity analysis are represented in the  figures in the Appendix. For each SMG we show the  
overdensity parameter $\delta_g$ as function of the projected radius $r$
measured from the central SMG, from 0.1' up to 5'. For each point we indicate the number of sources $N_r$ within the given radius, including the SMG. When the value of $\delta_g$ has a Poisson probability
$p(\geq N_r, n_r) \leq$ 0.05, that point is enclosed in a square.

A SMG is considered to reside in an overdensity given at least one the following conditions:  First, the probability of observing $\geq N_r$  objects defined analytically by a Poisson distribution within a radius $r$  is $p \leq$ 0.05 (see Sect.~\ref{subsec:falsedetrate}). Second, the false detection probability $P_{FD}$ (calculated numerically using mock catalogues 
at random SMG positions) of finding a number of sources $\geq N_r$ within a radius $r$ is $\leq$ 5\%  (see Sect.~\ref{subsec:falsedetrate}).

To evaluate the significance of $\delta_g$, we start by the smallest considered radius 0'.5.  If $p(\geq N_r, n_r)$ and $P_{FD} >$ 5\% for that radius, we continue to evaluate
 $\delta_g$ at the following values of the radius up to $r =$ 5'. Table~\ref{table:smallscale} shows those SMGs found in high-density peaks according to the above conditions. We find 31 SMGs out of the total 116 sample ($\sim$27\%) located in a significant overdensity, and nearly half of these (15 of 31) belong to the spectroscopic sample.
 
 %%%%%%%%%% TABLE small-scale OVERDENSITIES %%%%%%%%%%%%%%%%%%%%%%%%%%%
\begin{table}[htbp]
\renewcommand{\footnoterule}{}
\caption{SMGs found in an overdensity.}
{\small
\begin{minipage}{1\columnwidth}
\label{table:smallscale}
\begin{tabular}{llccc}
\hline
SMG  &  $r$ & $N_\mathrm{r}$ & Poisson probability &  False detection \\ 
name   &  ['] & & $p(\geq N_\mathrm{r}, n_\mathrm{r})$&  probability P$_\mathrm{FD}$ \\ 
\hline 
AzTEC/C5$^*$   &  0.5 &  2  &  0.055 & 0.031  \\
AzTEC/C6a$^*$  & 0.5 &  4  &  0.104 & 0.004  \\
AzTEC/C6b$^*$  &  0.5 &  4  &  0.106 & 0.002  \\
AzTEC/C9a      &  0.5 &  4  &  0.115 & 0.006  \\
AzTEC/C9b$^*$  &  0.5 &  3  &  0.112 & 0.028  \\
AzTEC/C9c$^*$  &  0.5 &  3  &  0.109 & 0.026  \\
AzTEC/C17$^*$  &  0.5 &  2  &  0.046 & 0.021  \\
AzTEC/C25$^*$  &  0.5 &  3  &  0.106 & 0.022  \\
AzTEC/C28a$^*$ &  0.5 &  5  &  0.012 & 0.005  \\
AzTEC/C28b     &  0.5 &  4  &  0.112 & 0.005  \\
AzTEC/C33a     &  0.5 &  3  &  0.112 & 0.032  \\
AzTEC/C34a     &  0.5 &  4  &  0.130 & 0.014  \\
AzTEC/C43b     &  5.0 & 148 & 0.034 & 0.063 \\
AzTEC/C45$^*$  &  0.5 &  3  &  0.112 & 0.031  \\
AzTEC/C48b     &  0.5 &  4  &  0.142 & 0.021  \\
AzTEC/C50      &  0.5 &  4  &  0.091 & 0.002  \\
AzTEC/C51b     &  0.5 &  4  &  0.154 & 0.033  \\
AzTEC/C52$^*$  &  2.5 & 67 & 0.020 & 0.069 \\
AzTEC/C55b     &  0.5 &  5  &  0.120 & 0.006  \\
AzTEC/C59$^*$  &  0.5 & 5 & 0.048 & 0.071 \\
AzTEC/C60b     &  0.5 &  2  &  0.029 & 0.008  \\
AzTEC/C61$^*$  &  0.5 &  4  &  0.088 & 0.001  \\
AzTEC/C65$^*$  &  1.0 & 12 & 0.013 & 0.085 \\
AzTEC/C71b$^*$ &  0.5 & 9 & 0.007 & 0.052 \\
AzTEC/C79      &  0.5 &  3  &  0.117 & 0.039  \\
AzTEC/C99      &  0.5 &  3  &  0.115 & 0.029  \\
AzTEC/C100a    &  2.5 & 48 & 0.040 & 0.093 \\
AzTEC/C101a   &  5.0 & 160 &  0.049 & 0.095 \\
AzTEC/C117     &  0.5 &  5  &  0.142 & 0.023  \\
AzTEC/C118$^*$ &  0.5 &  3  &  0.115 & 0.033  \\ 
AzTEC/C122a    &  1.0 & 12 & 0.023 & 0.067 \\ 
\hline 
\end{tabular} 
Column description: (1): SMG name according to \cite{2017A&A...608A..15B} nomenclature; (2): radius of $\delta_g$ statistically significant; (3): number of sources within $r$; (4) Poisson probability  $p(\geq N_r, n_r)$; (5) false detection probability $P_{FD}$.\\
 $^*$ SMGs with spectroscopic redshift.
\end{minipage} 
}
\end{table}

\begin{figure}[htbp]
\centering
\includegraphics[width=0.45 \textwidth]{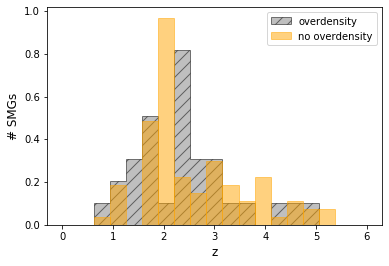}
    \caption{In grey (dashed) the redshift distribution of the SMGs found in a significant overdensity  and in orange (solid)
     those that are not found in a high-density environment. There is no
     significant difference between their redshift distribution at a 95\% confidence level according to the K-S test.} 
       \label{fig:figure2}
\end{figure}

We compare the redshift distributions of the sources located in an overdensity and non-overdensity environments in Fig.~\ref{fig:figure2}.  
A Kolmogorov-Smirnov (K-S) test \citep{1983MNRAS.202..615P,1987MNRAS.225..155F} shows no significant
difference at a 95\% confidence level.
 The physical distances corresponding to the radius of the overdensities as reported in the second column of Table~\ref{table:smallscale} vary  from 232 pkpc at the $z_{min} =$ 0.829,  251 pkpc at $z_{mean} =$ 2.410 and 199 pkpc at
$z_{max} =$ 4.772.
 
 %%%%%%%%%%%%%% Figure redshift histogram small-scale  %%%%%%%%%%%%%%

\subsection{Extended X-ray emission}
X-ray extended emission is an evidence of hot intergalactic medium in clusters of galaxies. Hence, we search for extended X-ray emission around the SMGs positions. 
The entire COSMOS region has been mapped through 54 overlapping {\it XMM-Newton} pointings and additional {\it Chandra} observations \citep{2009ApJS..184..158E,2016ApJ...819...62C} and
previous efforts have been made to search for X-ray emitting galaxy clusters in this field 
\citep{2007ApJS..172...29H,2007ApJS..172..182F,2011ApJ...742..125G}.

 In this work we first cross matched the positions of our SMGs with the revised catalogue of extended X-ray sources in the COSMOS field, which contains 247 X-ray groups with $M_{200c} =$ 8 $\times$ 10$^{12}$ - 3 $\times$ 10$^{14} M_{\odot}$ at a redshift range 0.08 $\leq z <$ 1.53. These sources were obtained from
the combined data from {\it Chandra} and {\it XMM-Newton} mosaic image in the 0.5 - 2 keV band  \citep{2019MNRAS.483.3545G}.
This catalogue has a total cluster flux depth of 3 $\times$ 10$^{-16}$ erg cm$^{-2}$ s$^{-1}$, covers an area of 2.1 deg$^2$, and the precision of the centres for extended sources goes down to 5". 
None of the positions of our SMGs  match any of the X-ray groups reported in the catalogue.

However, since only 11 of 116 SMGs from our sample are at the same redshift as the aforementioned X-ray group catalogue, 
  we then extracted the stacked X-ray flux outside of the already detected groups in the same  {\it Chandra} and {\it XMM-Newton} mosaic  image in the 0.5 - 2 keV band. If the SMG is at the centre of a cluster, it should be located near to the centre of a virialised dark matter halo. The  size of this halo at the typical redshift of SMGs is a few hundreds of pkpcs, so we used a 32" aperture size centred in the SMG, following
 \cite{2009ApJ...704..564F,2015A&A...576A.130F}.  
Prior to this we subtracted background and point-like sources such X-ray jets and cluster cores.
 We detected an average flux of 1.7 $\pm$ 0.9 $\times$ 10$^{-16}$ erg s$^{-1}$ cm$^{-2}$, which is a marginal detection
 so we ruled out the possibility of significant X-ray extended emission for our sample.
 Using the mean redshift of the sample of 2.3, the halo mas (M200) that corresponds to the marginal X-ray flux detection is $(2.5\pm0.7)\times 10^{13} M_\odot$.
We note that this mass is consistent with the typical SMG mass inferred in the clustering studies.

\section{Spectroscopic verification of overdensity candidates}
\label{sec:spectr}
As a secondary test, we now evaluate the overdensity parameter $\delta_g$ for those 15 SMGs having spectroscopic redshift  found in an overdensity 
(see Table~\ref{table:smallscale} in Sect.~\ref{sec:result} )
using spectroscopic redshift catalogues, to confirm what we previously found using photometry.
 For each SMG with a spectroscopic redshift in Table~\ref{table:smallscale}, 
 we calculated the overdensity parameter using the sources from the
  spectroscopic catalogue described in Sect.~\ref{sec:data}, within a radius of $r =$ 0'.5, 1', 2'.5, and 5' from the central SMG and in a redshift range of 
  $z_{spec} = z_{SMG} \pm 3 \sigma_{\Delta z / (1+z_{SMG})} (1+z_{SMG})$, being 
$\sigma_{\Delta z / (1+z_{spec})} =$ 0.0067 for $z \leq$ 3.5
and  $\sigma_{\Delta z / (1+z_{spec})} =$ 0.0155 for $z >$ 3.5, similar to the photometric analysis.

 In this case, $\Sigma_{bg}$ is calculated by randomly locating 1,000 SMGs in the non-uniform footprint of the COSMOS field covered by the spectroscopic catalogue, as reported in Sect.~\ref{sec:data}, 
 with a random redshift in the interval 0 $<z< 3.5+3 \sigma_{\Delta z / (1+z_{SMG})} (1+3.5)$ for $z\leq$ 3.5 within a radius of 5'. For $z>$ 3.5 the redshift is randomly selected
 within the range $3.5-3 \sigma_{\Delta z / (1+z_{SMG})} (1+3.5)< z <$ 6.0.
The value of the overdensity parameter $\delta_g$ along with its associated uncertainty
is calculated using a truncated Gaussian to the distribution of the number of galaxies generated in the 1,000 simulations at $z \leq$ 3.5. Since the number of recovered sources
is small at $z >$ 3.5 (35,964/36,274 sources in the spectroscopic catalogue lie at $z \leq$ 3.5), above this redshift the function fitted to evaluate  the overdensity parameter is instead Poissonian. 

 Out of the 15 SMGs analysed, we find 7 SMGs in environments that are significantly overdense, meaning $\delta_g \geq 3\sigma$, confirming the photometric identified overdensities for AzTEC/C5, AzTEC/C6a, AzTEC/C6b, AzTEC/C17, AzTEC/C52, AzTEC/C59, and AzTEC/C118.
We note that the compilation of the spectroscopic catalogue is strongly redshift dependent, where $\sim$ 99\% of the observed sources lie at $z \leq$ 3.5, which diminishes our capability 
to spectroscopically confirm sources at high redshift.
Additionally, since spectroscopic sources are taken from a compilation of different catalogues, the spectral coverage is not completely uniform.
These effects limit the conclusions that can be extracted from this analysis and can only be used as a confirmation for those cases showing positive results. 

\section{Discussion}
\label{sec:discussion}
 
 \subsection{Multi-component SMGs}
 \label{subsec:multicomponent}
  The sample in this paper contains 116 different SMGs, observed in 96 different  ALMA fields, meaning
19 out of 96 observations contain more than one SMG. These are the so-called multi-component systems and 39 of the 116 SMGs are one of them. In the following we unravel which SMGs belonging to multi-component systems are gravitationally bounded, and those for which this is simply due to chance alignment. 

The SMGs AzTEC/C6a and AzTEC/C6b are separated by $\Delta z =$ 0.023, a distance slightly higher than the threshold suggested by \cite{2013MNRAS.434.2572H}, which differentiates between chance and physical associations of ALMA sources. However, both SMGs reside in a high density environment, which points towards a physical association. 

Similarly, although each SMG of the triplet in the field AzTEC/C9 lies within 13'' of its closest neighbour, a bit higher than the threshold, all of these are found within an overdensity so they are very likely physically associated.
The redshifts for both AzTEC/C28a and  AzTEC/C28b are compatible with being identical as is their separation; they are physically associated and are found in an overdensity.

Although redshifts in AzTEC/C43a and AzTEC/C43b are consistent with being identical, we find only one SMG in an overdensity, that is AzTEC/C43b.  This could happen if SMG is not located at the centre of the overdensity, so only one of the components would be found using our method. The same occurs for the field AzTEC/C48, 
while AzTEC/C48a and AzTEC/C48b are physically related, only AzTEC/C48b is found in a dense environment.

 Although photometric redshifts for AzTEC/C55a and AzTEC/C55b are compatible with being identical 
($z_{AzTEC/C55a} =$ 2.49$^{+0.33}_{-0.45}$ and $z_{AzTEC/C55b} =$ 2.77$^{+0.32}_{-0.41}$),
considering their physical separation (17.2", 41 kpc at the measured redshift) it is very unlikely that they are physically associated and only AzTEC/C55b is located in an overdensity.
AzTEC/C34a and AzTEC/C34b are not physically related since their redshifts differ strongly
 ($z_{AzTEC/C34a} =$ 3.53$^{+0.02}_{-0.52}$ and $z_{AzTEC/C34b} =$ 2.49$^{+0.26}_{-0.50}$), 
 their association is only due to chance alignment and we only find AzTEC/C34a in an overdensity.
 AzTEC/C60a and AzTEC/C60b are not physically associated ($z_{AzTEC/C60a} =$ 0.96$^{+0.14}_{-0.40}$ and $z_{AzTEC/C60b} =$ 4.77$^{+0.14}_{-0.75}$) and only AzTEC/C60b is found in an overdensity. The same situation is found for the field AzTEC/C100; only AzTEC/C100a is found in a dense environment since
AzTEC/C100a and AzTEC/C100b ($z_{AzTEC/C100a} =$ 1.63$^{+0.17}_{-0.44}$ and $z_{AzTEC/C100b} =$ 2.68$^{+0.42}_{-0.63}$) results from a chance alignment. 

\subsection{Comparison with previous results}
Early attempts to measure the environment surrounding SMGs involved their projected
two-dimensional distribution of the sky using projected
two-dimensional angular correlation function (ACF). The results for these early efforts are ambiguous, mostly because of the lack of redshifts to trace the three-dimensional structure \citep[see ][]{2002MNRAS.331..817S,2003MNRAS.344..385B,2003MNRAS.338..303A,2004MNRAS.355..485B,2004ApJ...611..725B,2009ApJ...707.1201W,2012MNRAS.421..284H}.
\cite{2011ApJ...733...92W} analysed several SMGs in the COSMOS field, assuming various redshift distributions to estimate their de-projected ACF. These authors could only set upper limits to the correlation length. Nonetheless, ACF studies have the important limitation that they are only able to measure the average clustering properties of a population, missing the individual differences between its components.

We do not find a significant redshift difference between those SMGs lying in density peaks and those located in environments that are indistinguishable from field galaxies. The correlation between redshift and clustering in SMGs is yet unclear. For instance, studying the clustering of galaxies selected in the IRAC bands, \cite{2006ApJ...641L..17F} did not find any strong redshift evolution in their sample.  \cite{2017MNRAS.464.1380W} however find that on average, SMGs at $z >$ 2.5 occupy high-mass dark matter haloes.   

\cite{2010ApJ...708L..36A} find three SMGs embedded in compact groups (with a typical radial extent of 5" - 10") centred at the positions of the SMGs by constructing number
 density maps of high-redshift {\textit BzK} galaxies in the COSBO field, the inner 20'$\times$ 20' region of the COSMOS field \citep{2007ApJS..172..132B}. Two of the sources in their sample, AzTEC/C6a and AzTEC/C7, also appear in ours. We find  AzTEC/C6a in a overdensity too, while for  AzTEC/C7 the overdensity is not significant in both studies.
 Additionally, we find a similar percentage of SMGs in overdensities, since they find 30\% of their sample in overdensities and we find $\sim$ 27\% and similar physical sizes, although our sample spans a higher redshift interval.
AzTEC/C6 belongs to a proto-cluster, it has been previously found within an X-ray emitting region with 17 spectroscopically confirmed member galaxies 
 \citep{2015ApJ...808L..33C,2016ApJ...828...56W,2018A&A...619A..49C}.

 \cite{2017A&A...597A...4S} performed an analysis using a similar method over a smaller sample of SMGs in the COSMOS field, and ten of our sources overlap (it is important to note they use a different nomenclature; the list of corresponding names is given in Table A.1 of  \cite{2017A&A...608A..15B}. Nonetheless a direct comparison can be performed only for the "small overdensities" part of their study, where the physical distances analysed are similar to those of this study.  Considering only their kpc-scale study, the fraction of SMGs found in a significant overdensity (i.e. $\sim$27\%) is the same as this study.  
The sources AzTEC/C3a, AzTEC/C14, AzTEC/C22a, and AzTEC/C22b (AzTEC2, AzTEC9, AzTEC11S, and AzTEC11N, respectively in \cite{2017A&A...597A...4S}) are not found in an overdensity in both studies. However, for AzTEC/C2a and AzTEC/C18 (AzTEC8, AzTEC12, respectively in their nomenclature), neither of us find an overdensity at small scales, but they find high galactic density when looking at Mpc scales.
The SMGs AzTEC/C5, AzTEC/C6a, and AzTEC/C17 (AzTEC1, Cosbo 3, and J1000+0234 for these authors) are found in a small-scale overdensity and confirmed spectroscopically by both studies. The only difference we encounter is with the source AzTEC/C42 (AzTEC5), which they find in a significantly dense environment while we do not. This could be because the redshift value they use is $z_{phot} = 3.05^{+0.33}_{0.28}$ \citep{2012A&A...548A...4S}, while  a more up-to-date value is available for our study $z_{phot} = 3.63^{+0.56}_{0.37}$ \citep{2017A&A...597A...4S}.

\section{Summary and conclusions}
\label{sec:conclusion}
We explore the clustering properties of a sample of 116 SMGs described as the "strict" sample in \cite{2017A&A...608A..15B},
 drawn from a S/N-limited sample  
initially discovered with the AzTEC
camera on ASTE, and identified with high-resolution
1.25 mm ALMA imaging.
Their redshifts lie within the interval 0.829 $< z <$ 5.152 and they are located within the COSMOS field. We evaluate the overdensity parameter $\delta_g$ in an interval $\Delta z$ and in steps of $r=$ 0'.5, 1', 2'.5, and 5'
from the central SMG, 
using the latest version of the COSMOS photometric catalogue (COSMOS2015). The accuracy is  $\sigma_{\Delta z / (1+z_{spec})} =$ 0.0067 for $z_{phot} \leq$ 3.5
and  $\sigma_{\Delta z / (1+z_{spec})} =$ 0.0155 for $z_{phot} >$ 3.5. Our main results can be summarised as follows:

\begin{enumerate}[label=(\alph*)]
\item  Thirty-one out of 116 ($\sim$ 27\%) of  our sample of AZTEC/ASTE sources selected within the COSMOS survey field are located 
in high-density environments, suggesting that a fraction of the SMGs are linked to  formation of structures.
\item For those SMGs found lying in overdensities with spectroscopic redshifts (15 out of 31), the photometric approach is tested using spectroscopically verified overdensities, which are able to confirm 7 of these SMGs high-density peaks. However, because of the lack of completeness of the spectroscopic catalogue used for this analysis, this approach can only be used as a lower limit. 
\item We search for extended X-ray emission around SMGs via matching the positions of our SMGs to those
 of the revised catalogue of extended X-ray sources in the COSMOS field, which contains combined {\it XMM-Newton} and {\it Chandra} data  in the 0.5 - 2 keV band, with negative results.
 Moreover, we perform an X-ray stacking analysis in the 0.5 - 2 keV band using a 32" aperture size, but
 the average flux found is 1.7 $\pm$ 0.9 $\times$ 10$^{-16}$ erg s$^{-1}$ cm$^{-2}$; this contribution is smaller than 1$\sigma$ so
not statistically significant. 
\end{enumerate}

This is consistent with previous results that show a similar fraction of SMGs in high-density environments at these angular scales \citep{2010ApJ...708L..36A,2017A&A...597A...4S}.
About one-third of our sources are related to the formation of structures at high redshift since they are located in regions with enhanced galaxy density, a fraction that is similar to previous studies \citep{2010ApJ...708L..36A,2017A&A...597A...4S}. 
We do not appreciate a correlation between redshift and clustering strength.

The reason why we do not find a higher fraction of SMGs 
associated with strong galaxy overdensities could be due to
  biases that affect our data. It is a known effect that using photometric redshifts leads to a weaker value of $\delta_g$ \citep{2013ApJ...779..127C}, therefore our results might be biased towards the most prominent overdensities and we could be missing the not-so-dense environments.
To overcome all the caveats present in this work and analyse the environments of SMGs in a robust way, further theoretical and observational efforts are needed, such as a dedicated spectroscopic campaign of the galaxies in the area surrounding the SMGs such as that performed for AzTEC/C6a.

\begin{acknowledgements}
        We thank the anonymous referee for her/his insightful comments that led to improvements in the paper.
        This research was funded by the European Union's Seventh Framework programme under grant agreement 337595 (ERC Starting Grant, "CoSMass"). 
        NAC and LB are supported by European Space Agency (ESA) Research Fellowships.
        We   thank B. C. Lemaux and G. Zamorani for helpful discussion and comments that led to improvements in this paper.
        We gratefully acknowledge the contributions of the entire COSMOS collaboration consisting of more than 100 scientists. 
        More information on the COSMOS survey is available at \url{http://www.astro.caltech.edu/~cosmos}.
        We thank the VUDS team for making the data in the COSMOS field available prior to public release.

\end{acknowledgements}

%-------------------------------------------------------------------
\bibliographystyle{bibtex/aa} % style aa.bst

\appendix
\section*{Appendix A: Figures}

\begin{figure*}[b]
\begin{center}
\includegraphics[width=0.23\textwidth]{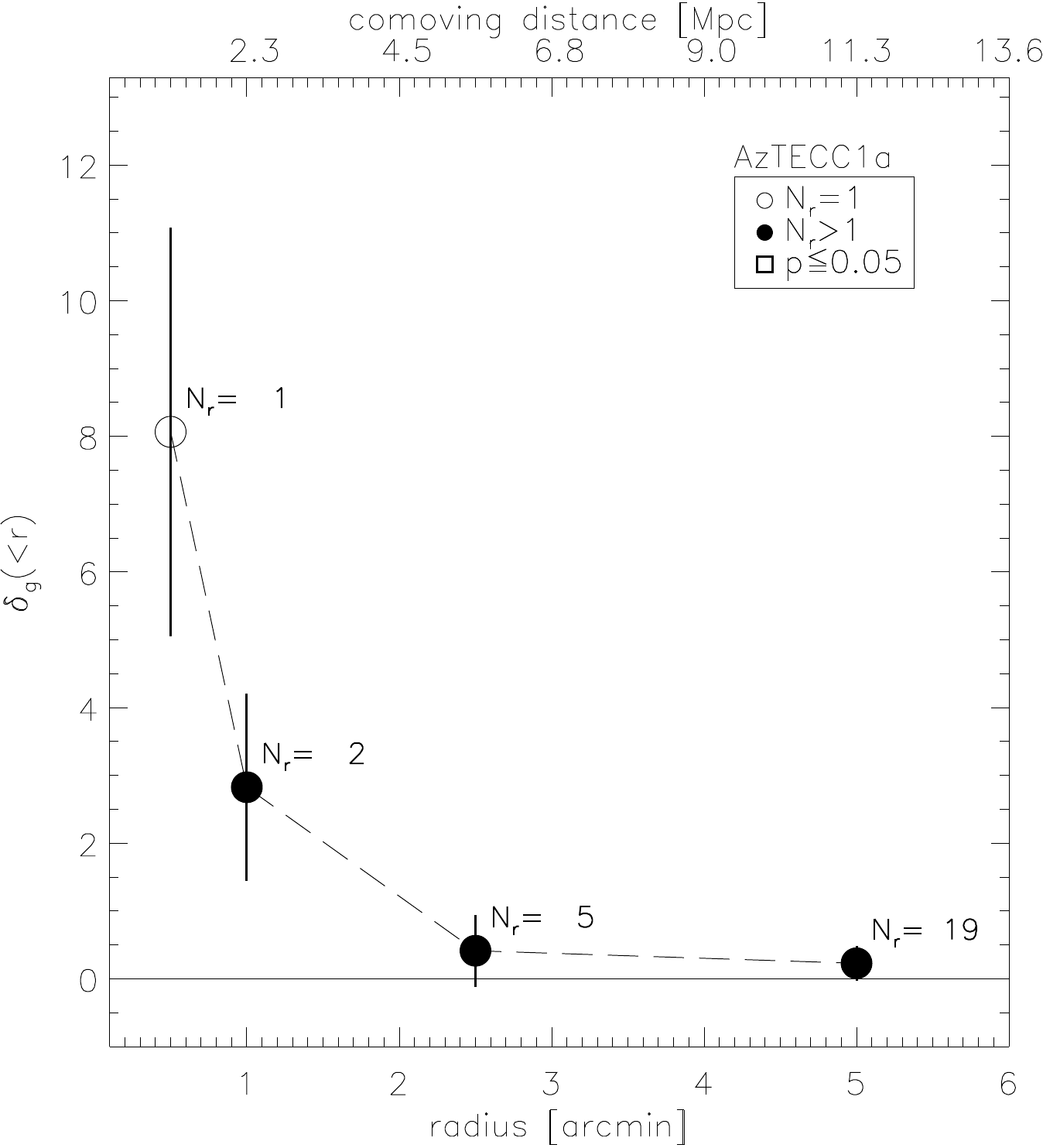}
\includegraphics[width=0.23\textwidth]{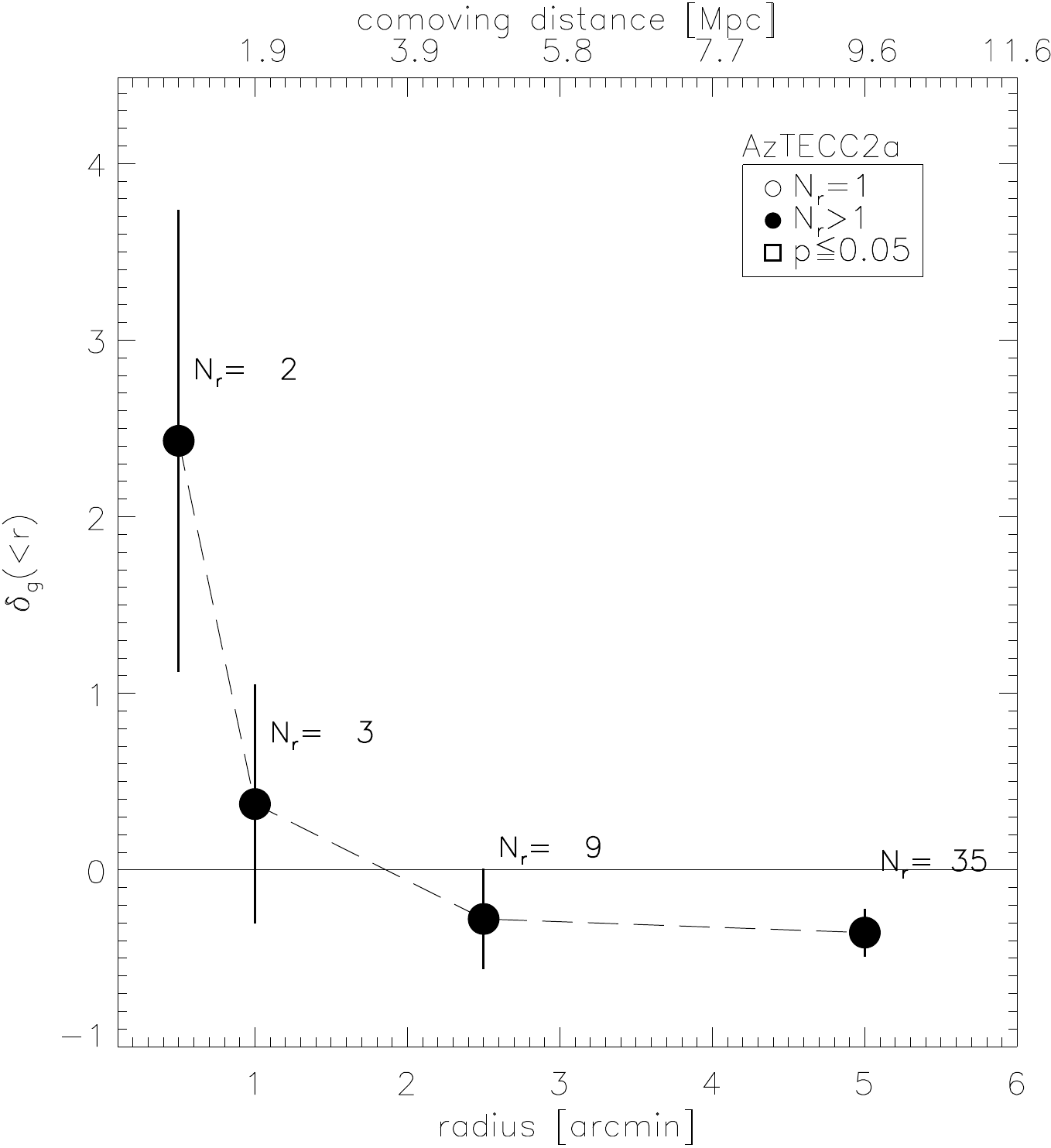}
\includegraphics[width=0.23\textwidth]{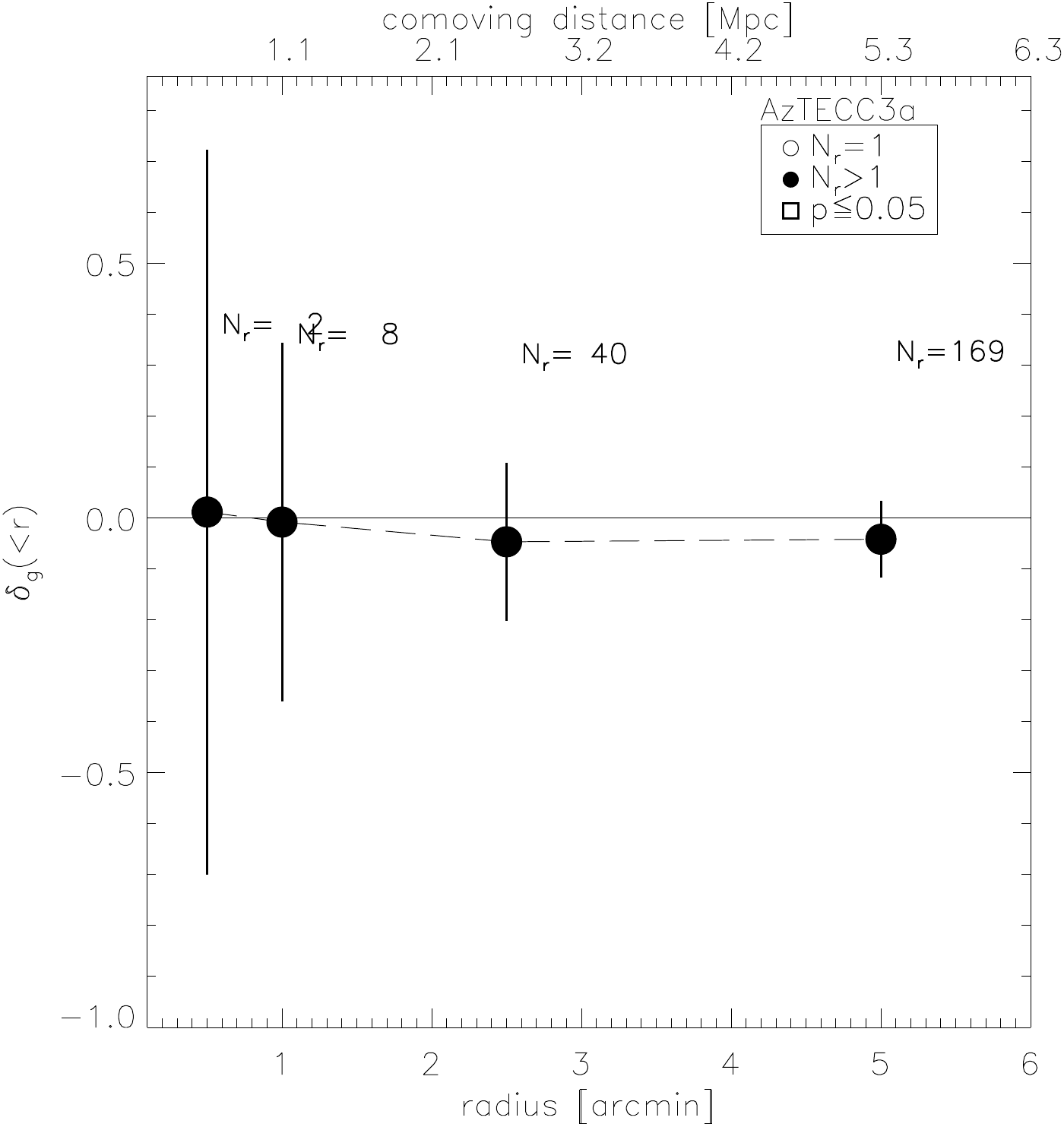}
\includegraphics[width=0.23\textwidth]{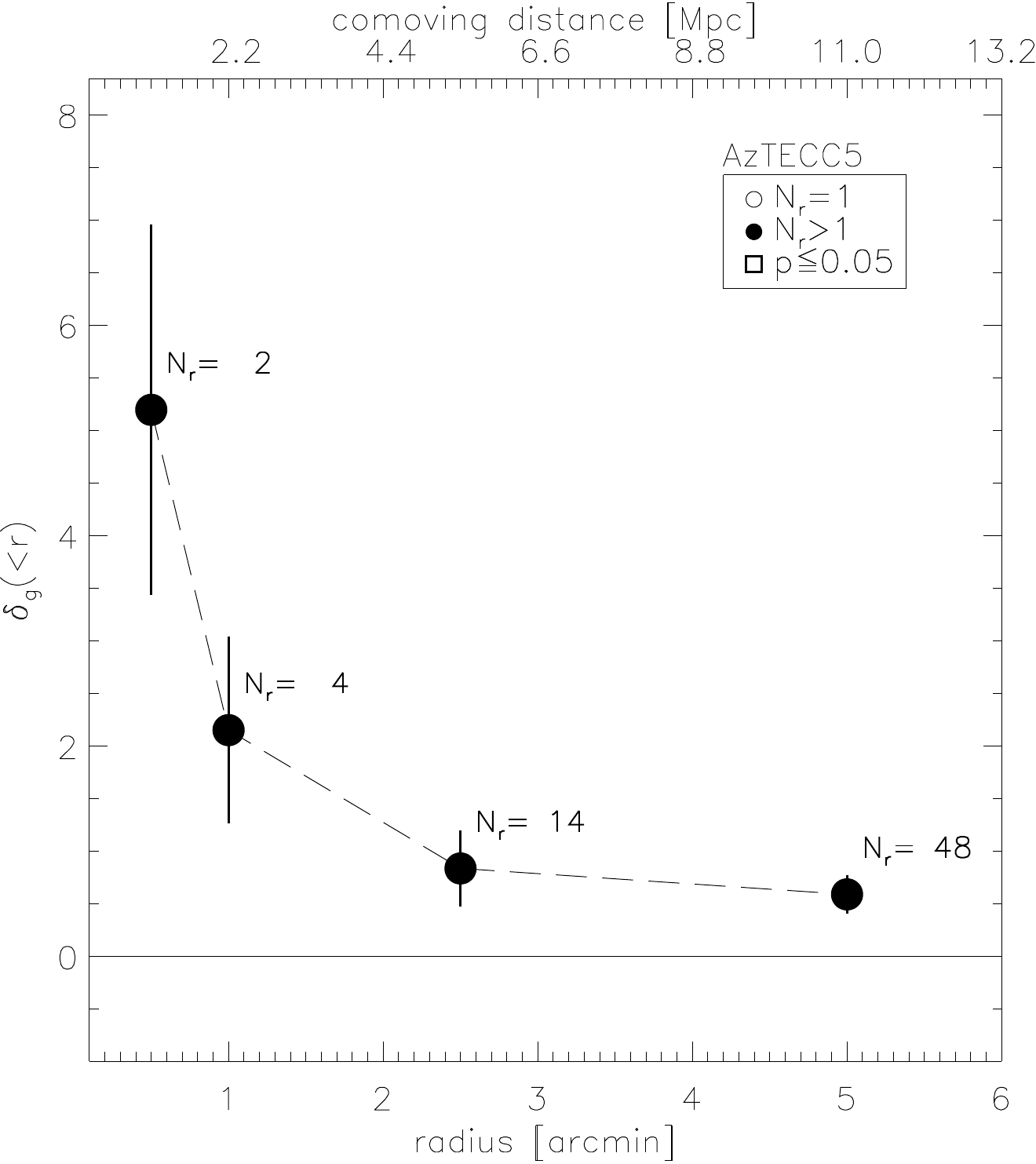}
\includegraphics[width=0.23\textwidth]{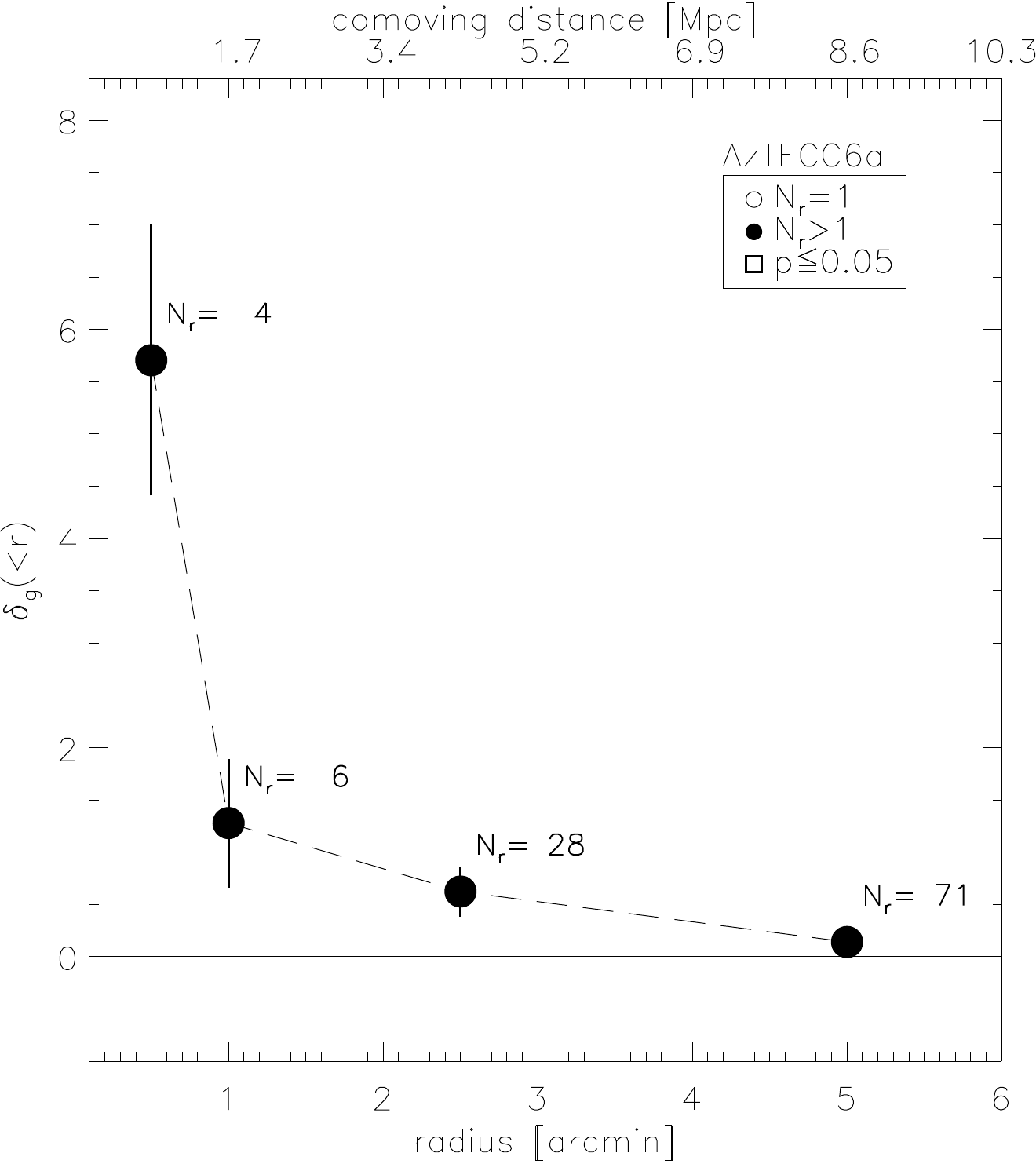}
\includegraphics[width=0.23\textwidth]{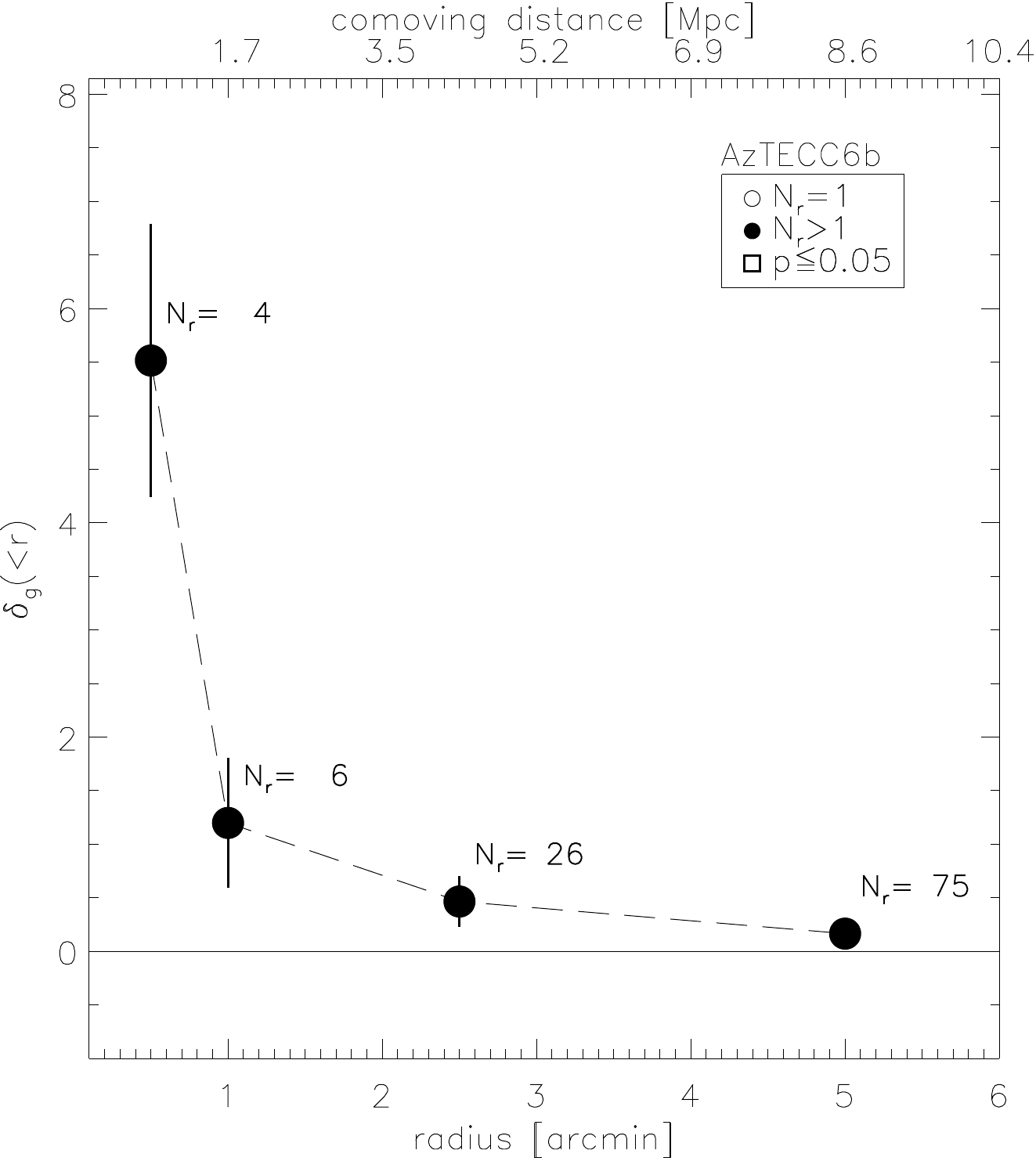}
\includegraphics[width=0.23\textwidth]{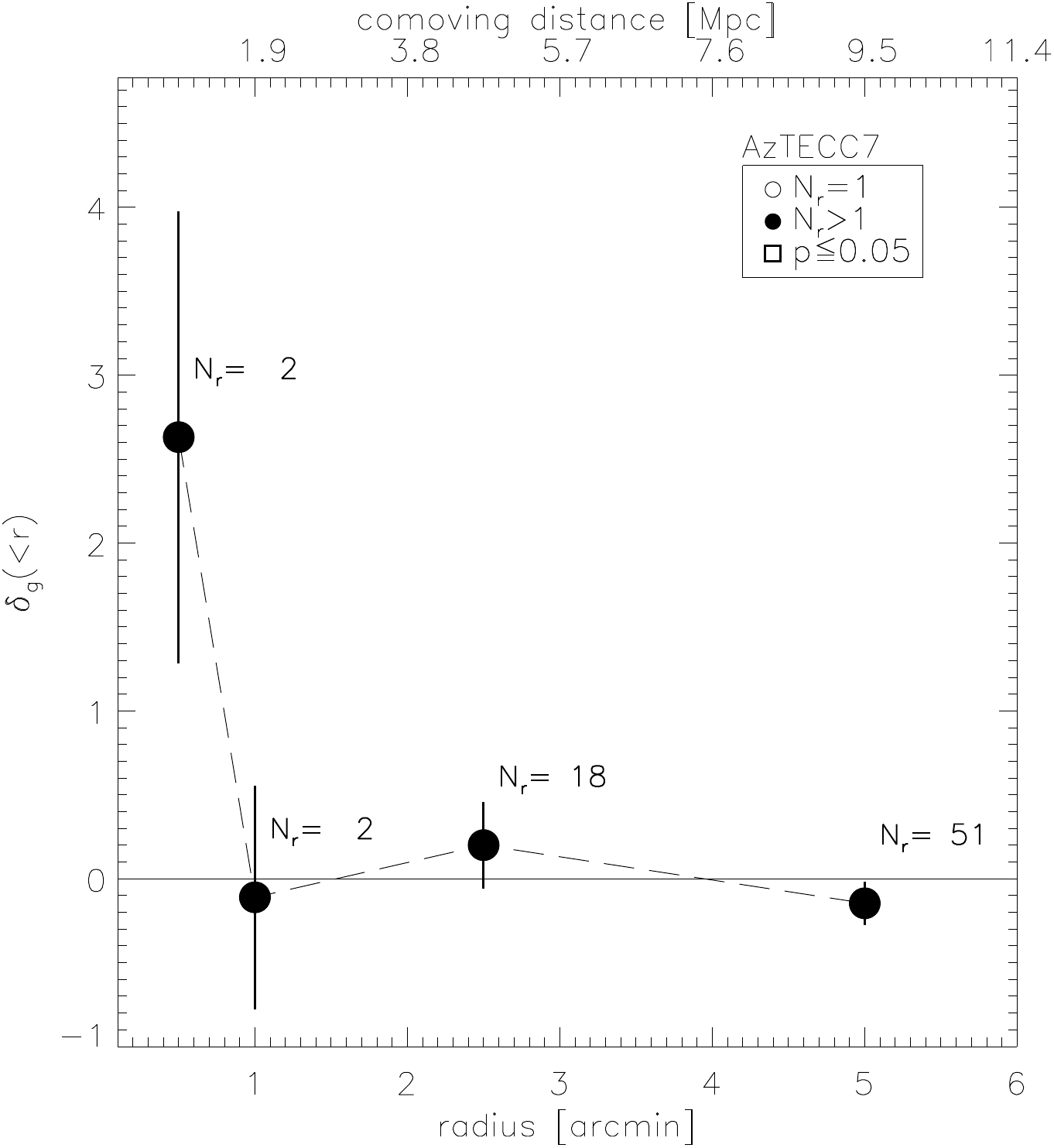}
\includegraphics[width=0.23\textwidth]{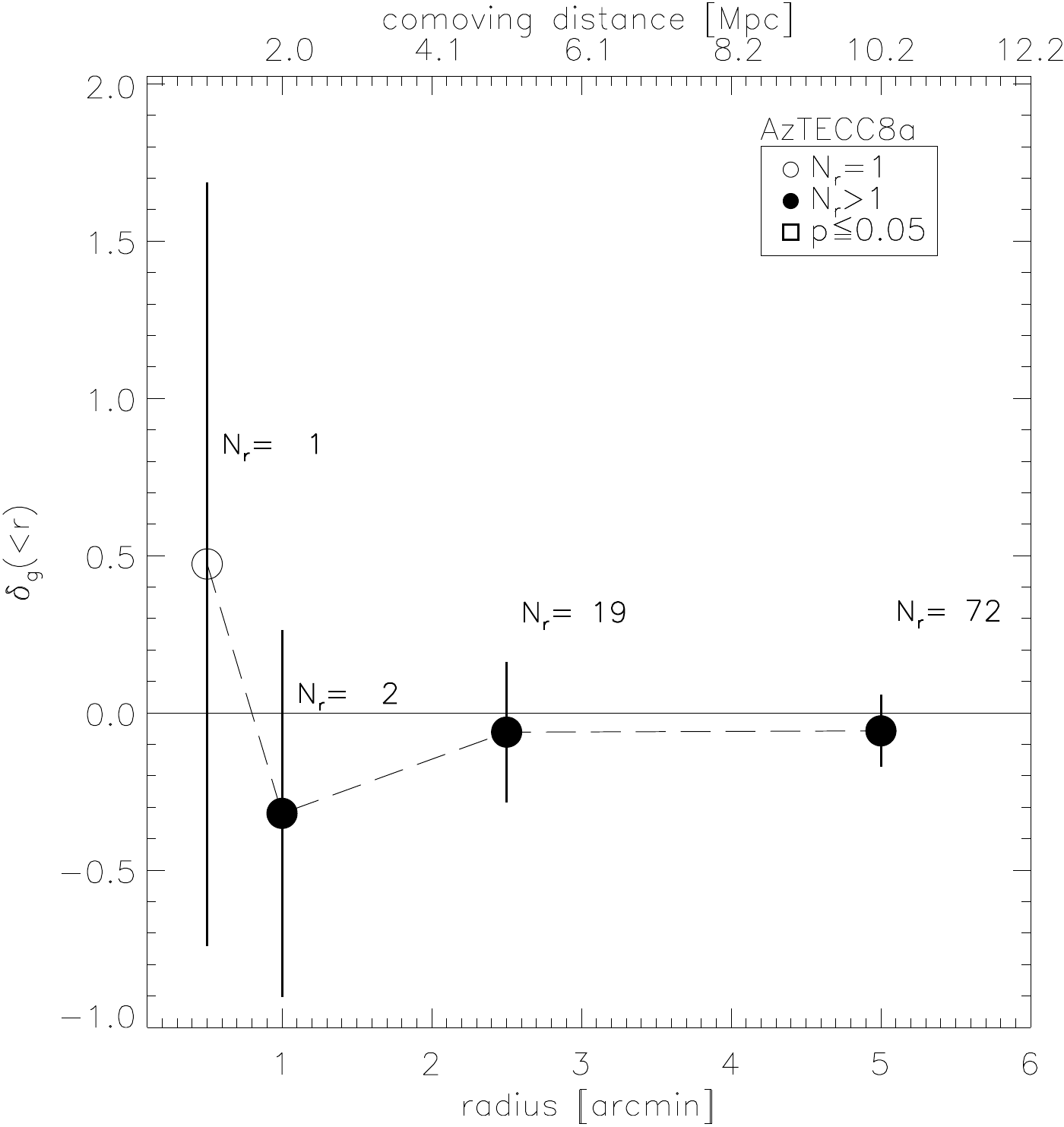}
\includegraphics[width=0.23\textwidth]{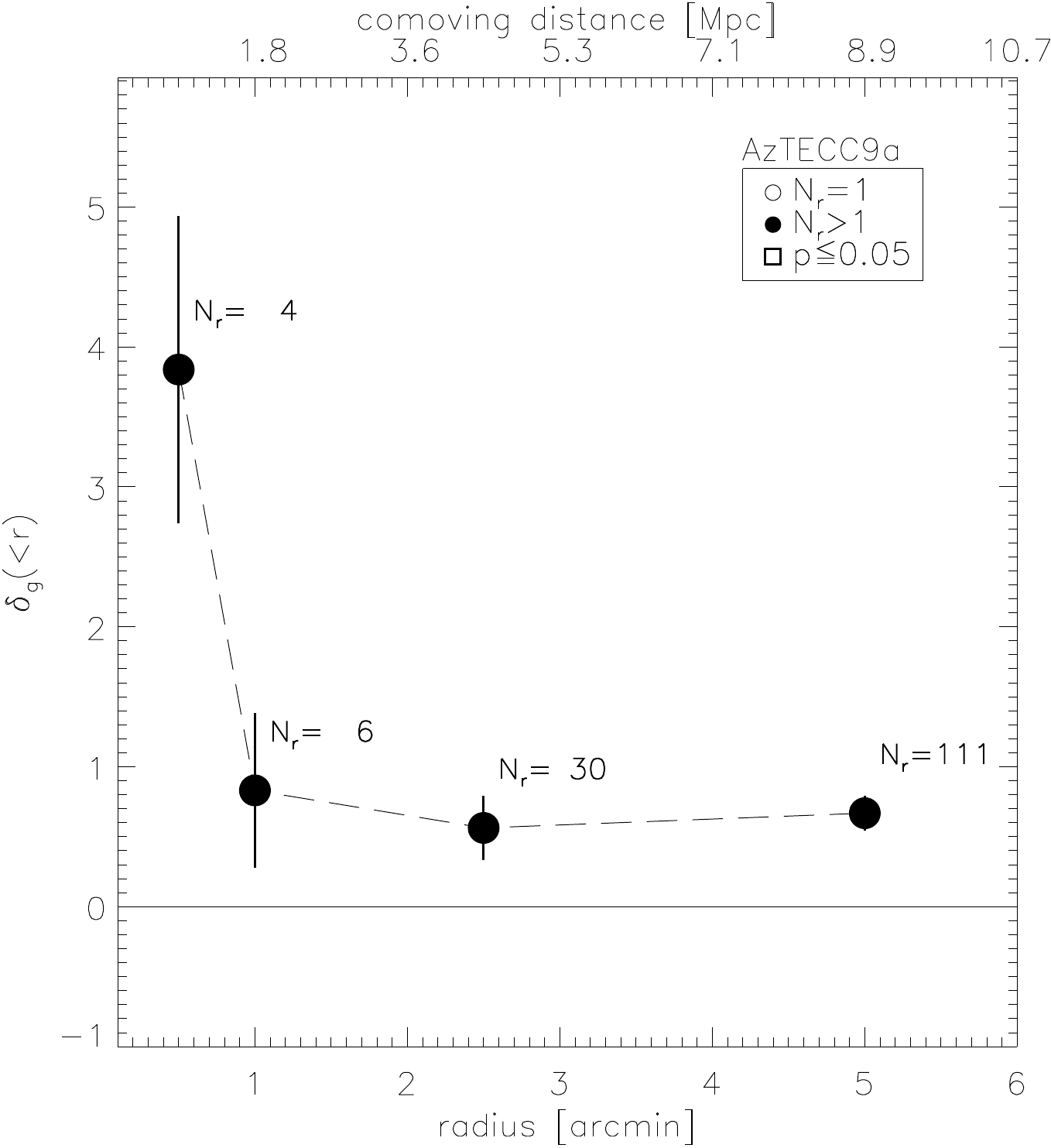}
\includegraphics[width=0.23\textwidth]{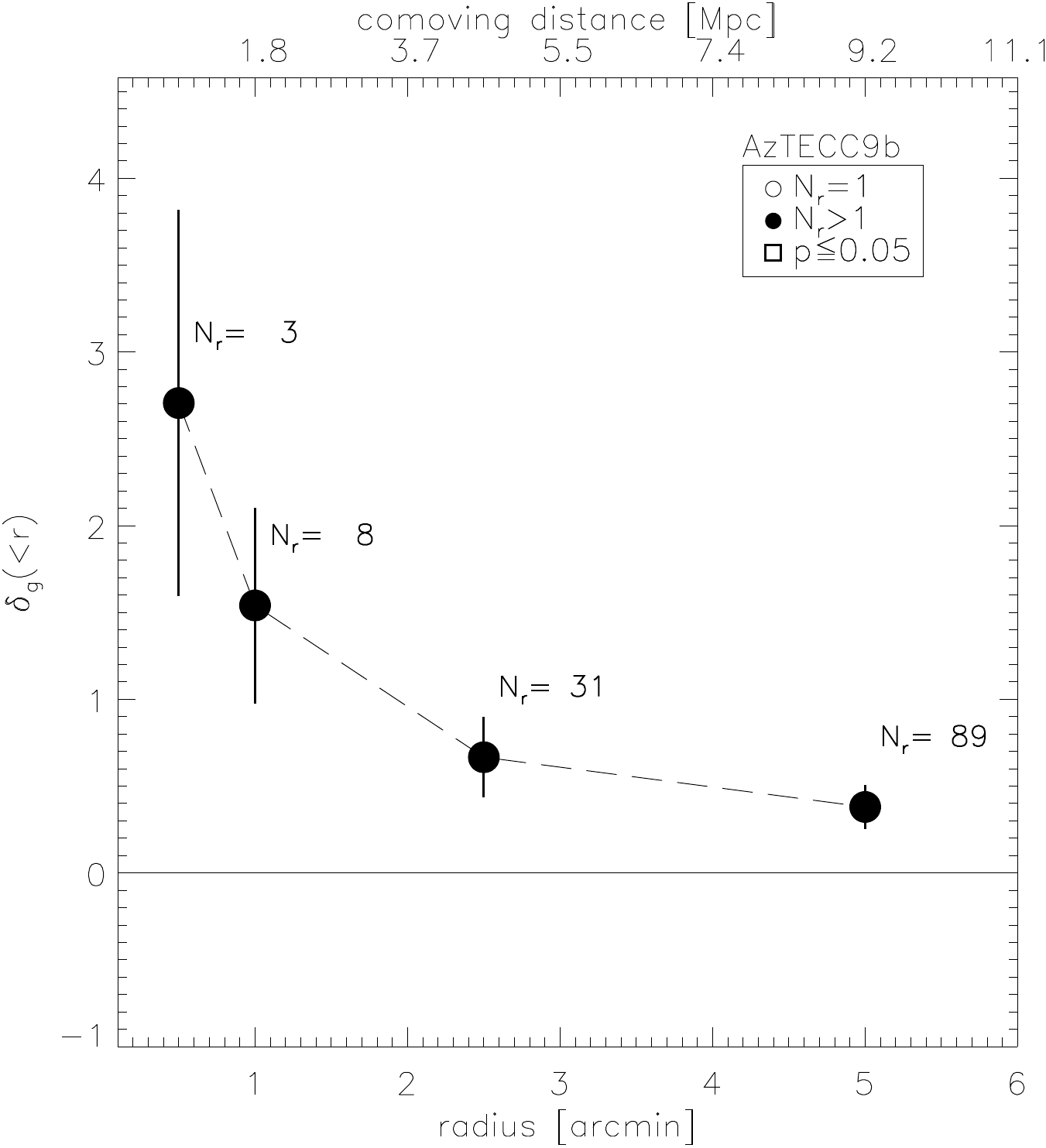}
\includegraphics[width=0.23\textwidth]{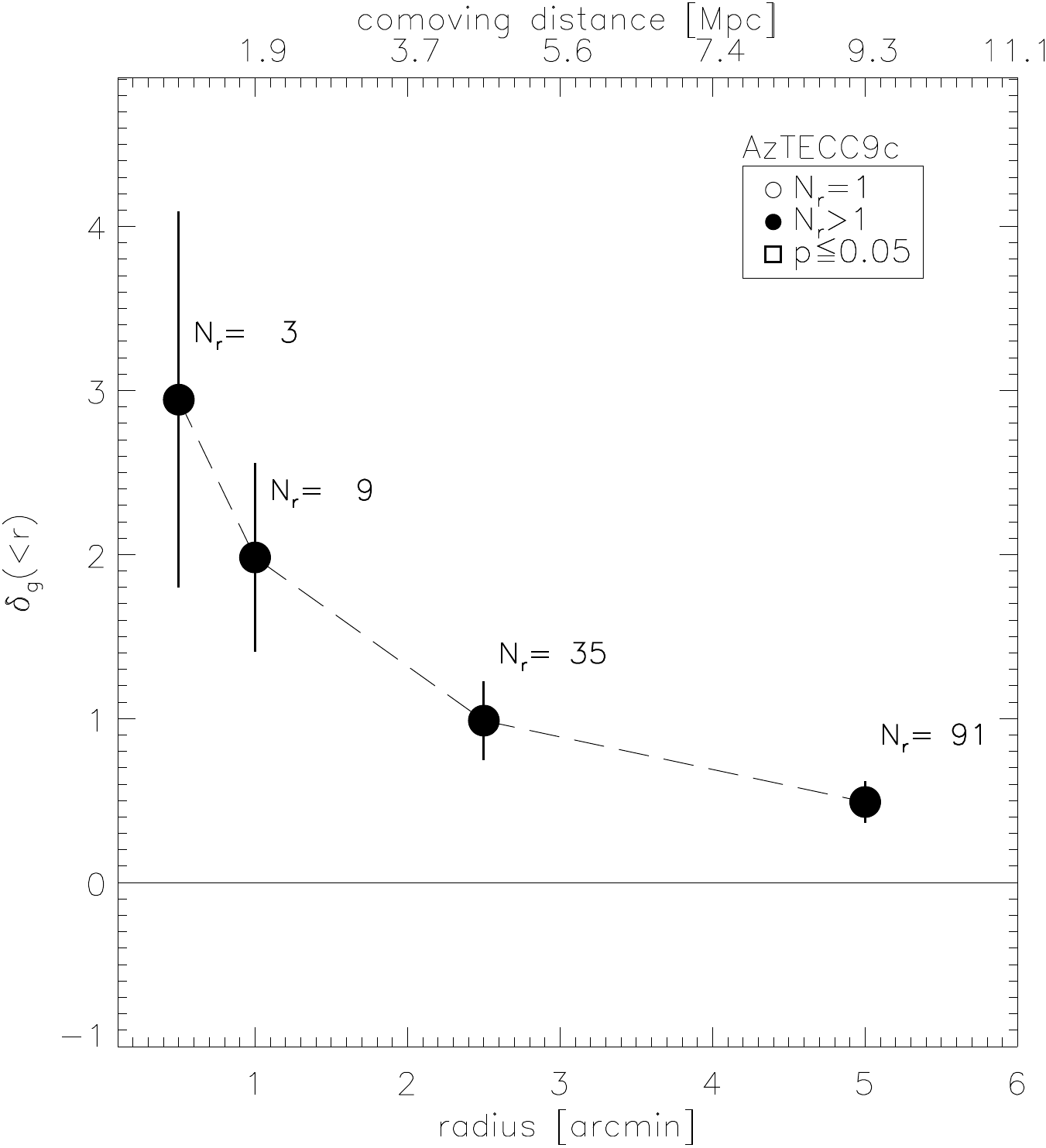}
\includegraphics[width=0.23\textwidth]{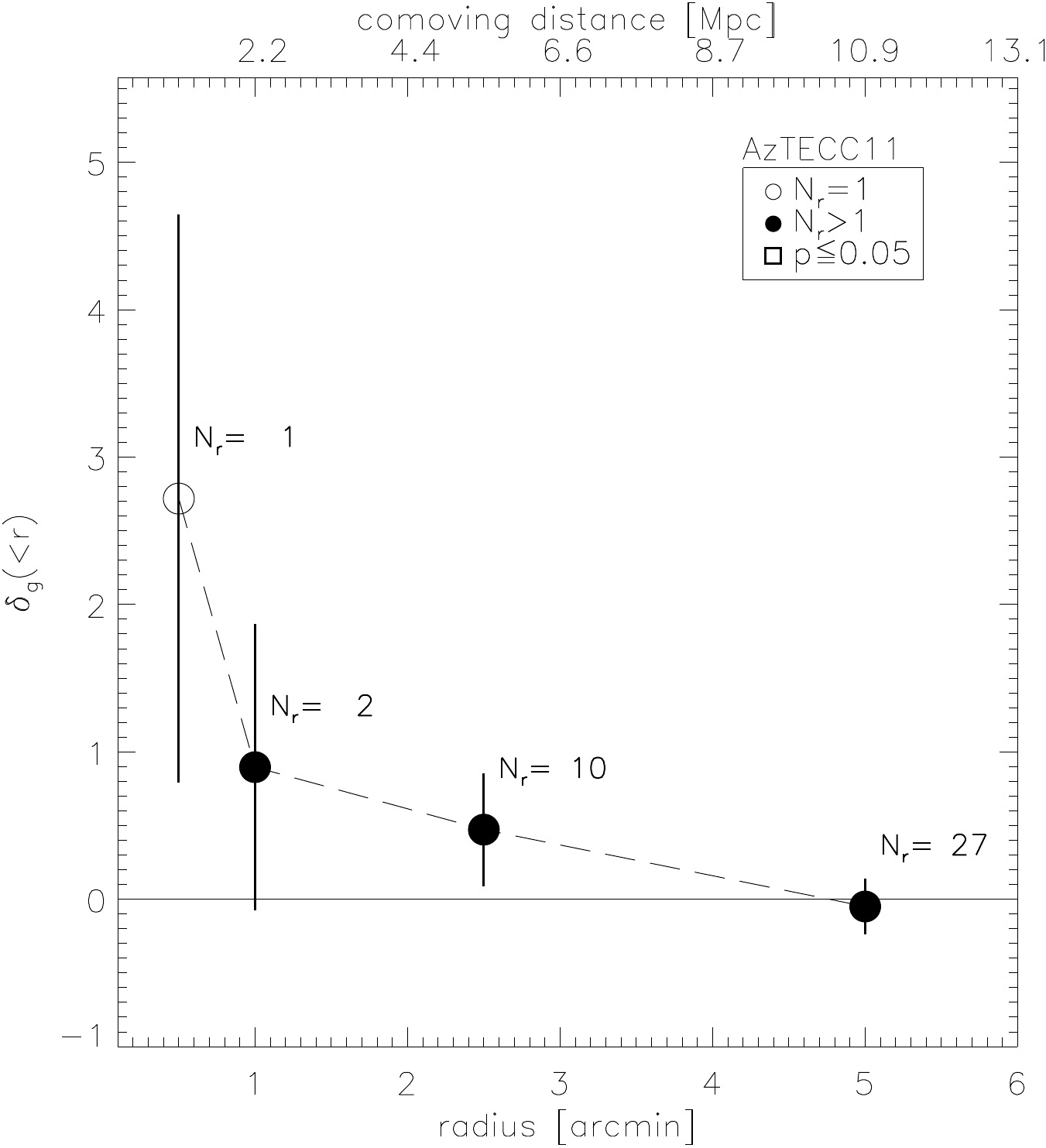}
\includegraphics[width=0.23\textwidth]{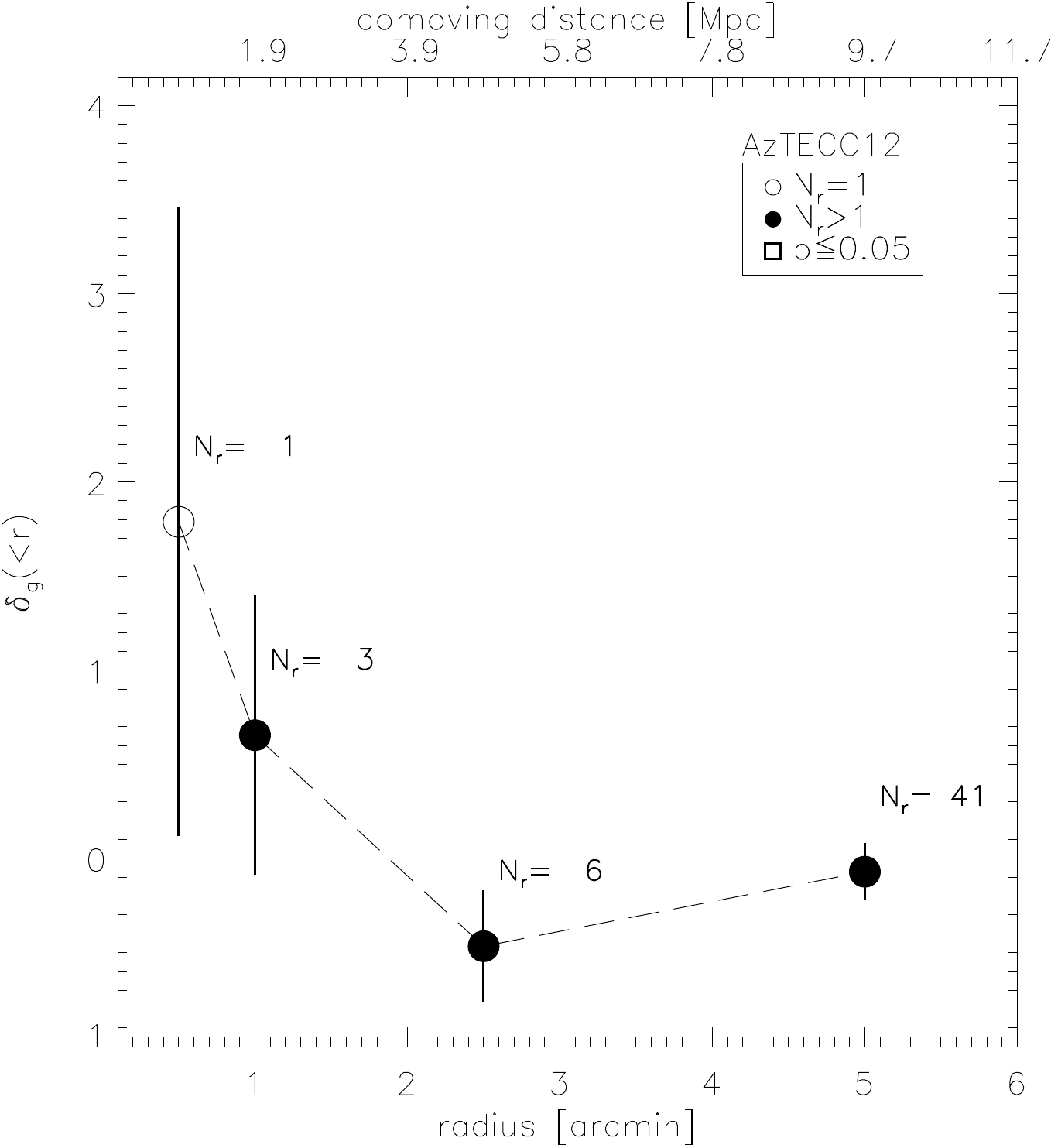}
\includegraphics[width=0.23\textwidth]{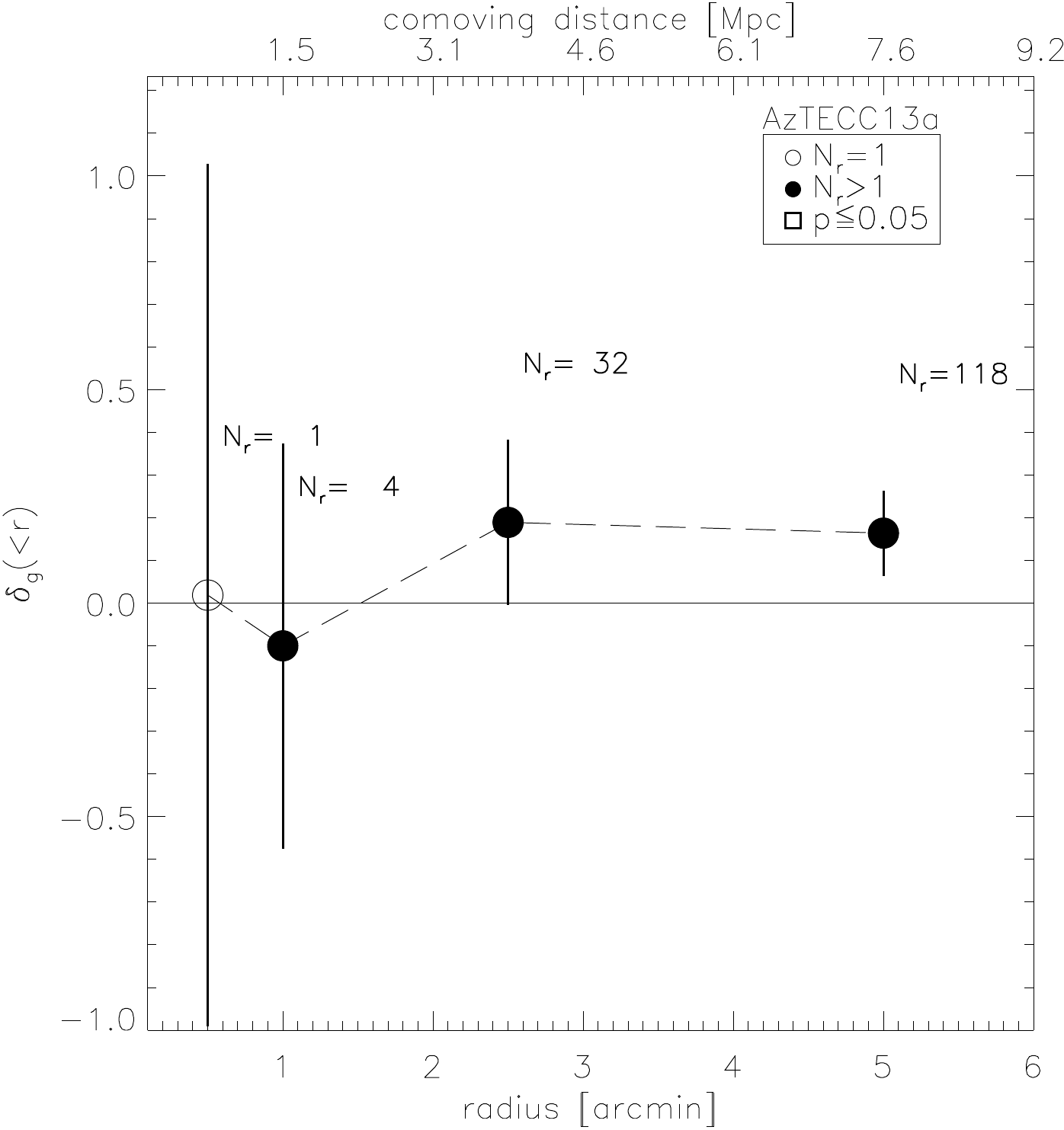}
\includegraphics[width=0.23\textwidth]{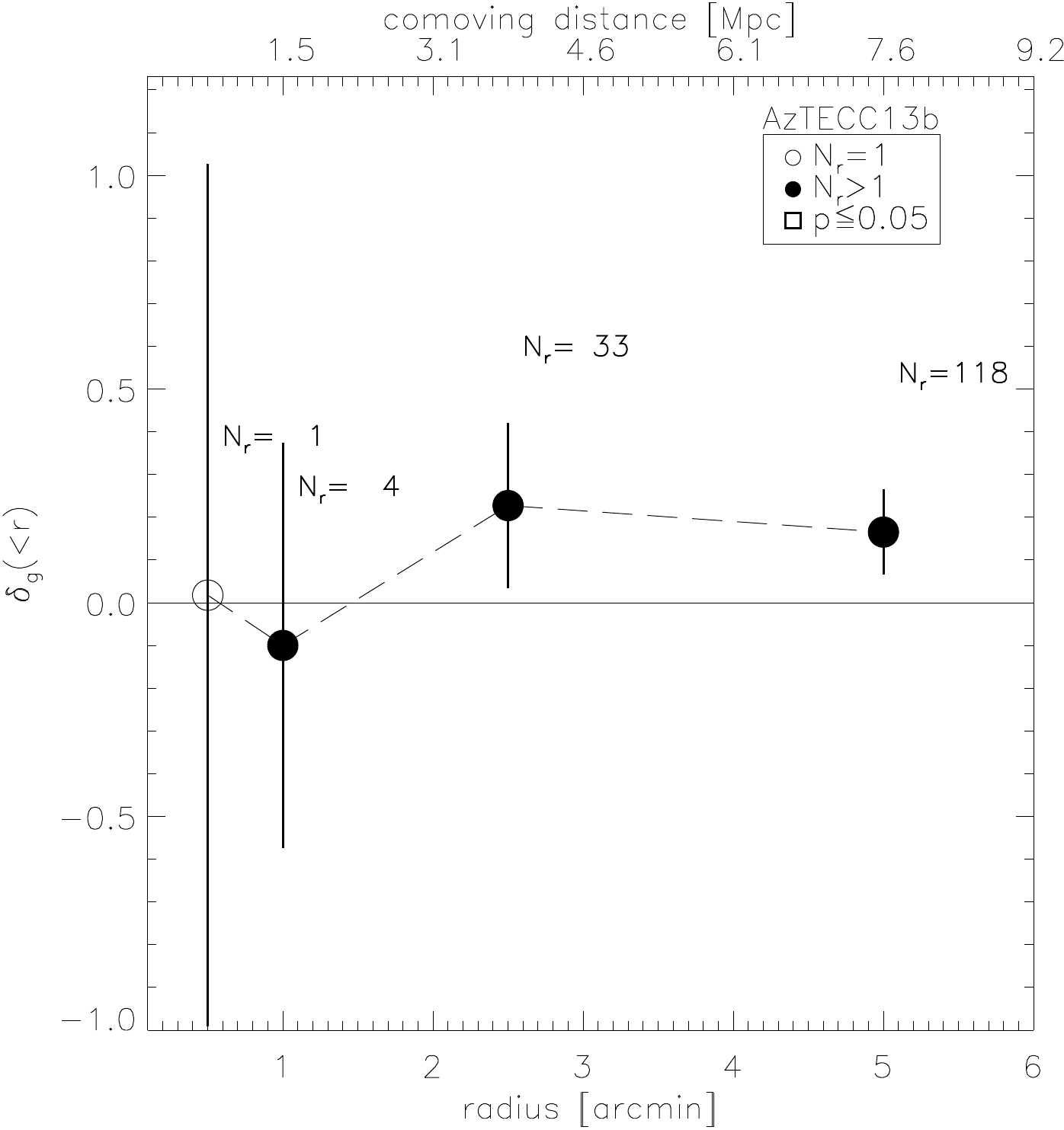}
\includegraphics[width=0.23\textwidth]{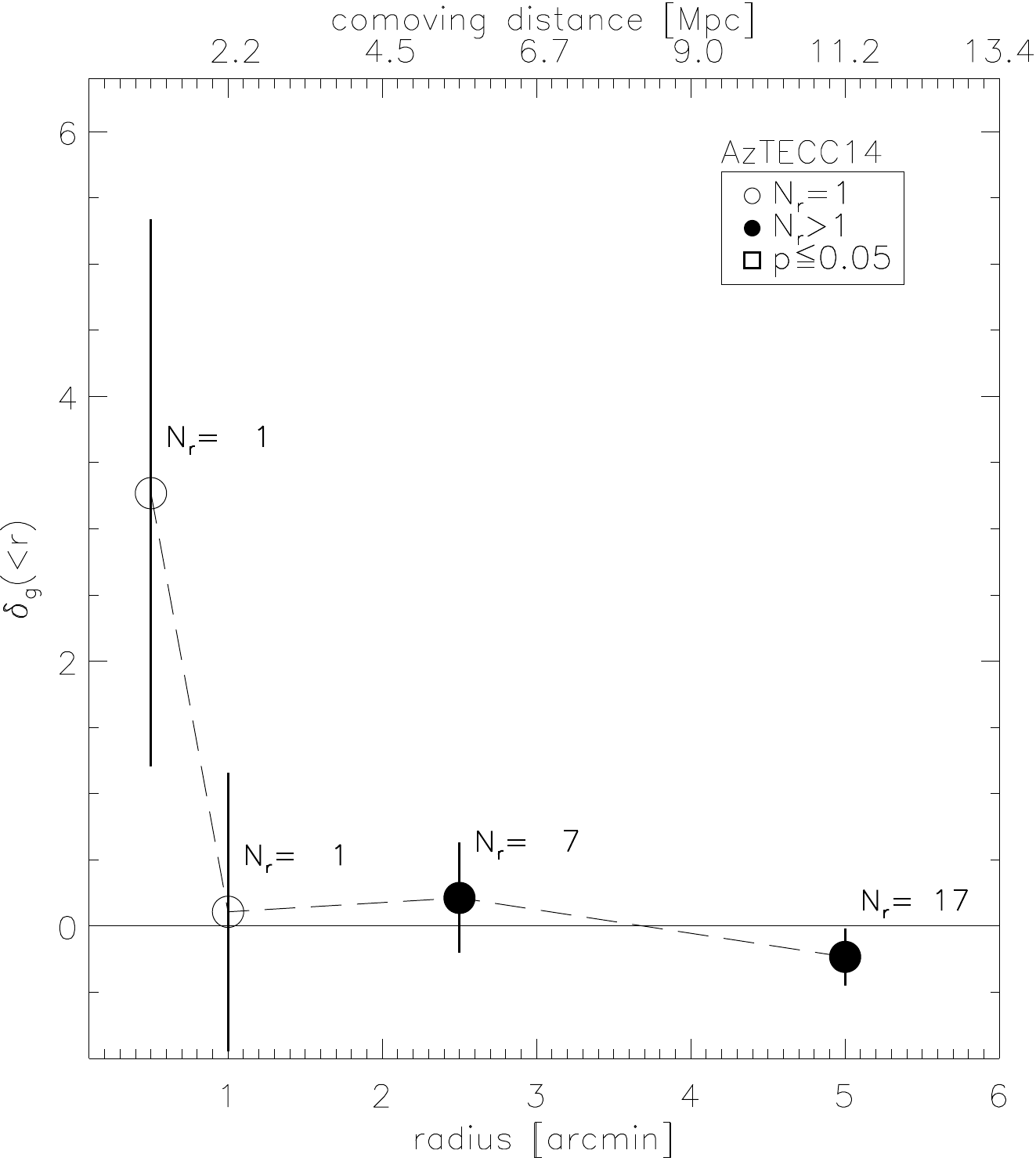}

\caption{
Overdensity parameter $\delta_g$ vs. radius $r$ measured from the central SMG.  Open dots represent when the only source measured within that radius is the target SMG, and black dots represent when at least one more source than the central SMG is observed.
 Error bars represent Poissonian errors. The dot is enclosed in a square when Poisson probability is $p \leq$ 0.05. The name of each SMG is indicated in the legend and
the number of sources found within a radius $r$ is indicated next to each point. A horizontal line represents $\delta_g =$ 0.
}
\label{figure:voronoi}
\end{center}
\end{figure*}

\addtocounter{figure}{-1}
\begin{figure*}
\begin{center}
\includegraphics[width=0.23\textwidth]{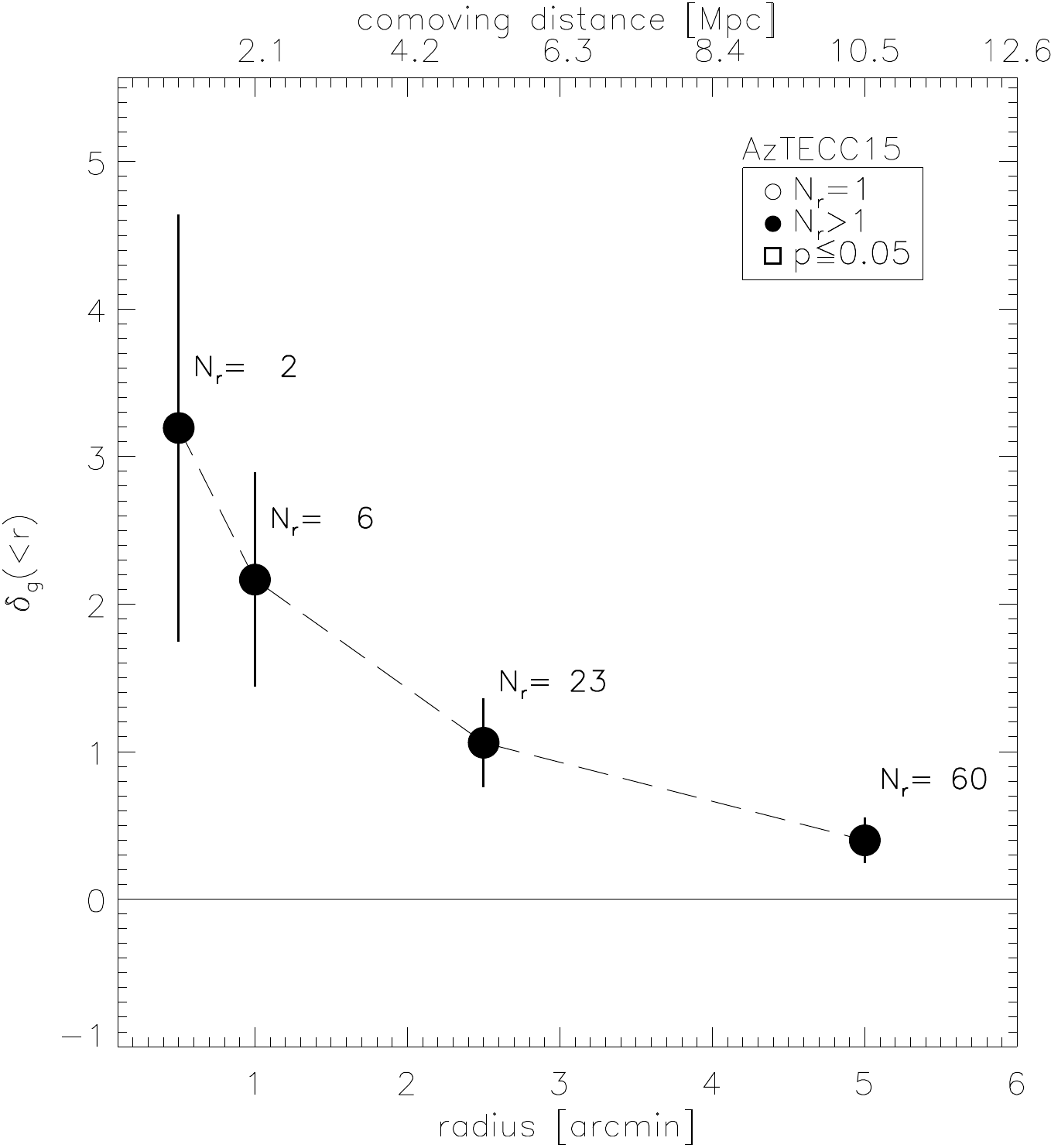}
\includegraphics[width=0.23\textwidth]{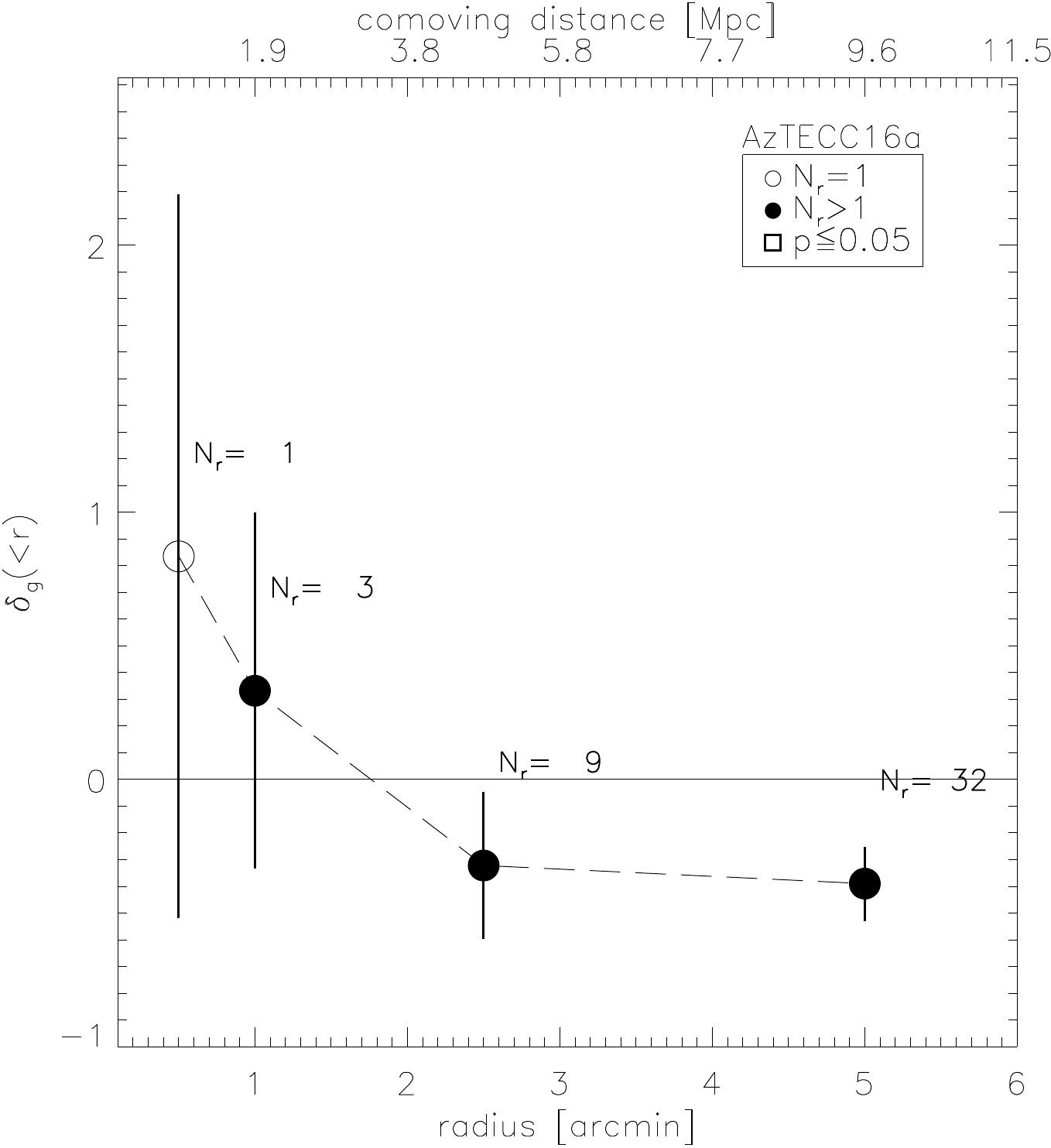}
\includegraphics[width=0.23\textwidth]{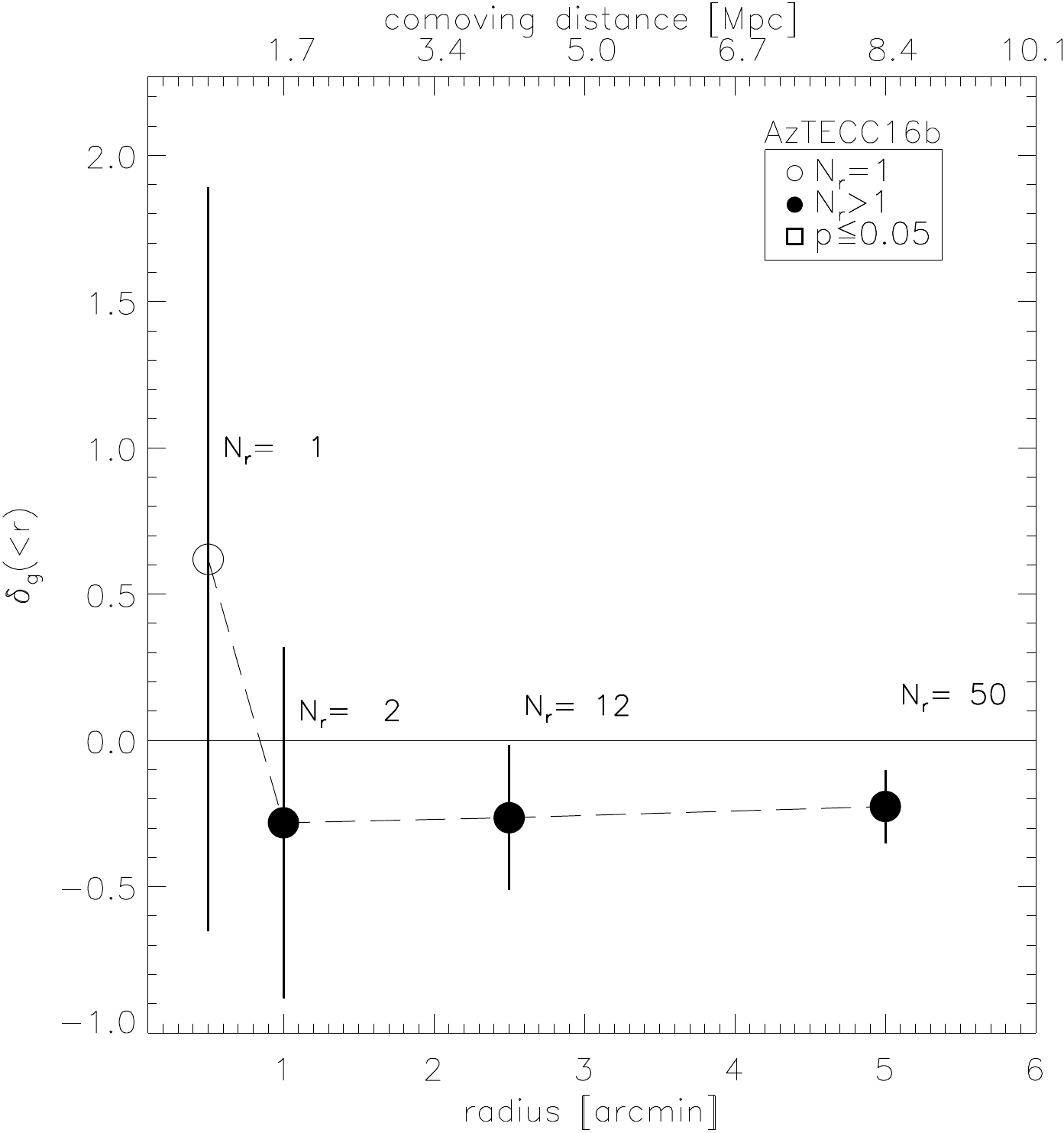}
\includegraphics[width=0.23\textwidth]{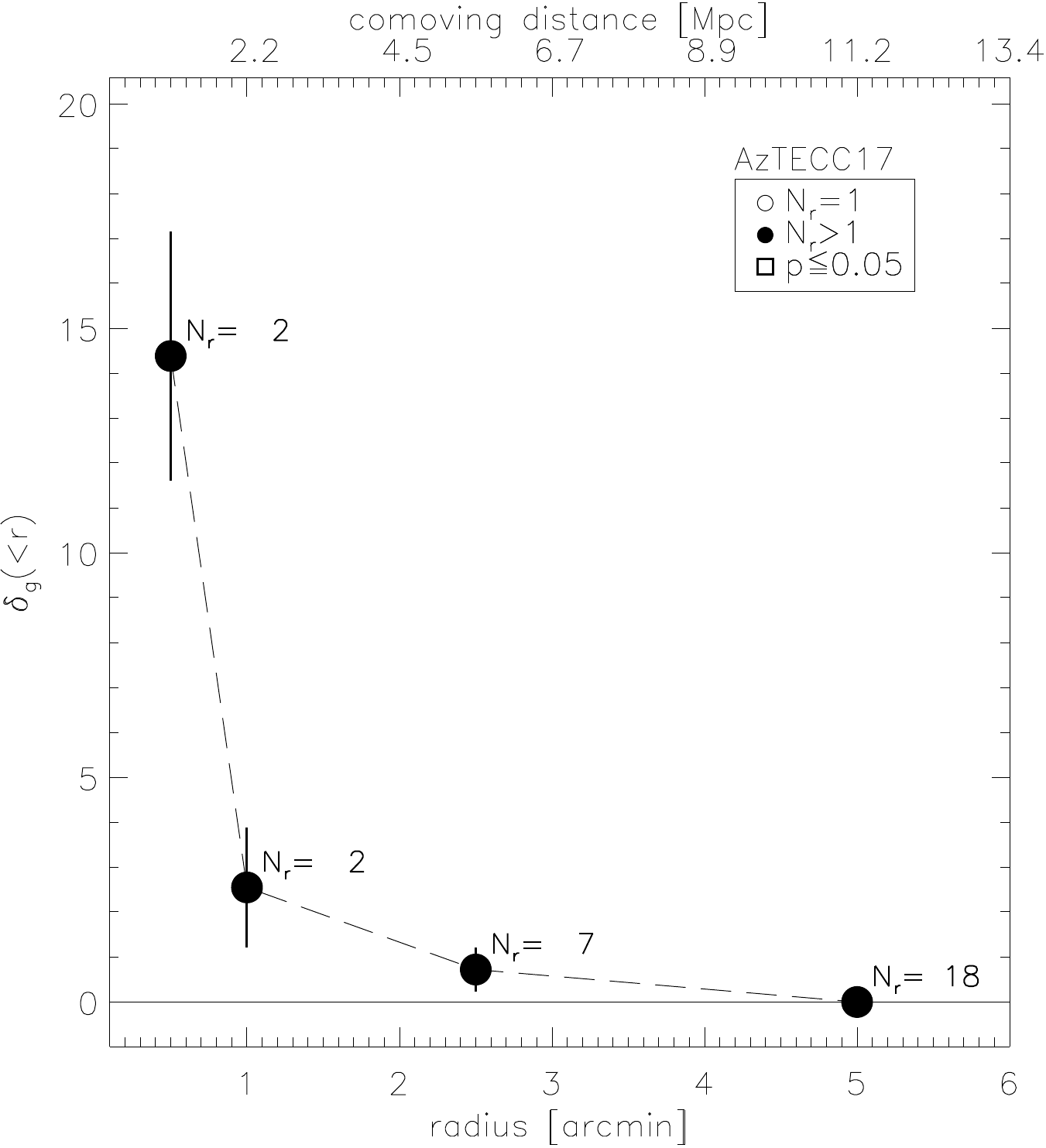}
\includegraphics[width=0.23\textwidth]{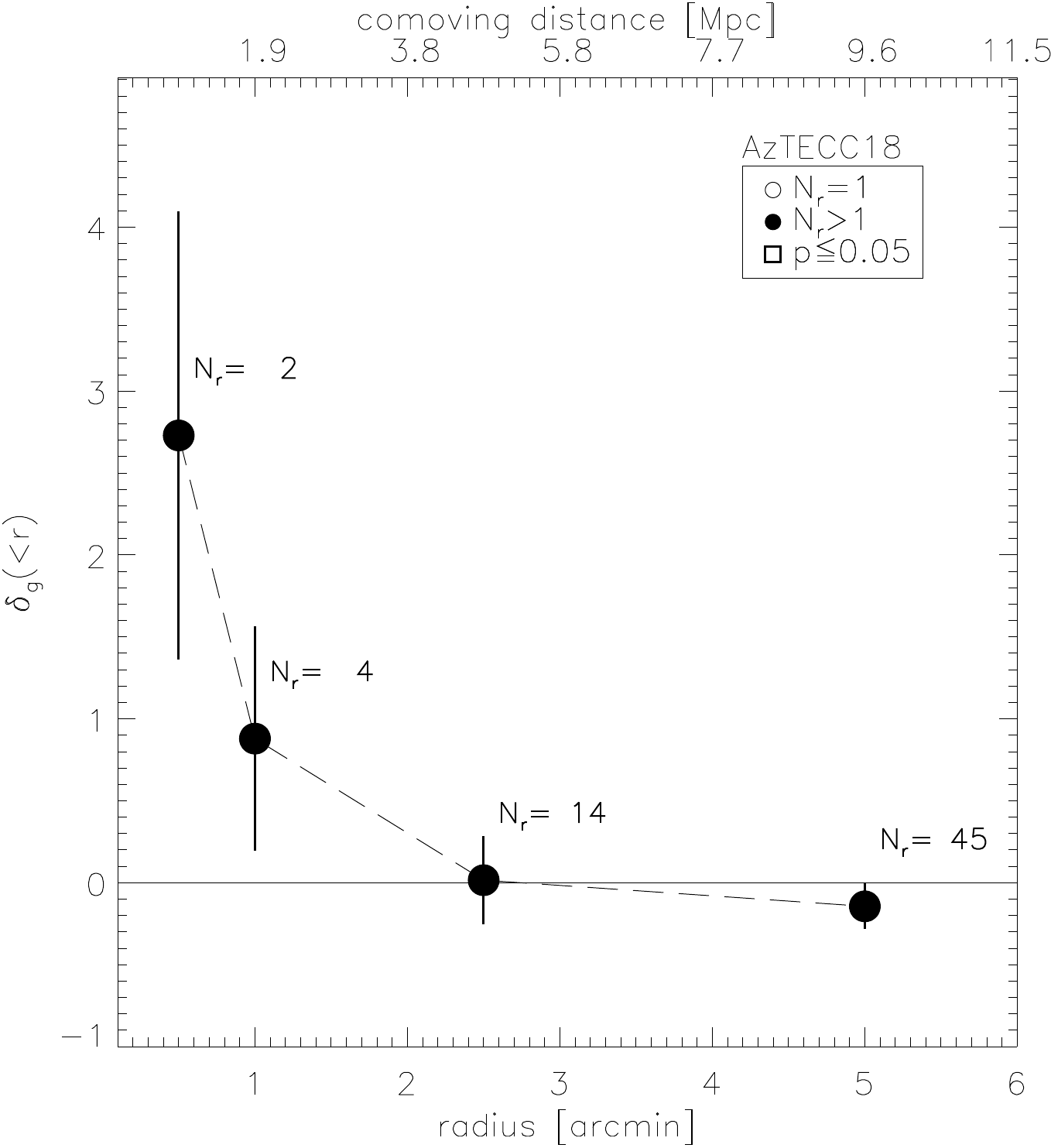}
\includegraphics[width=0.23\textwidth]{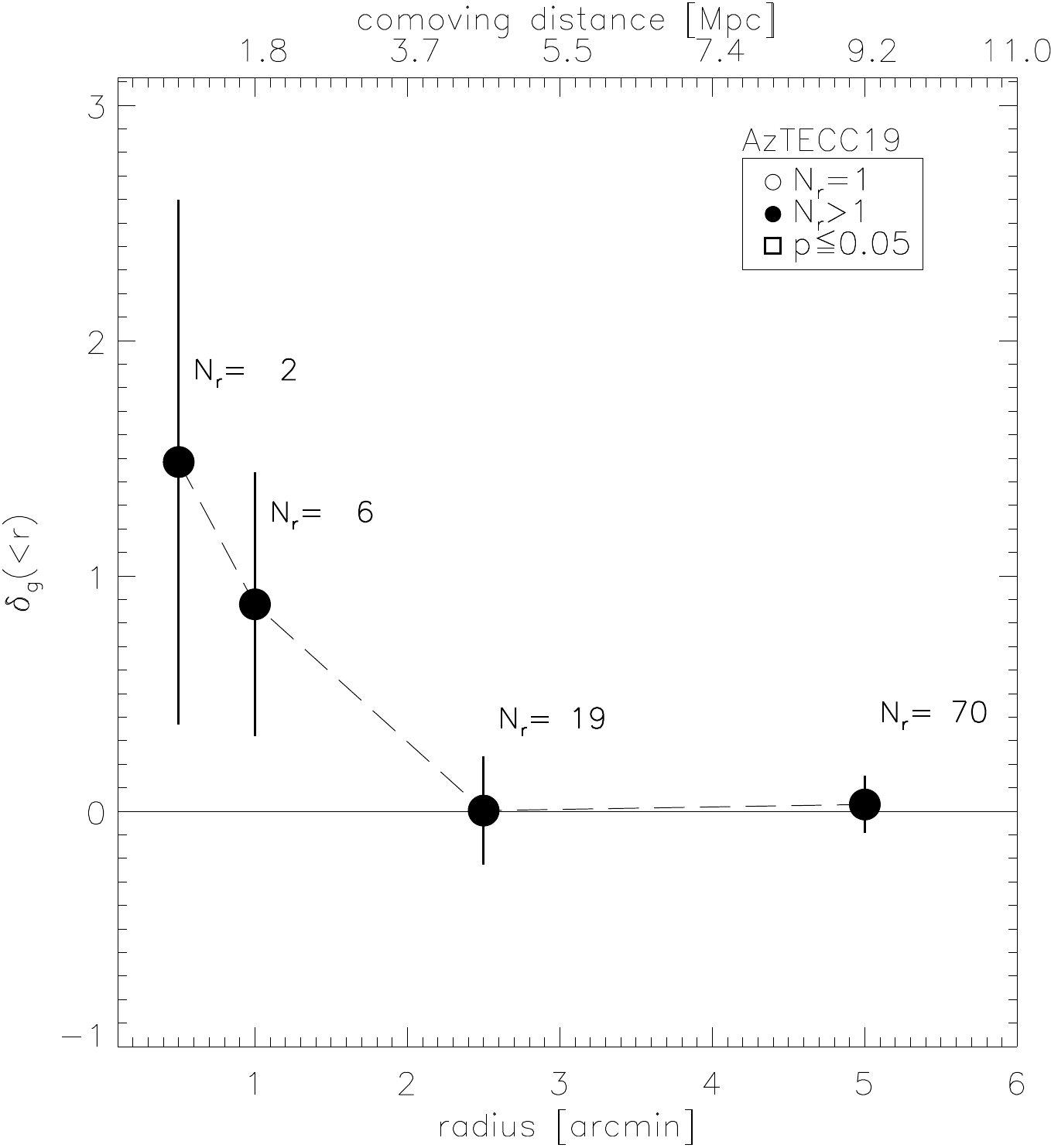}
\includegraphics[width=0.23\textwidth]{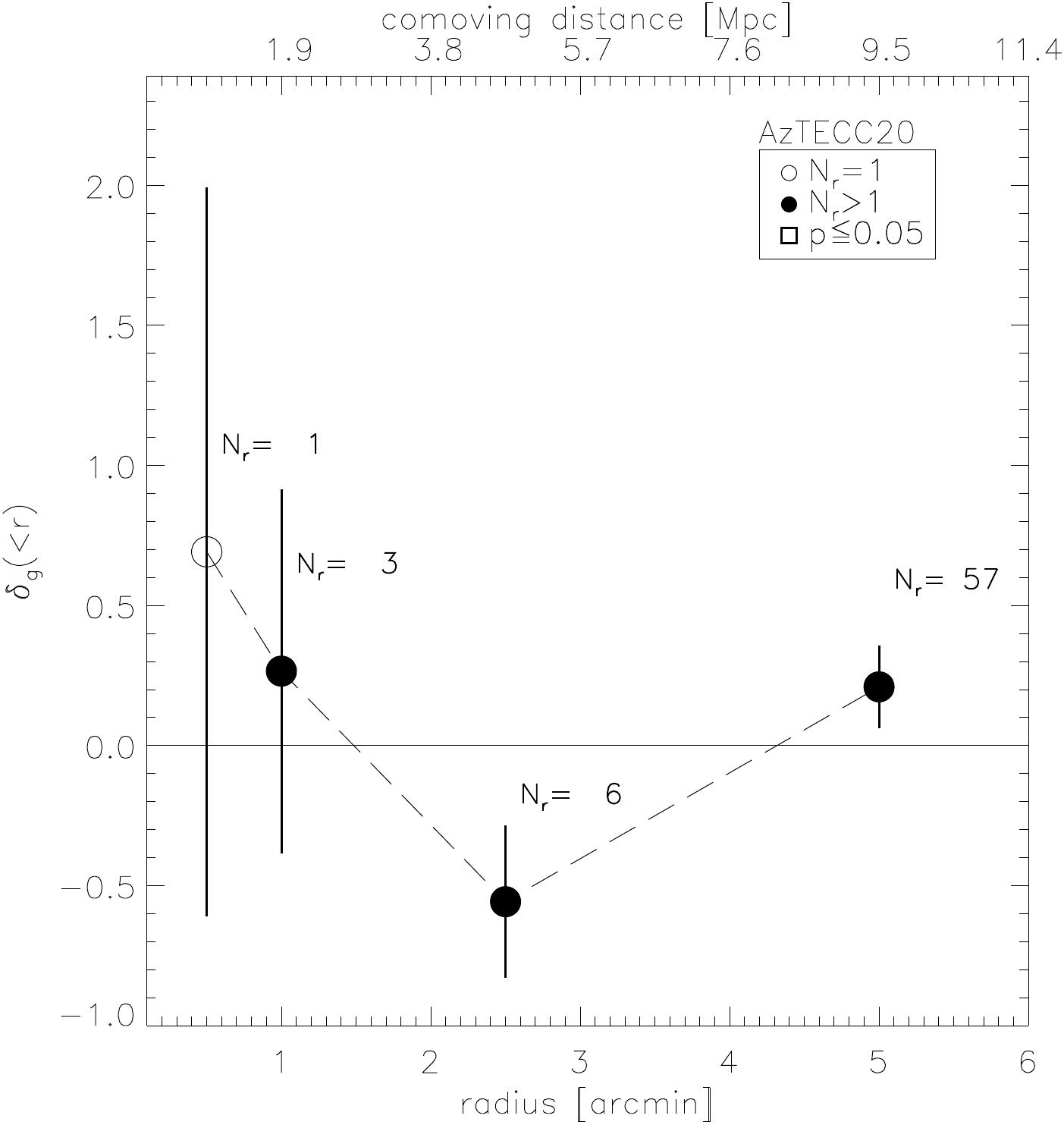}
\includegraphics[width=0.23\textwidth]{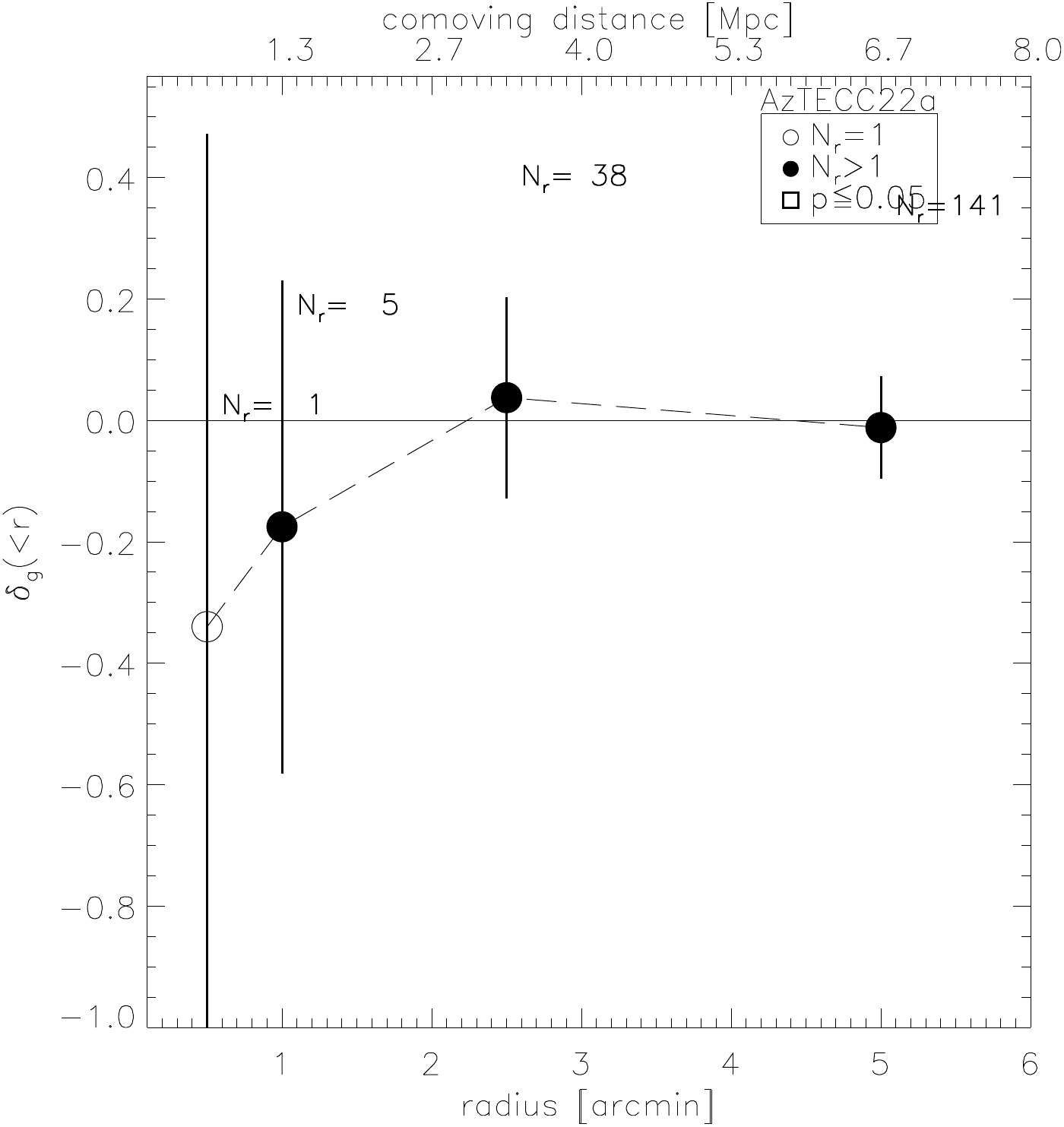}
\includegraphics[width=0.23\textwidth]{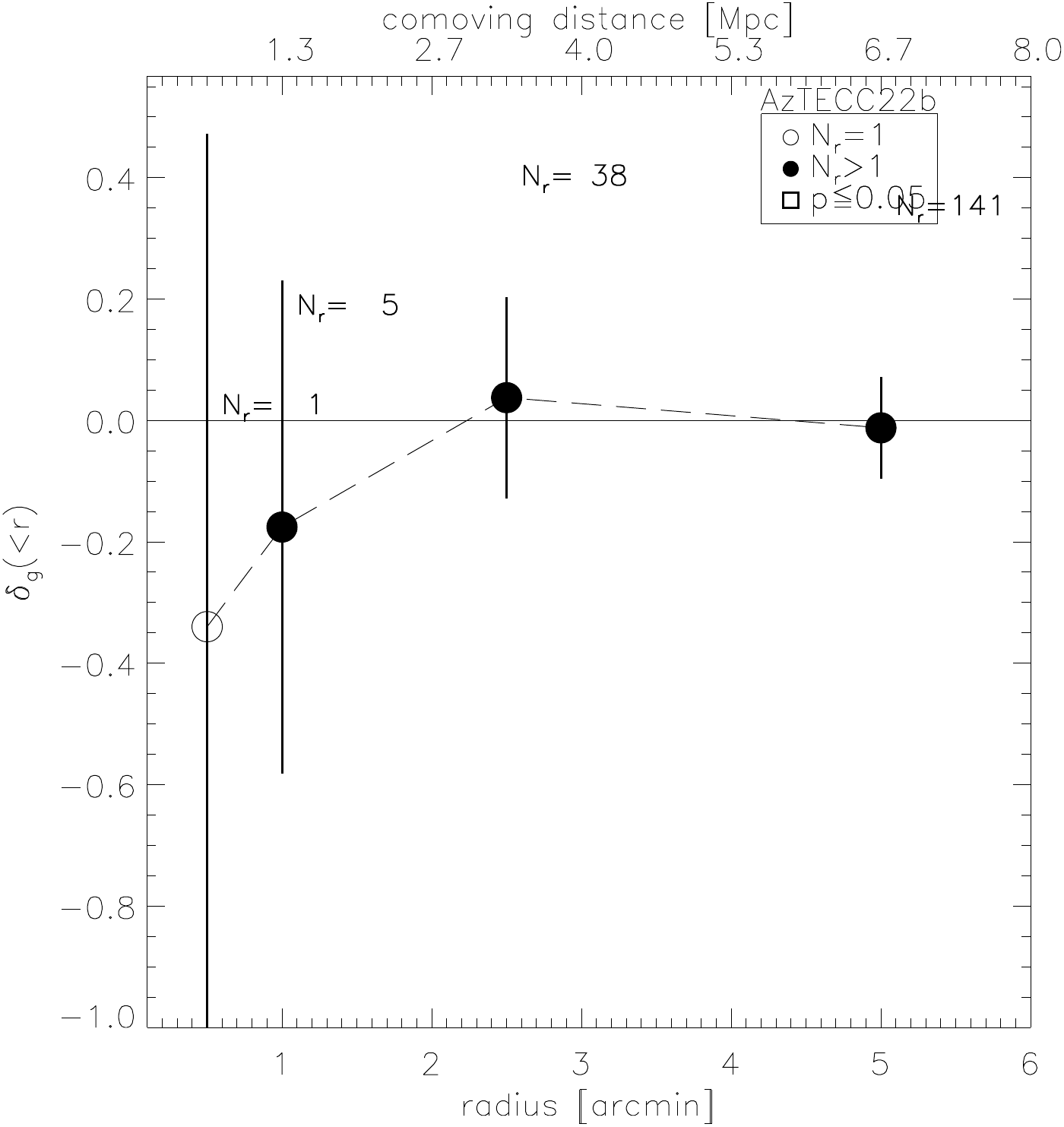}
\includegraphics[width=0.23\textwidth]{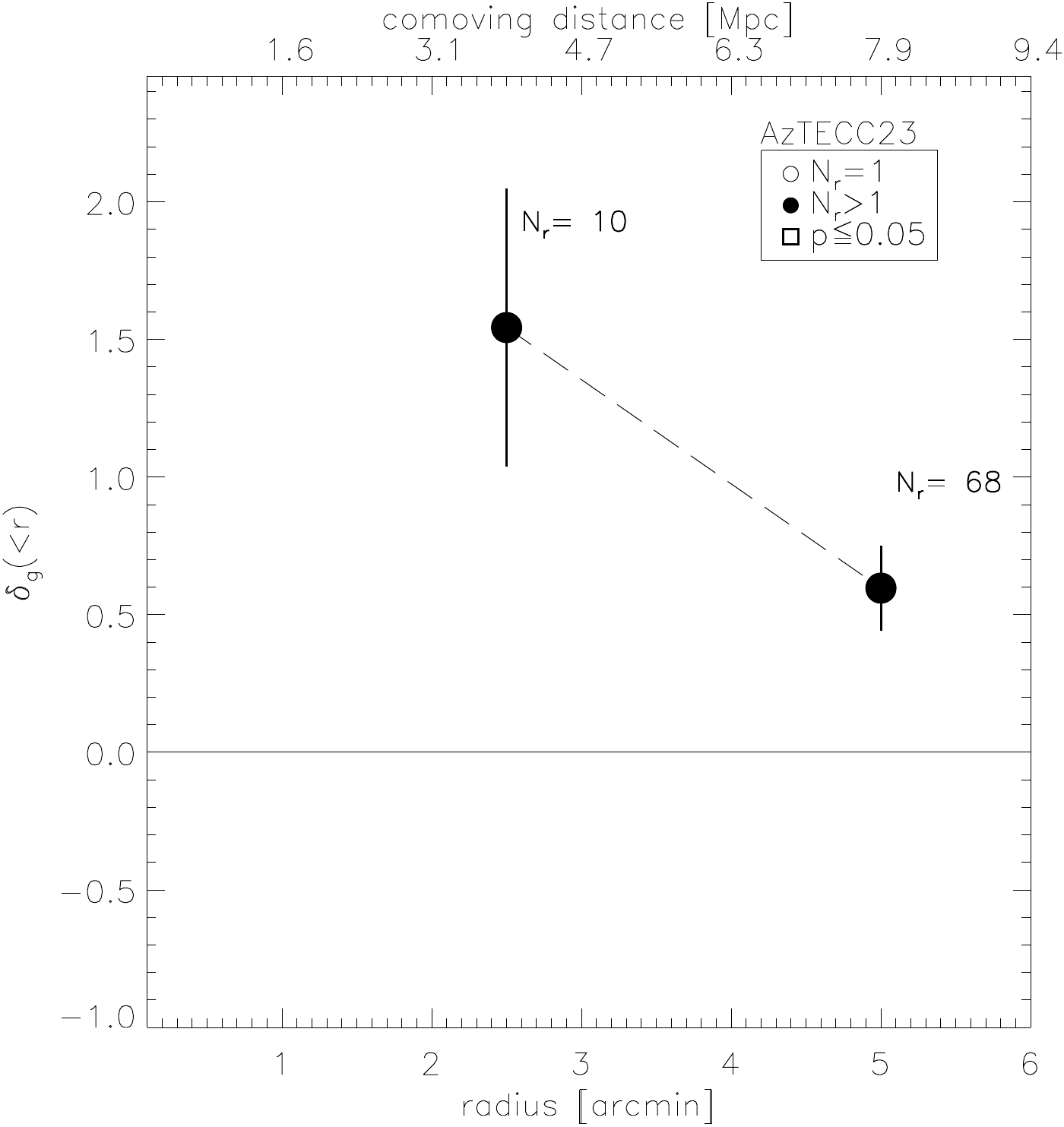}
\includegraphics[width=0.23\textwidth]{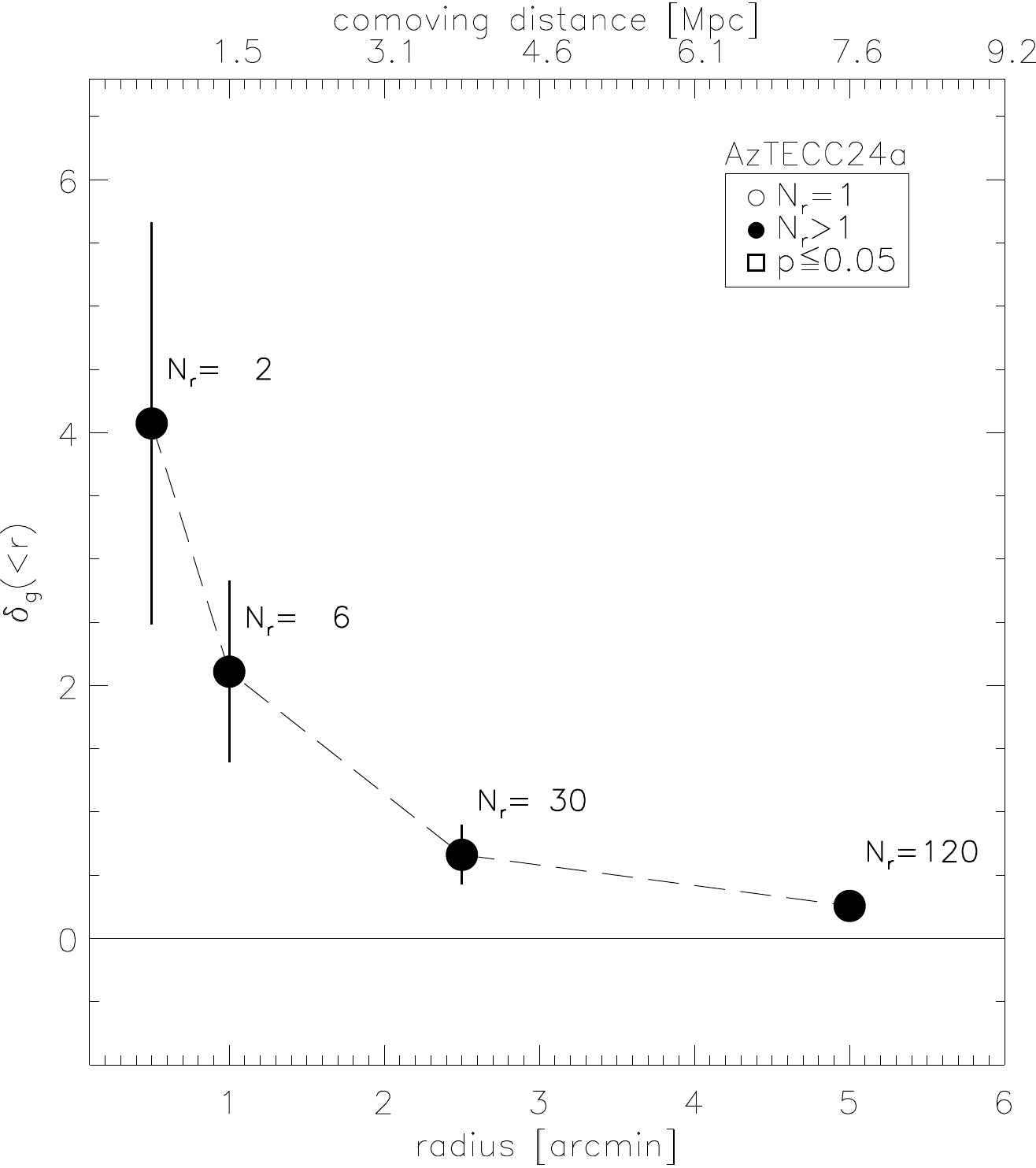}
\includegraphics[width=0.23\textwidth]{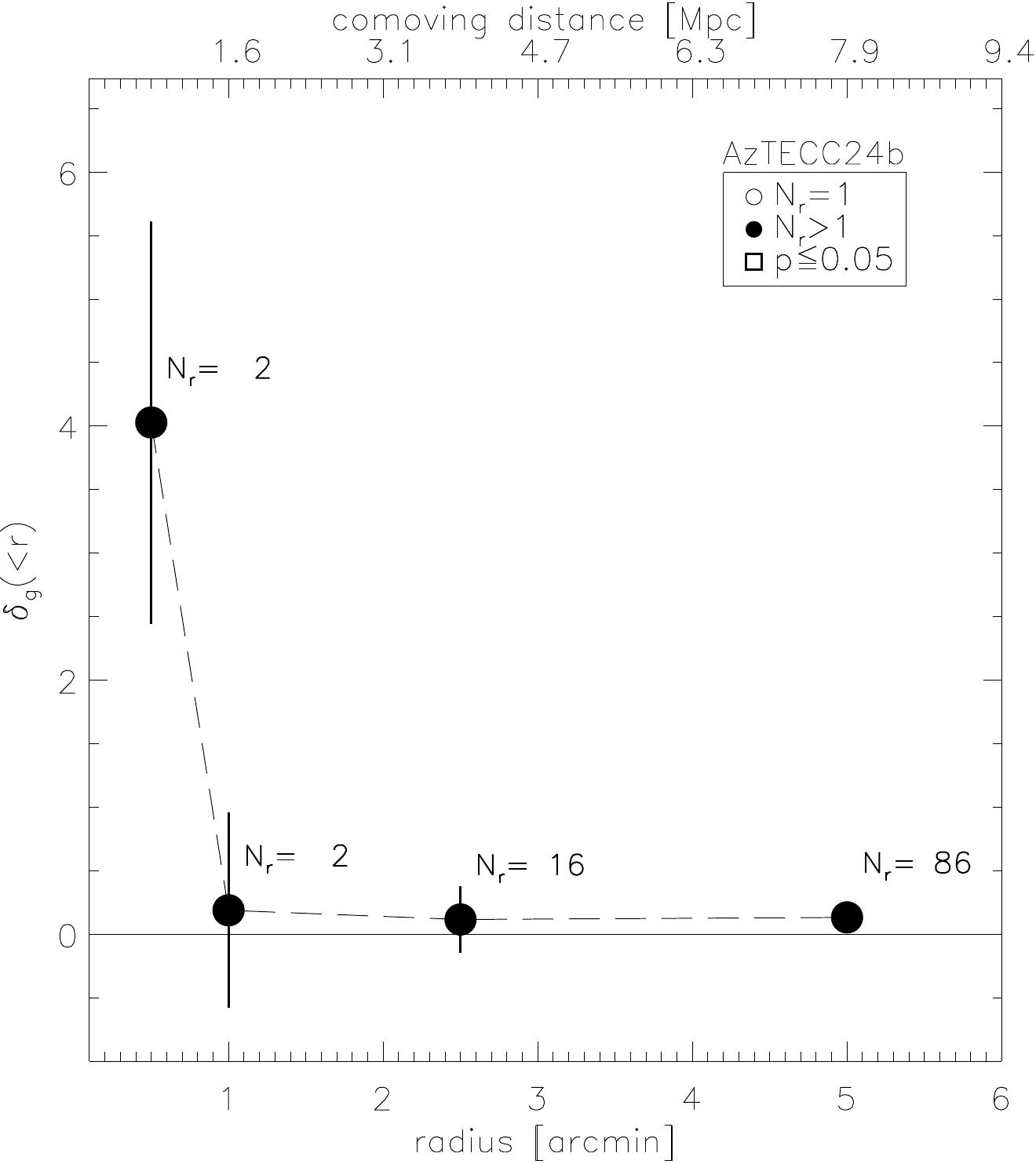}
\includegraphics[width=0.23\textwidth]{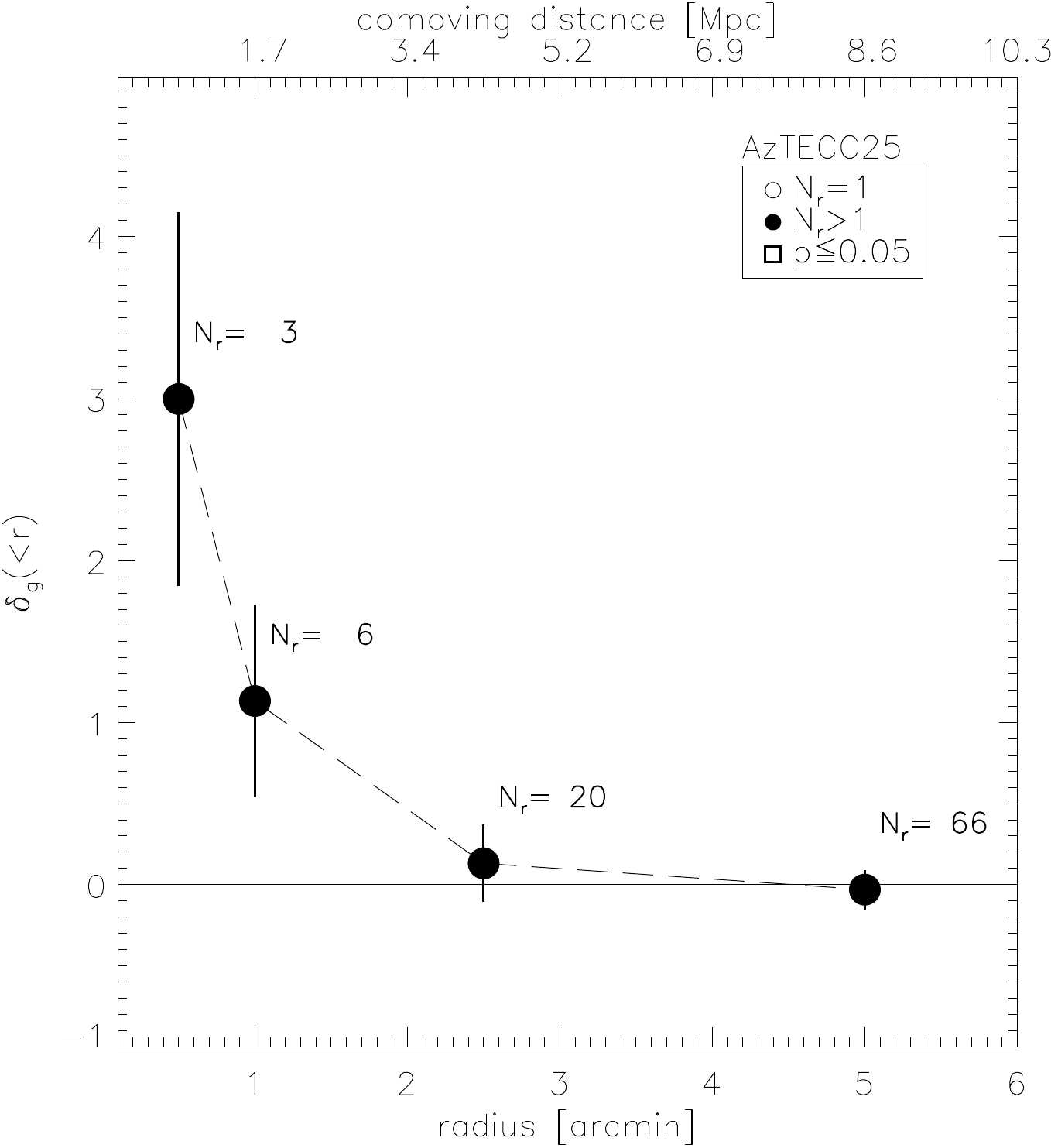}
\includegraphics[width=0.23\textwidth]{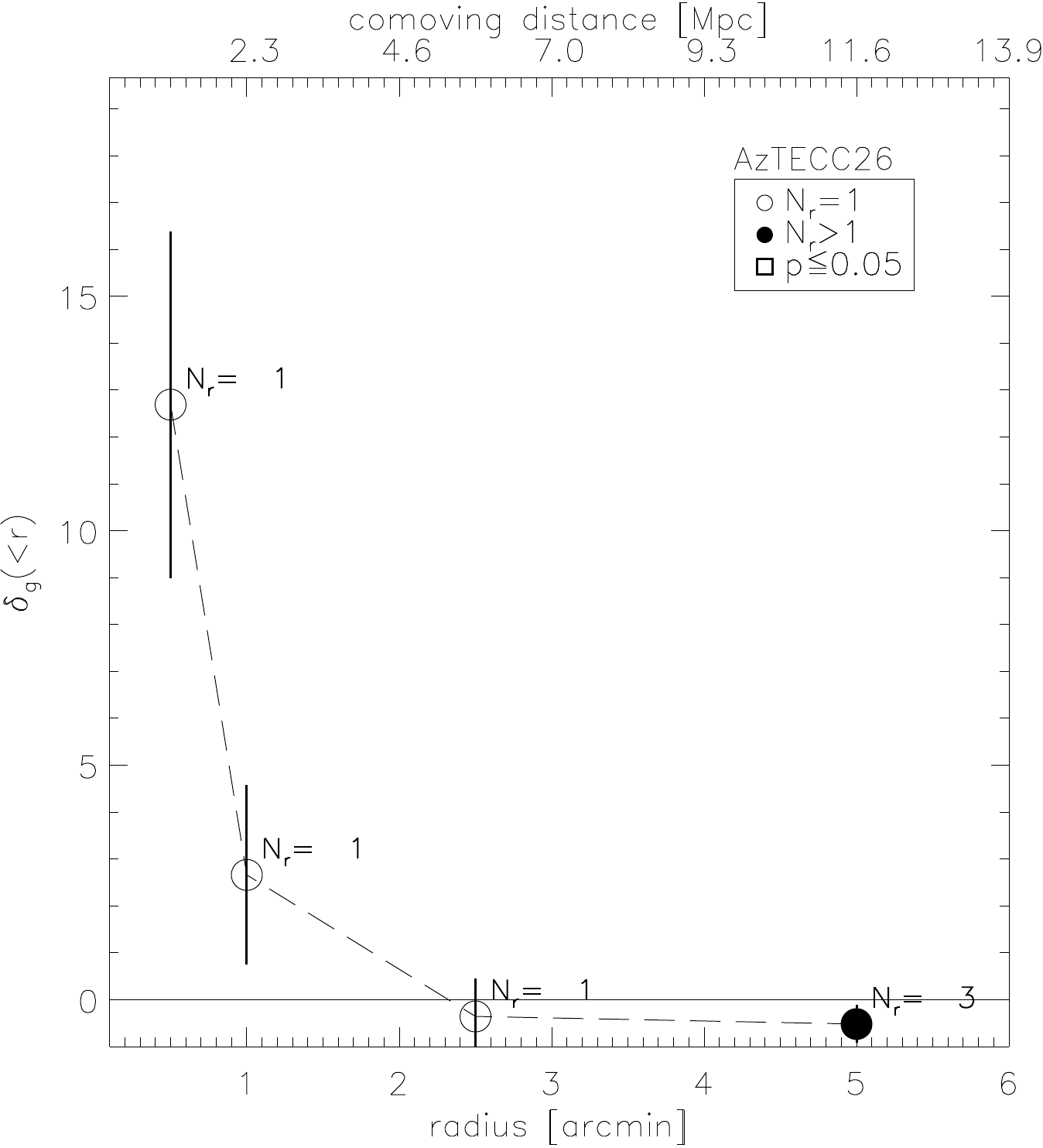}
\includegraphics[width=0.23\textwidth]{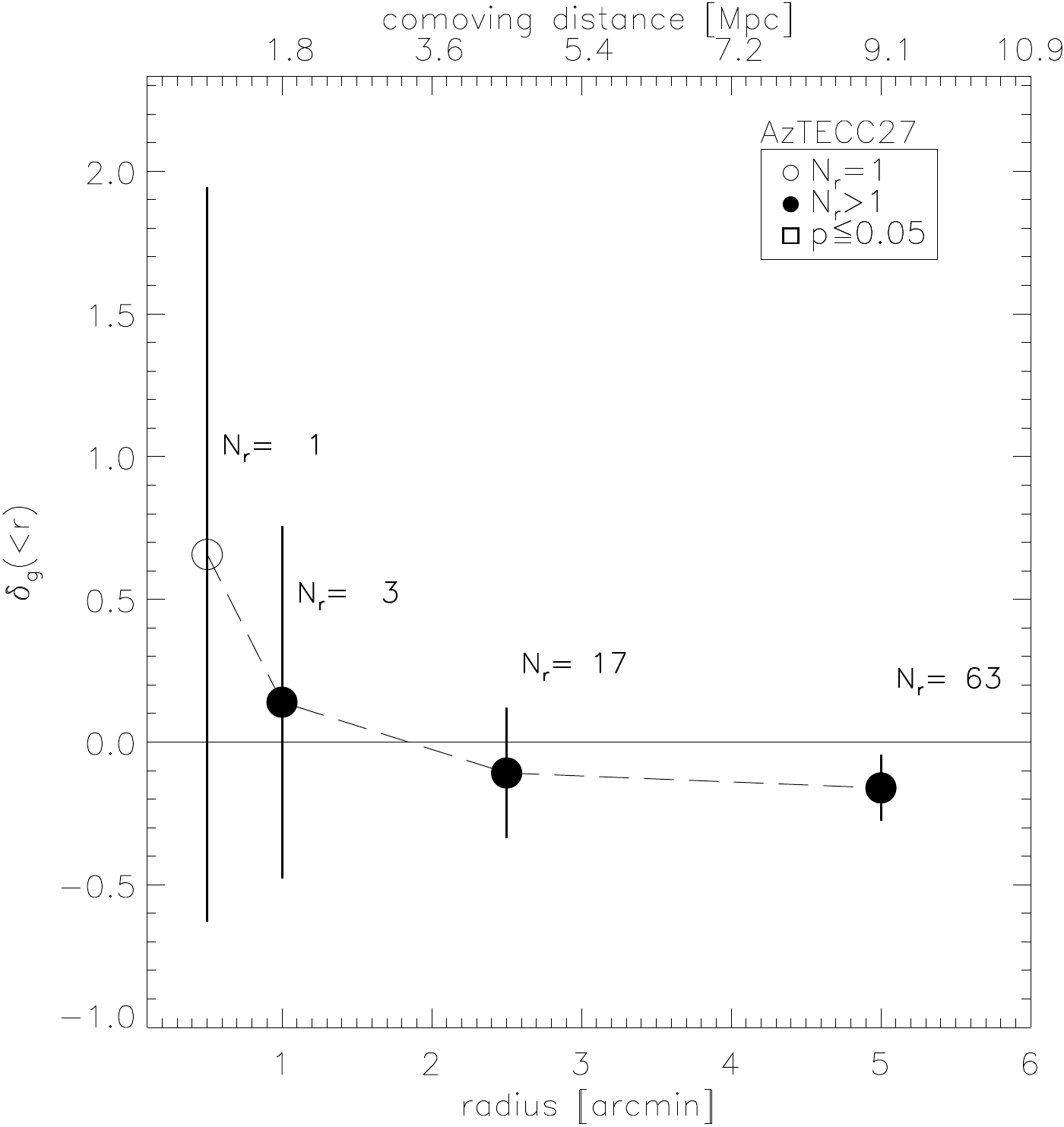}
\includegraphics[width=0.23\textwidth]{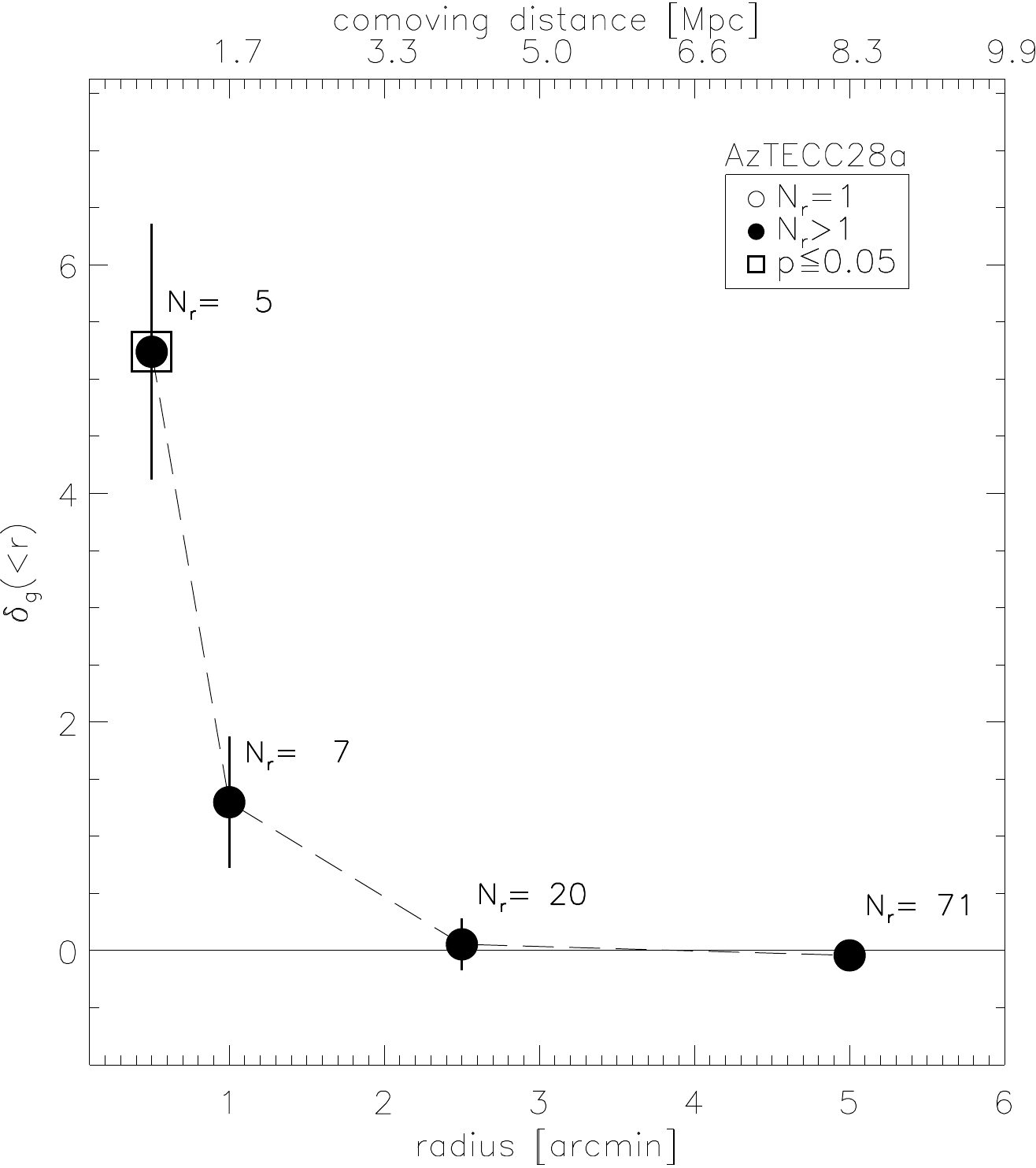}
\includegraphics[width=0.23\textwidth]{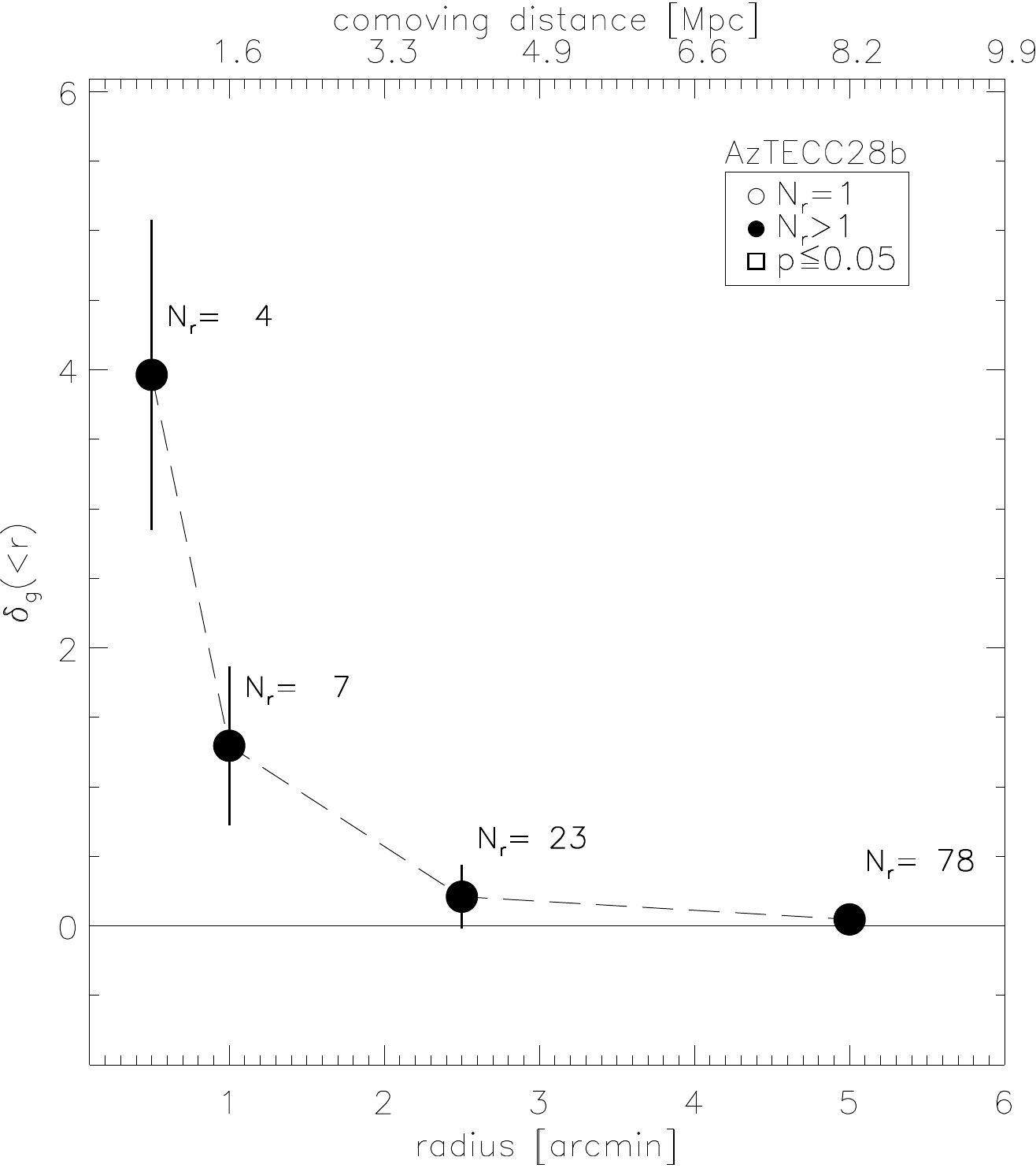}
\includegraphics[width=0.23\textwidth]{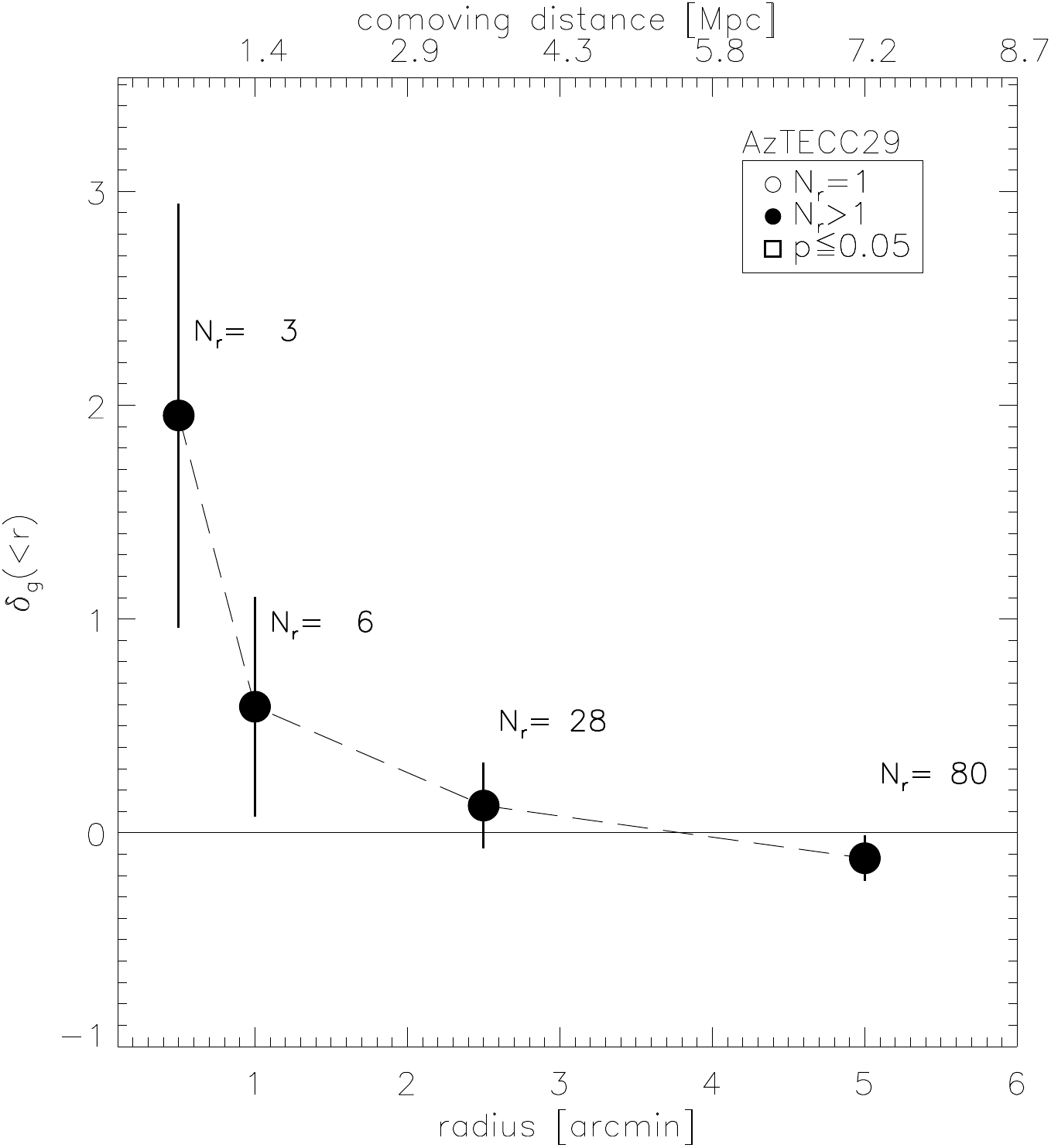}
\includegraphics[width=0.23\textwidth]{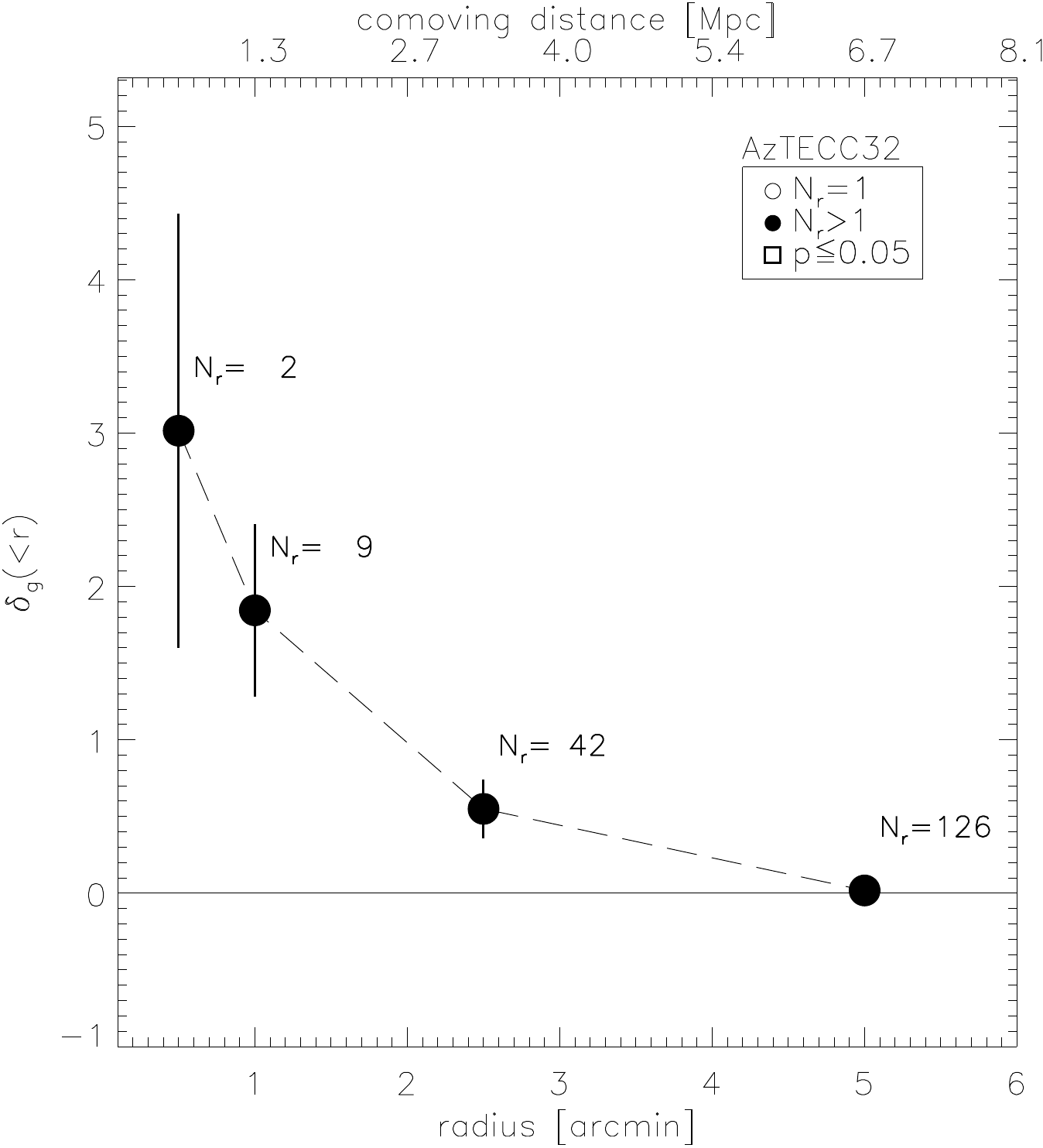}
\includegraphics[width=0.23\textwidth]{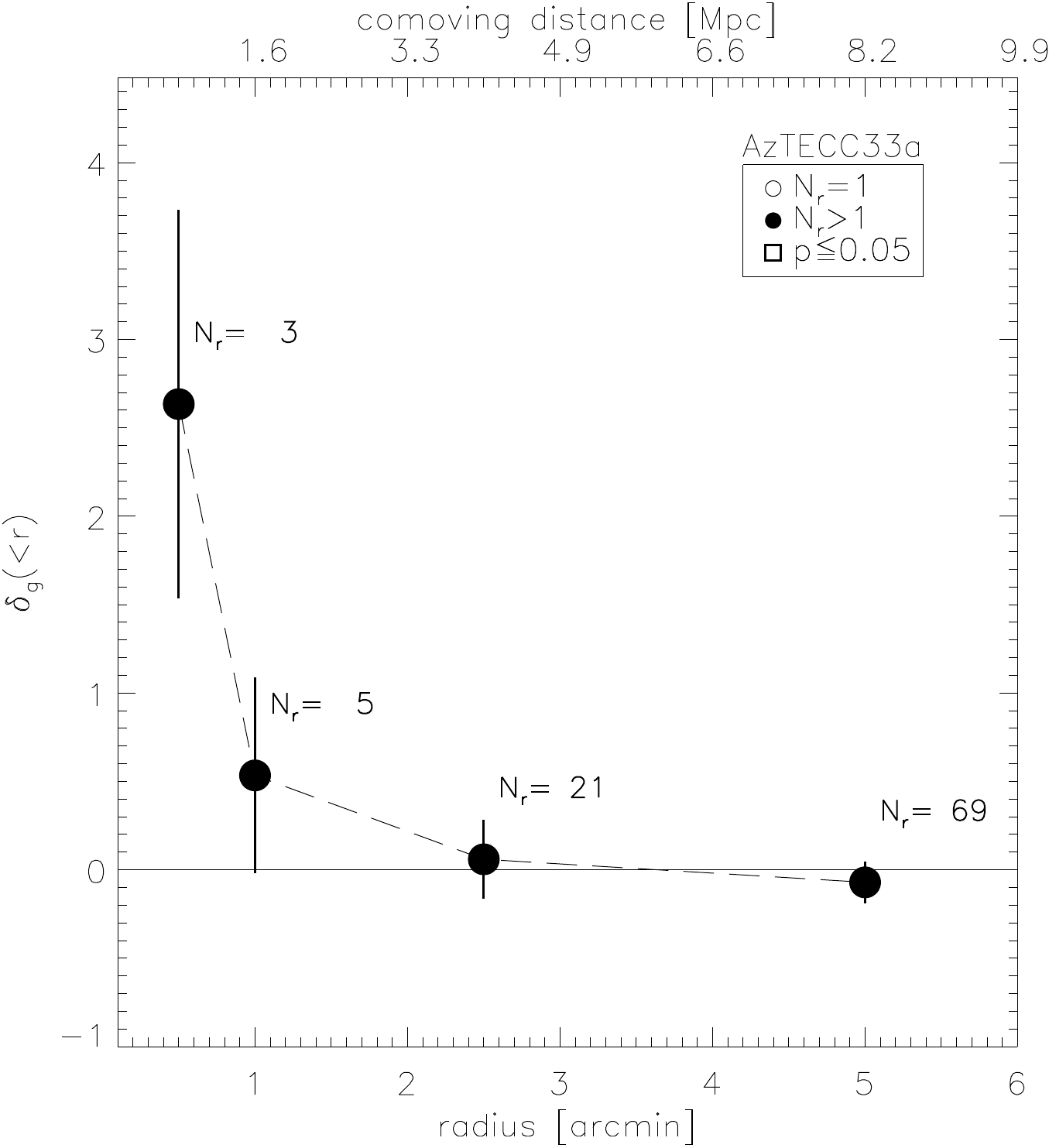}

\caption{continued.}
\end{center}
\end{figure*}

\addtocounter{figure}{-1}
\begin{figure*}
\begin{center}
\includegraphics[width=0.23\textwidth]{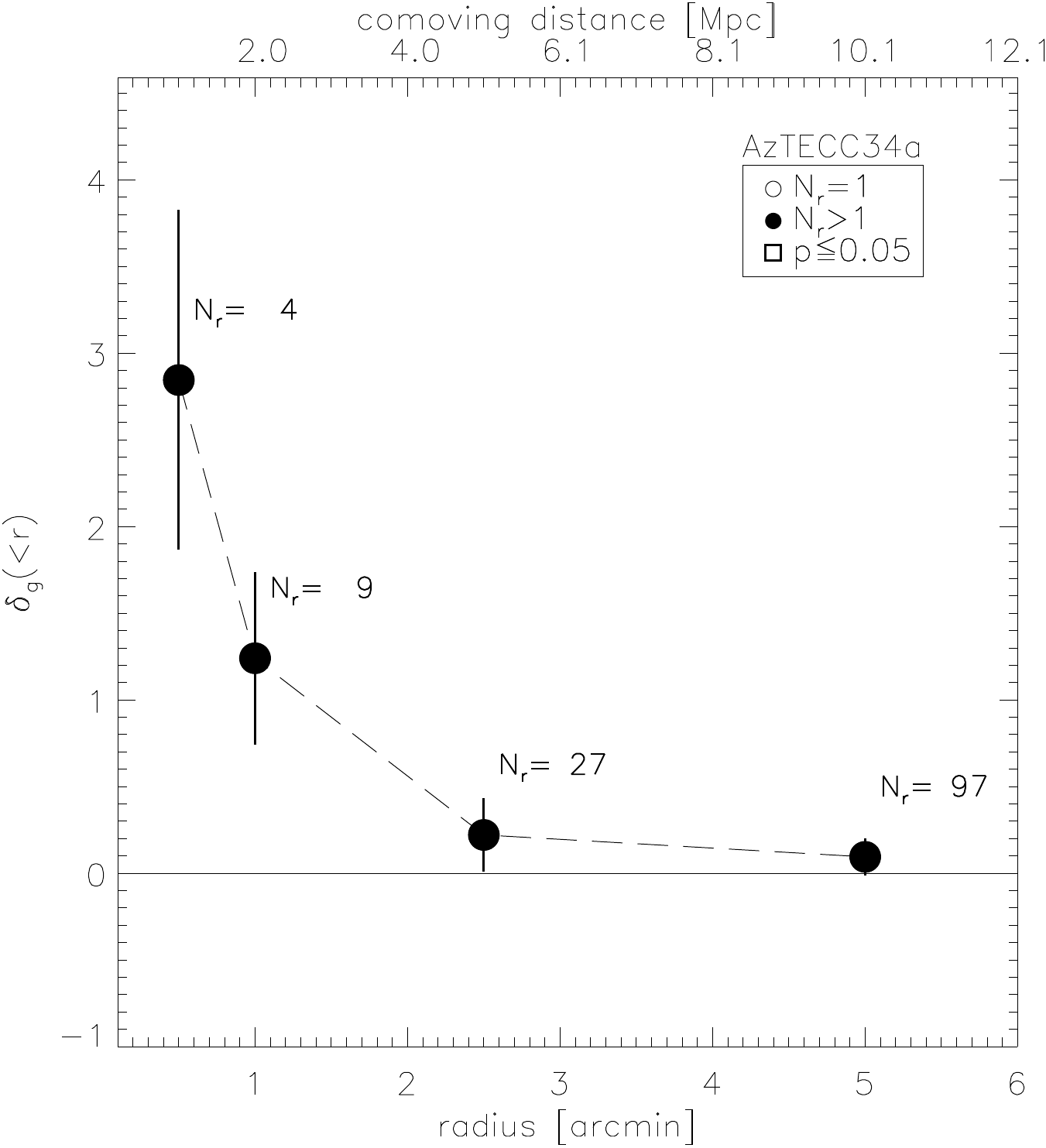}
\includegraphics[width=0.23\textwidth]{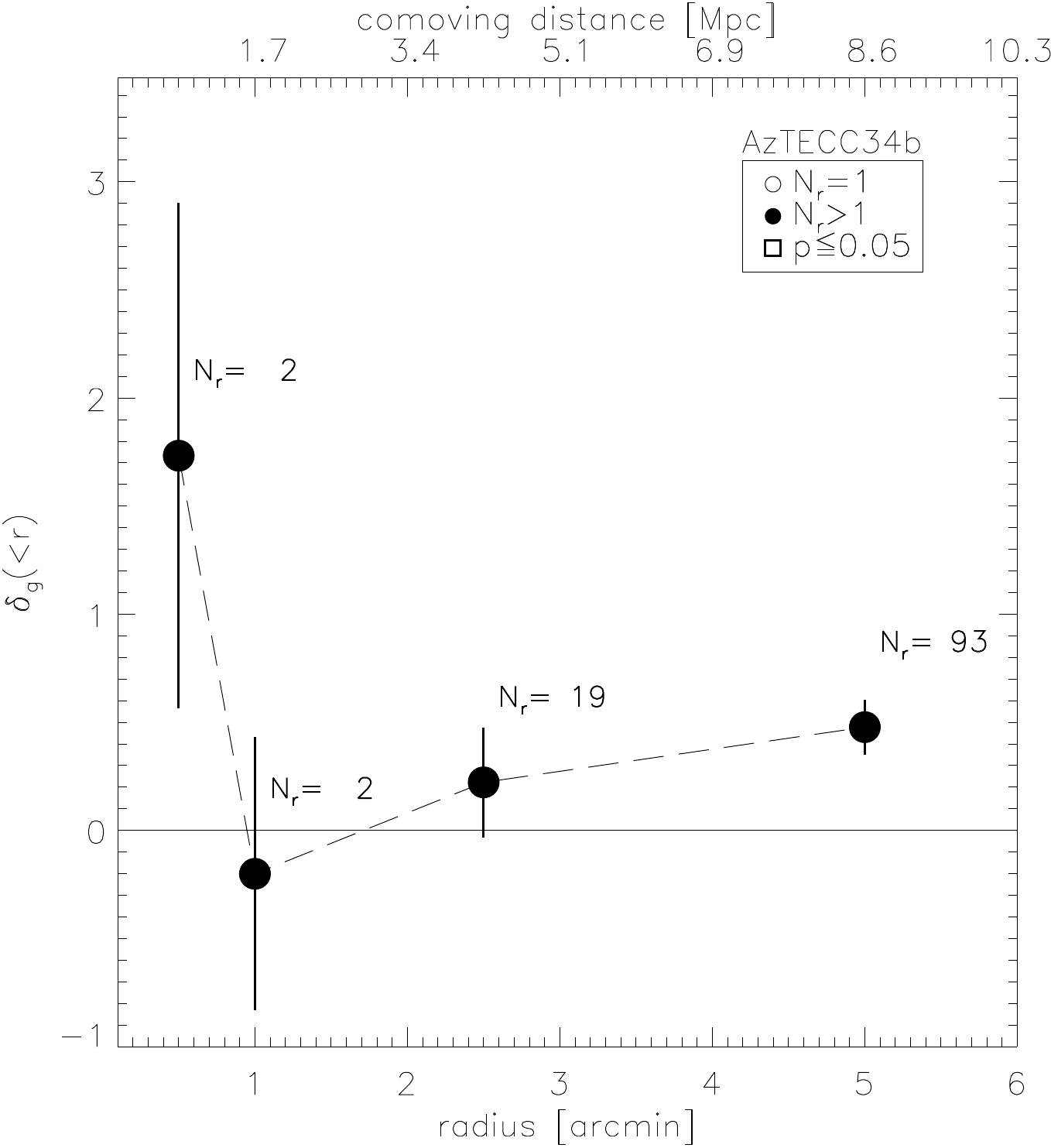}
\includegraphics[width=0.23\textwidth]{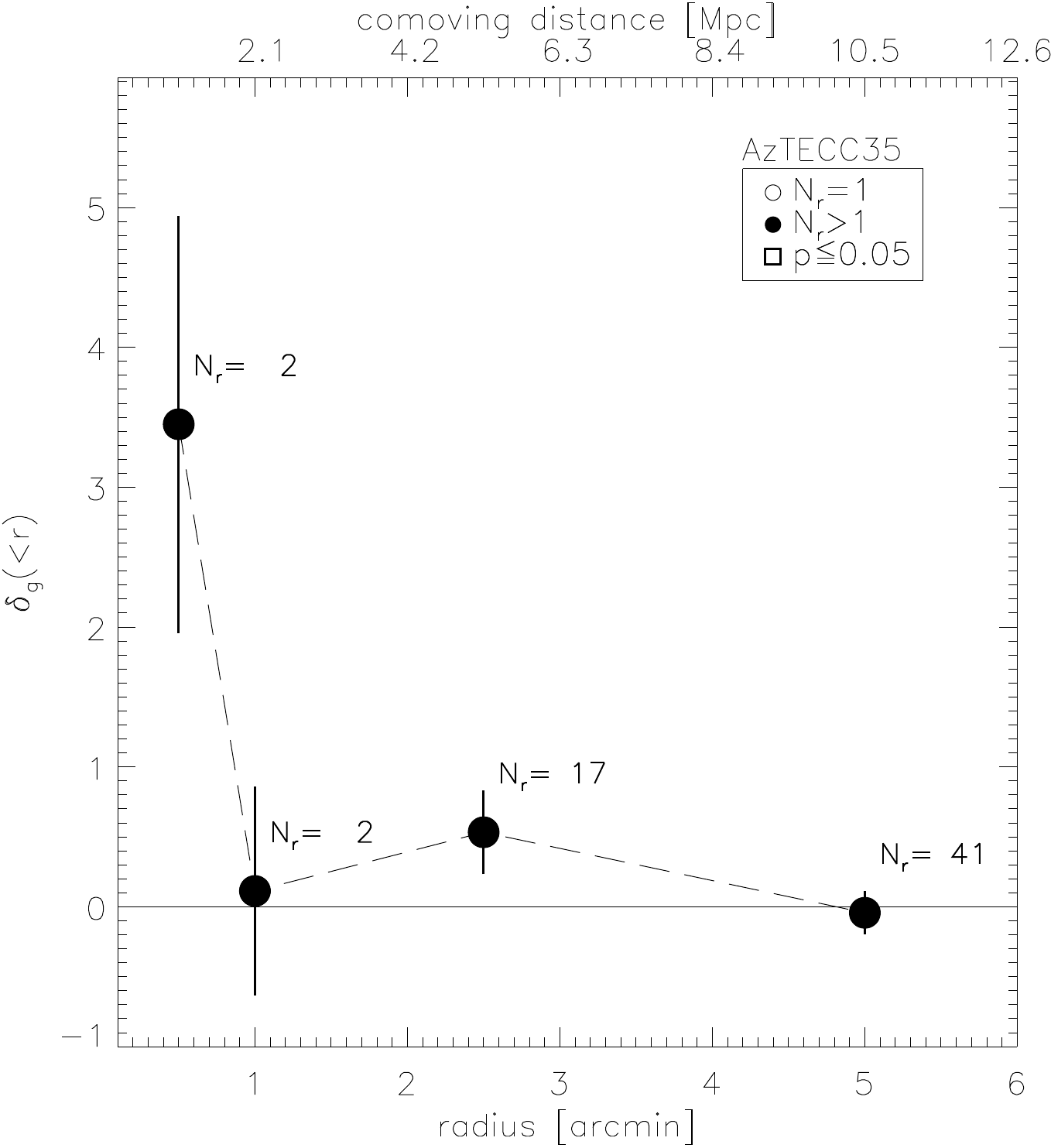}
\includegraphics[width=0.23\textwidth]{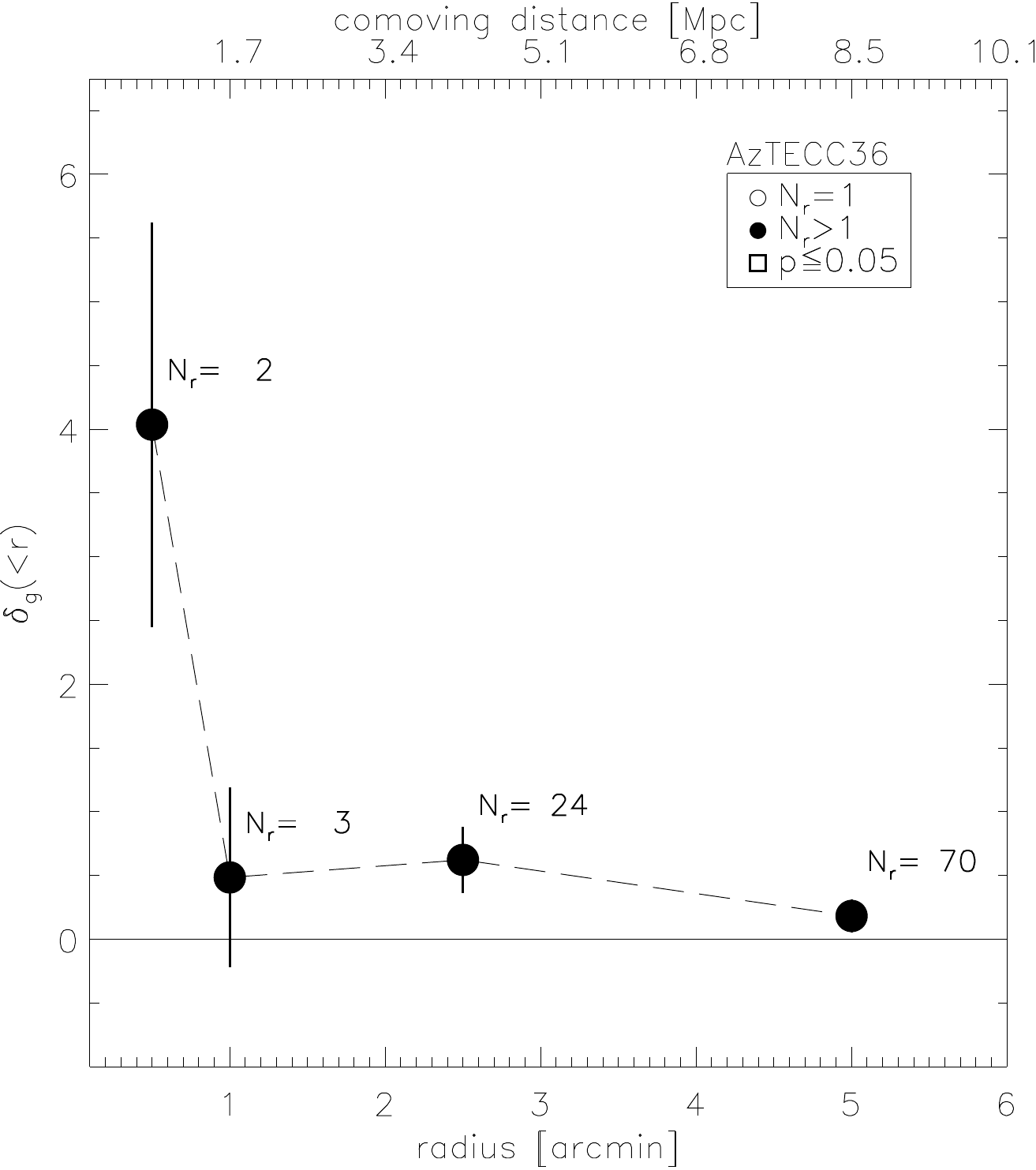}
\includegraphics[width=0.23\textwidth]{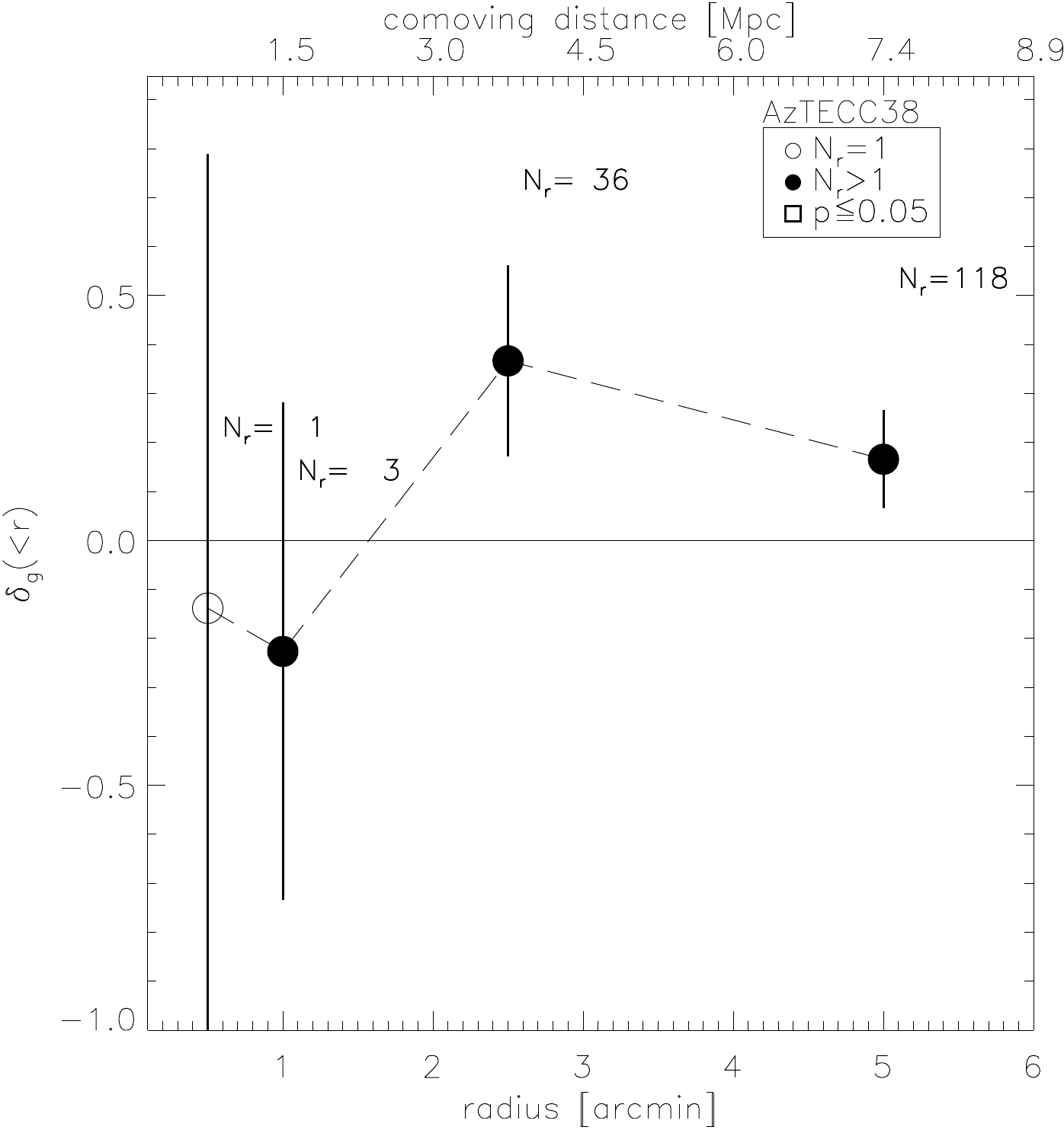}
\includegraphics[width=0.23\textwidth]{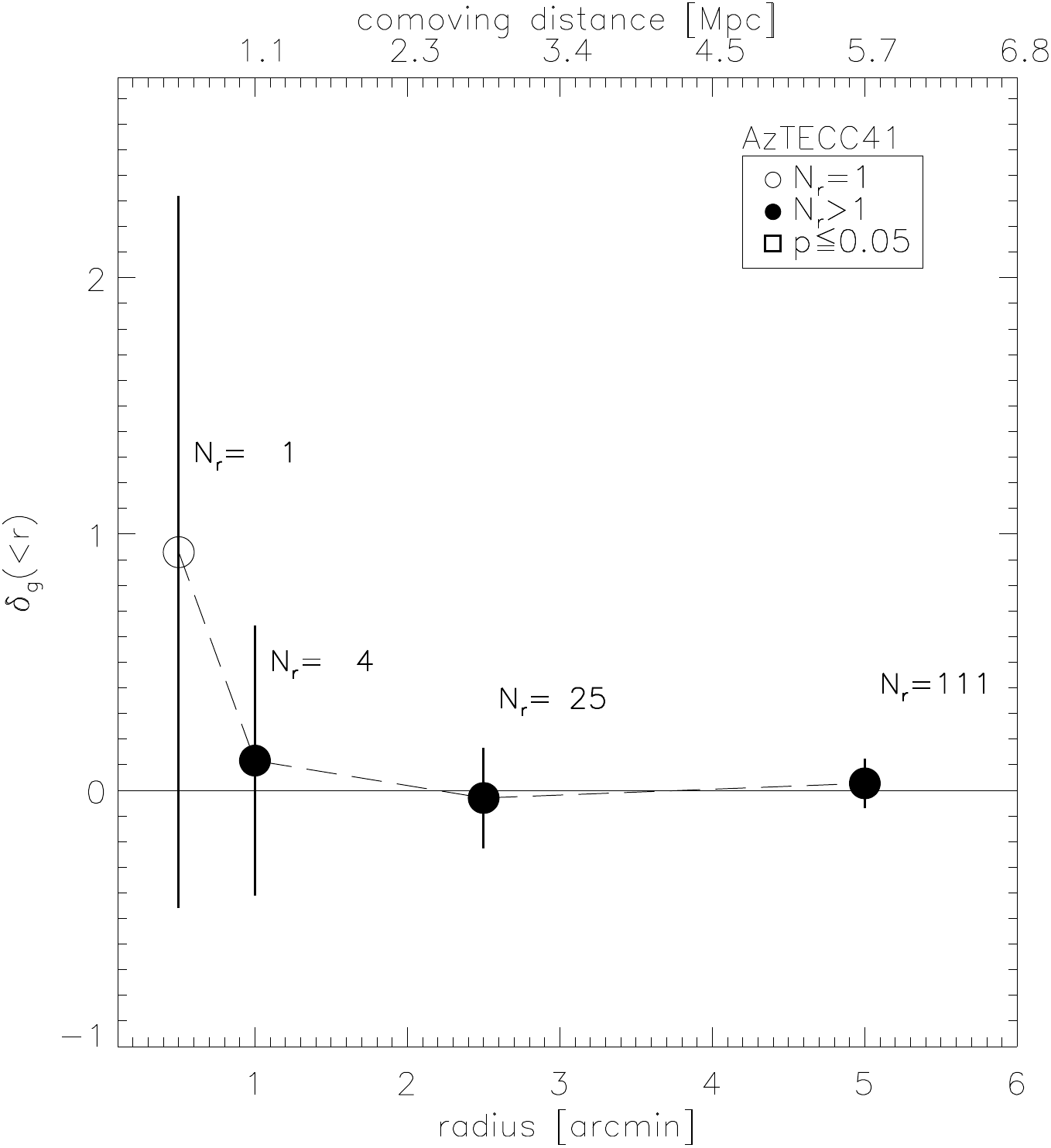}
\includegraphics[width=0.23\textwidth]{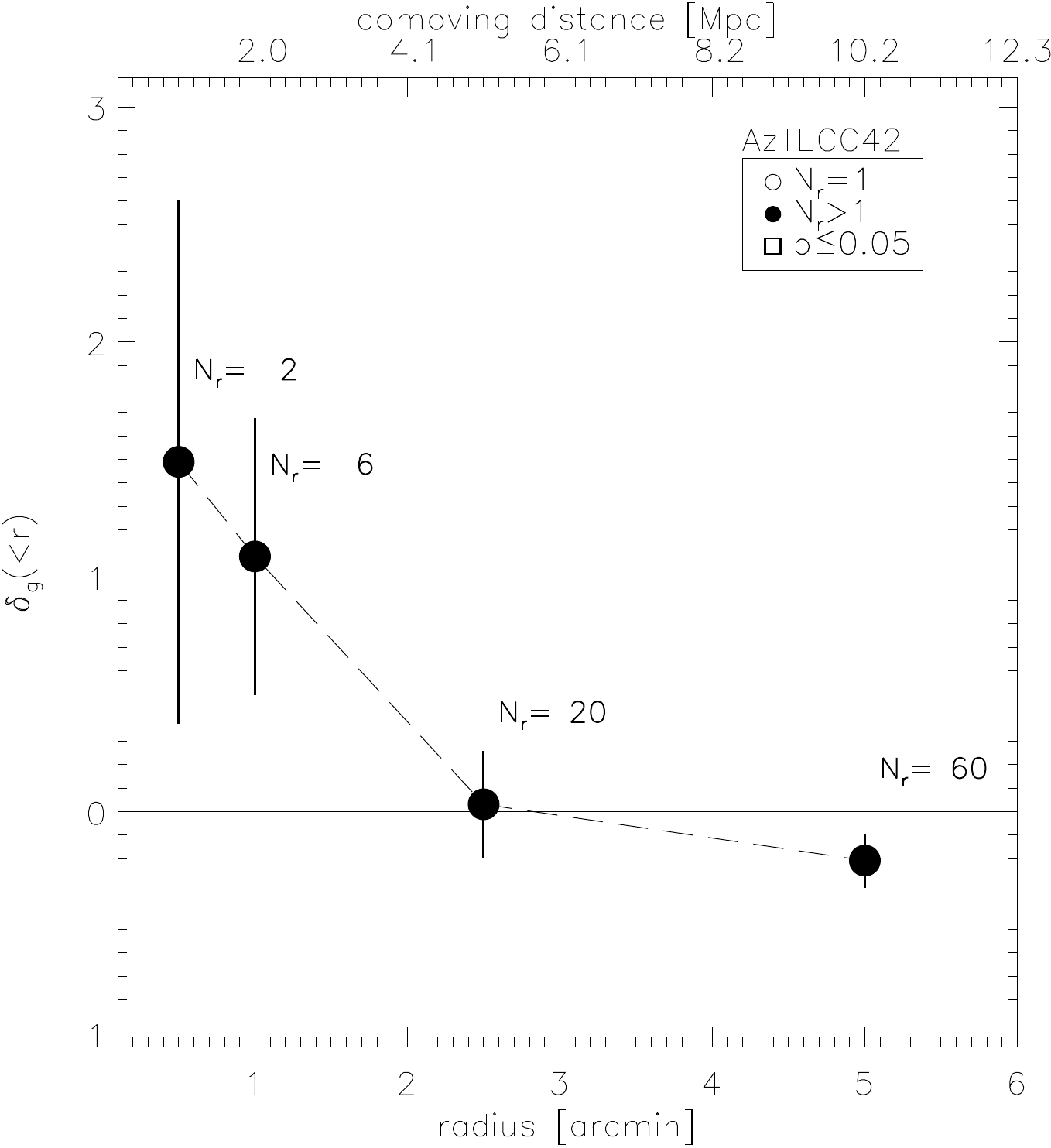}
\includegraphics[width=0.23\textwidth]{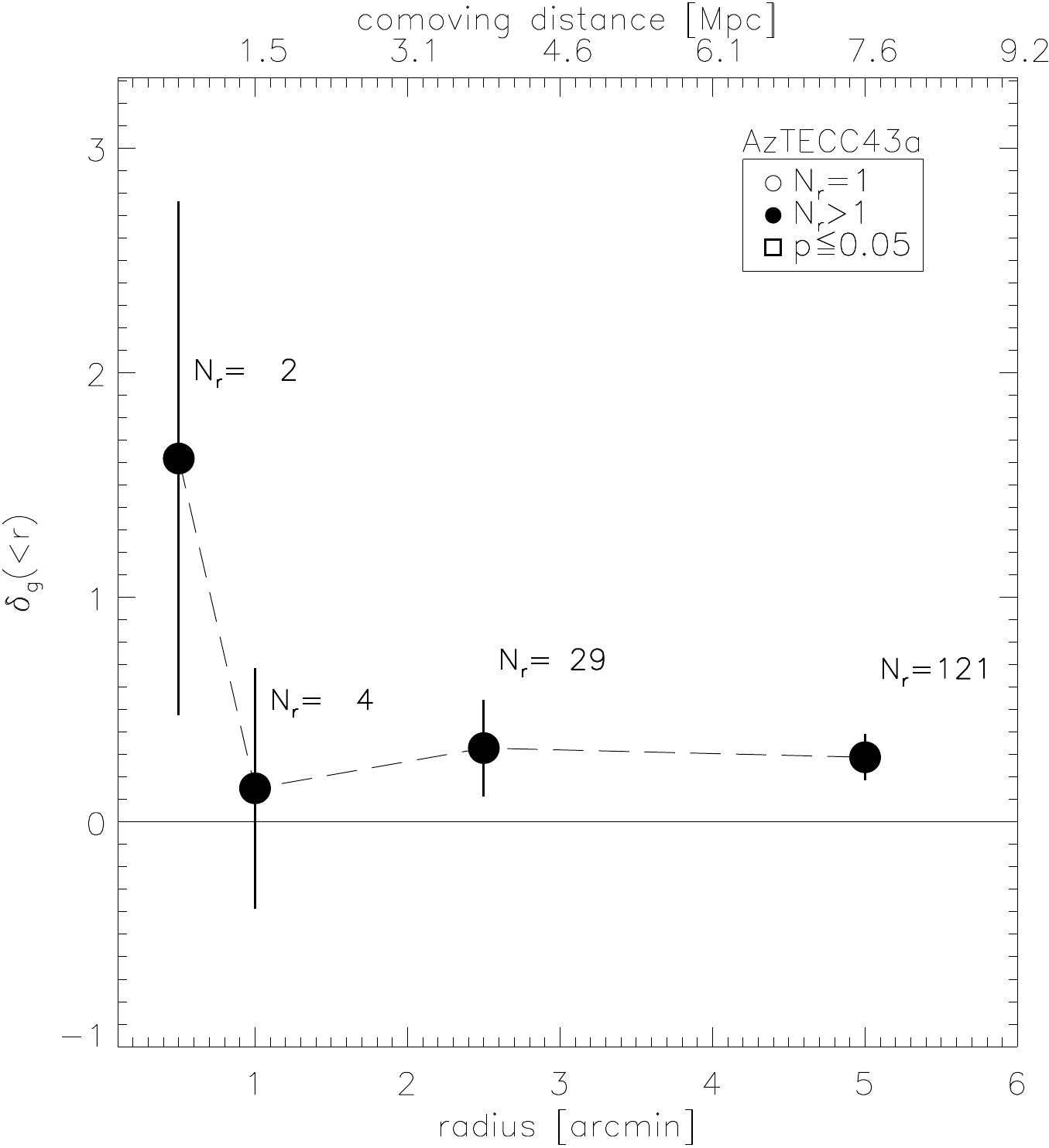}
\includegraphics[width=0.23\textwidth]{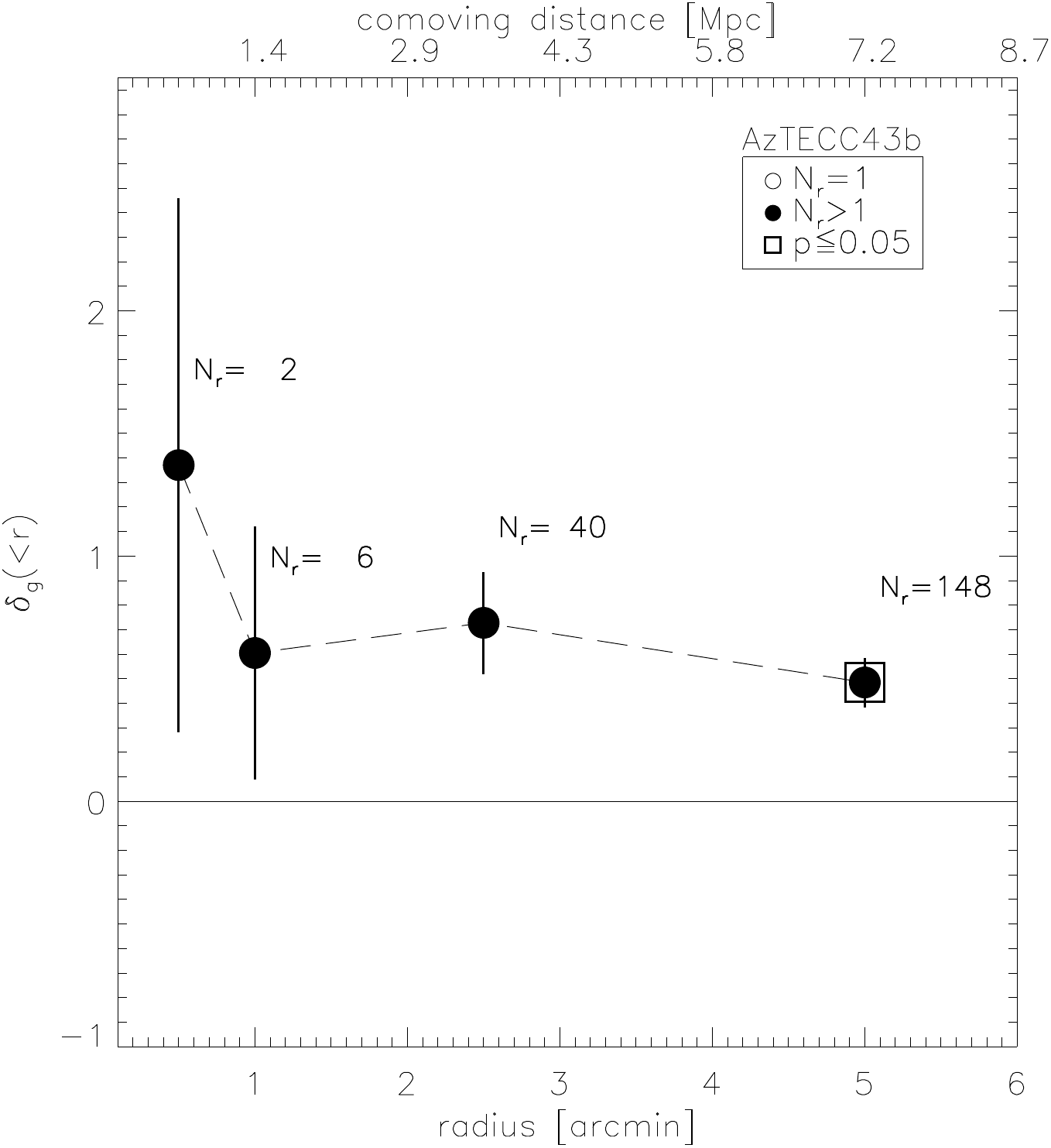}
\includegraphics[width=0.23\textwidth]{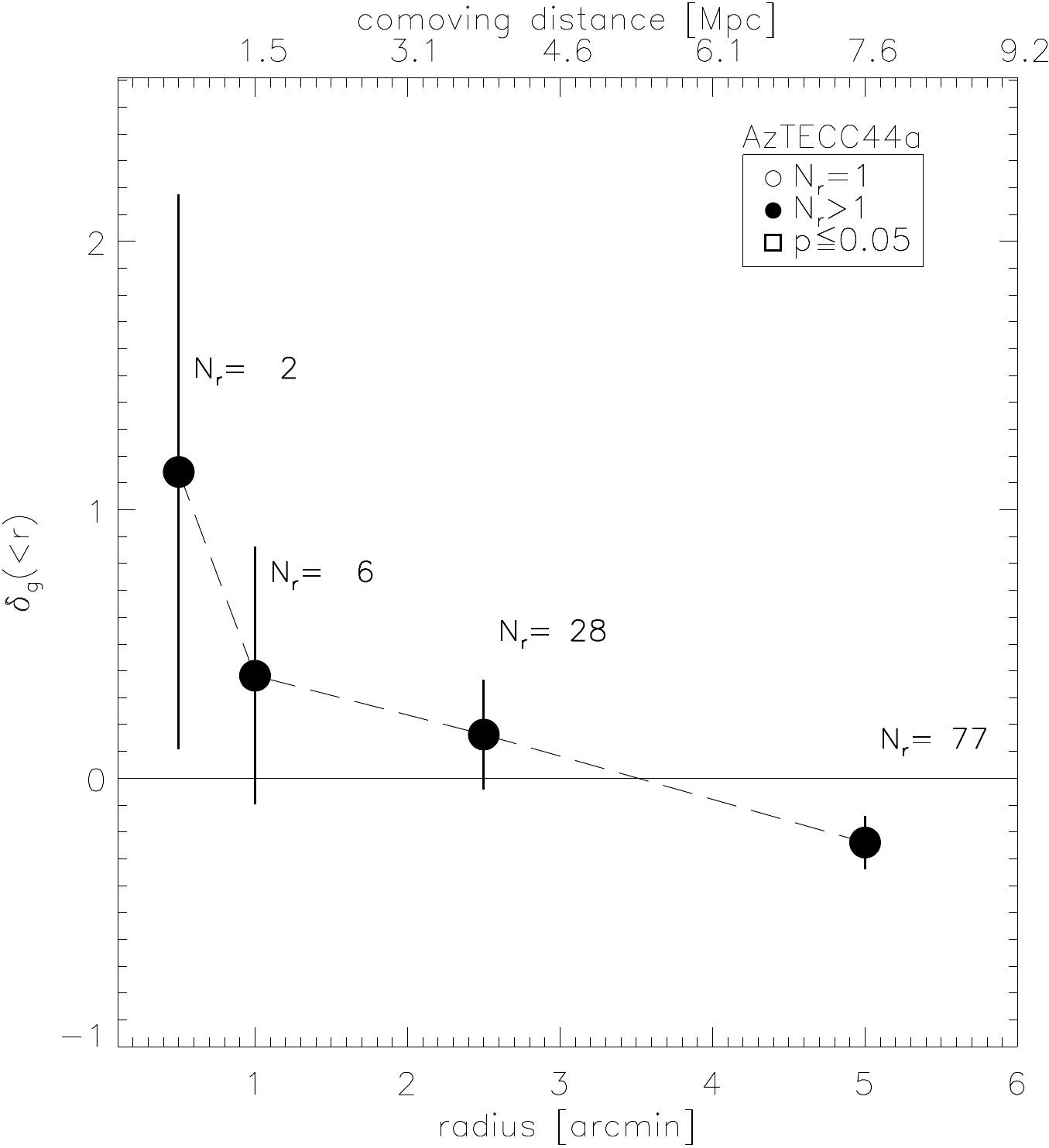}
\includegraphics[width=0.23\textwidth]{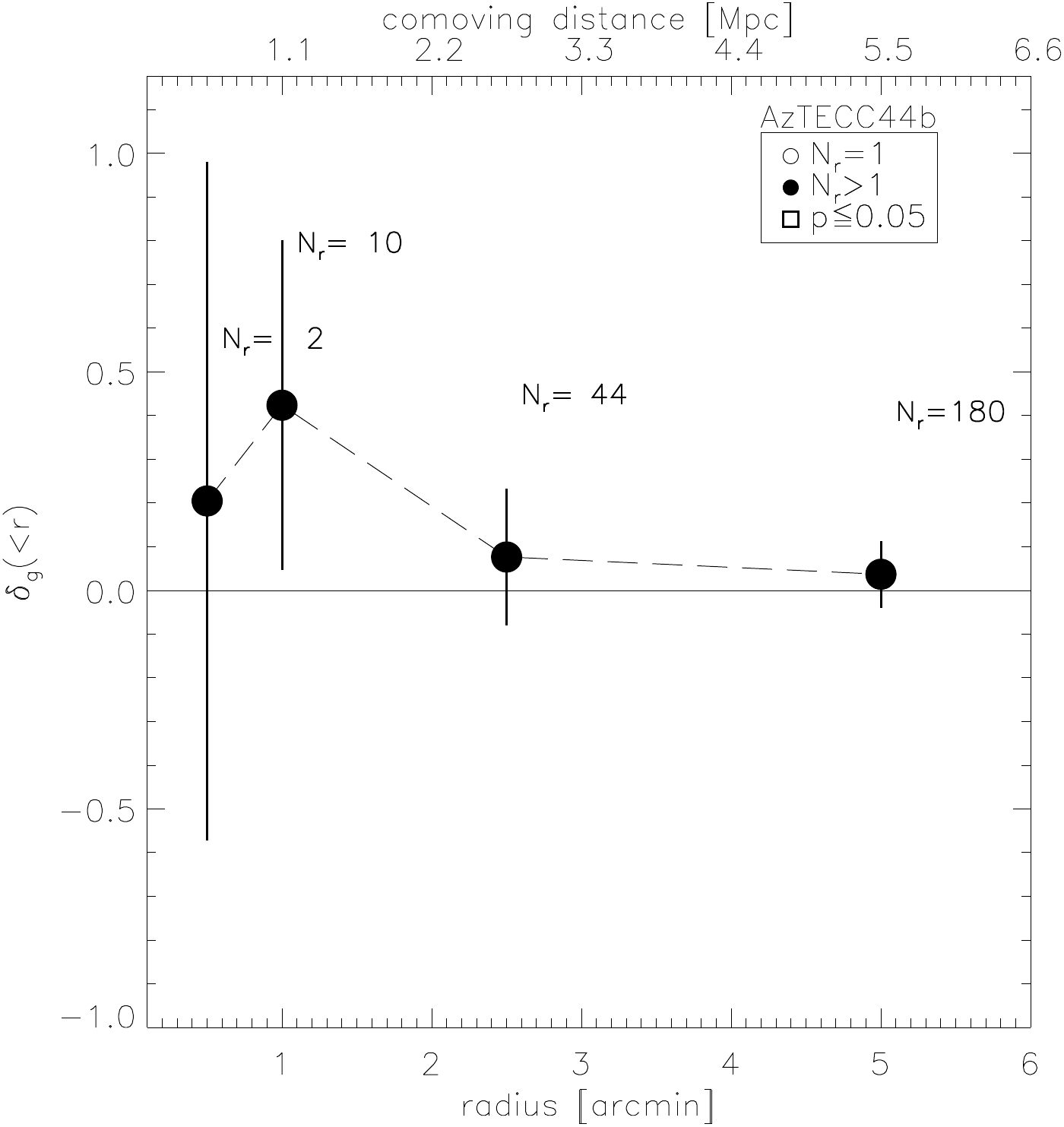}
\includegraphics[width=0.23\textwidth]{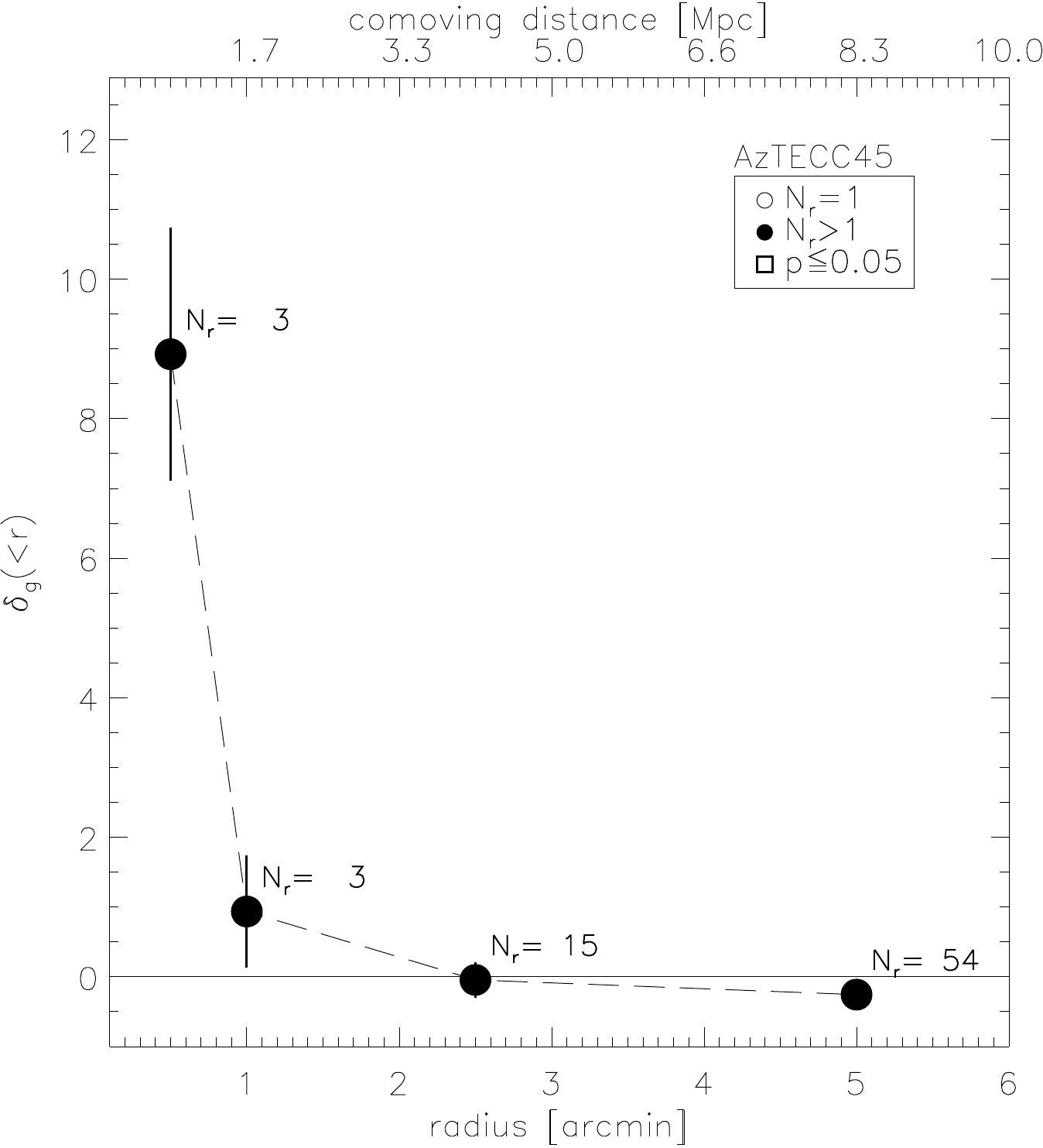}
\includegraphics[width=0.23\textwidth]{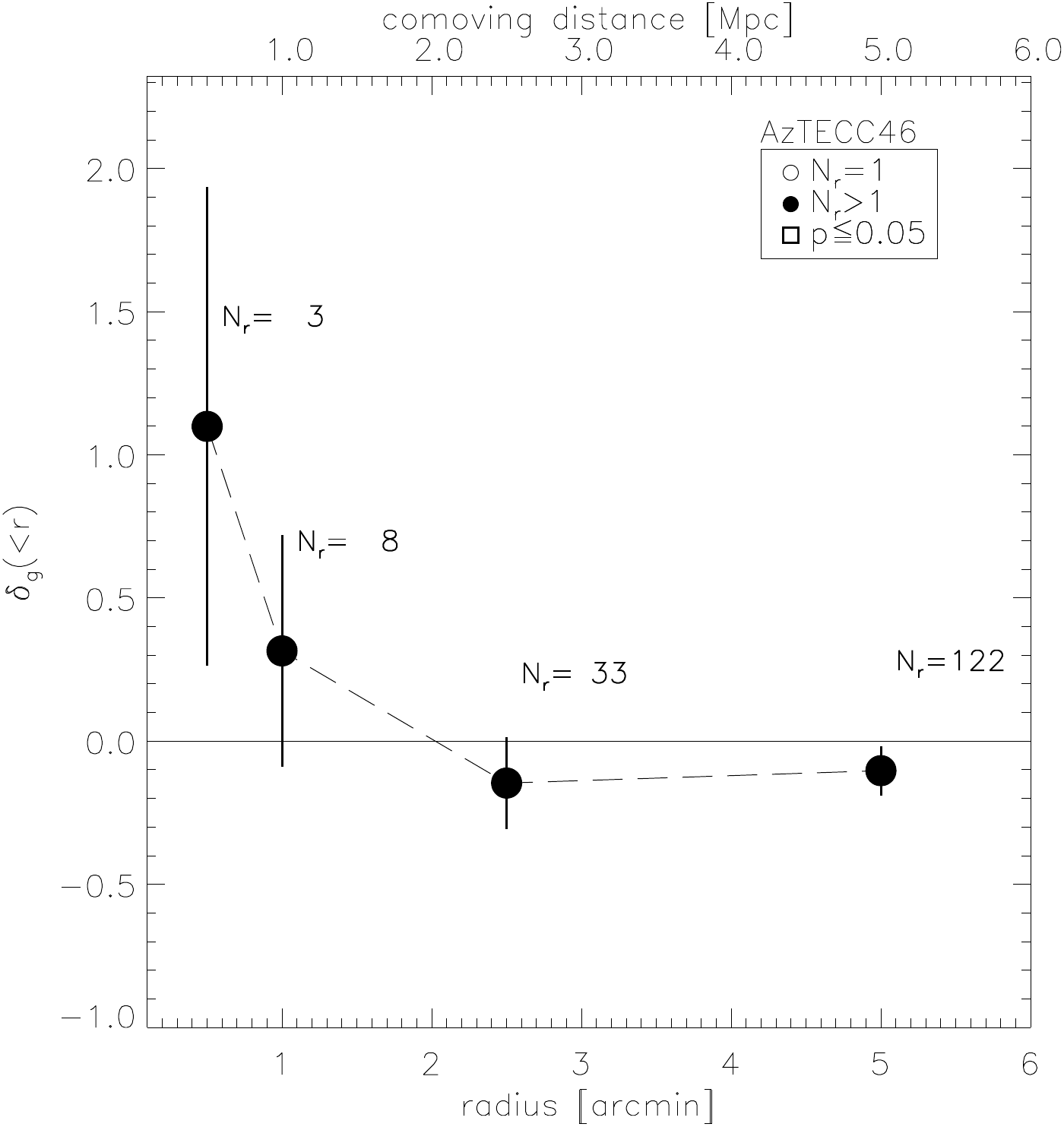}
\includegraphics[width=0.23\textwidth]{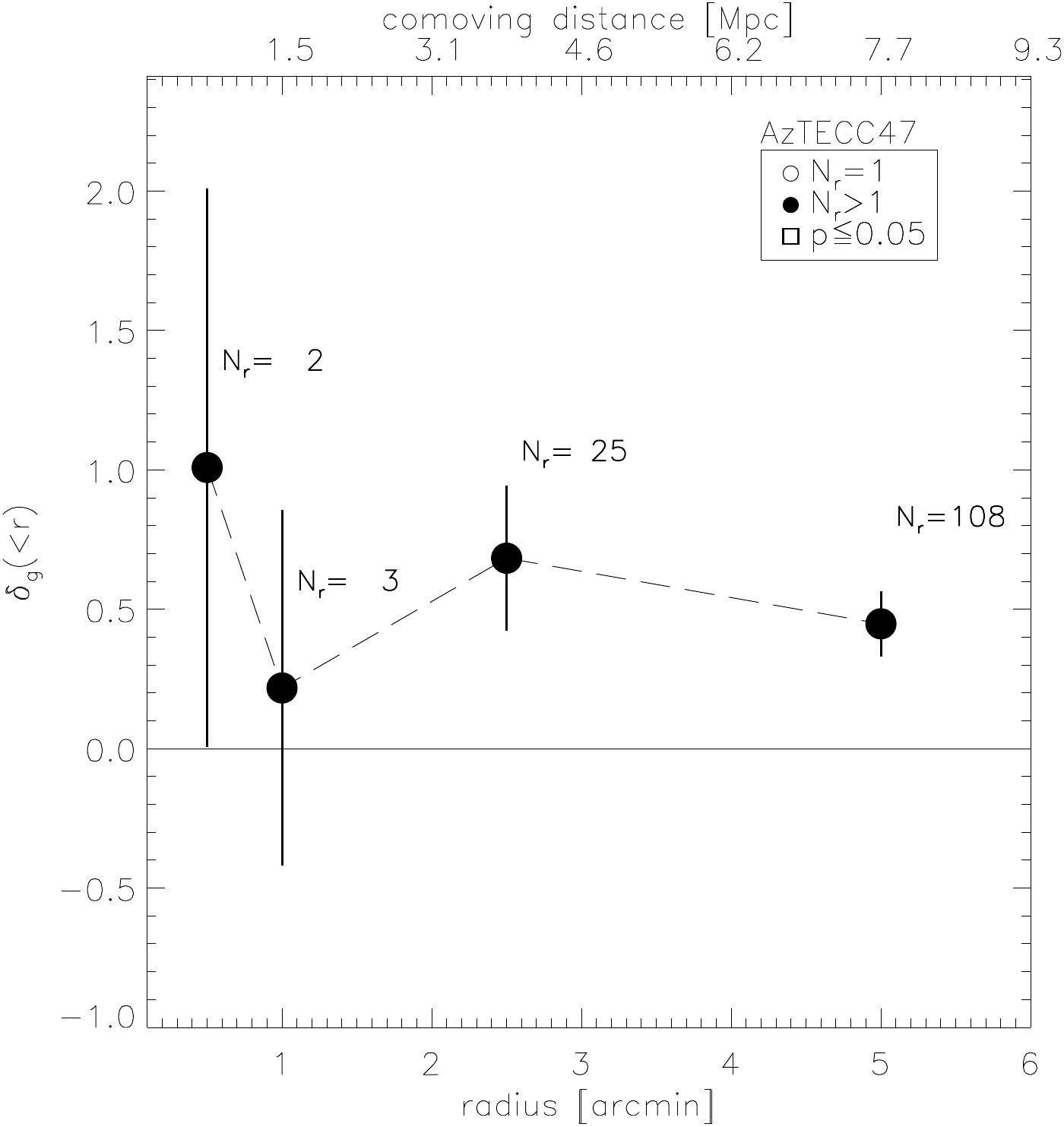}
\includegraphics[width=0.23\textwidth]{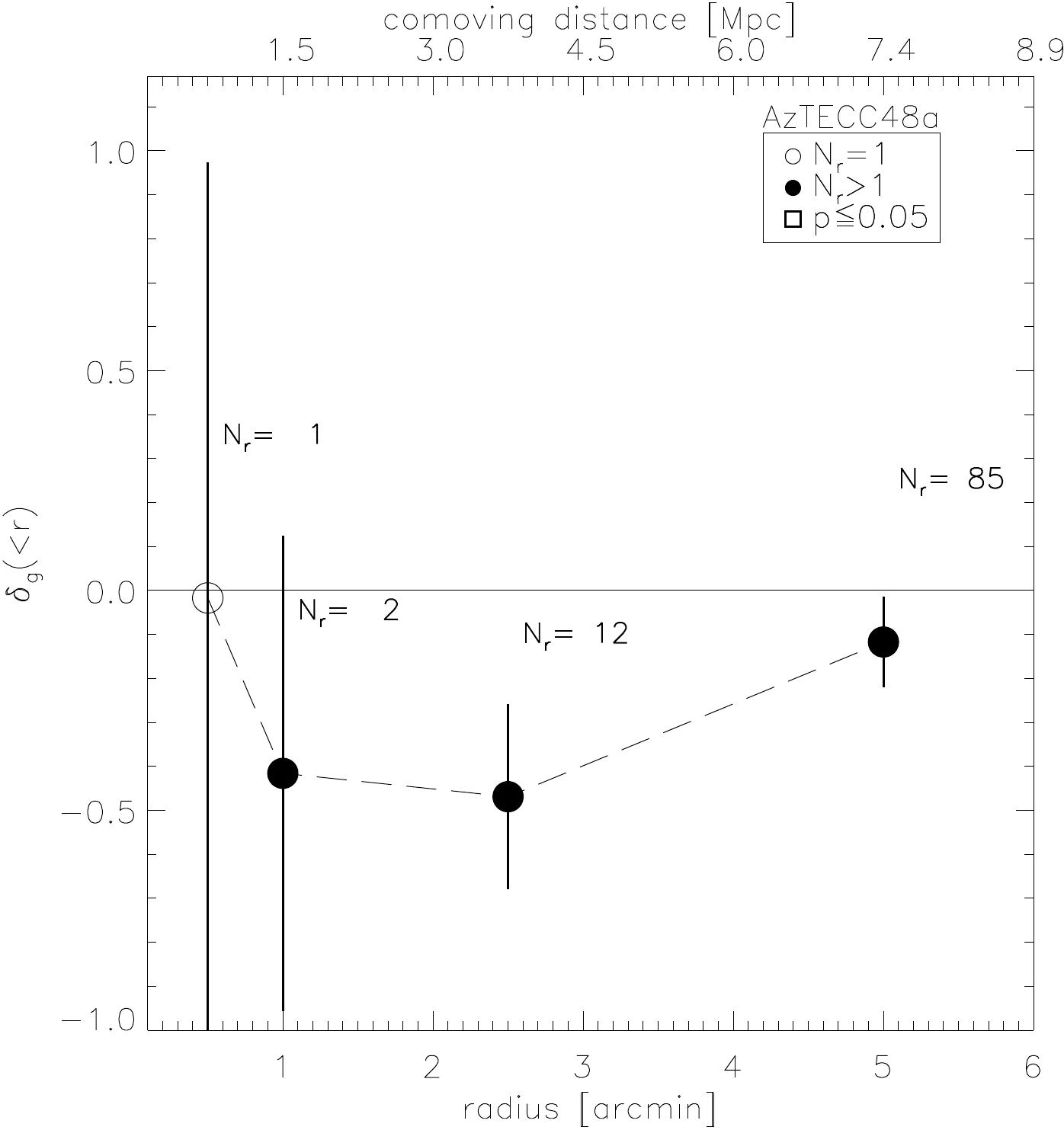}
\includegraphics[width=0.23\textwidth]{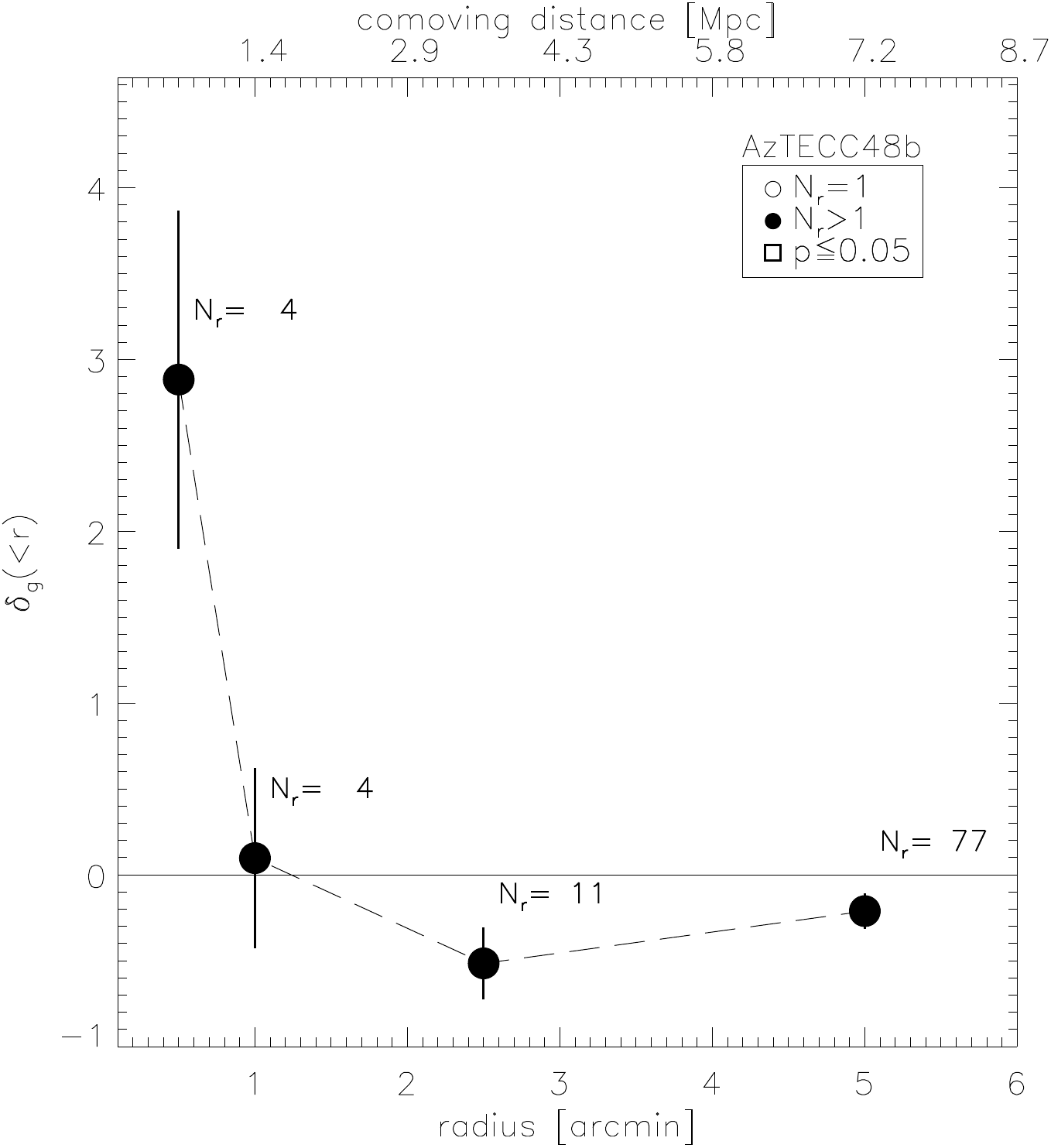}
\includegraphics[width=0.23\textwidth]{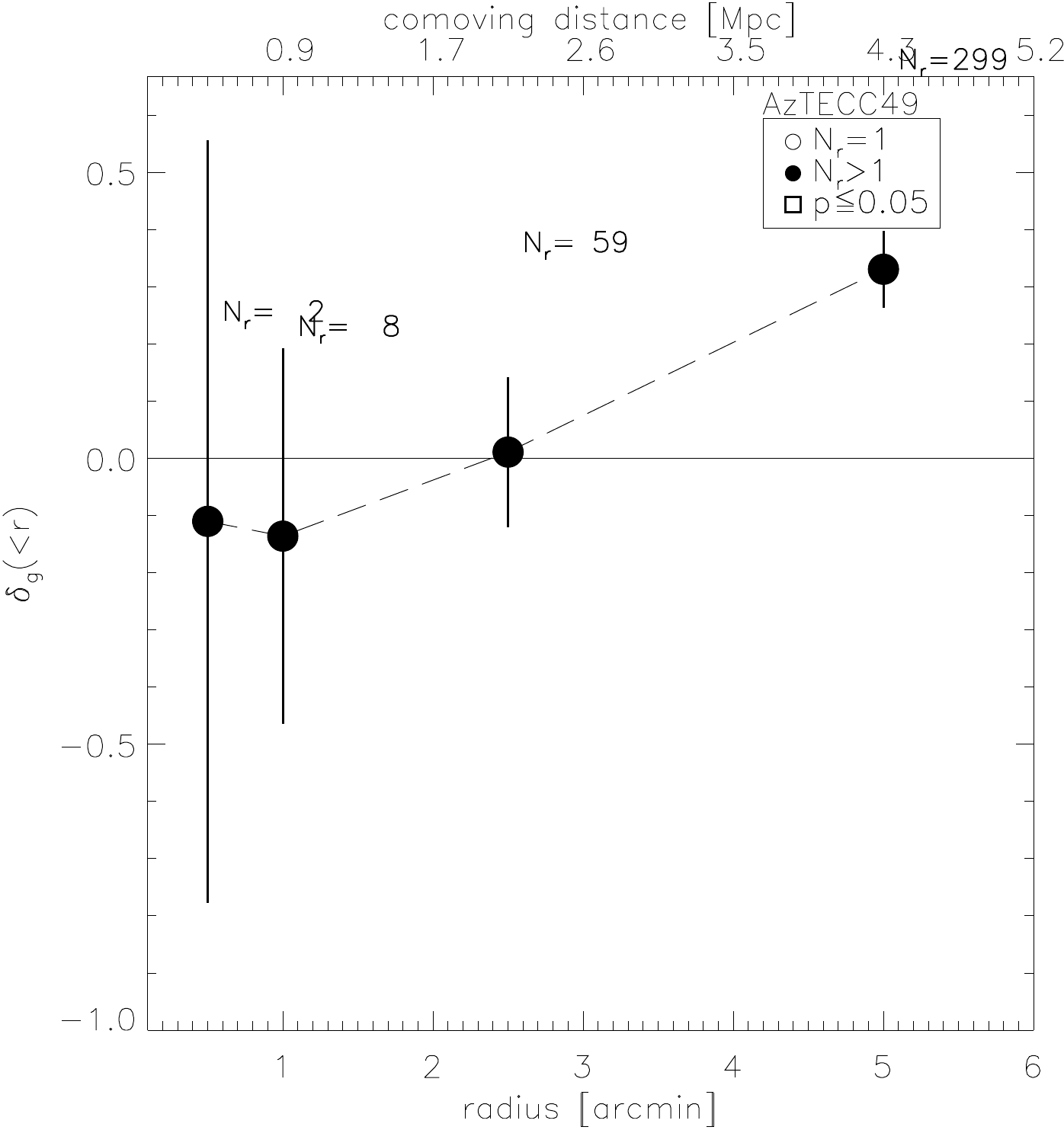}
\includegraphics[width=0.23\textwidth]{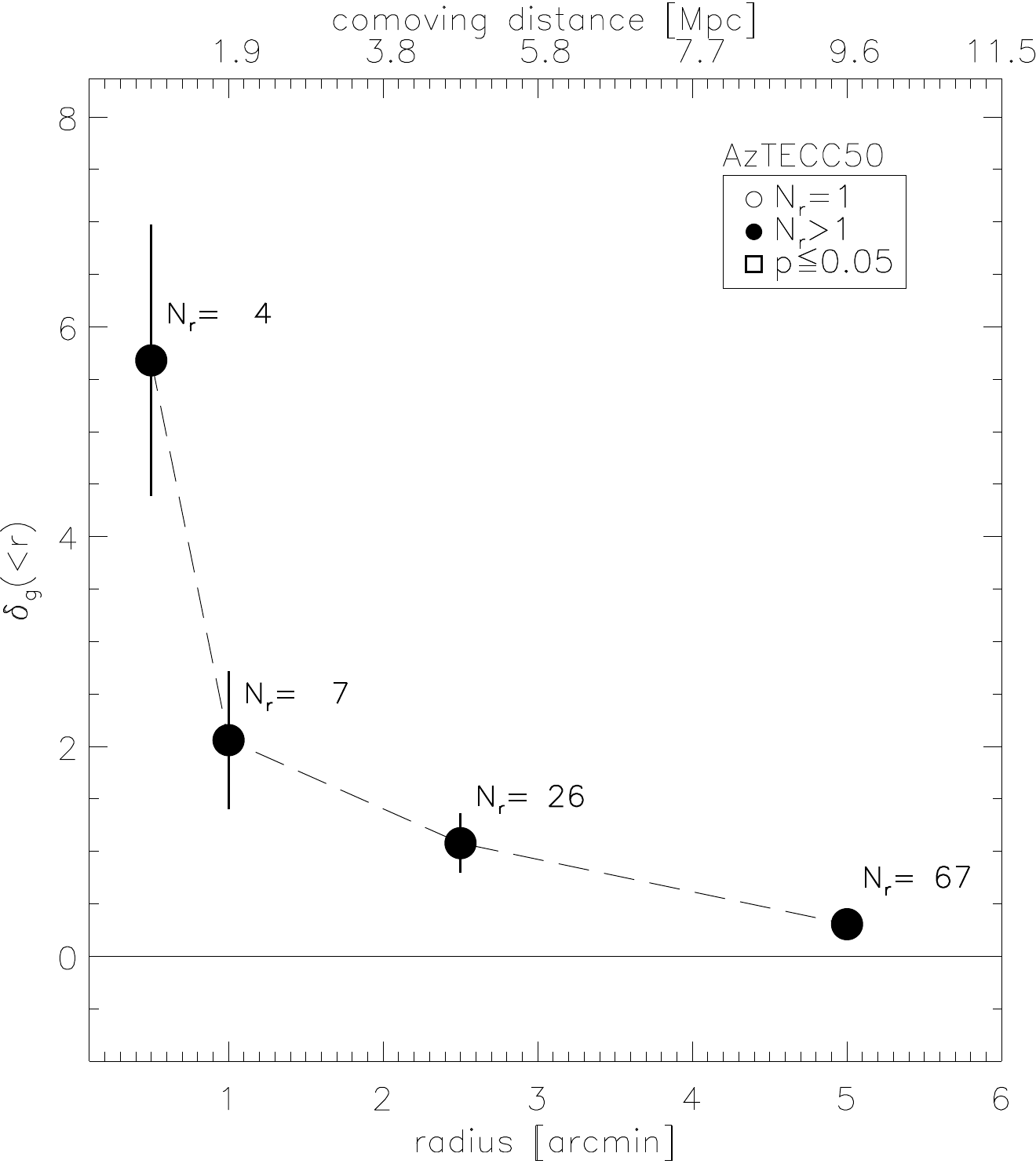}
\includegraphics[width=0.23\textwidth]{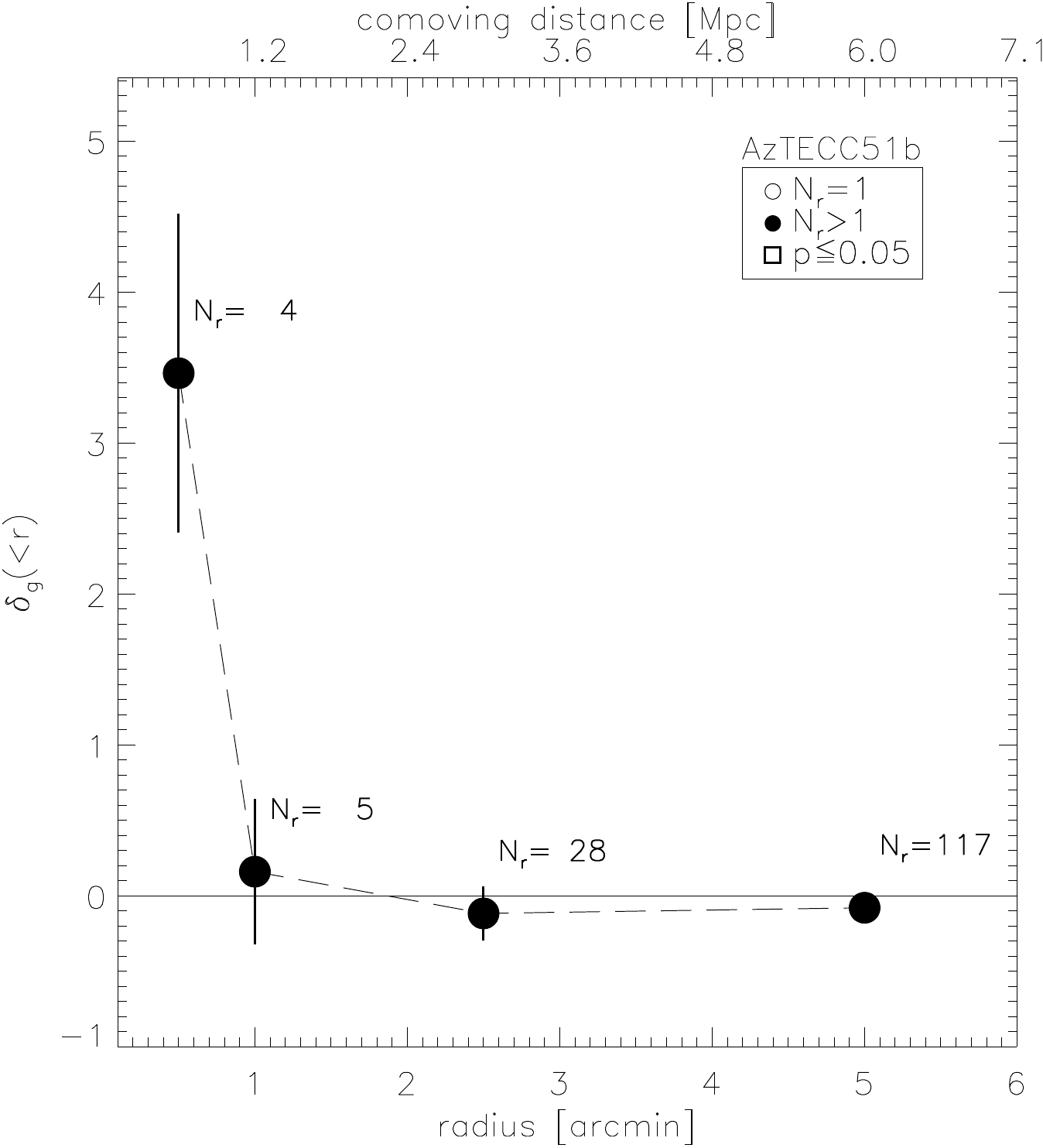}
\includegraphics[width=0.23\textwidth]{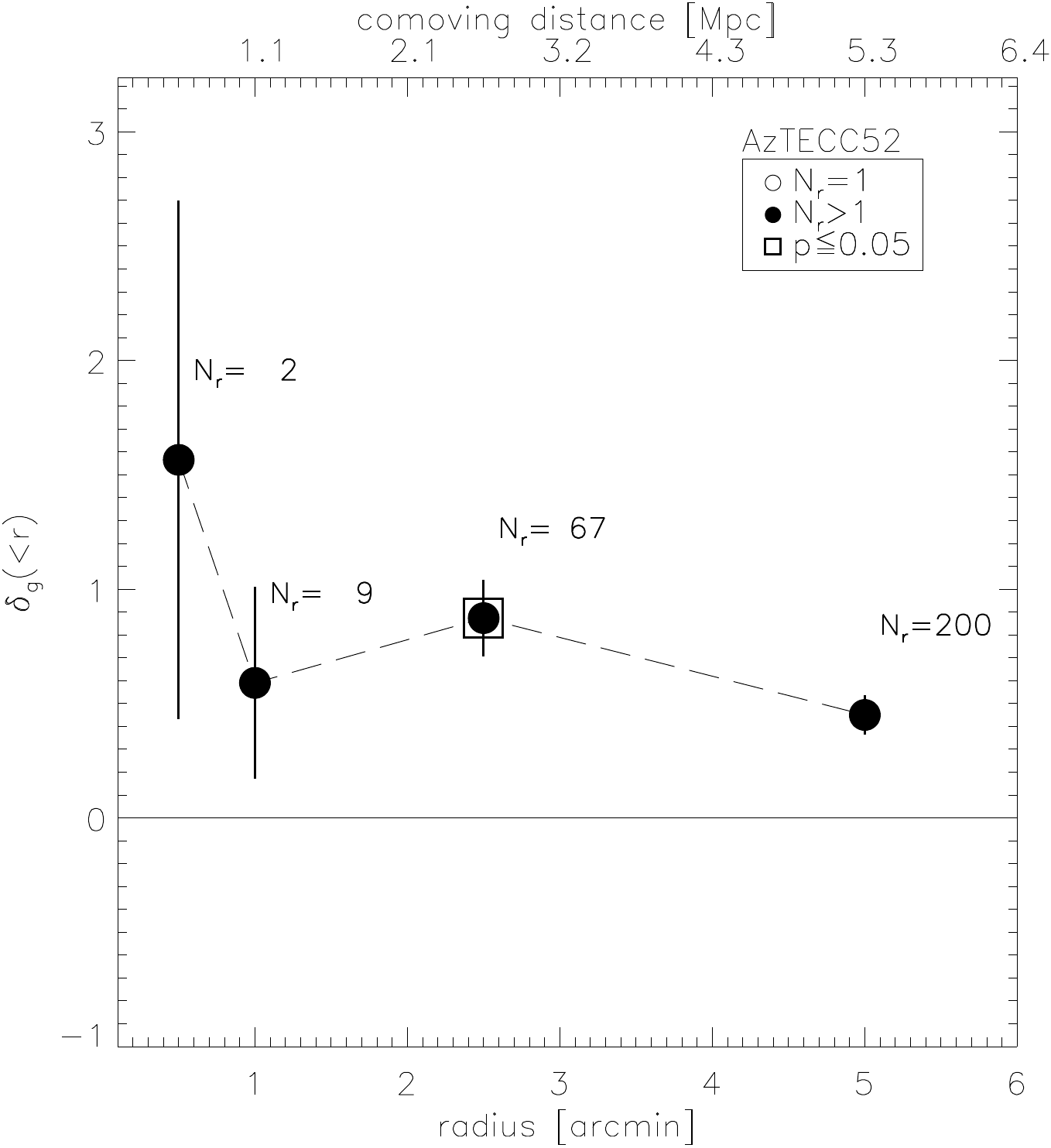}

\caption{continued.}
\end{center}
\end{figure*}

\addtocounter{figure}{-1}
\begin{figure*}
\begin{center}
\includegraphics[width=0.23\textwidth]{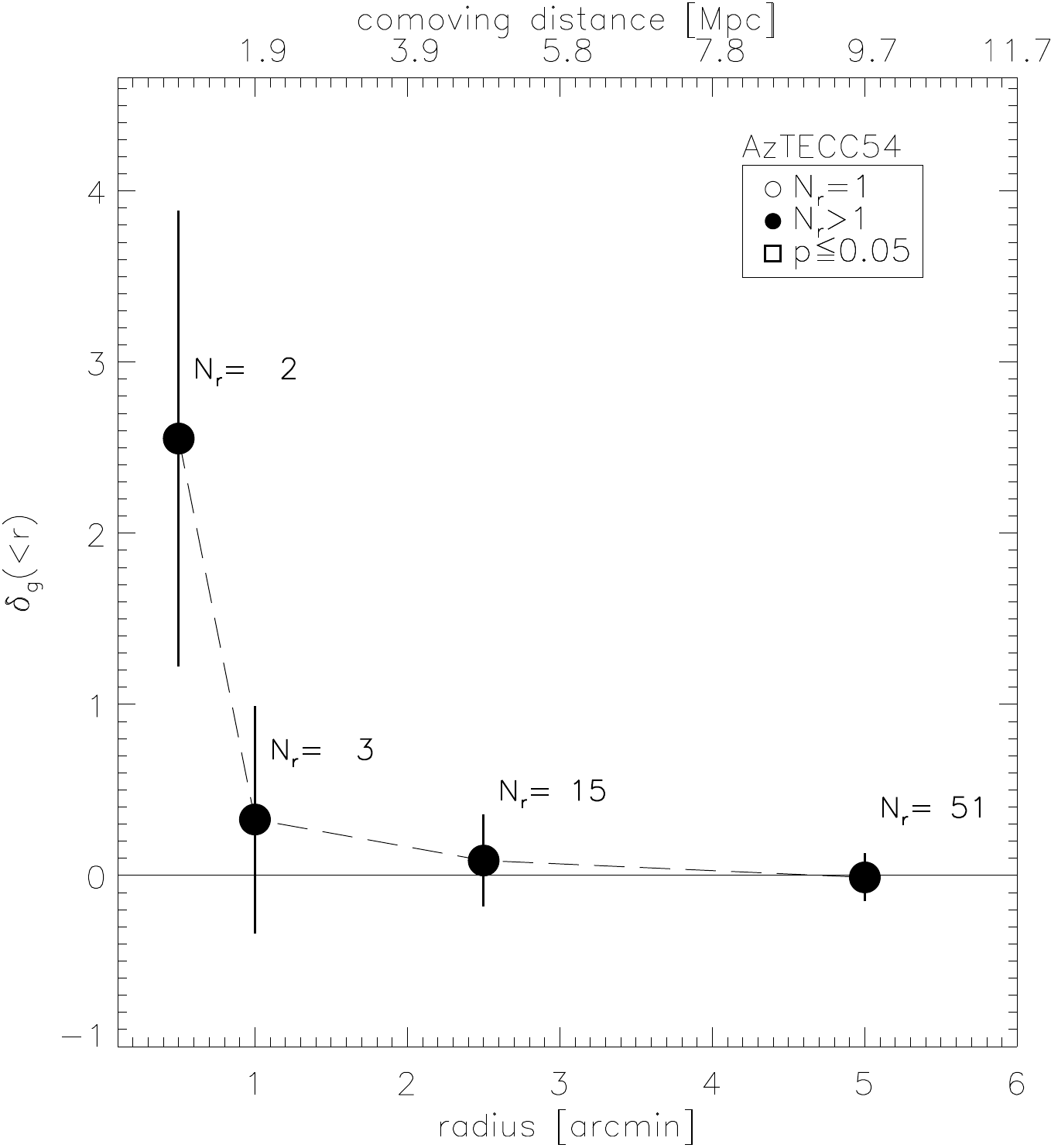}
\includegraphics[width=0.23\textwidth]{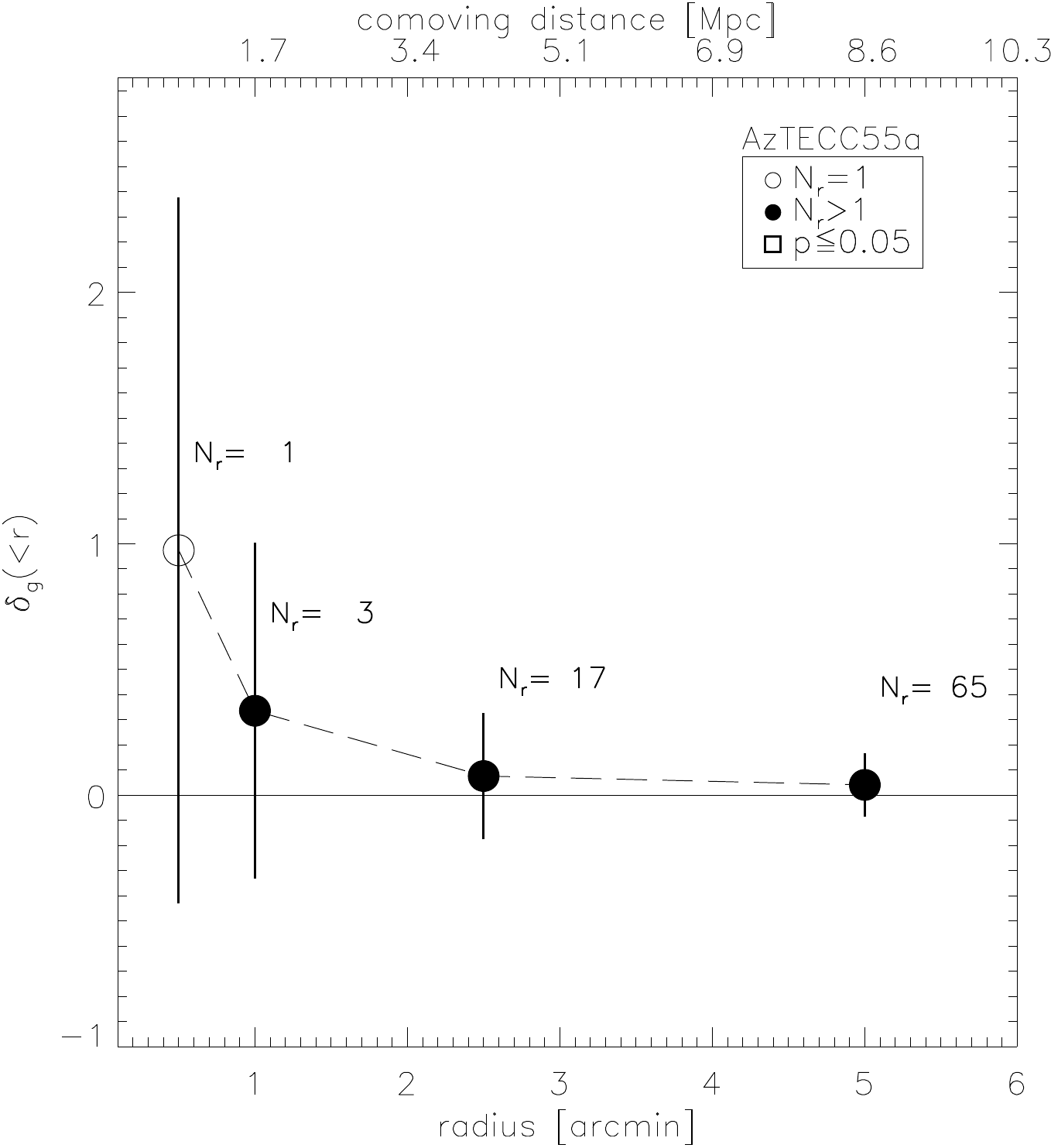}
\includegraphics[width=0.23\textwidth]{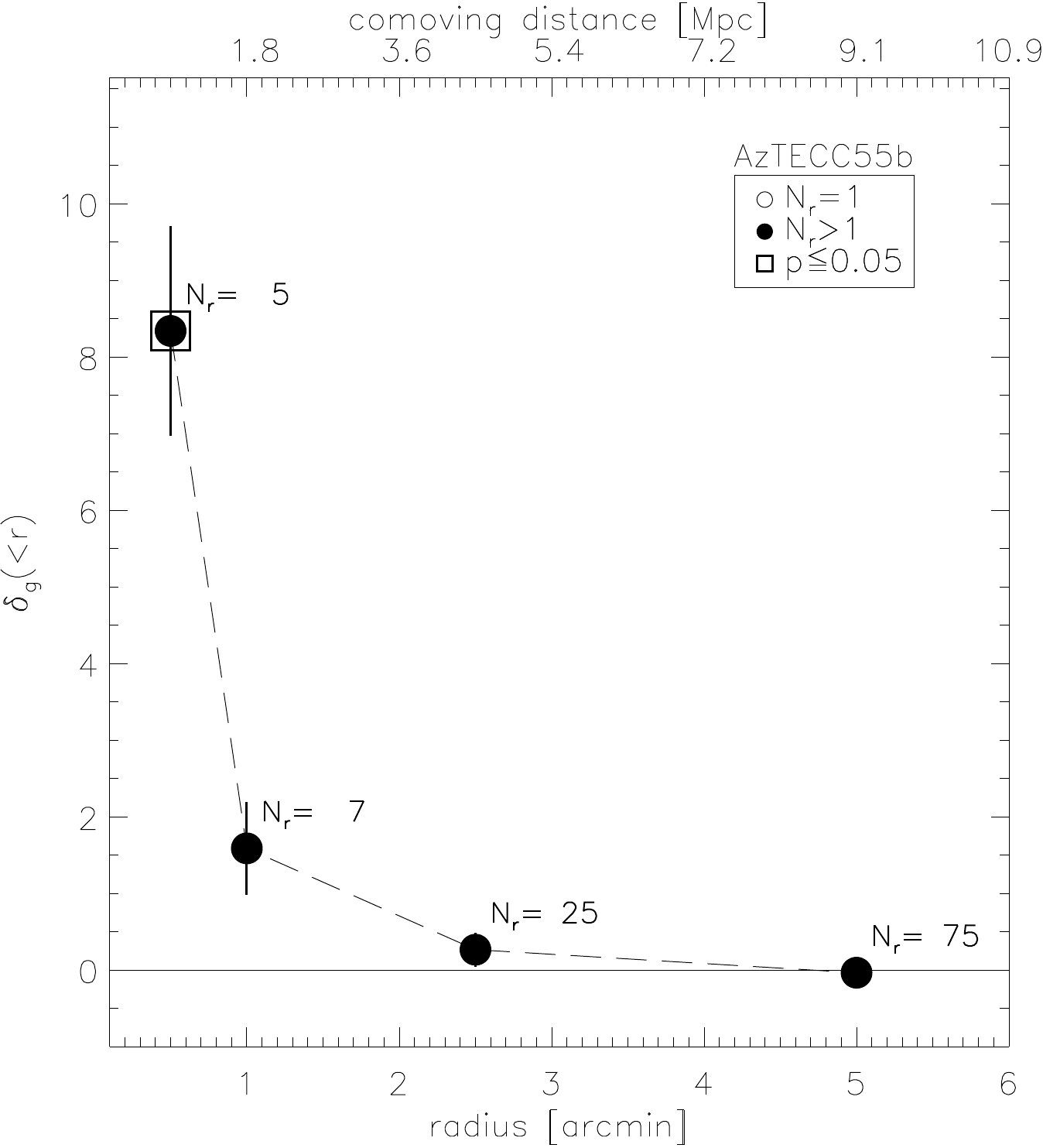}
\includegraphics[width=0.23\textwidth]{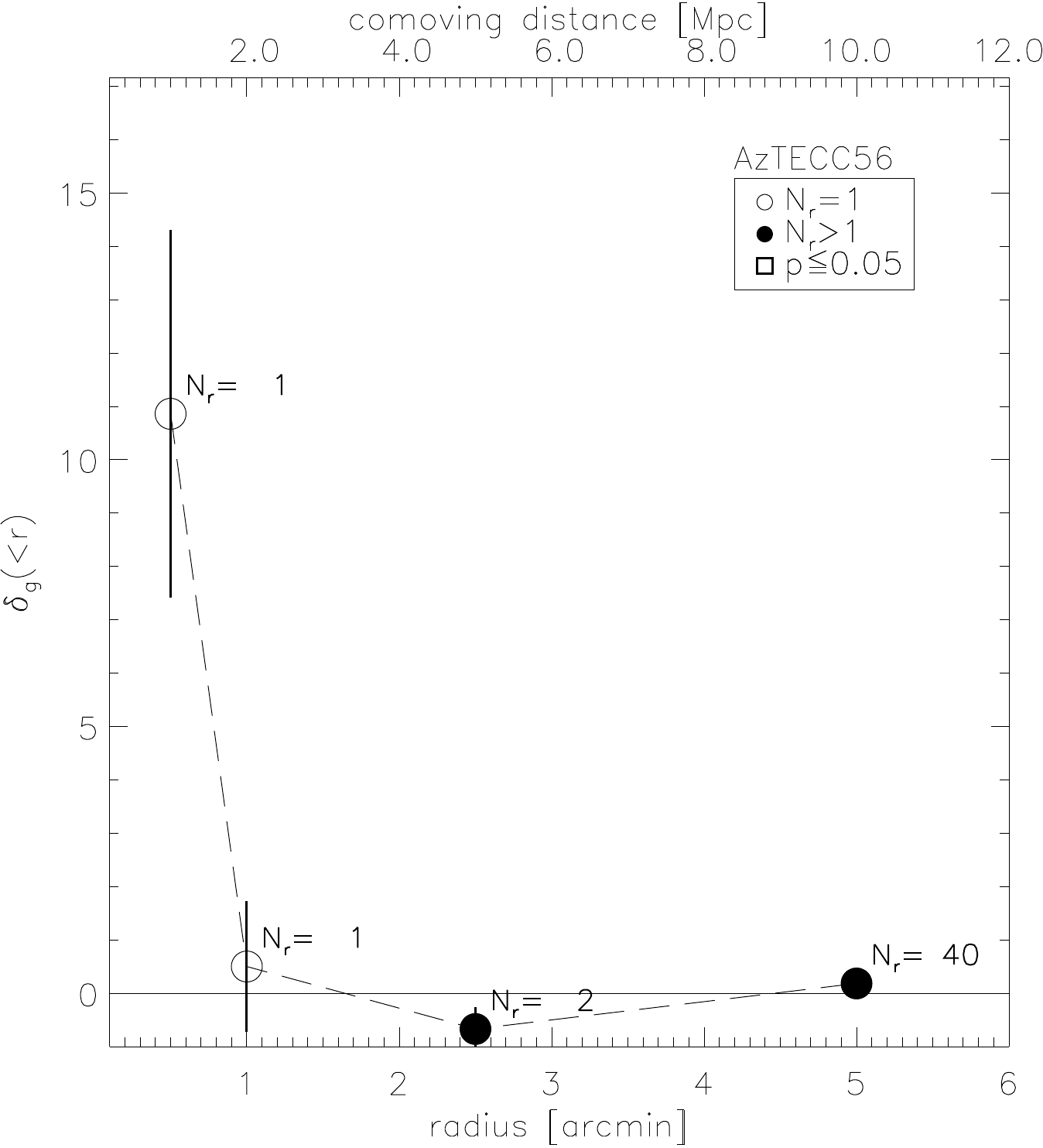}
\includegraphics[width=0.23\textwidth]{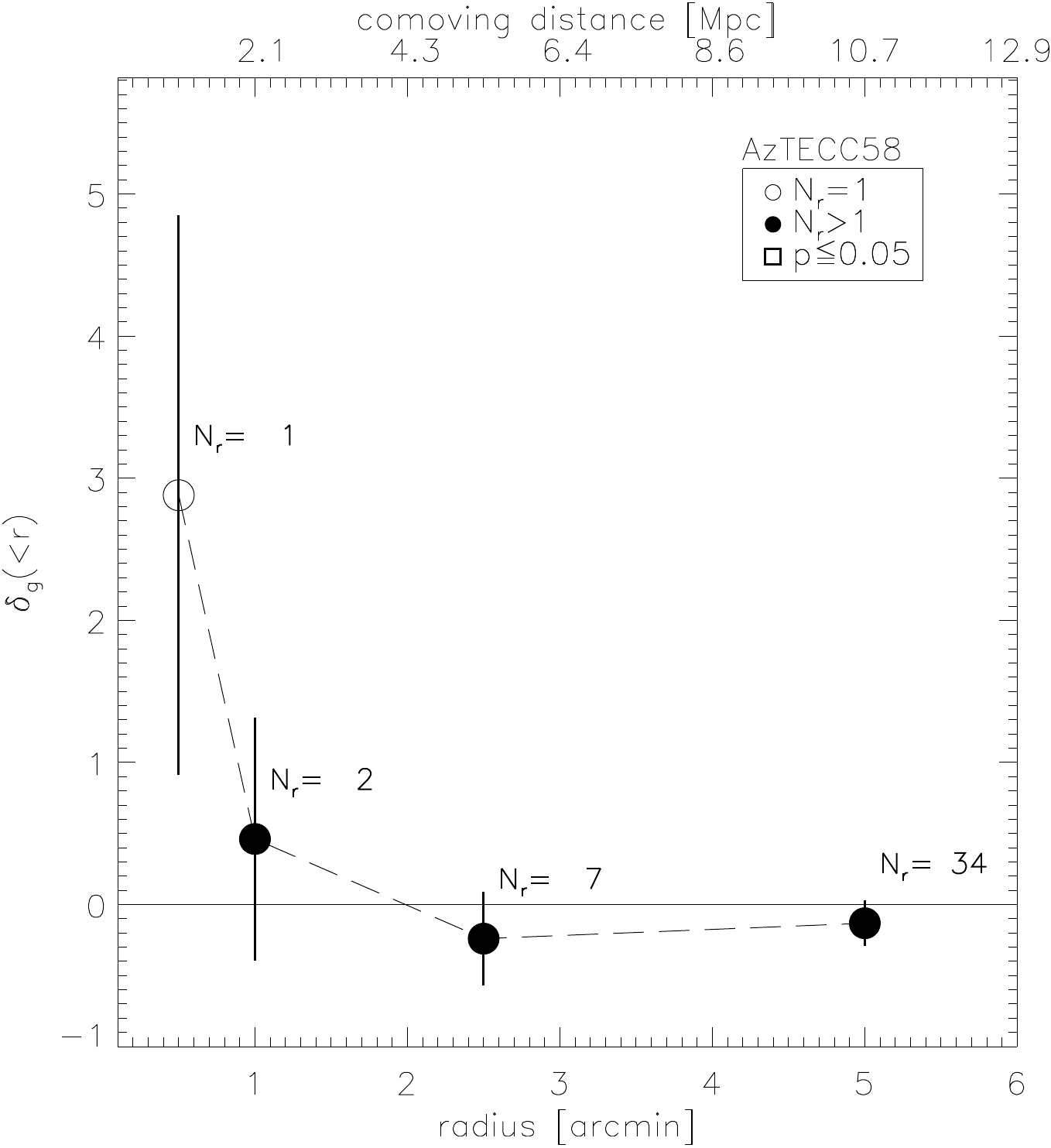}
\includegraphics[width=0.23\textwidth]{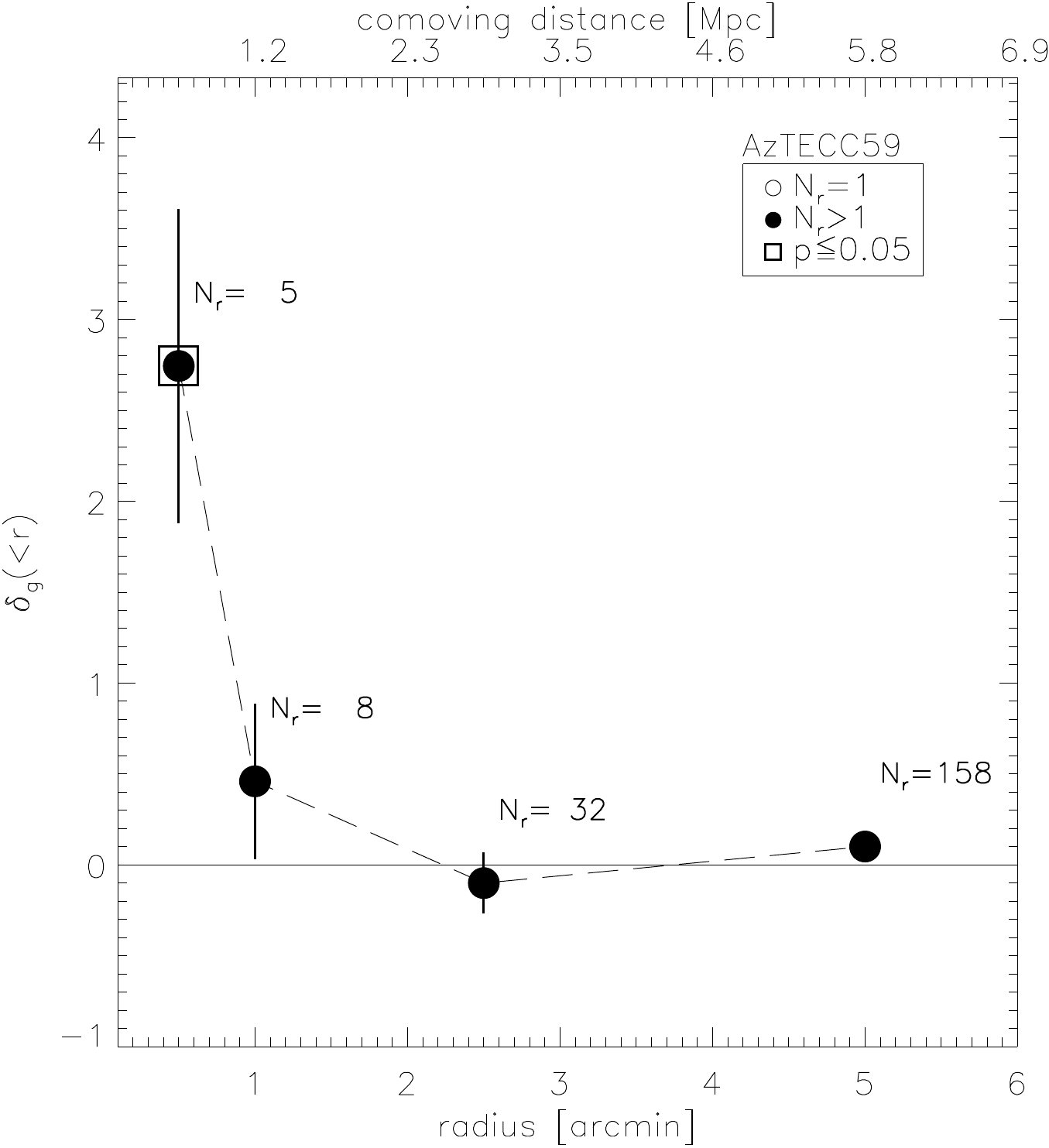}
\includegraphics[width=0.23\textwidth]{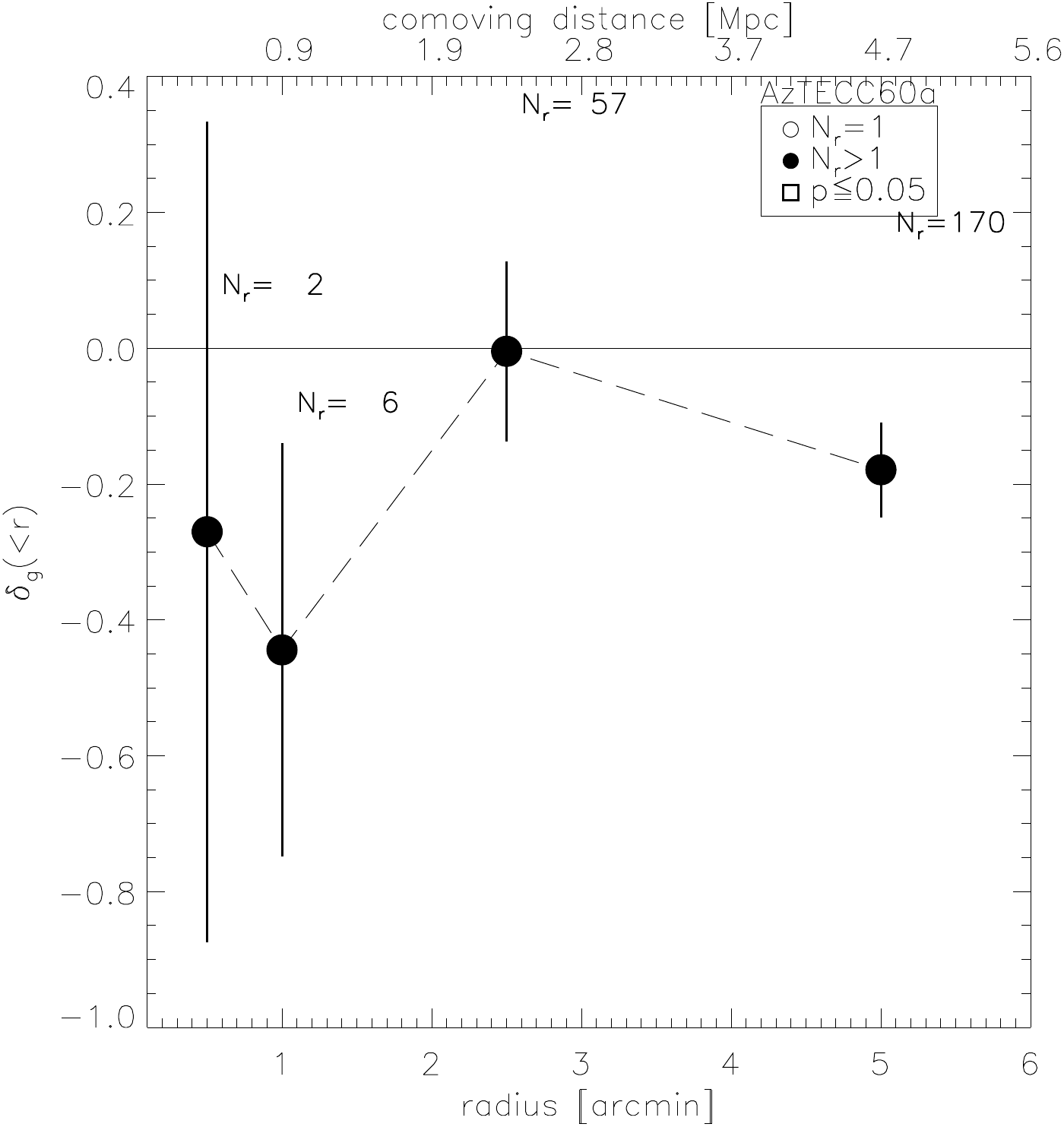}
\includegraphics[width=0.23\textwidth]{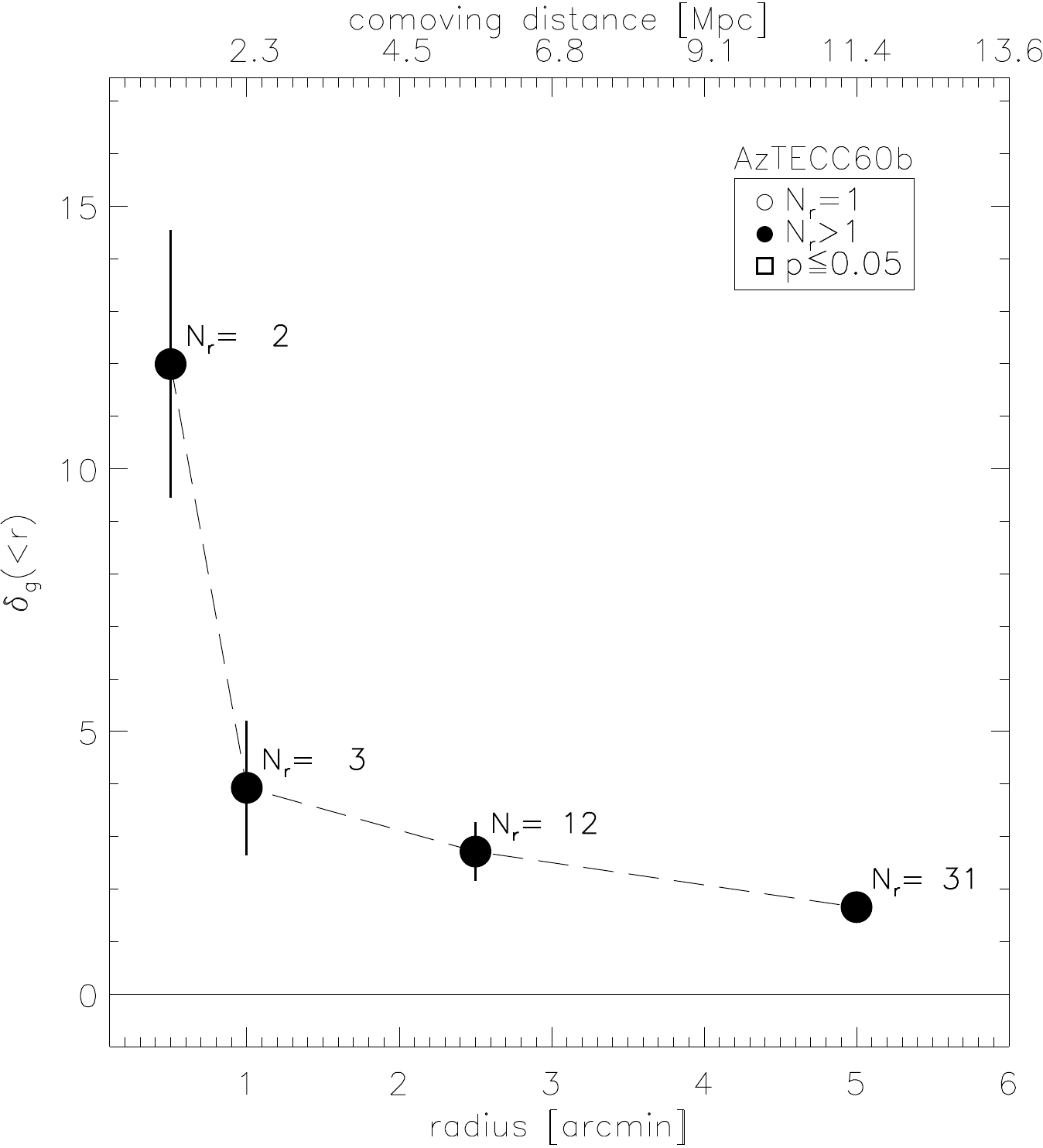}
\includegraphics[width=0.23\textwidth]{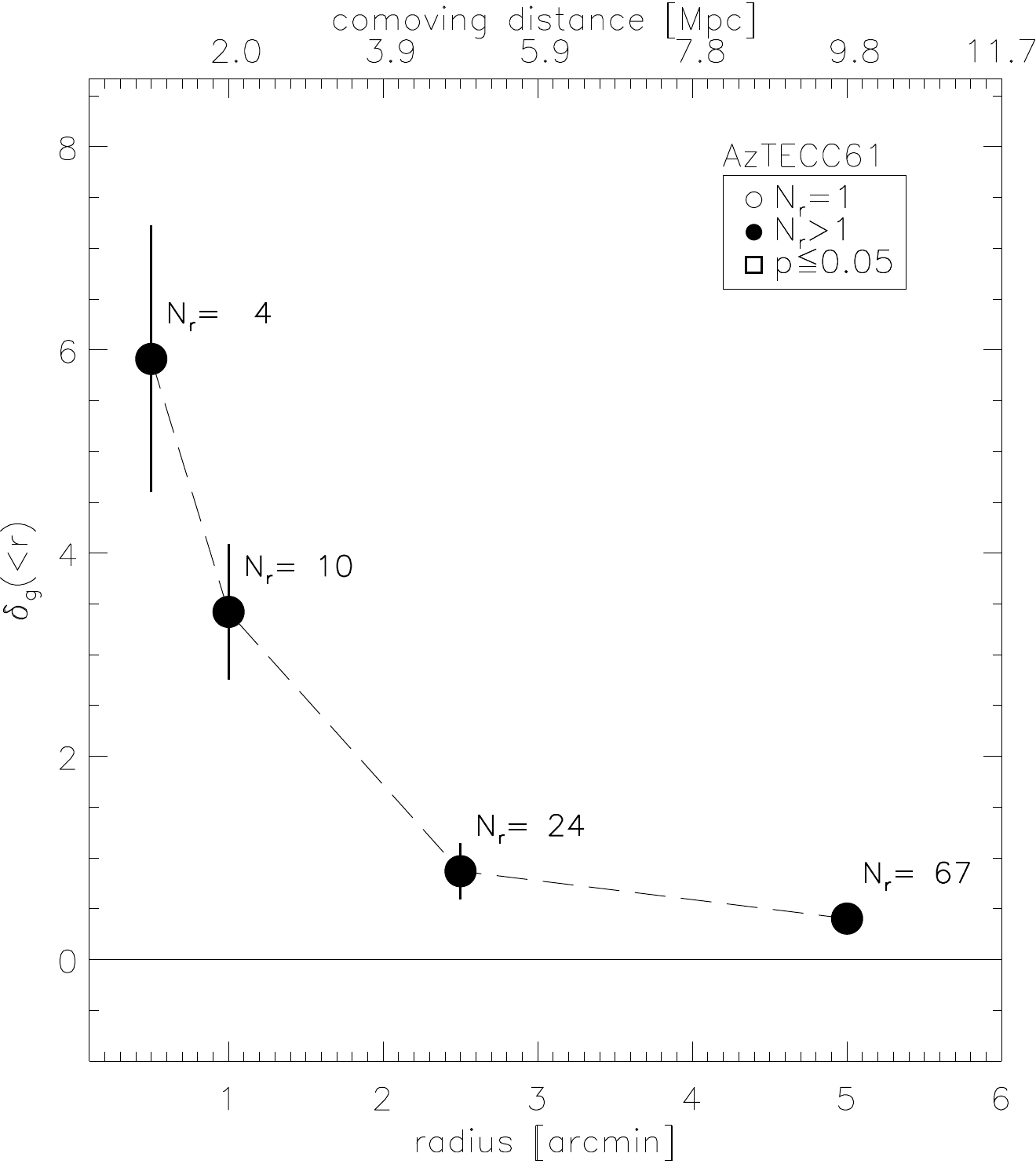}
\includegraphics[width=0.23\textwidth]{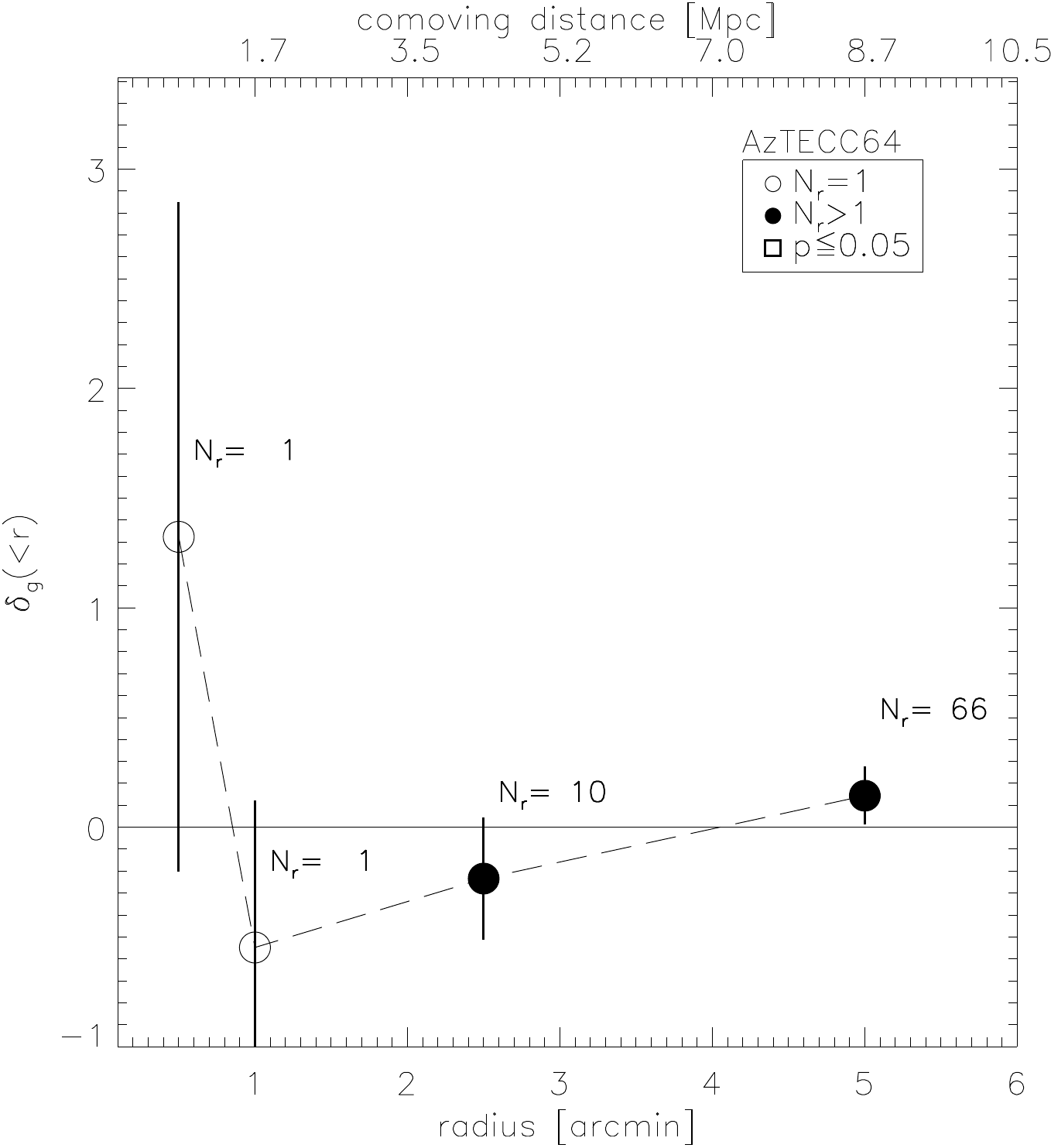}
\includegraphics[width=0.23\textwidth]{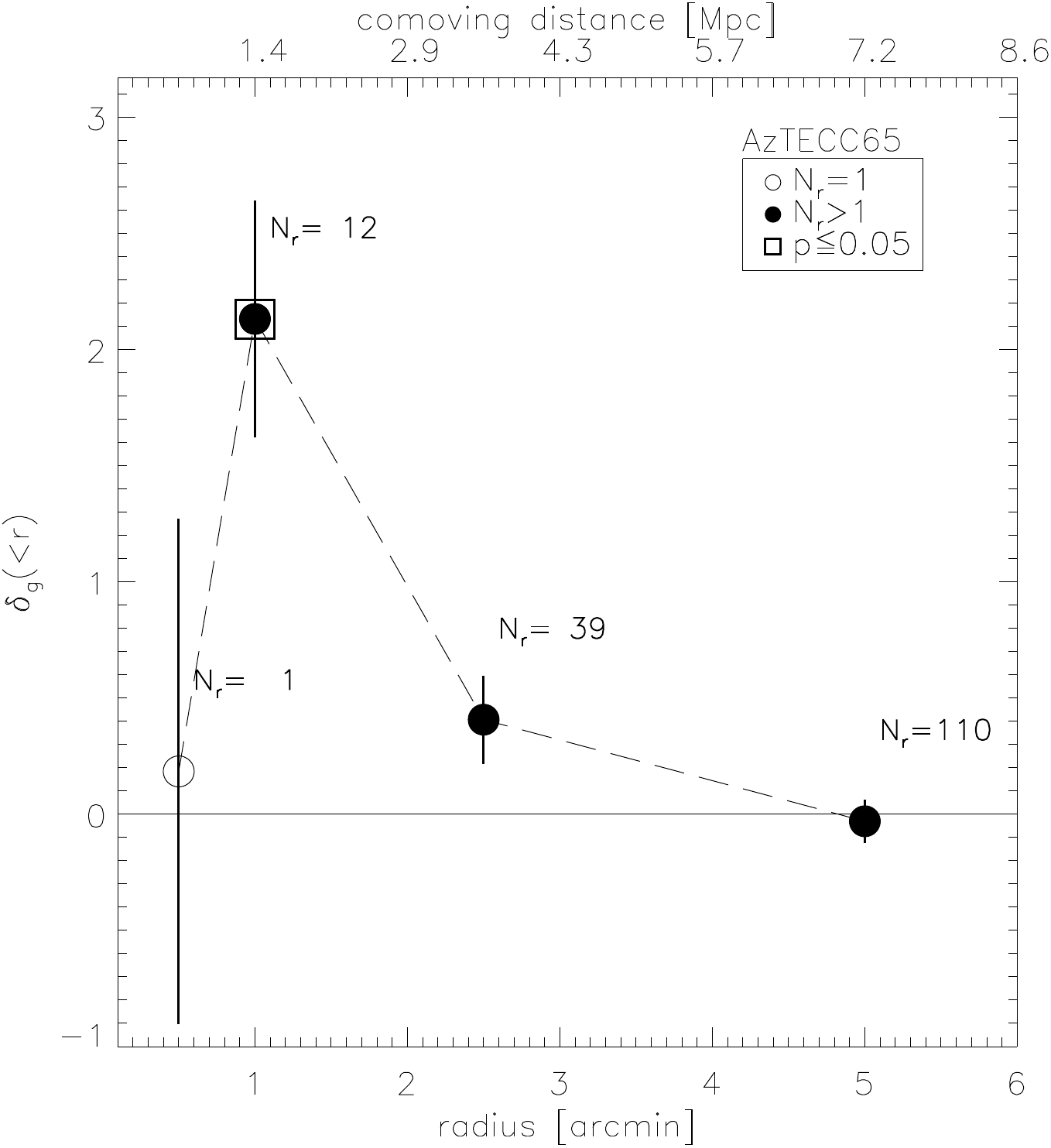}
\includegraphics[width=0.23\textwidth]{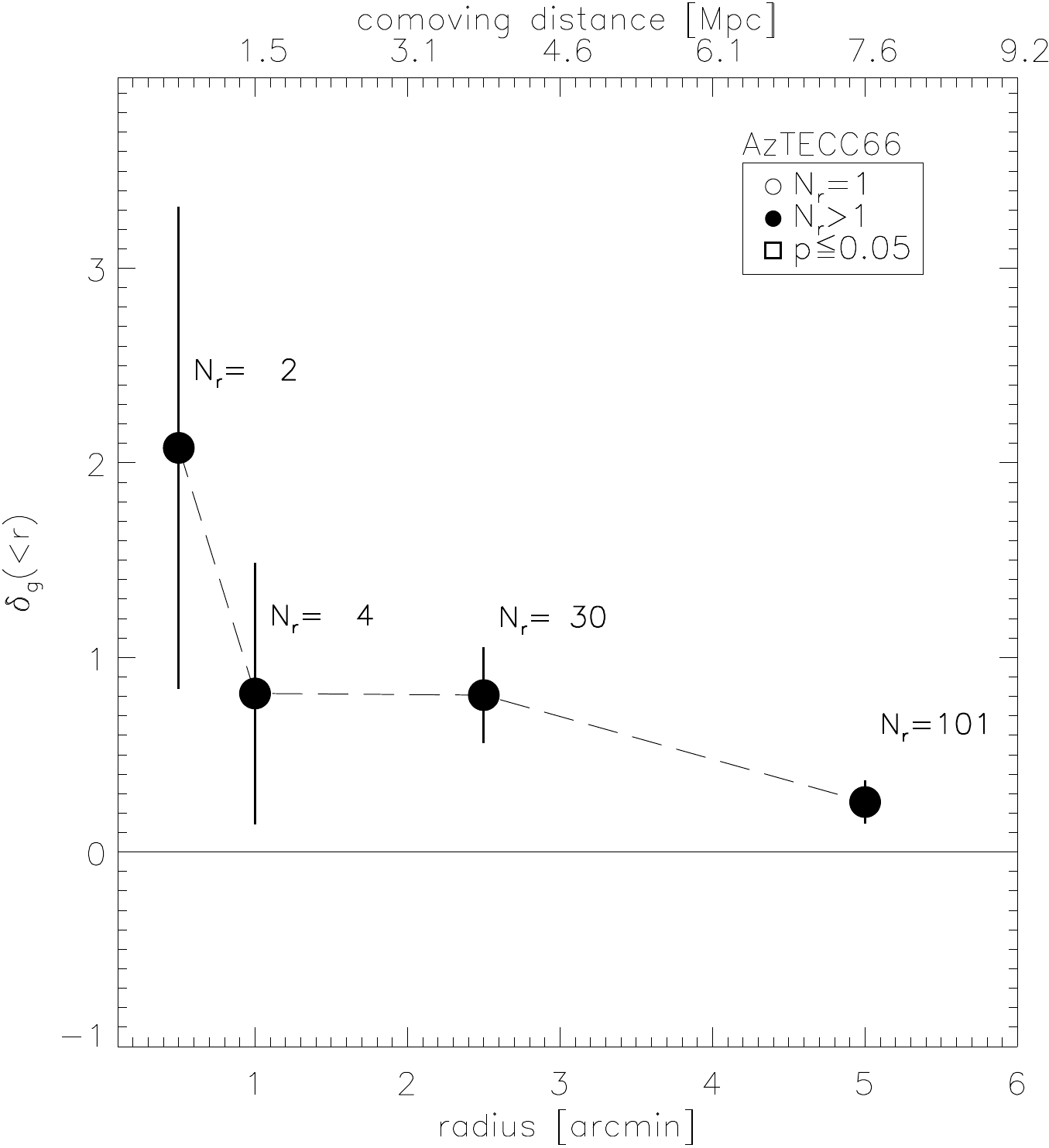}
\includegraphics[width=0.23\textwidth]{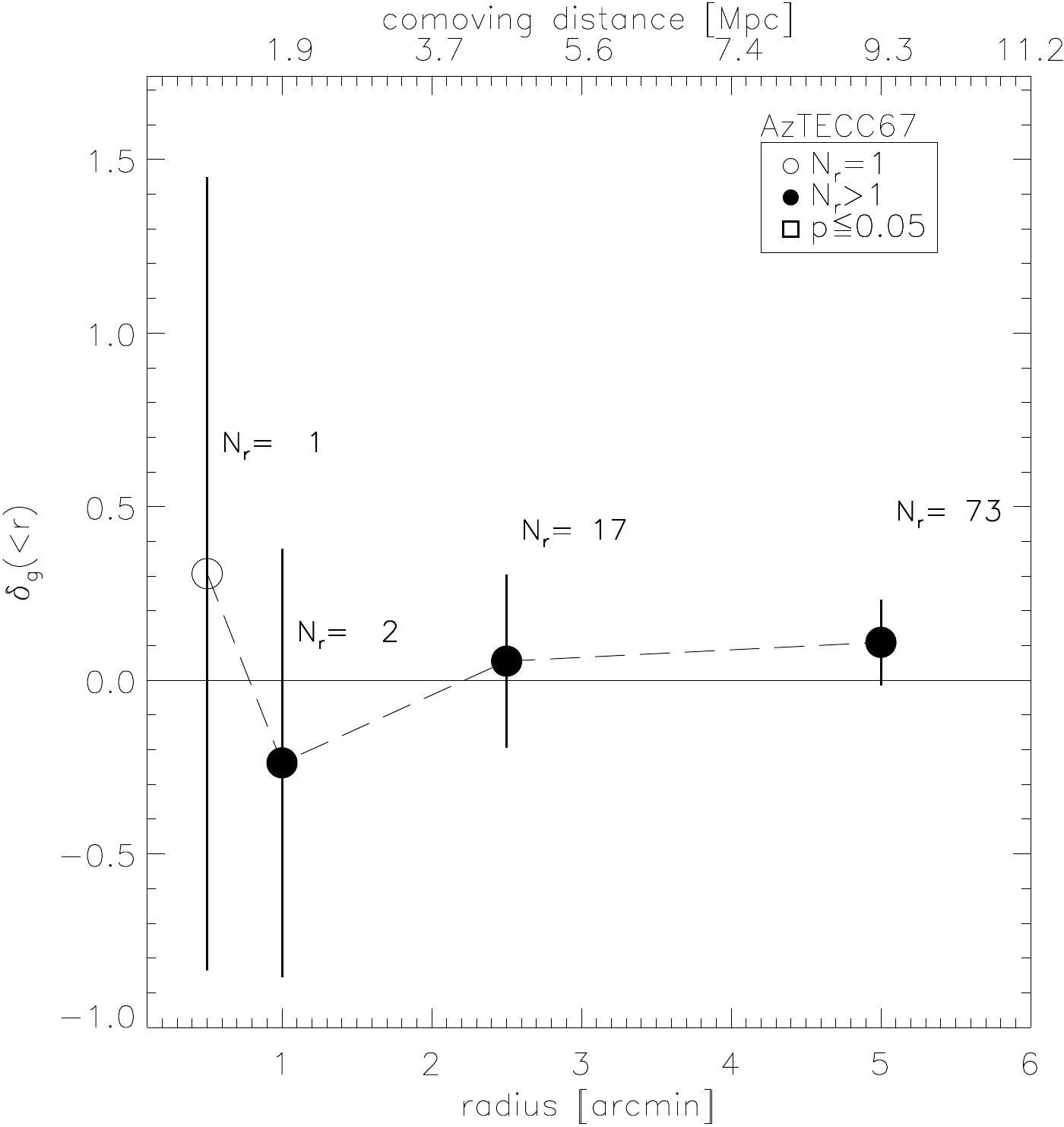}
\includegraphics[width=0.23\textwidth]{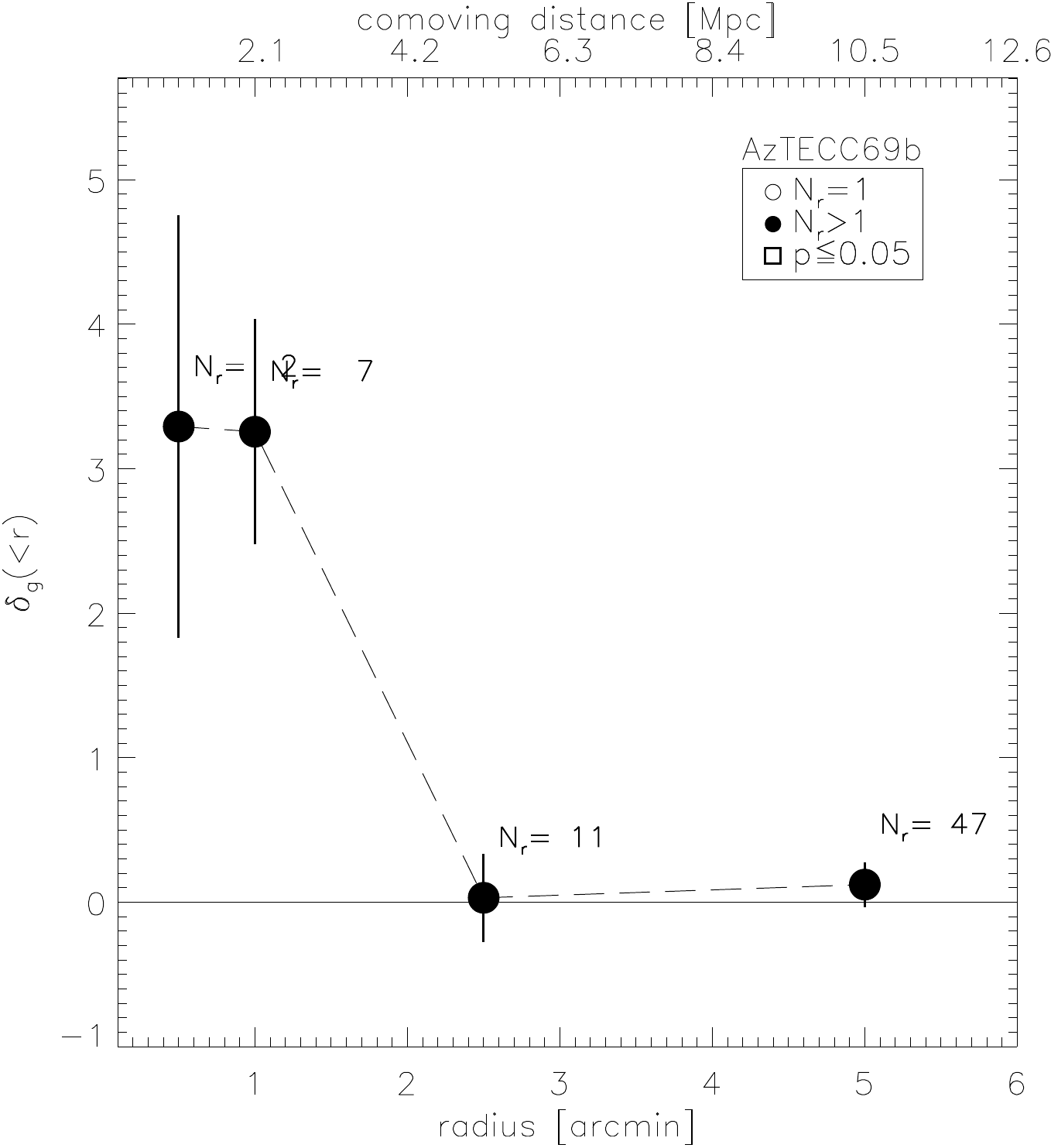}
\includegraphics[width=0.23\textwidth]{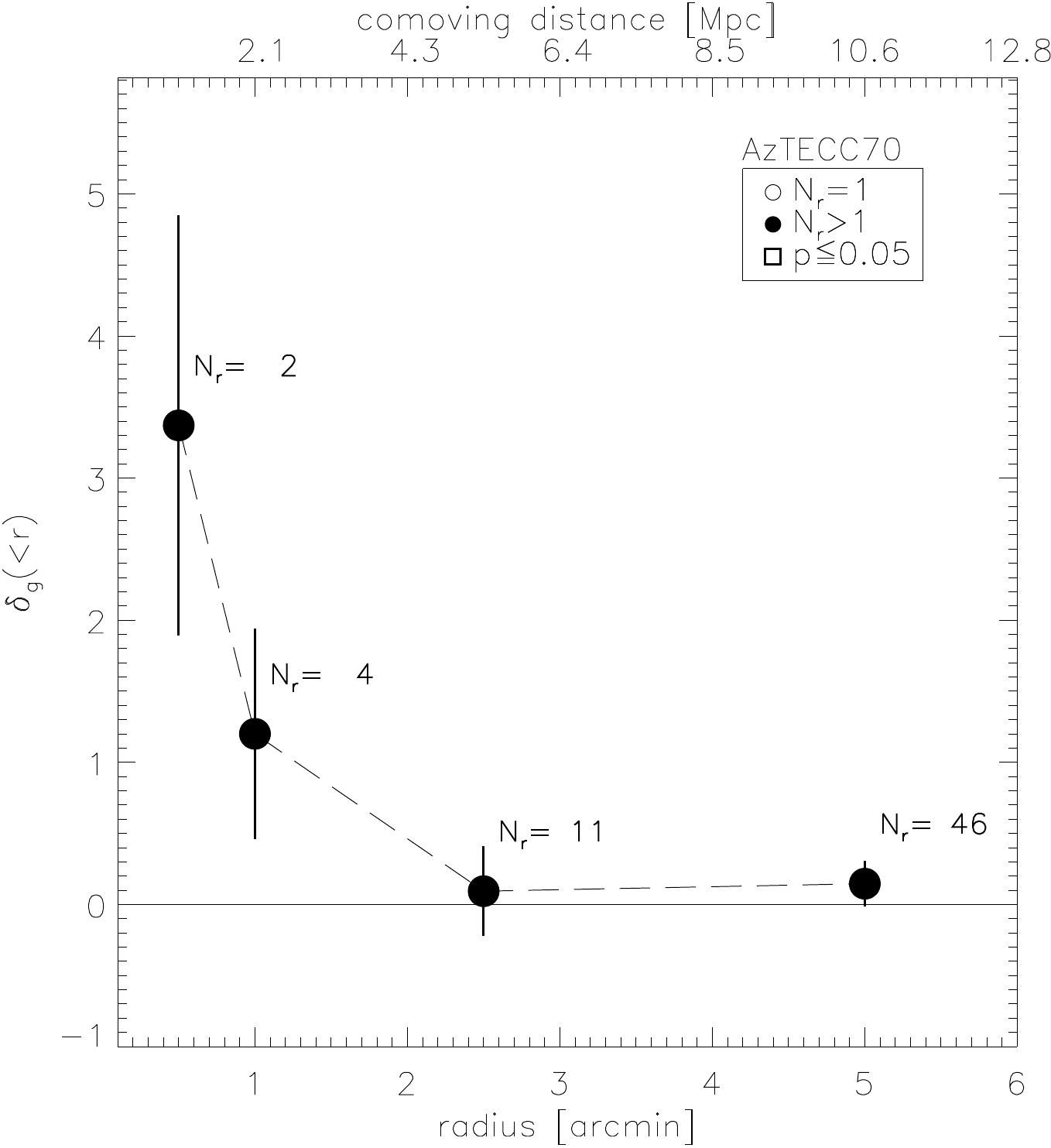}
\includegraphics[width=0.23\textwidth]{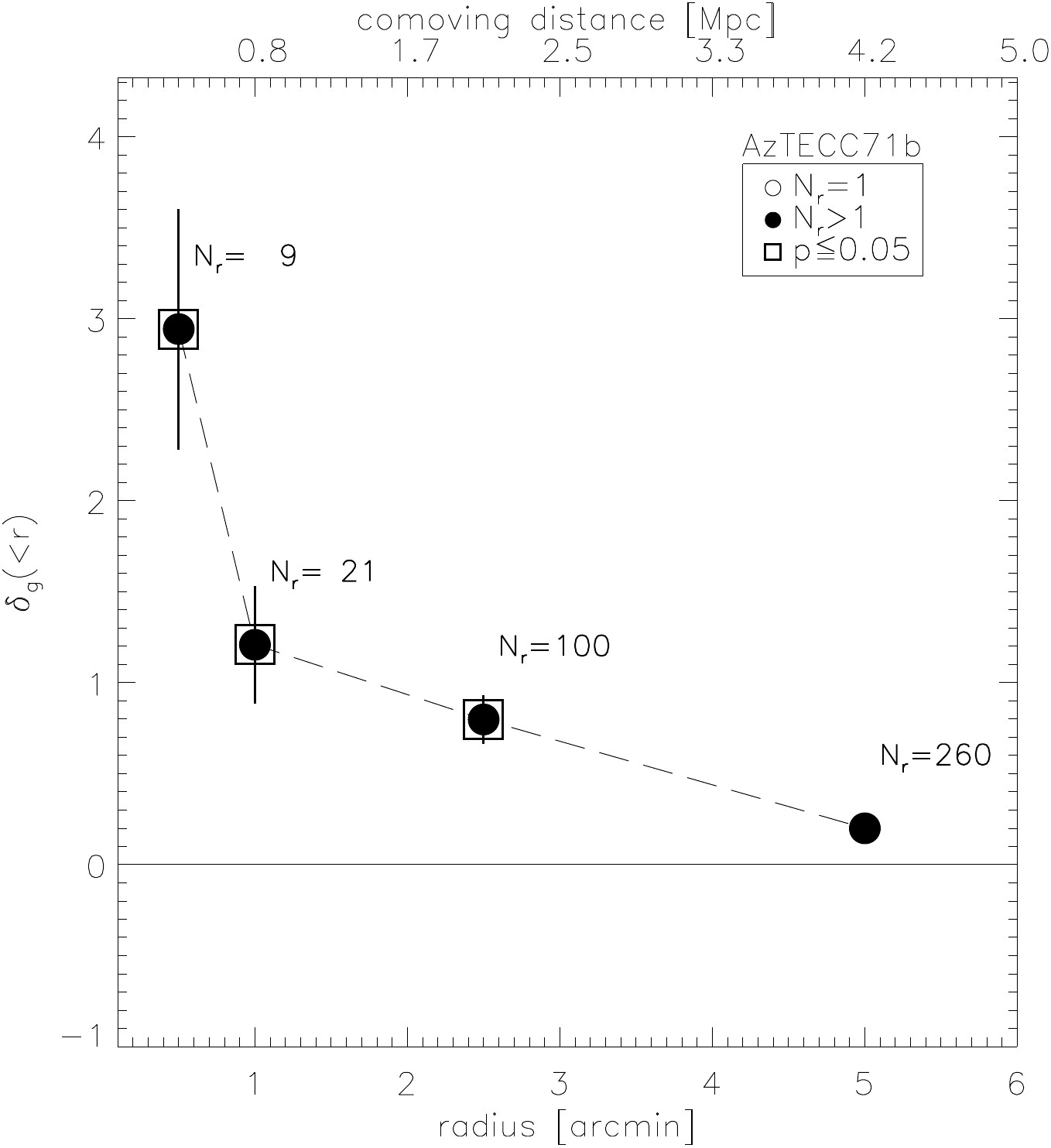}
\includegraphics[width=0.23\textwidth]{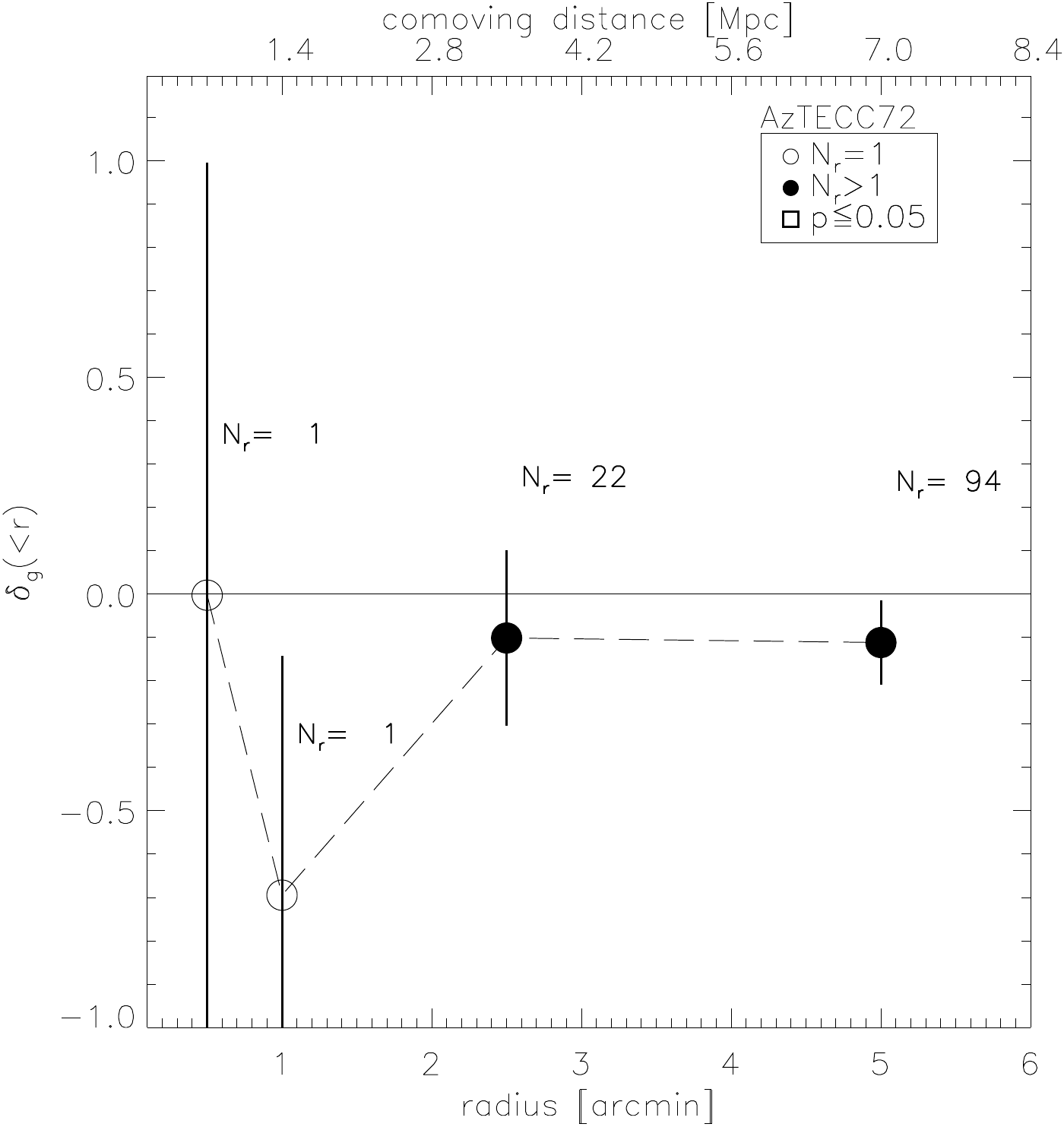}
\includegraphics[width=0.23\textwidth]{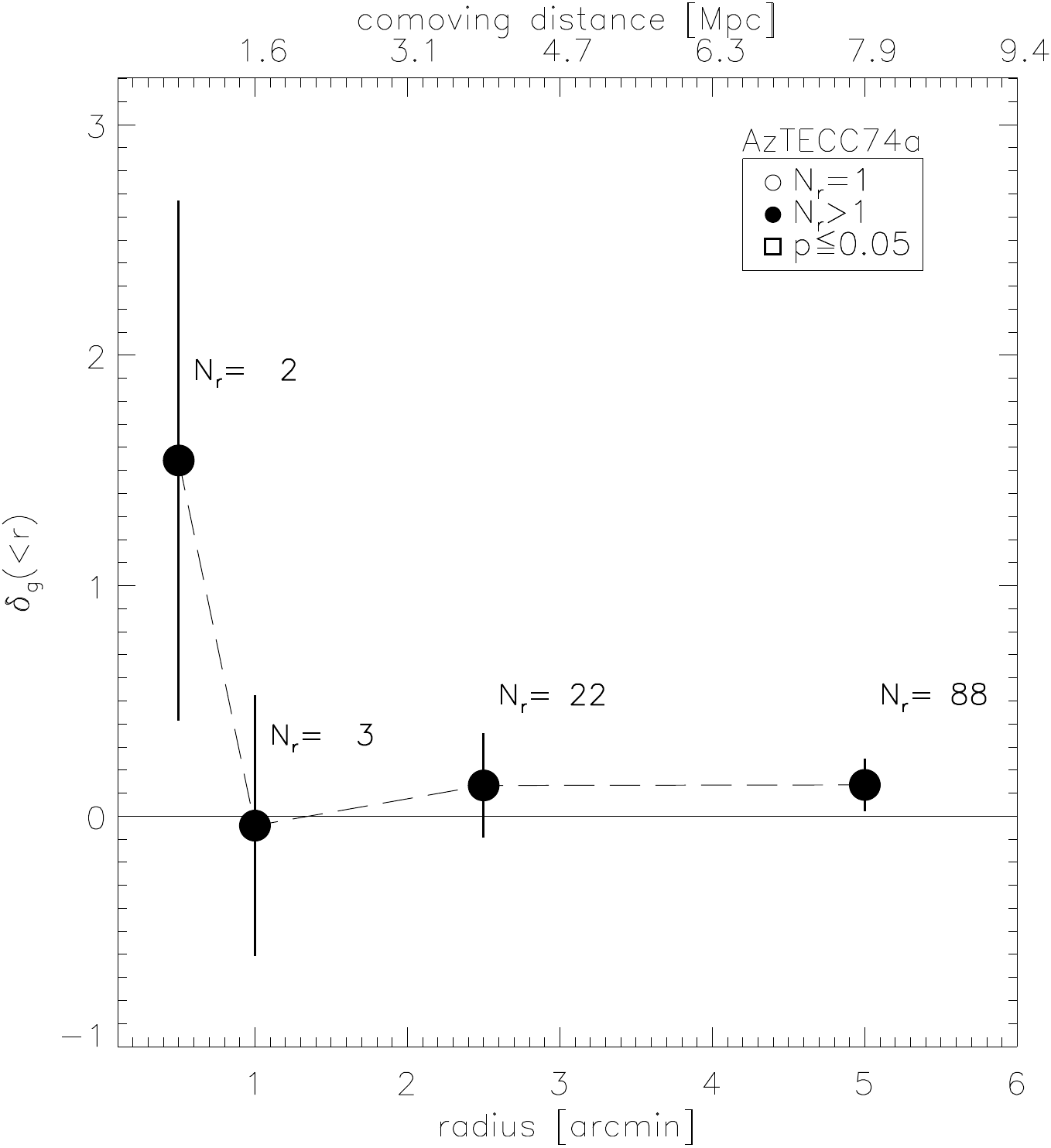}
\includegraphics[width=0.23\textwidth]{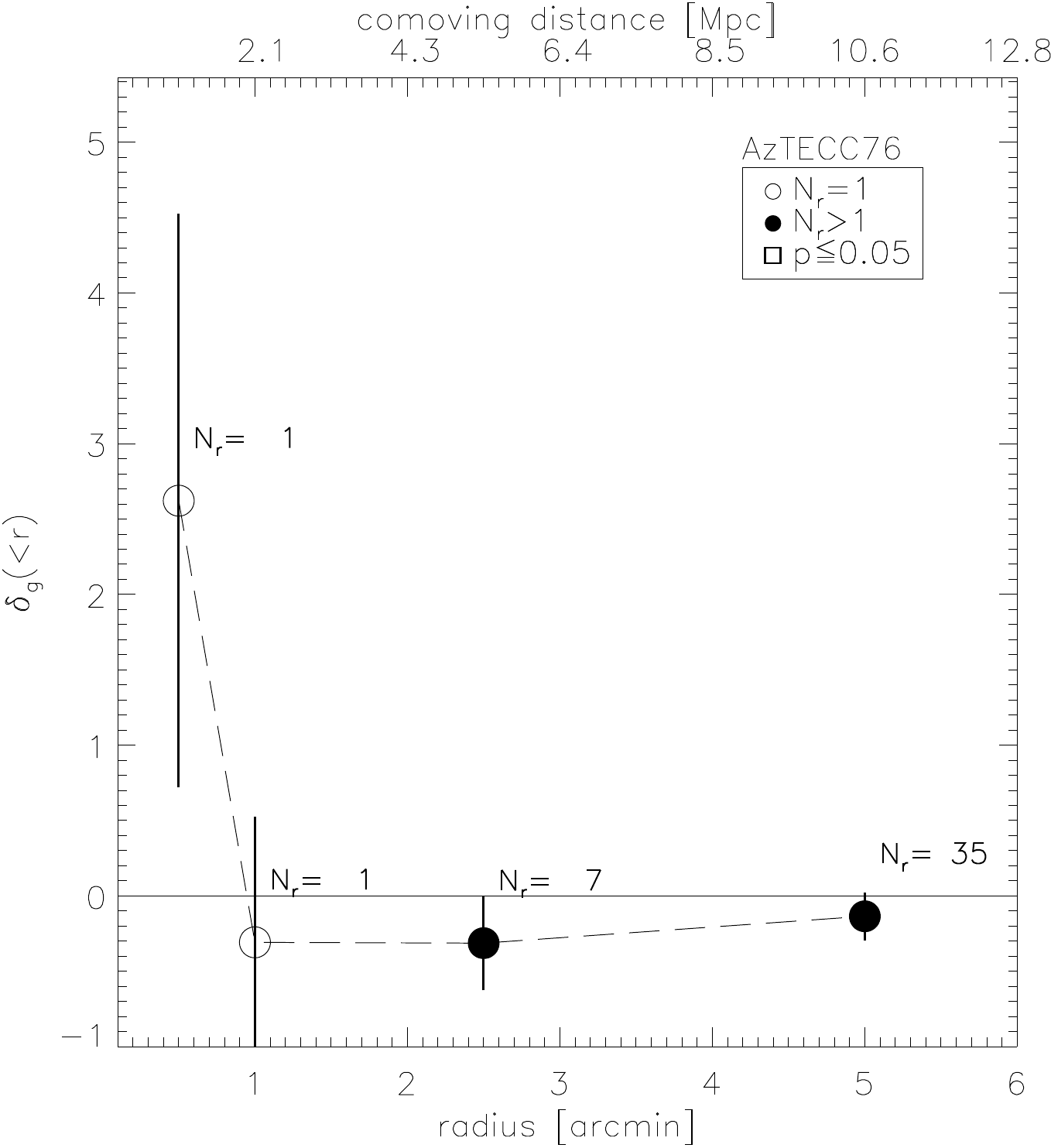}
\includegraphics[width=0.23\textwidth]{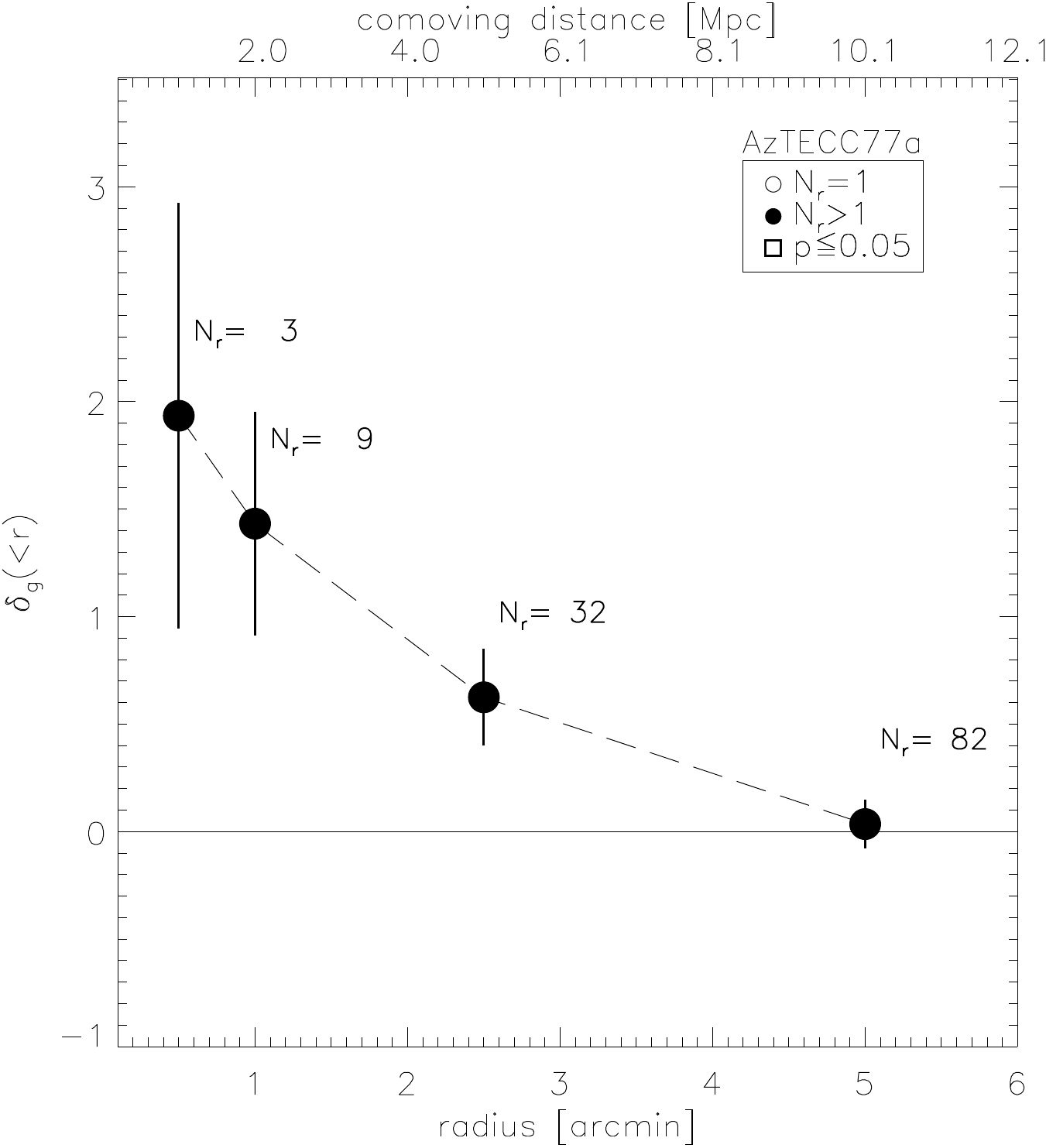}

\caption{continued.}
\end{center}
\end{figure*}

\addtocounter{figure}{-1}
\begin{figure*}
\begin{center}
\includegraphics[width=0.23\textwidth]{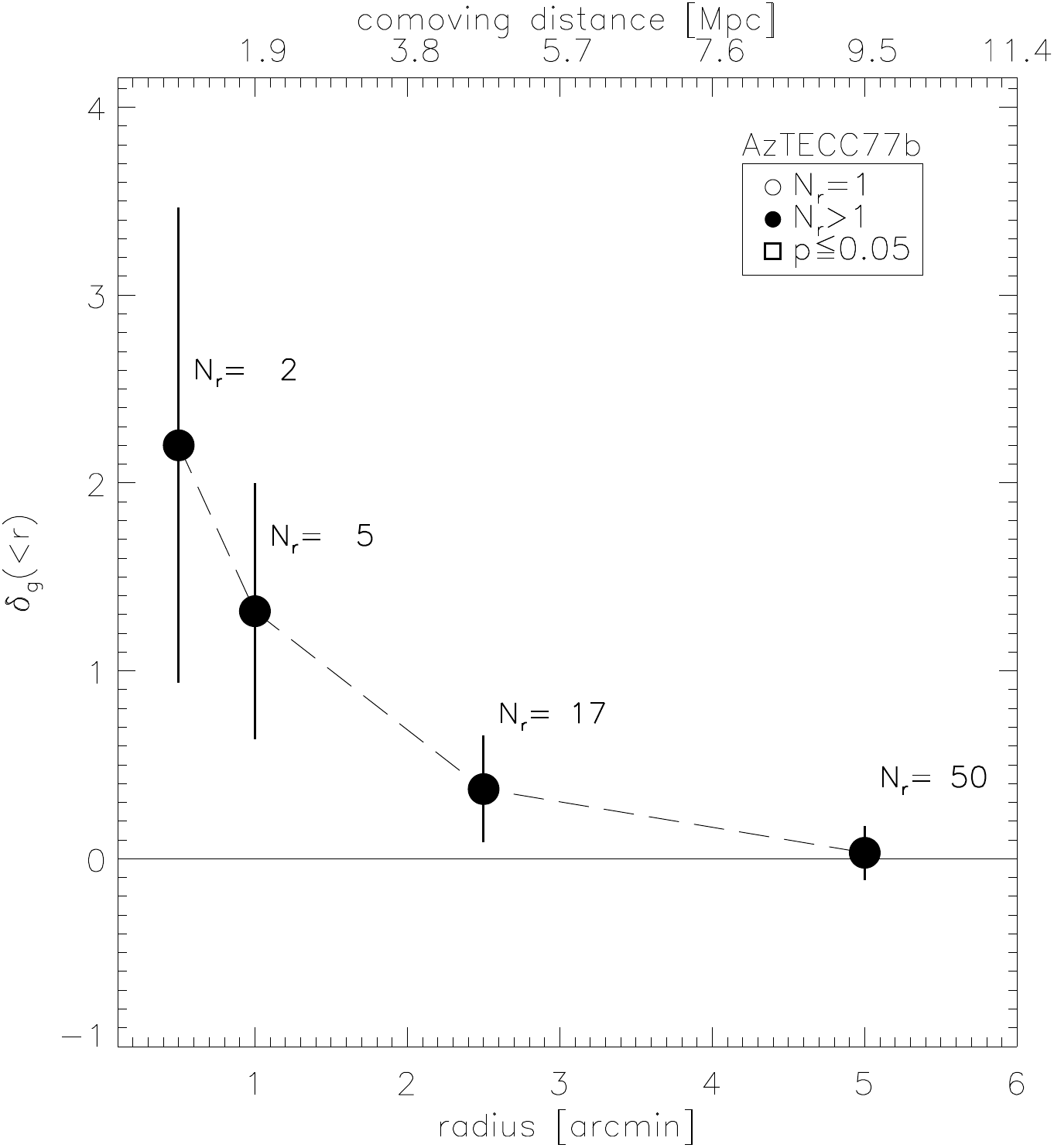}
\includegraphics[width=0.23\textwidth]{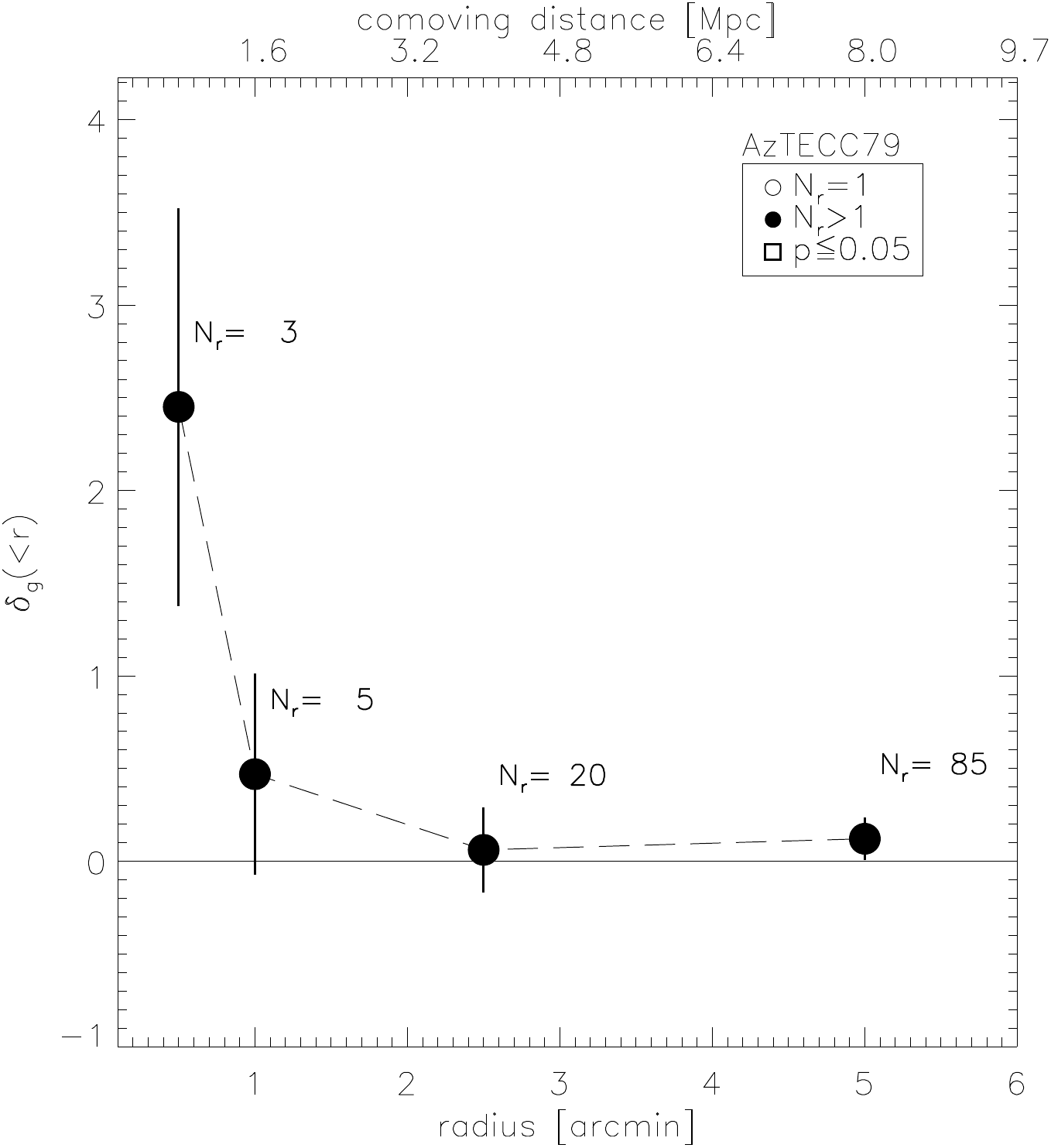}
\includegraphics[width=0.23\textwidth]{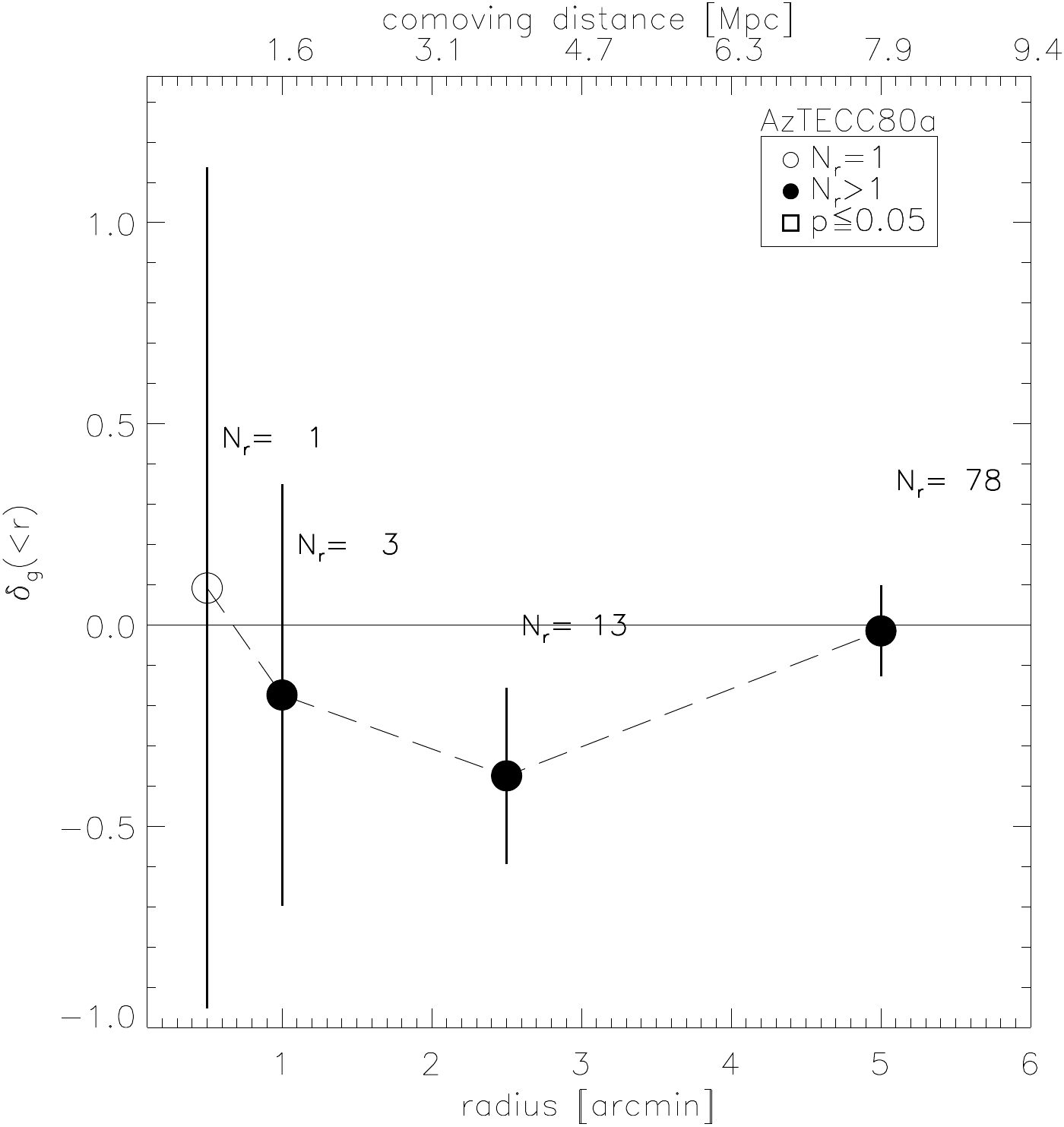}
\includegraphics[width=0.23\textwidth]{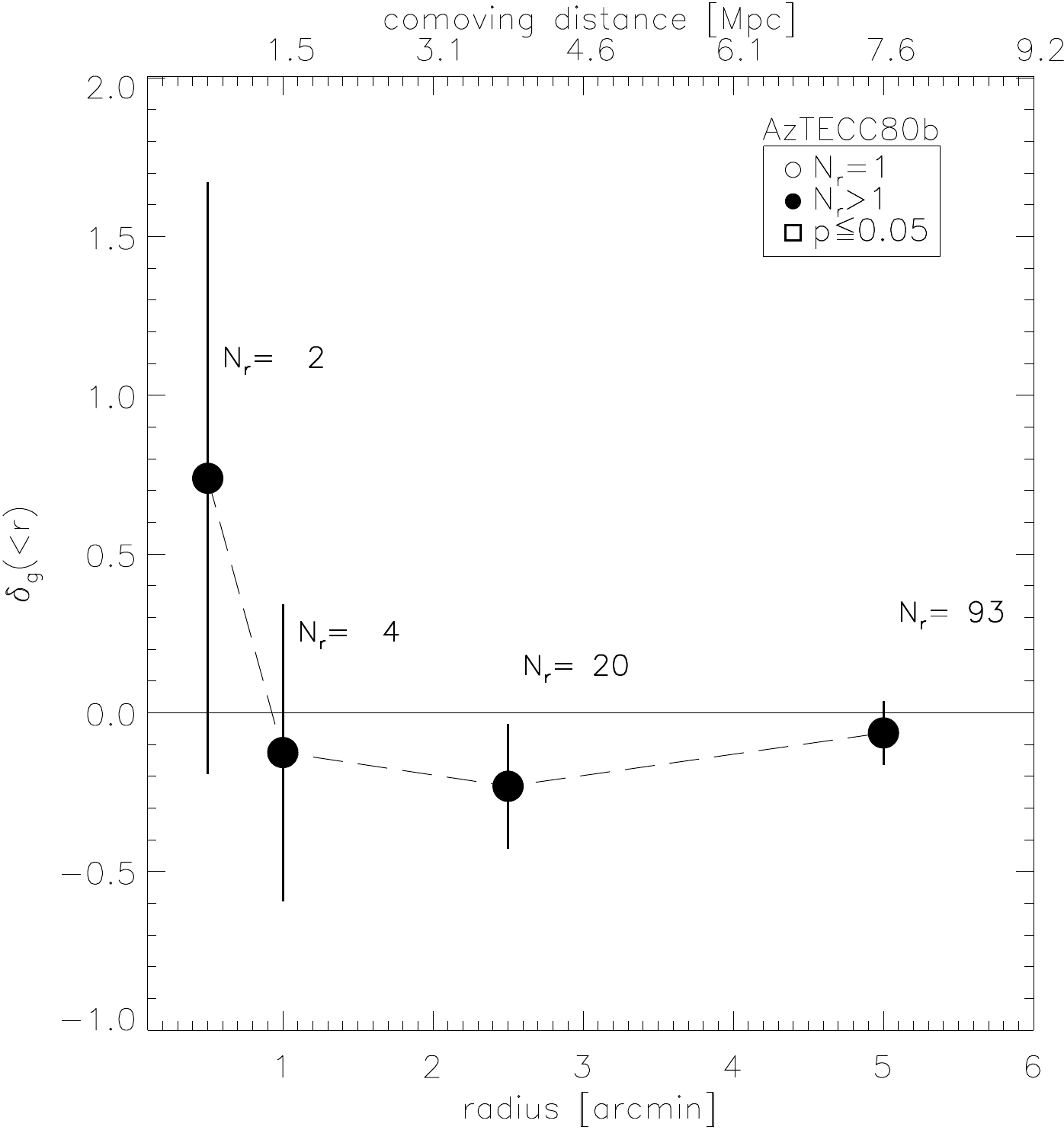}
\includegraphics[width=0.23\textwidth]{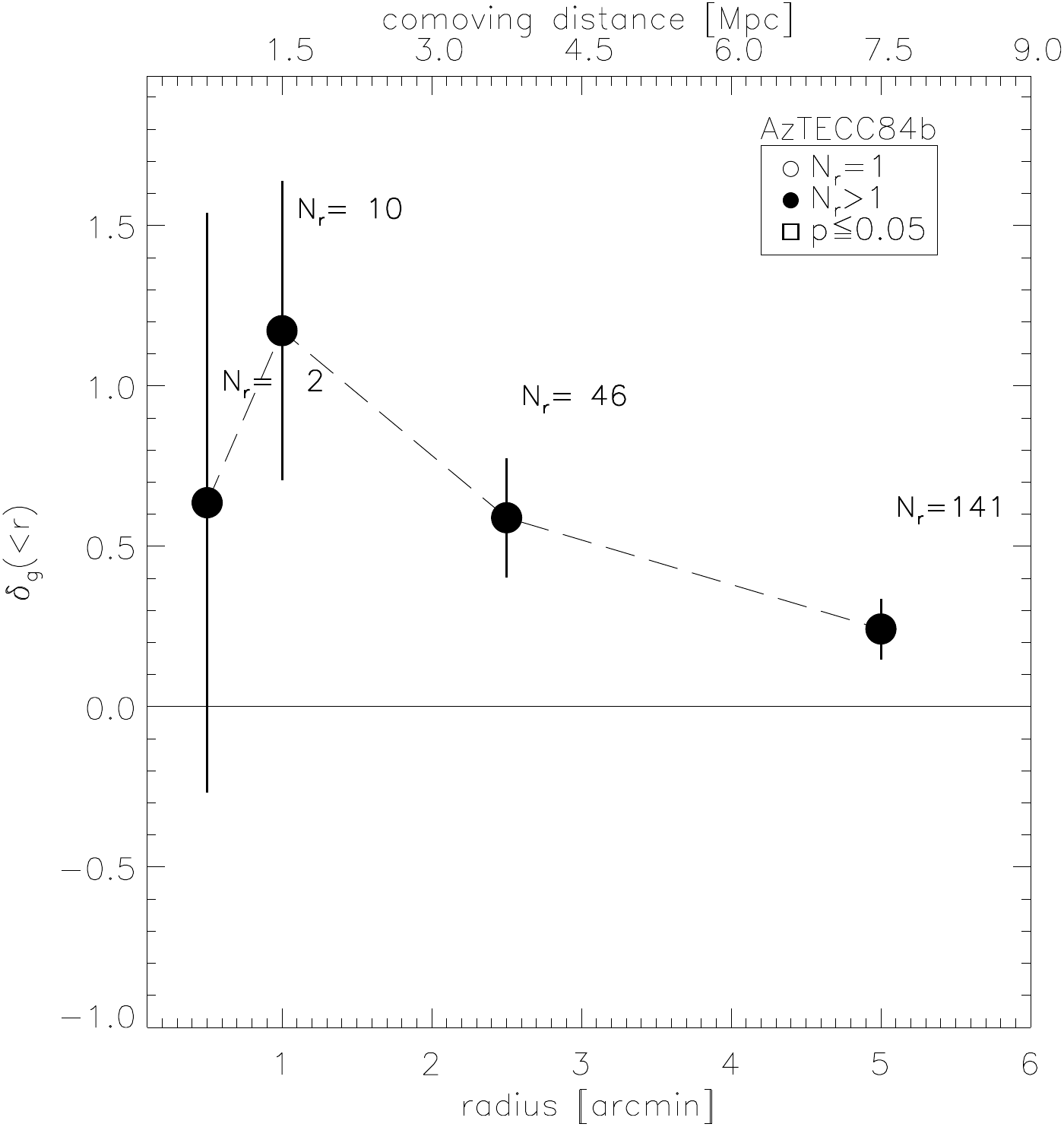}
\includegraphics[width=0.23\textwidth]{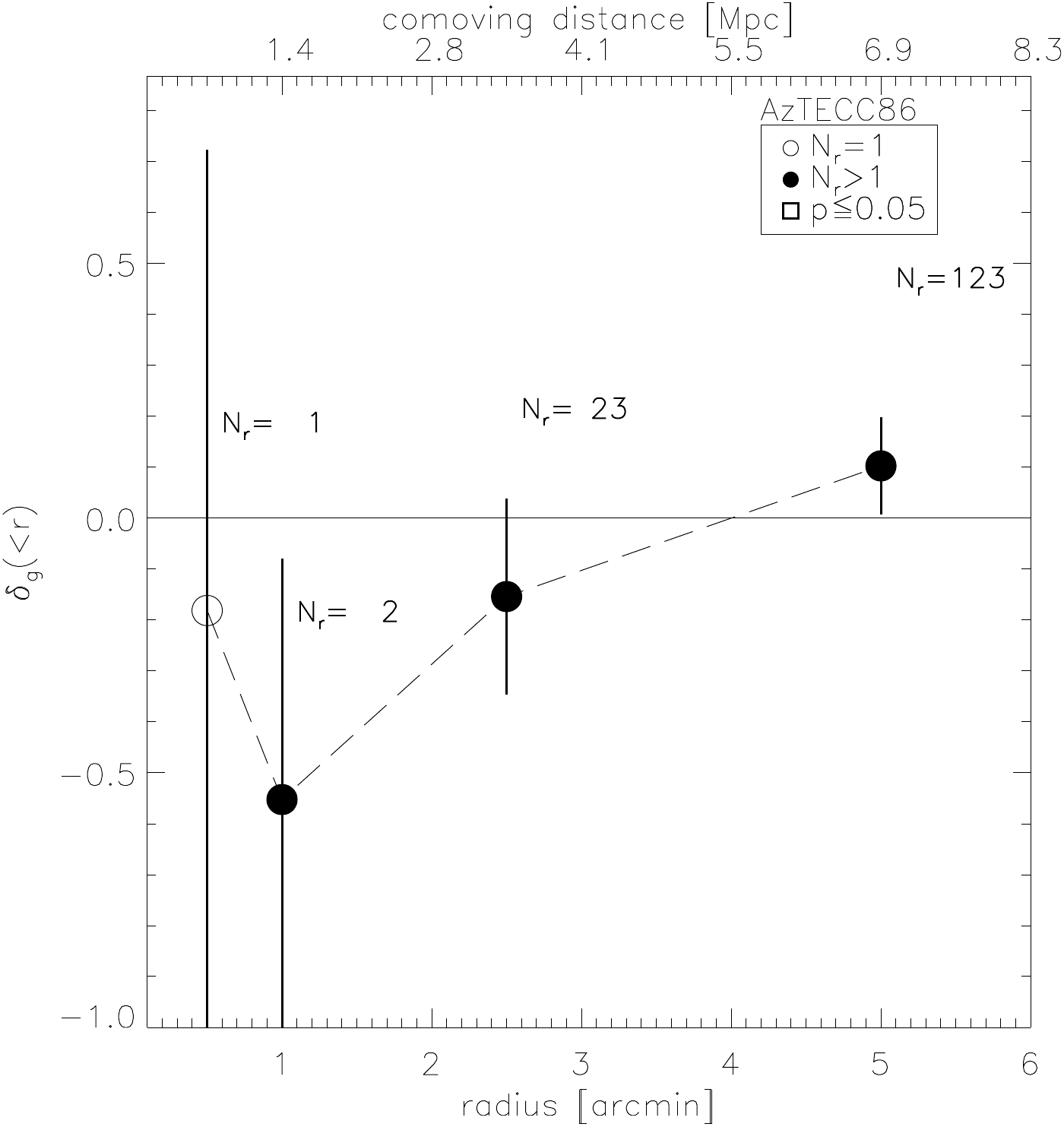}
\includegraphics[width=0.23\textwidth]{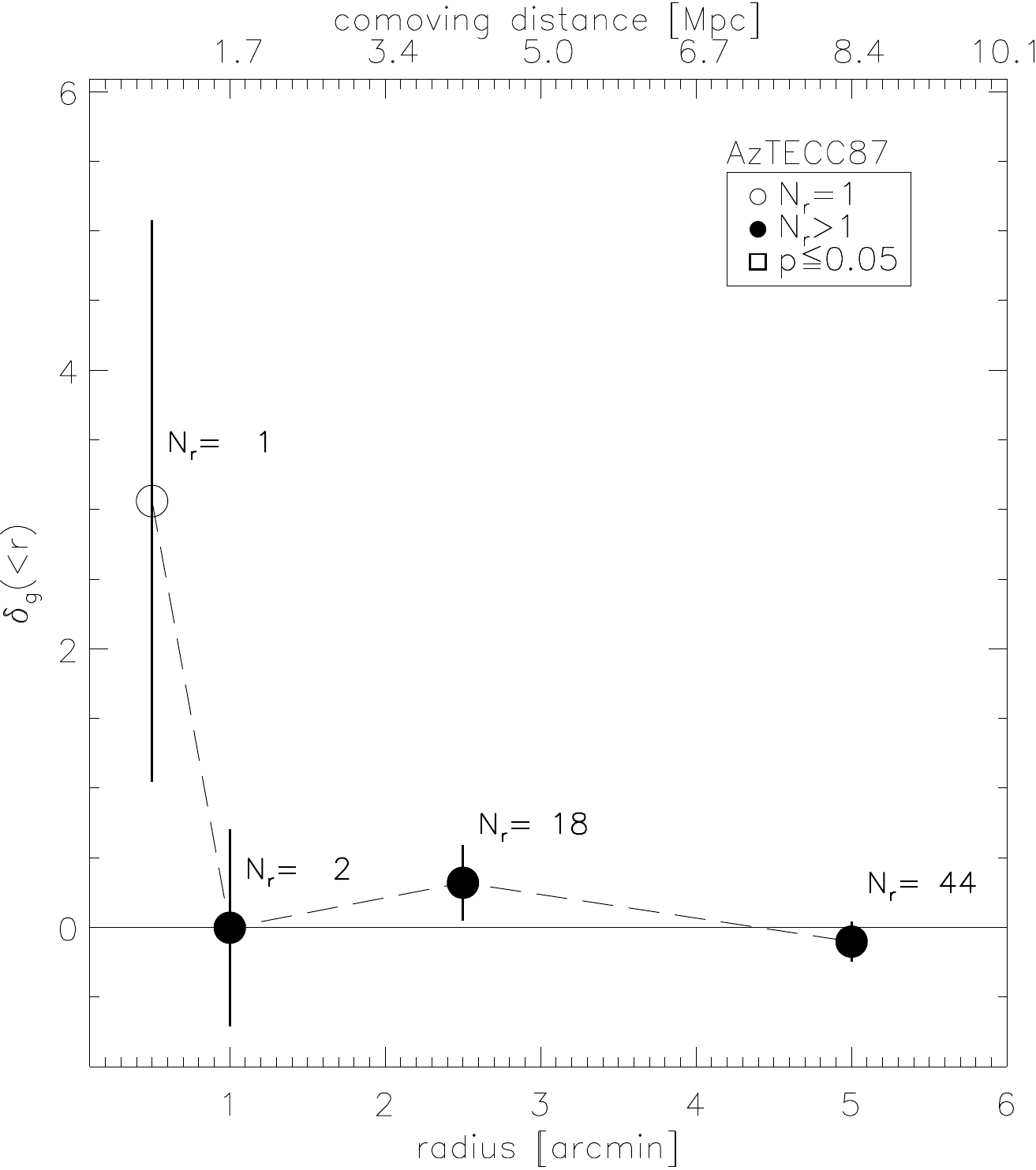}
\includegraphics[width=0.23\textwidth]{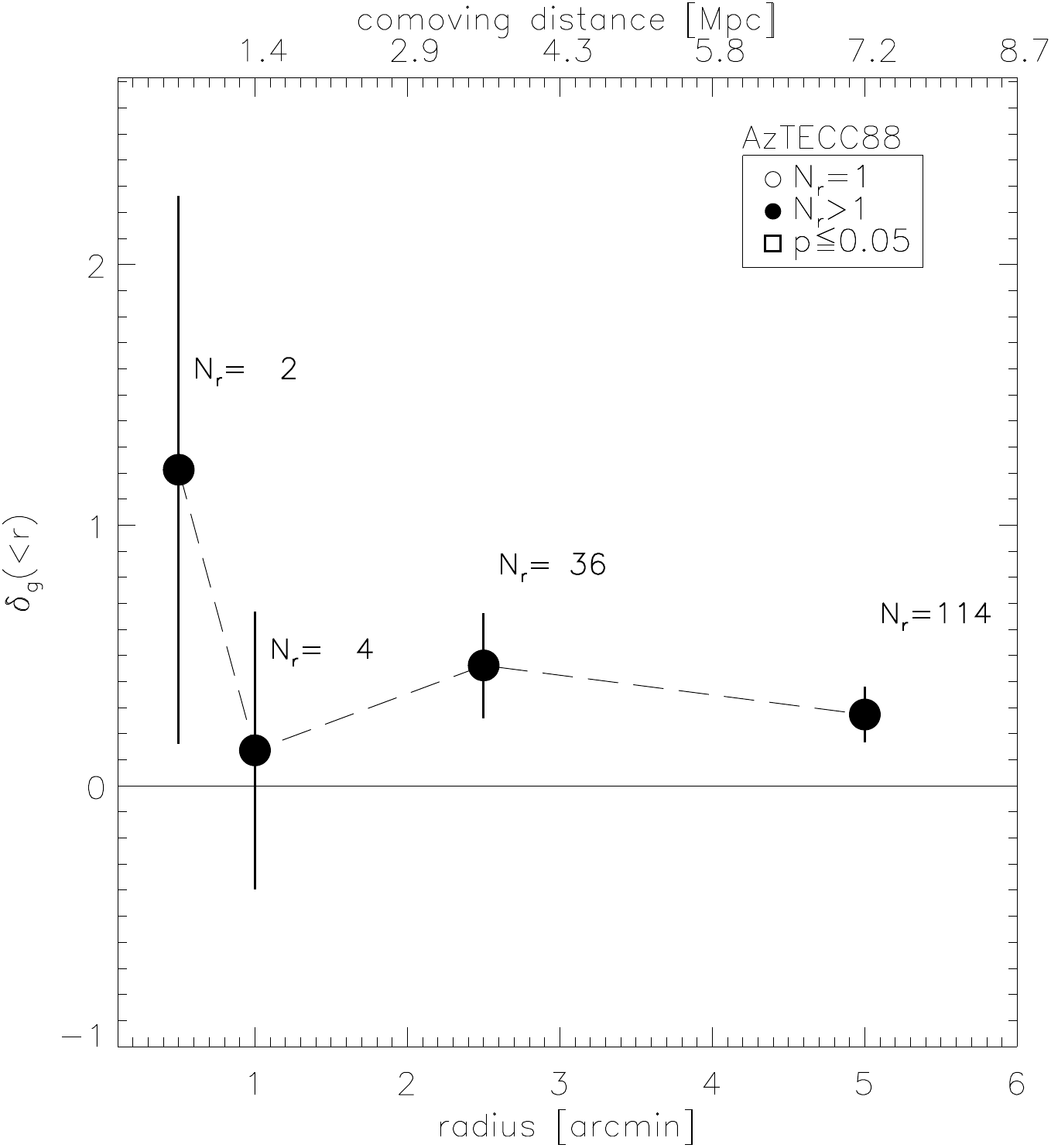}
\includegraphics[width=0.23\textwidth]{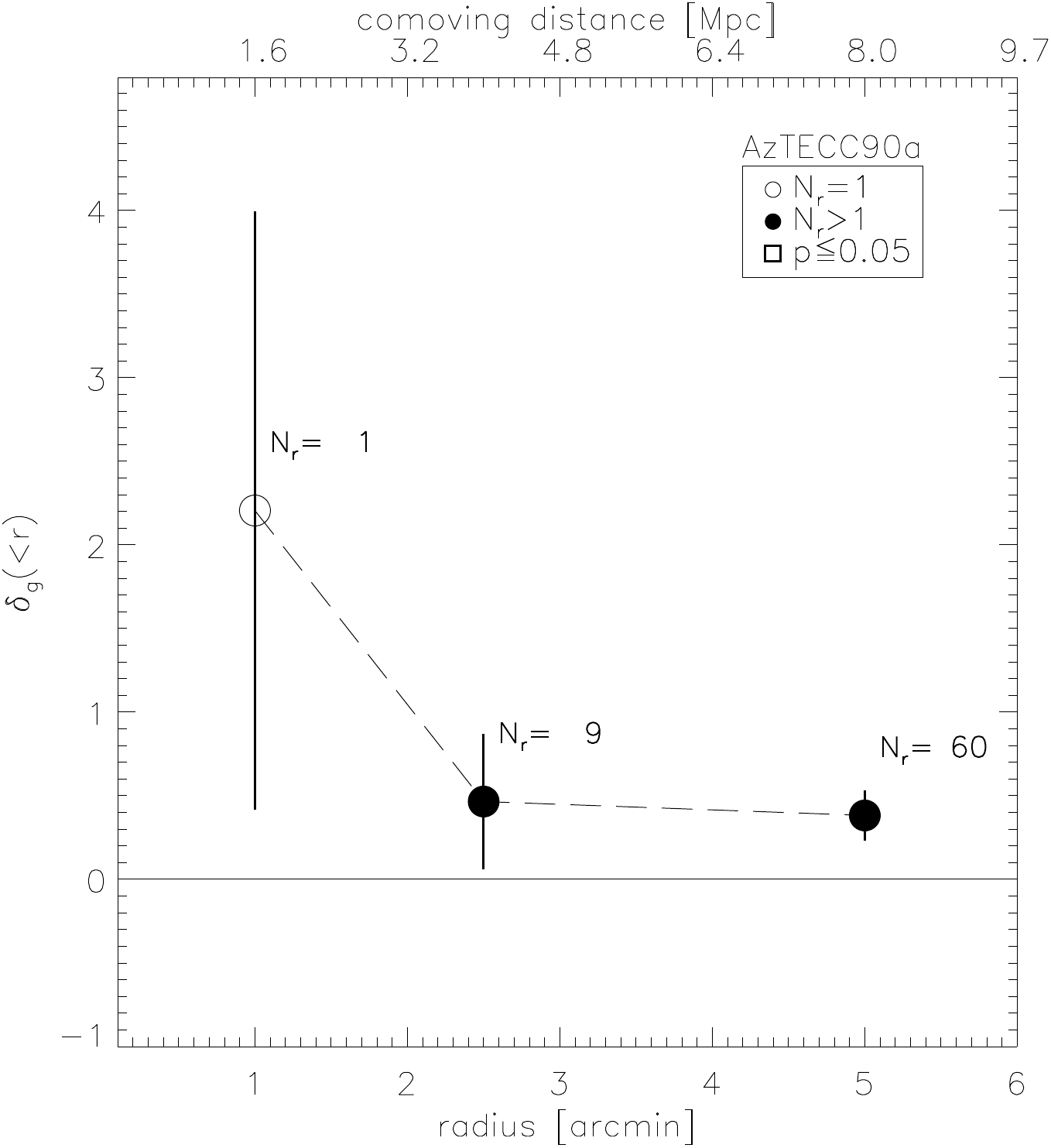}
\includegraphics[width=0.23\textwidth]{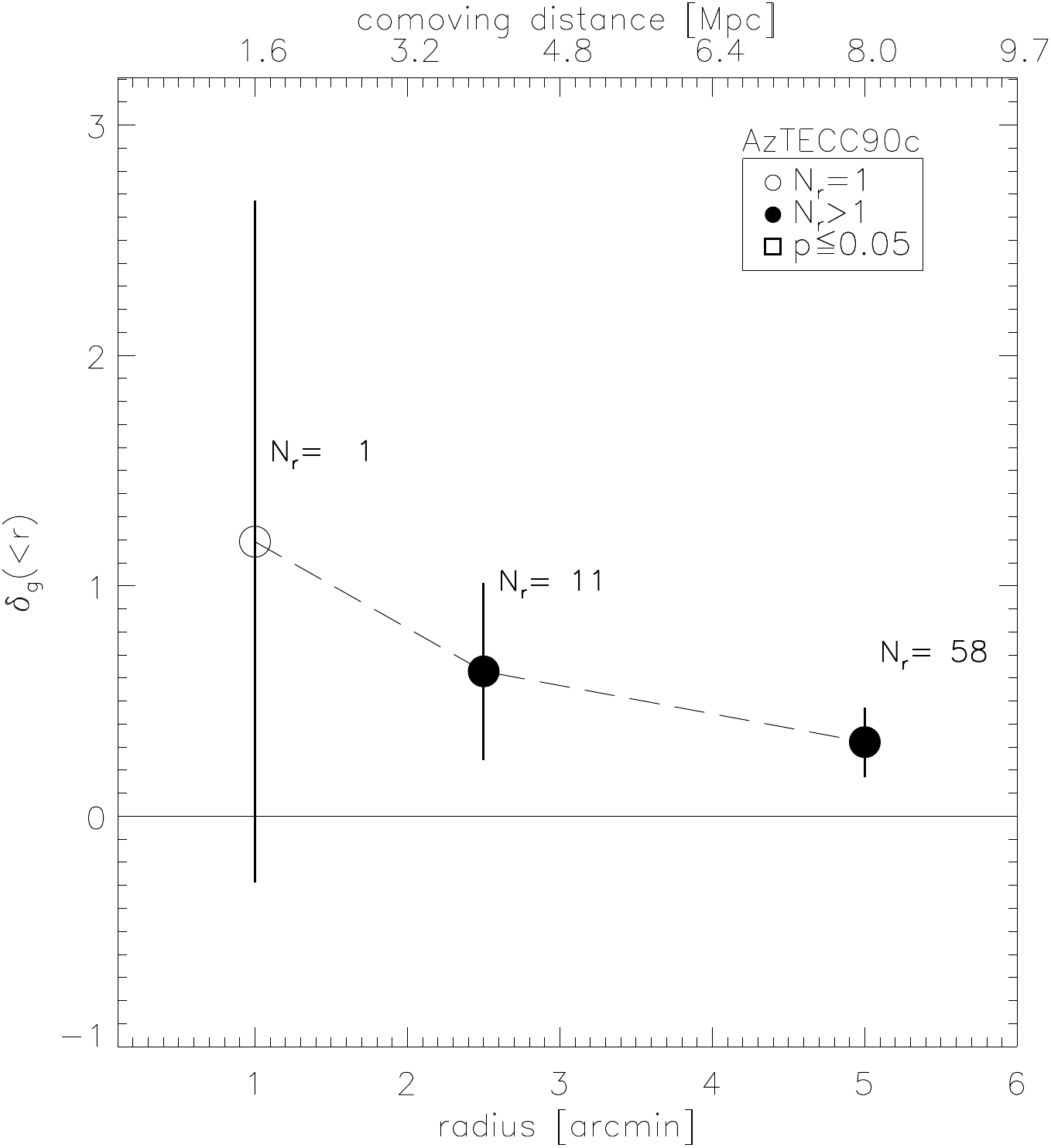}
\includegraphics[width=0.23\textwidth]{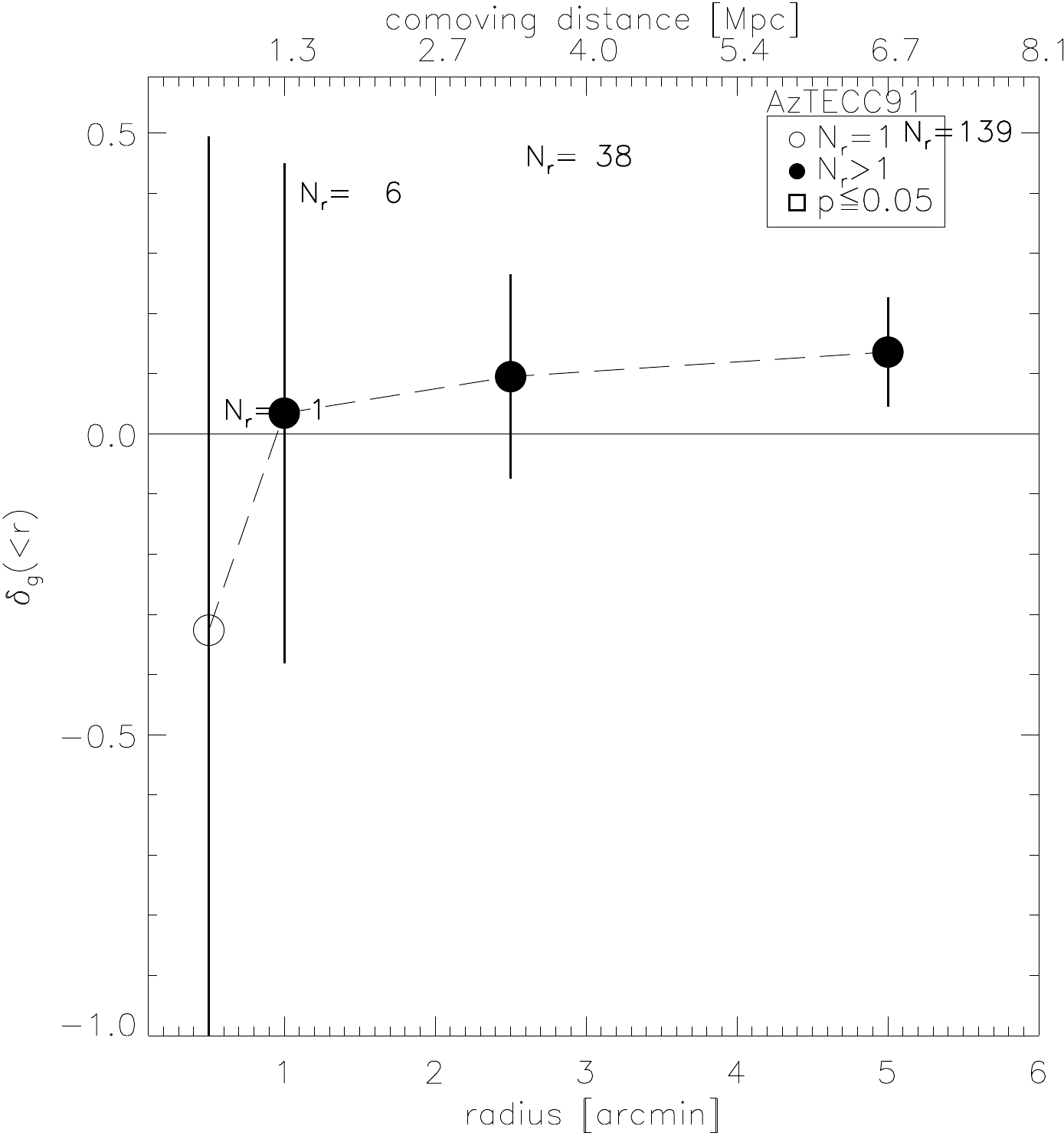}
\includegraphics[width=0.23\textwidth]{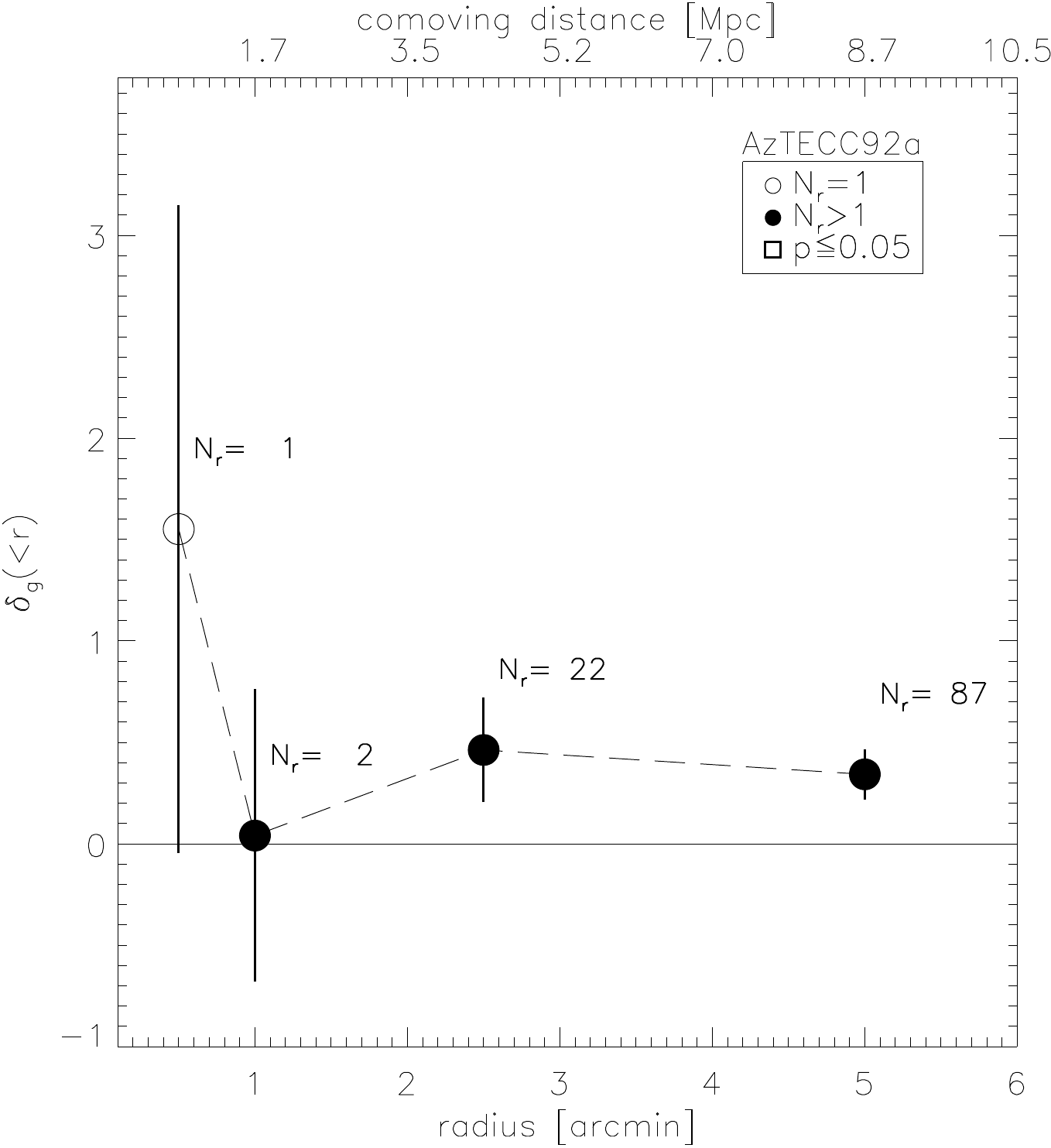}
\includegraphics[width=0.23\textwidth]{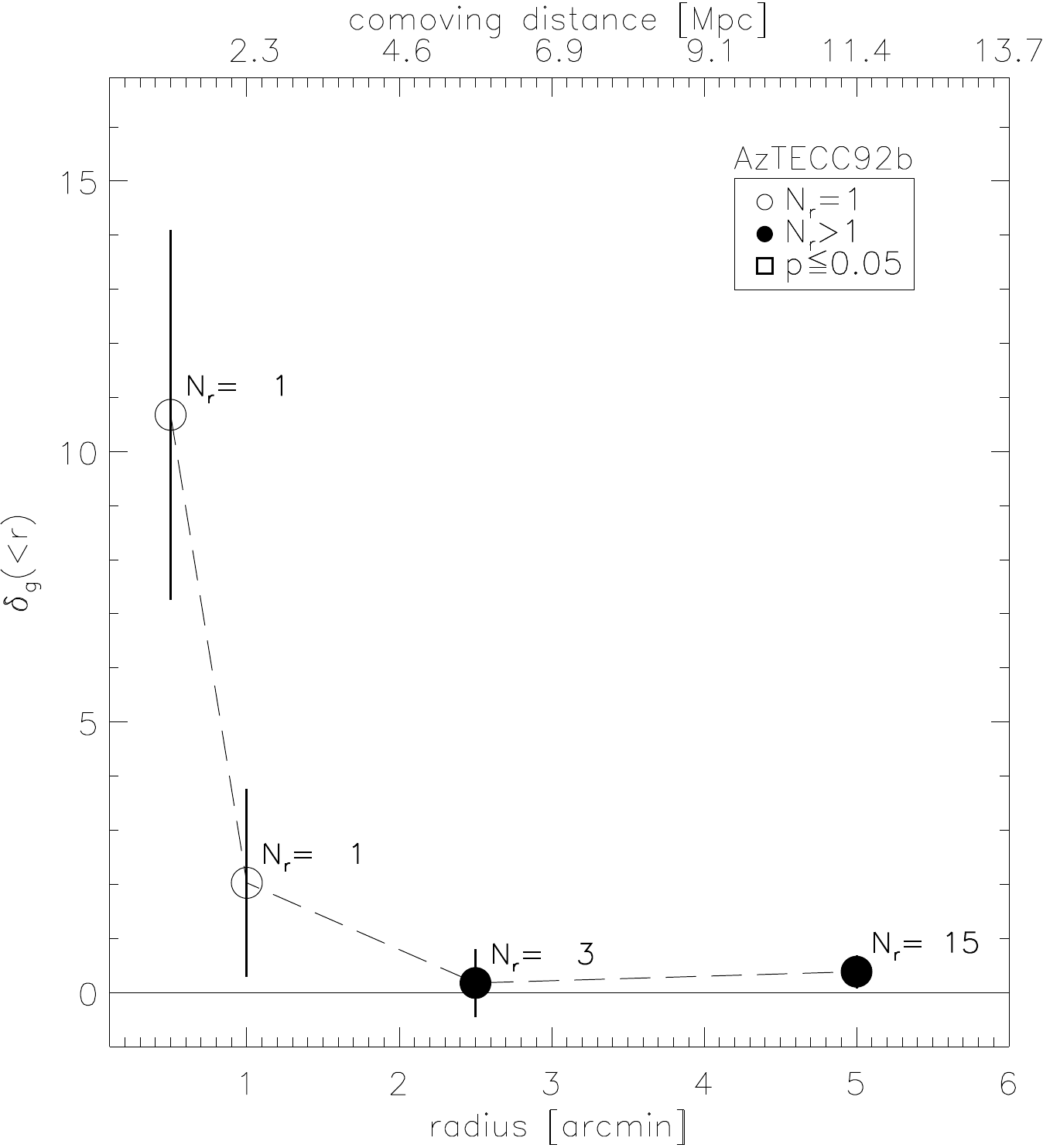}
\includegraphics[width=0.23\textwidth]{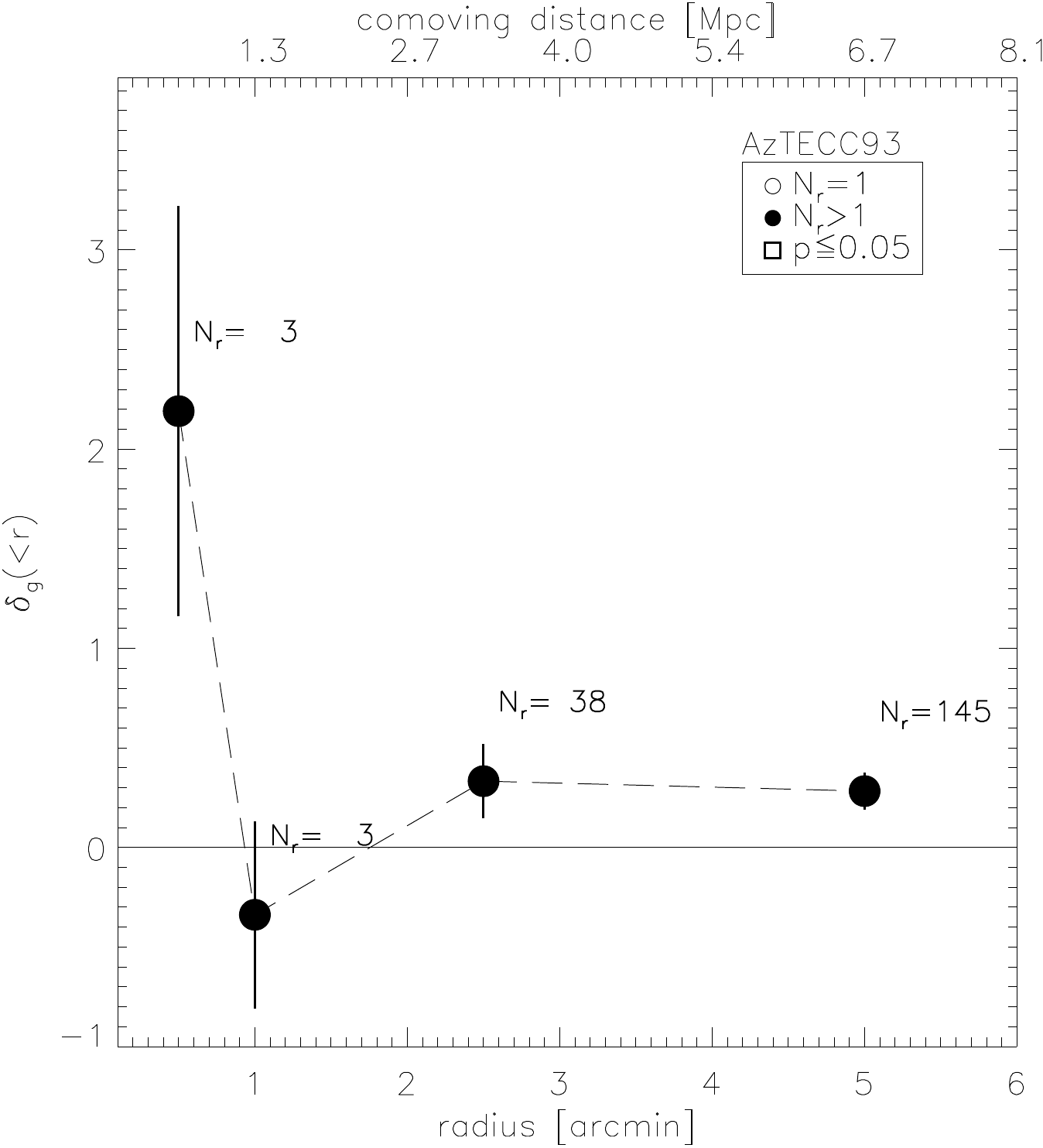}
\includegraphics[width=0.23\textwidth]{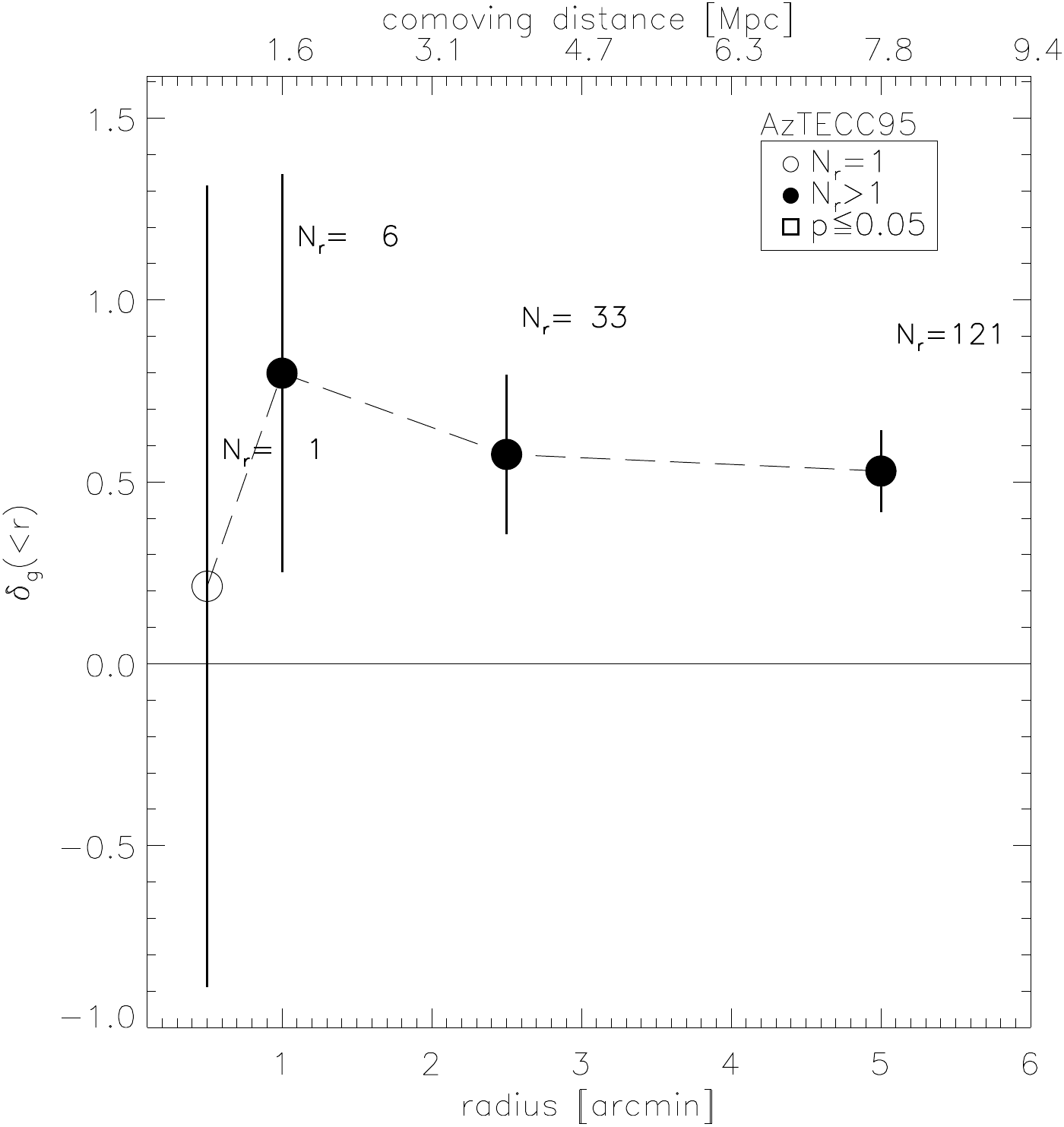}
\includegraphics[width=0.23\textwidth]{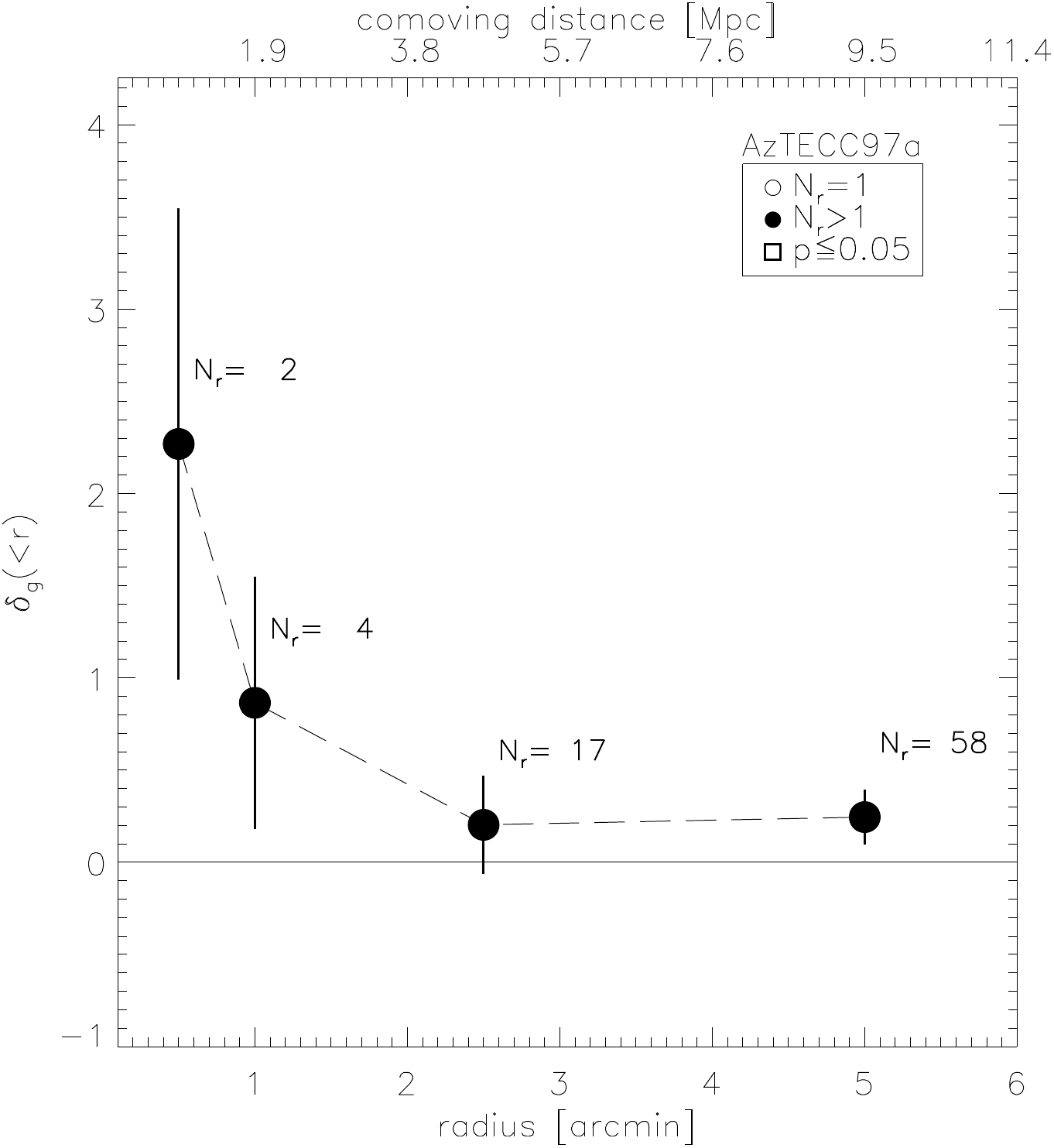}
\includegraphics[width=0.23\textwidth]{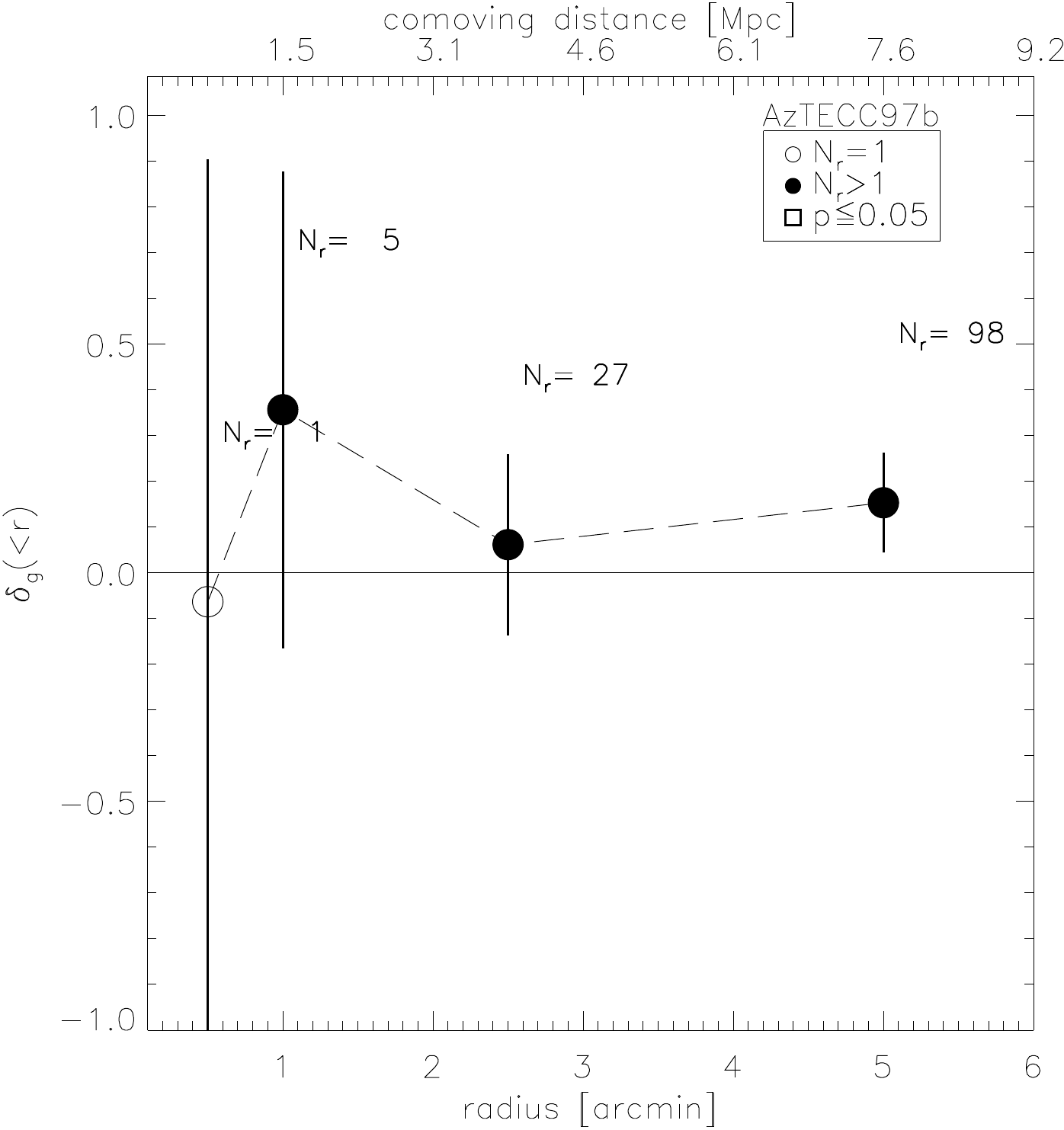}
\includegraphics[width=0.23\textwidth]{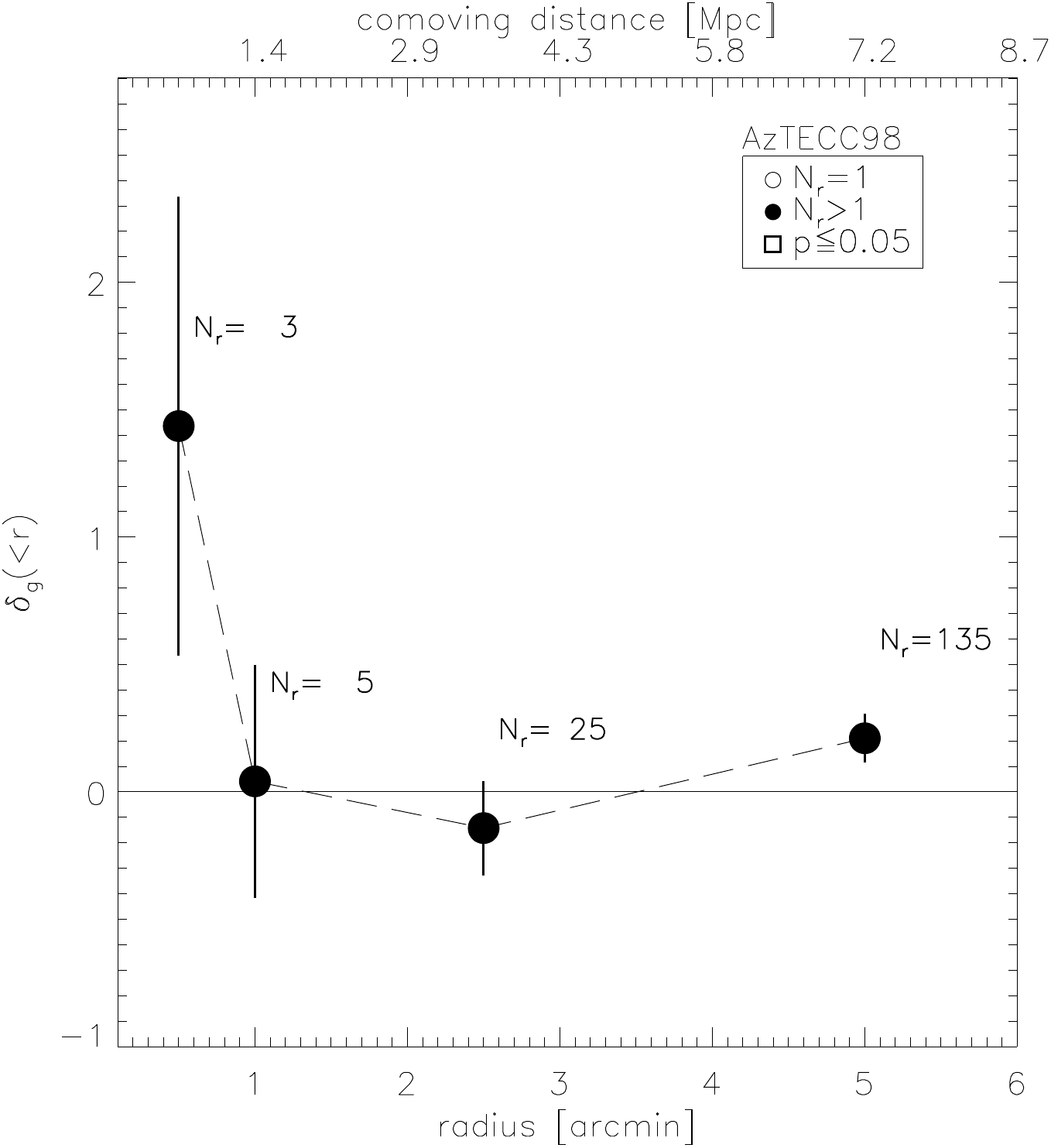}
\includegraphics[width=0.23\textwidth]{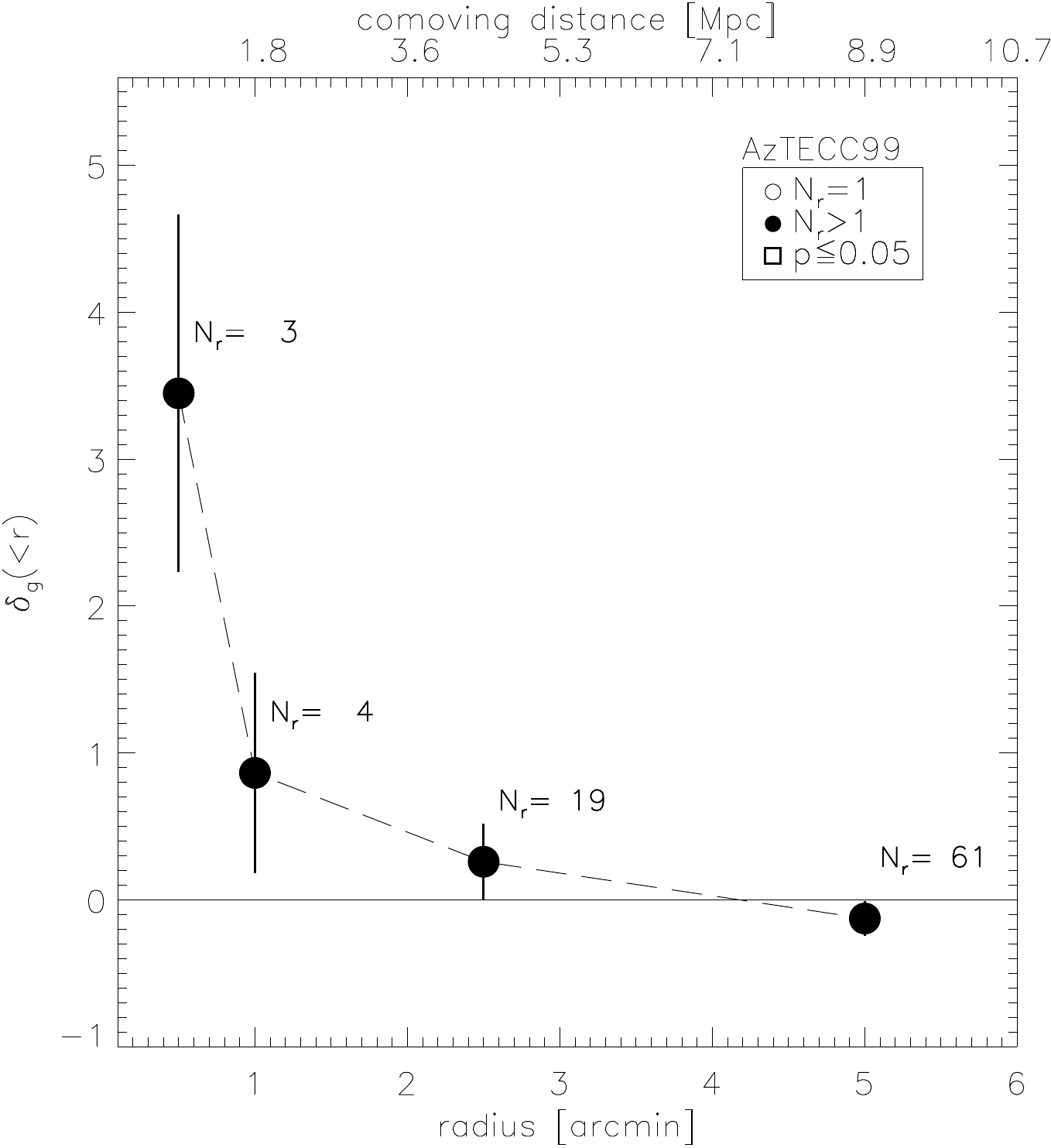}
\includegraphics[width=0.23\textwidth]{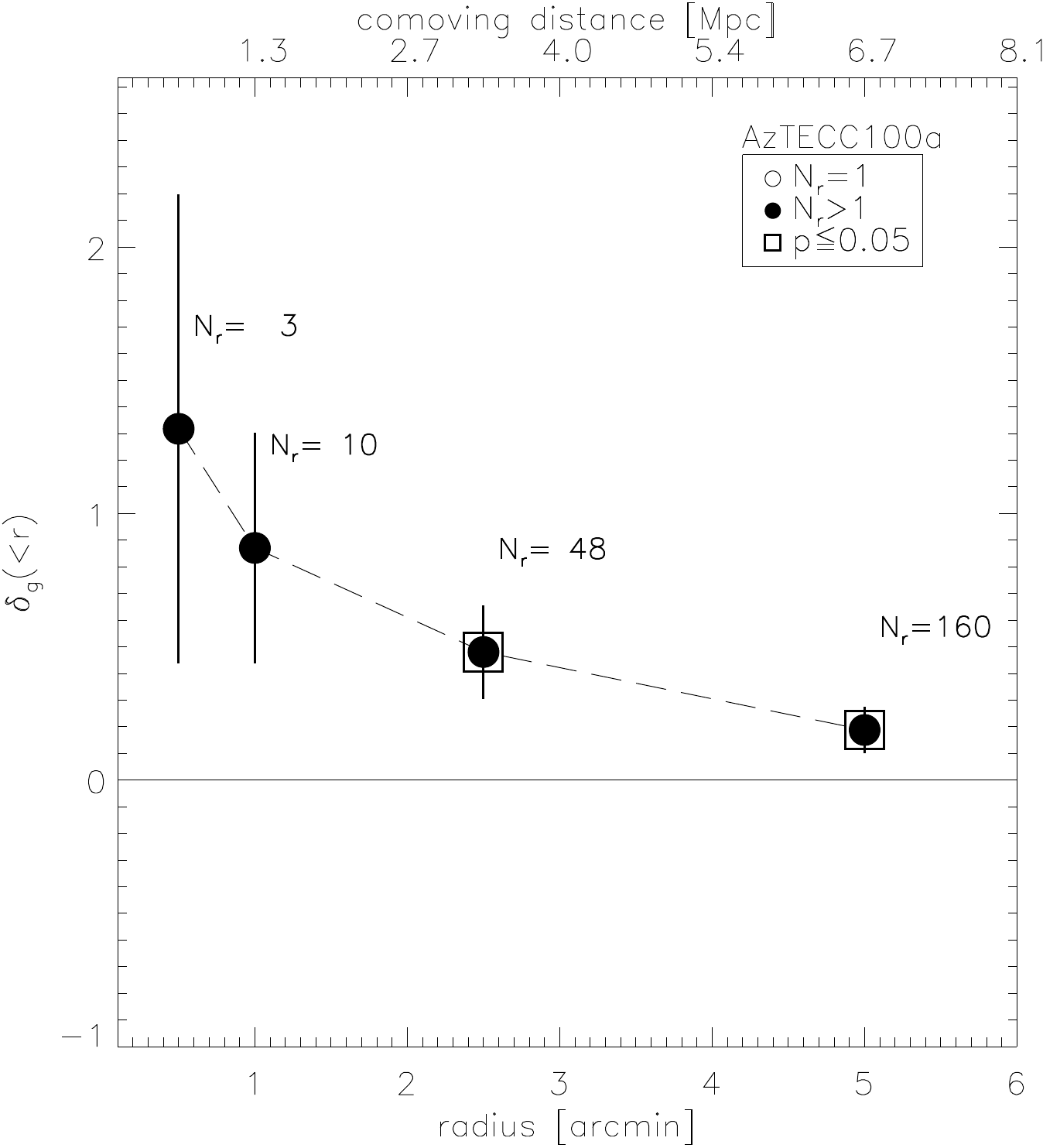}

\caption{continued.}
\end{center}
\end{figure*}

\addtocounter{figure}{-1}
\begin{figure*}
\begin{center}
\includegraphics[width=0.23\textwidth]{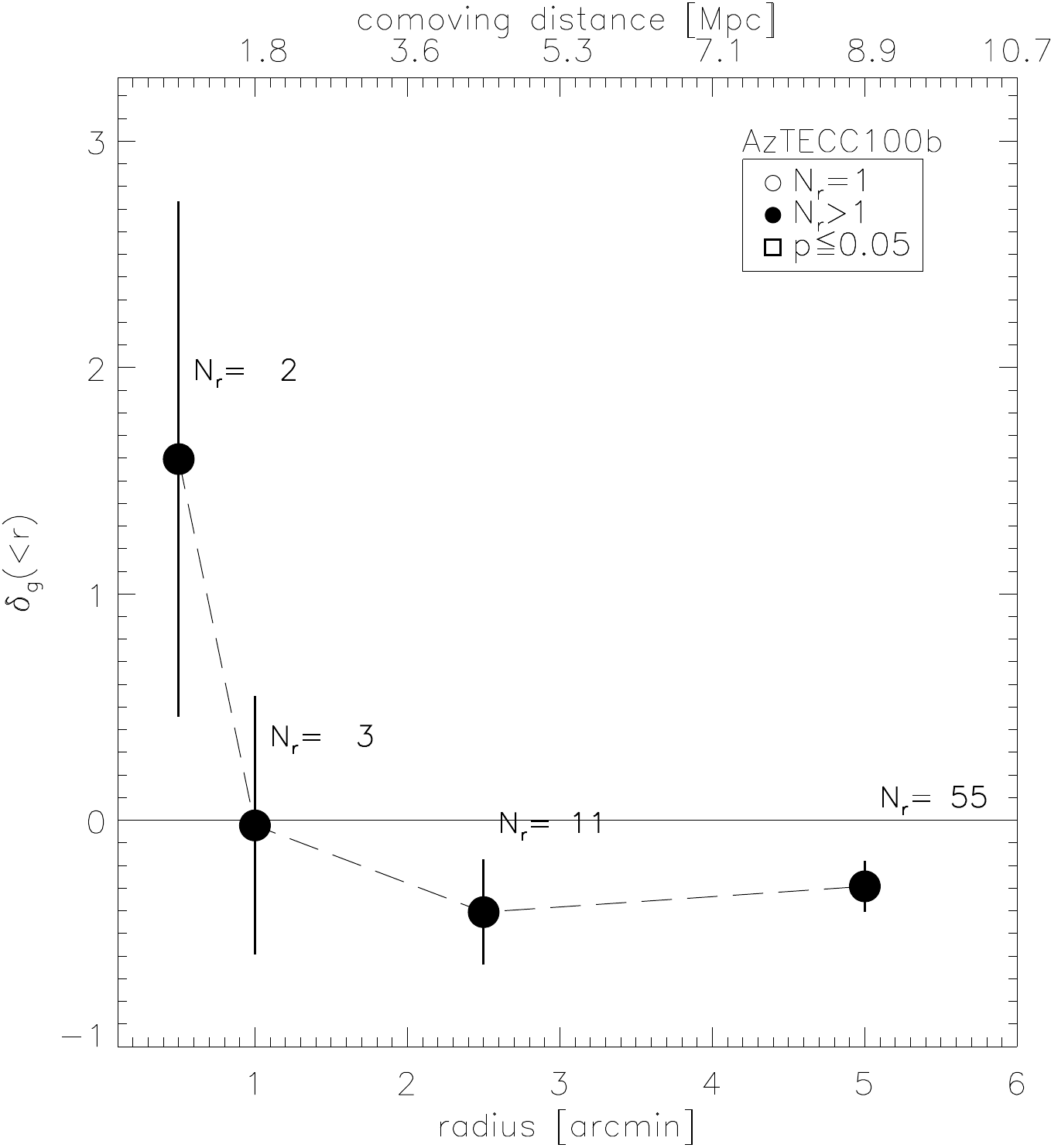}
\includegraphics[width=0.23\textwidth]{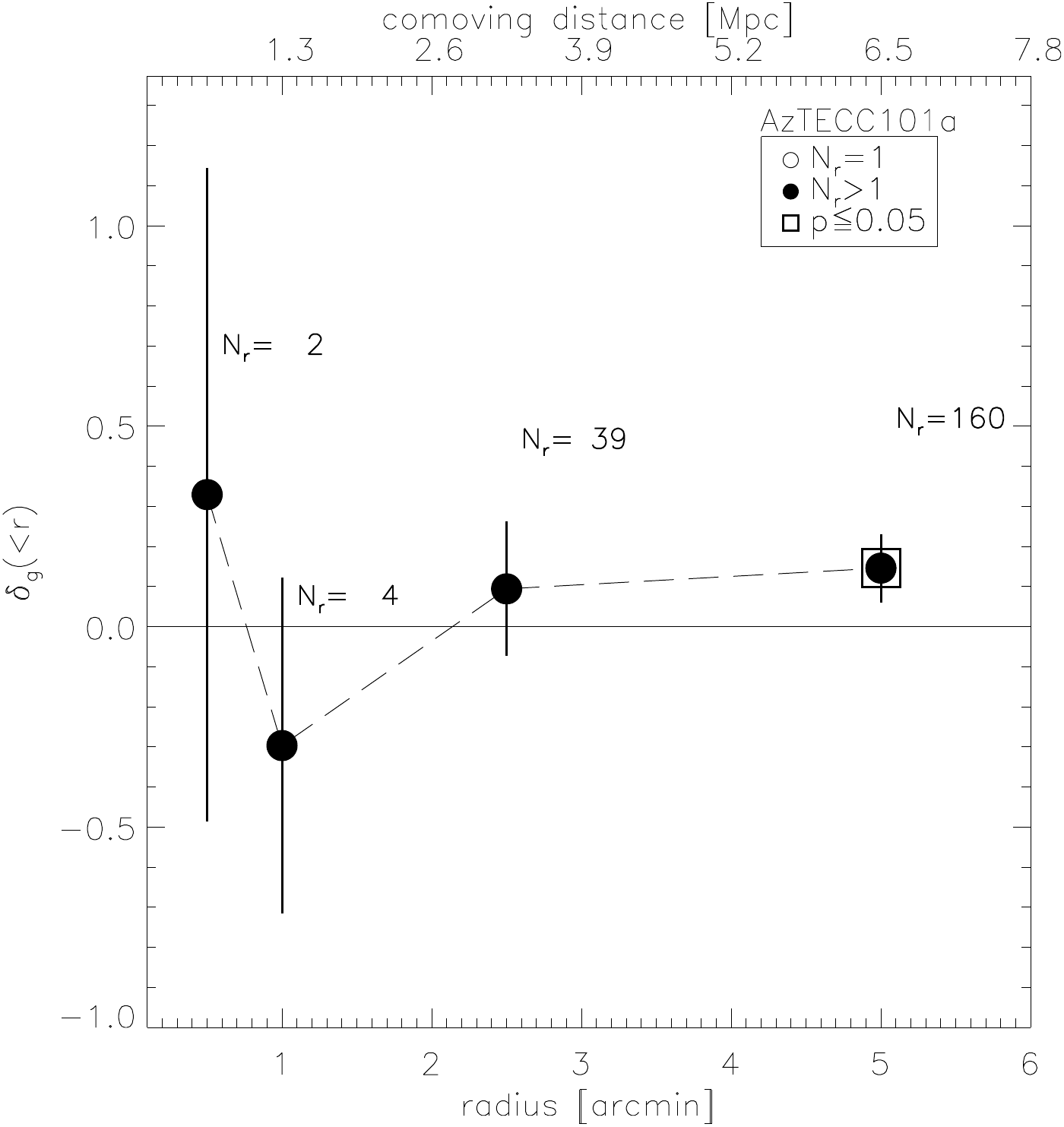}
\includegraphics[width=0.23\textwidth]{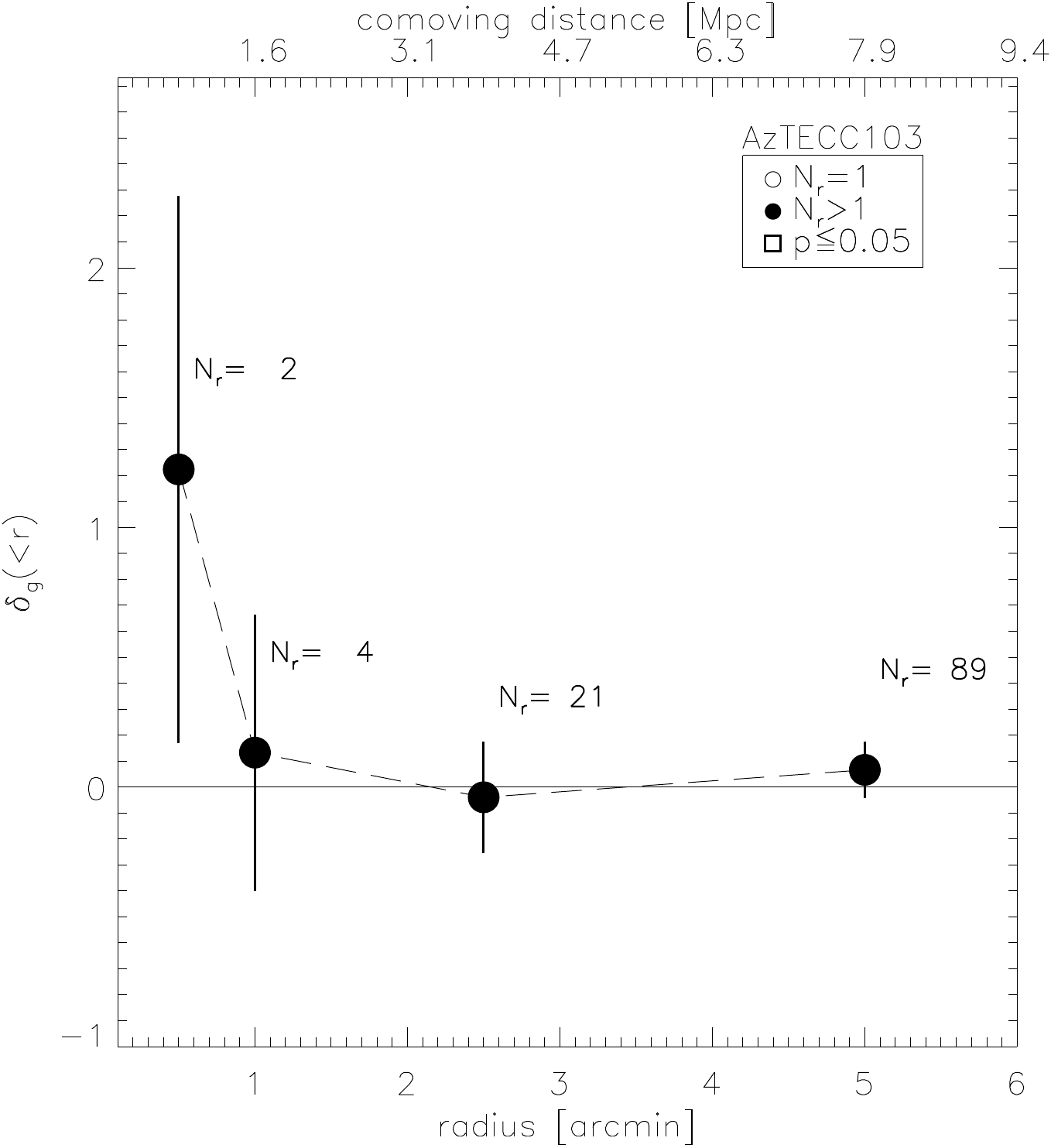}
\includegraphics[width=0.23\textwidth]{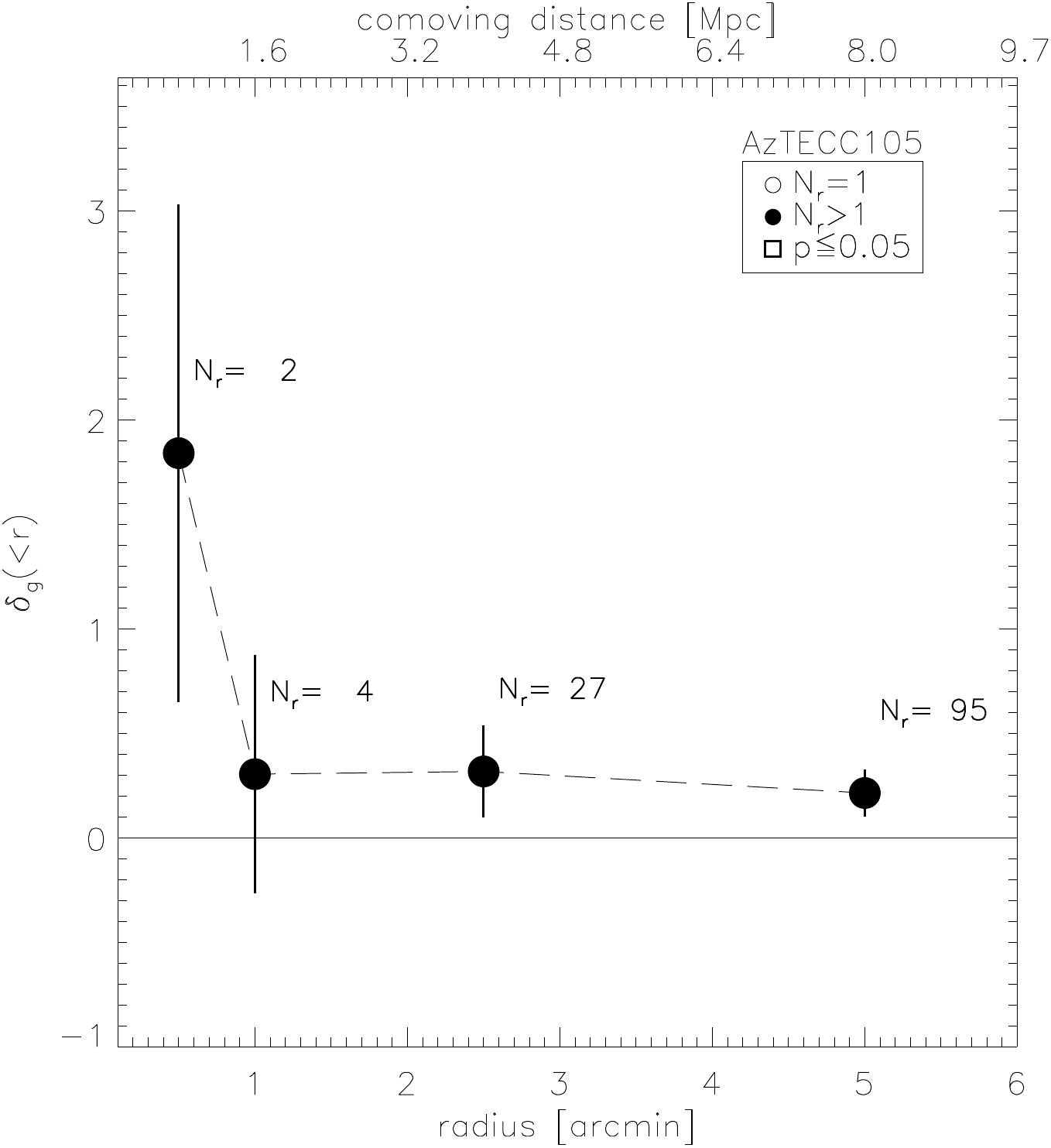}
\includegraphics[width=0.23\textwidth]{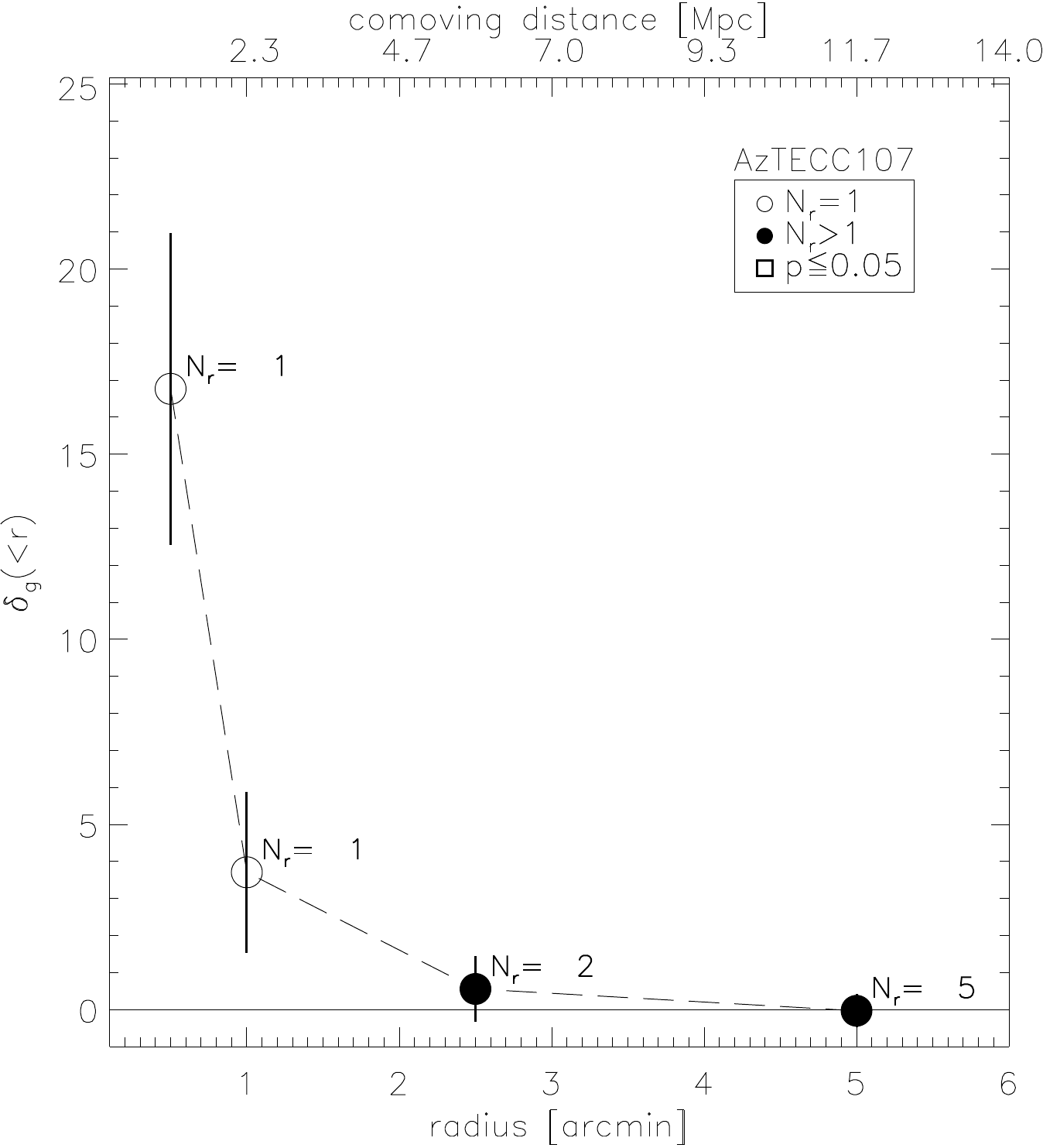}
\includegraphics[width=0.23\textwidth]{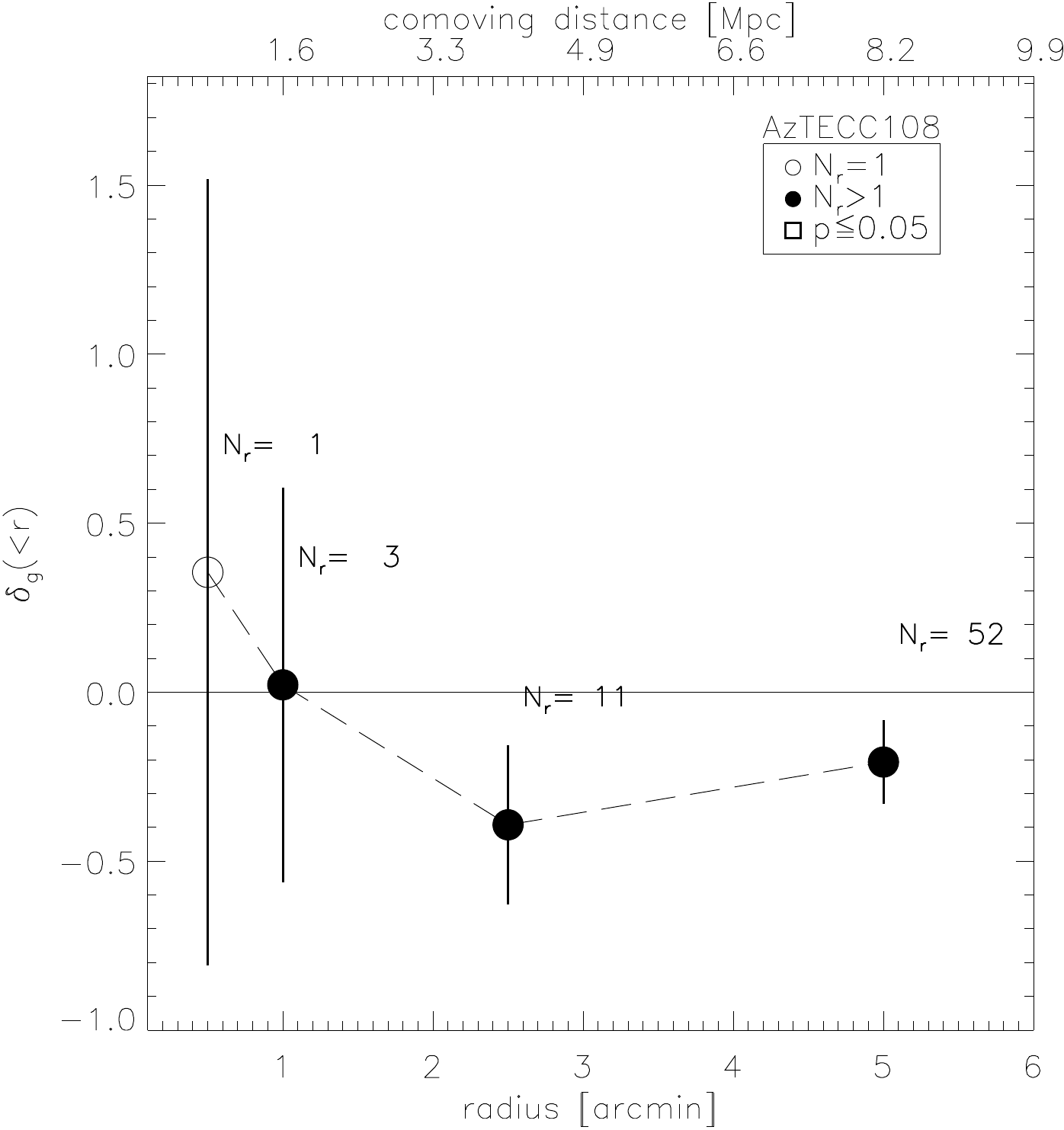}
\includegraphics[width=0.23\textwidth]{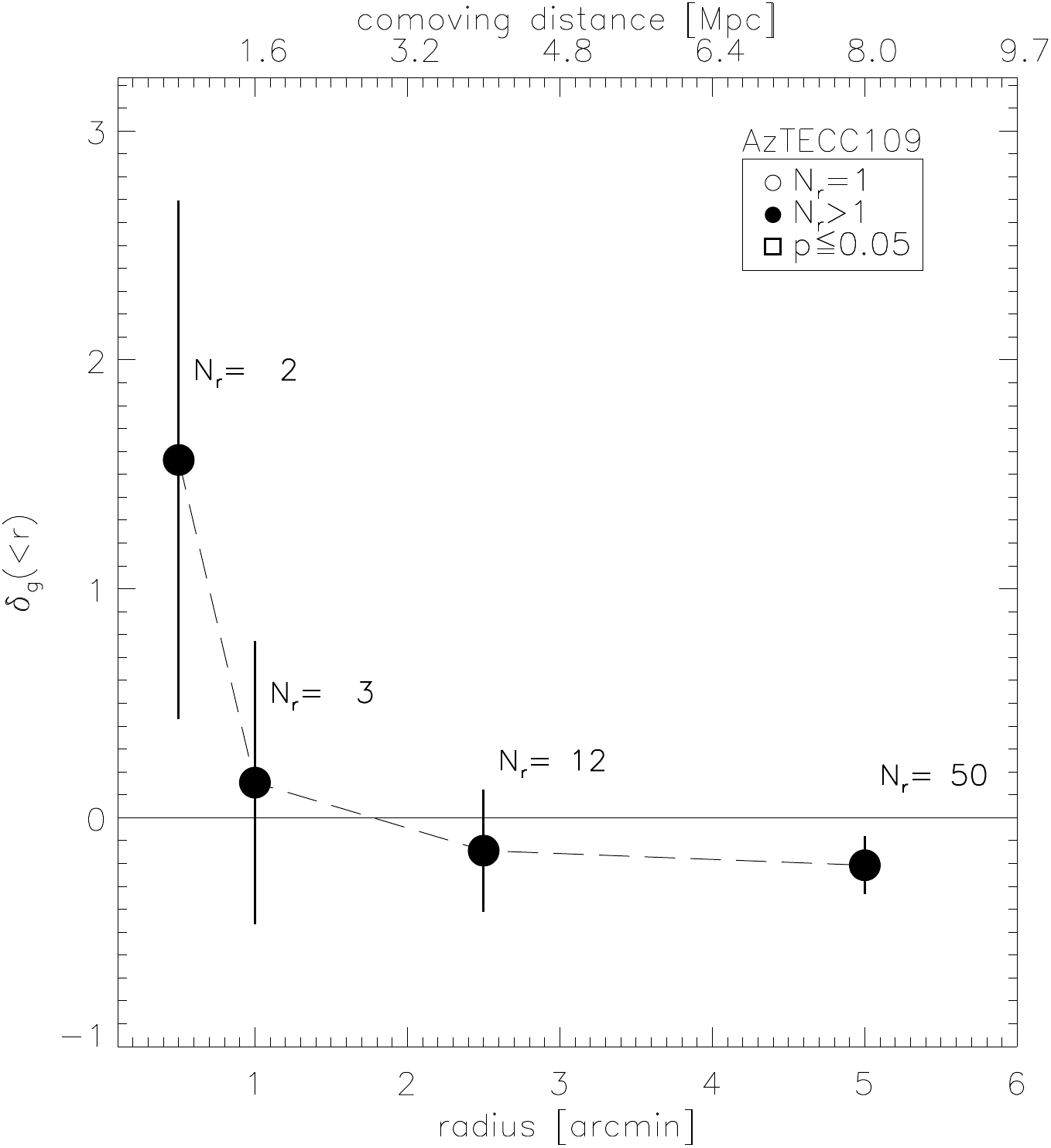}
\includegraphics[width=0.23\textwidth]{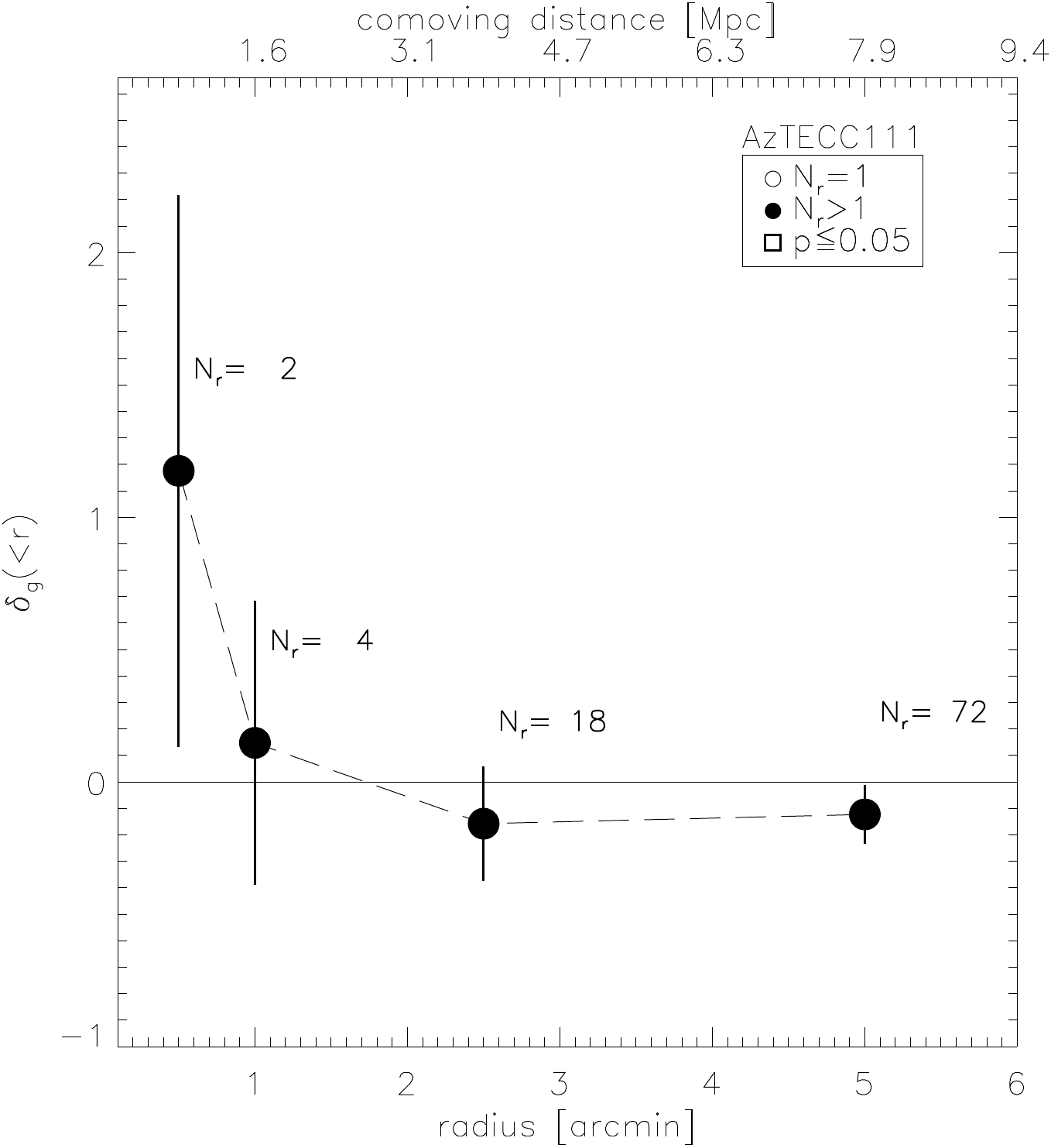}
\includegraphics[width=0.23\textwidth]{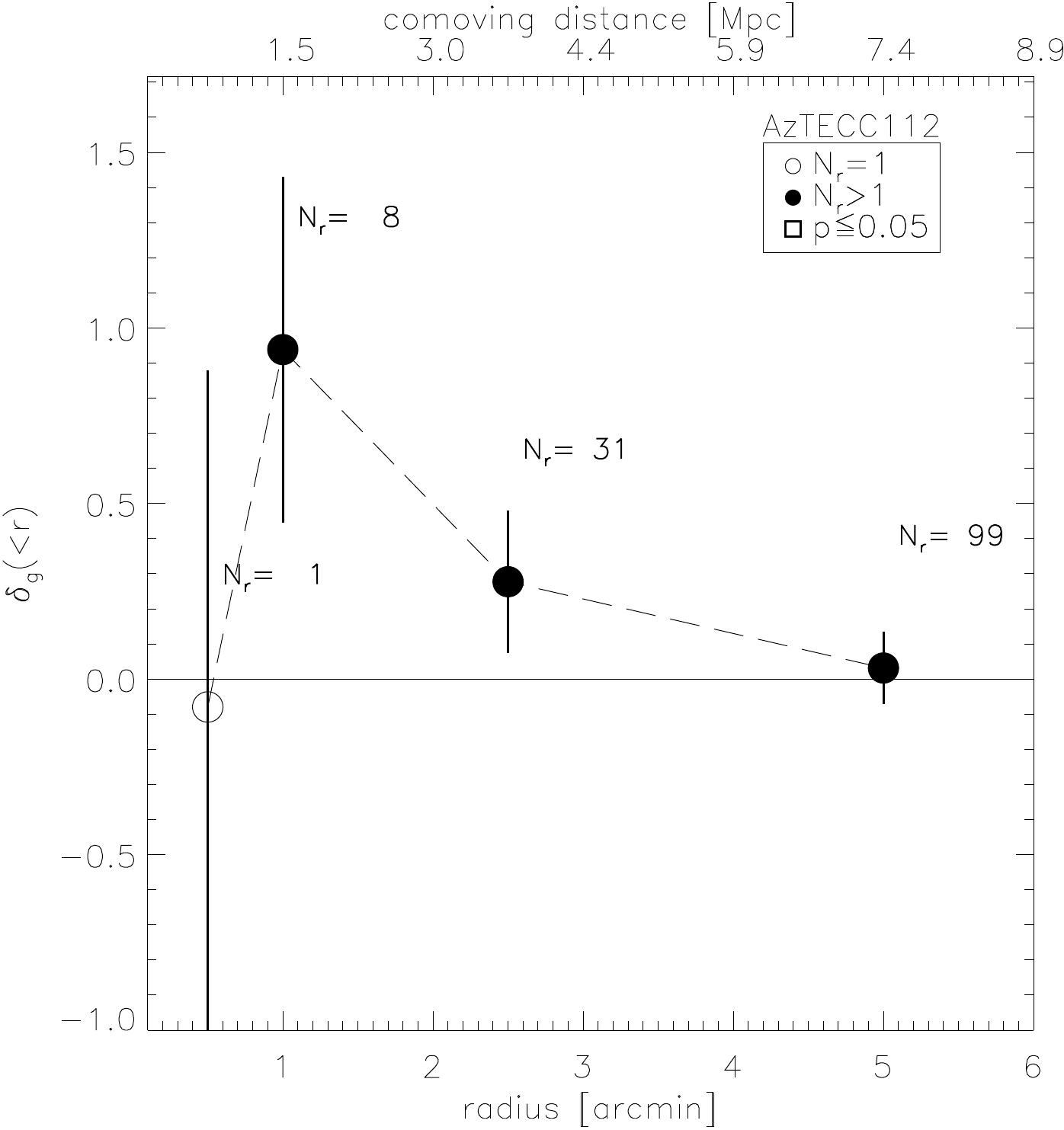}
\includegraphics[width=0.23\textwidth]{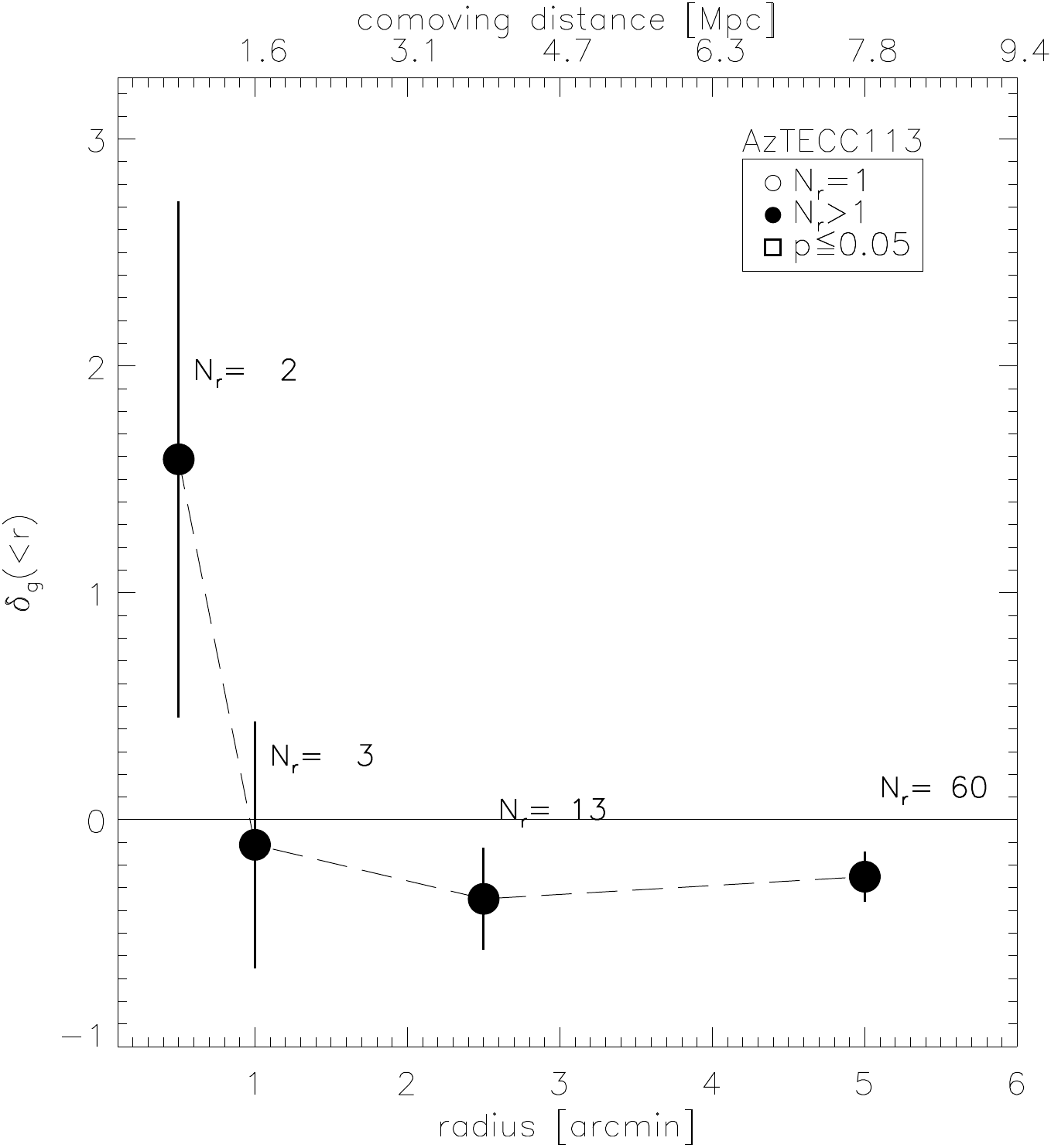}
\includegraphics[width=0.23\textwidth]{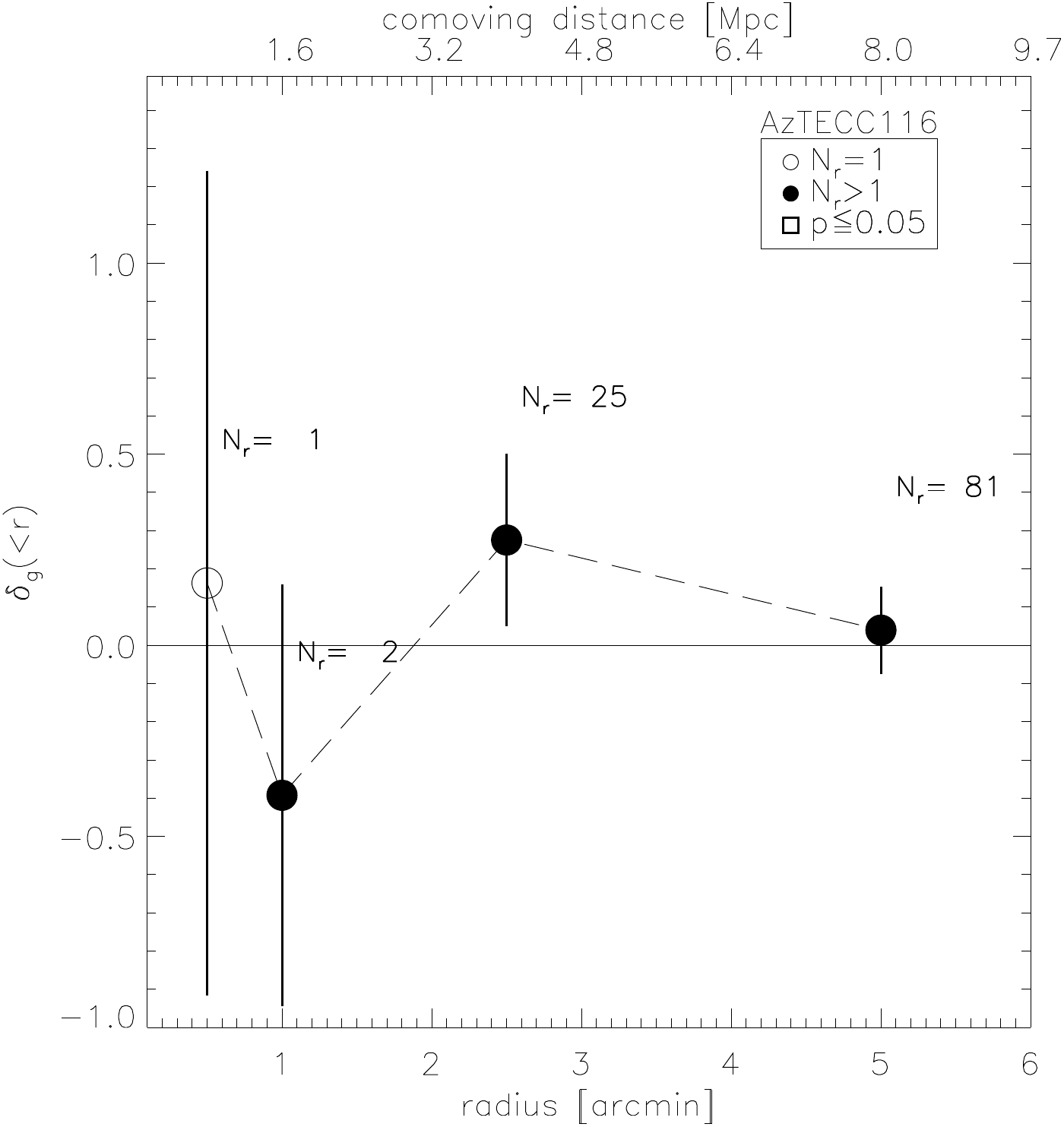}
\includegraphics[width=0.23\textwidth]{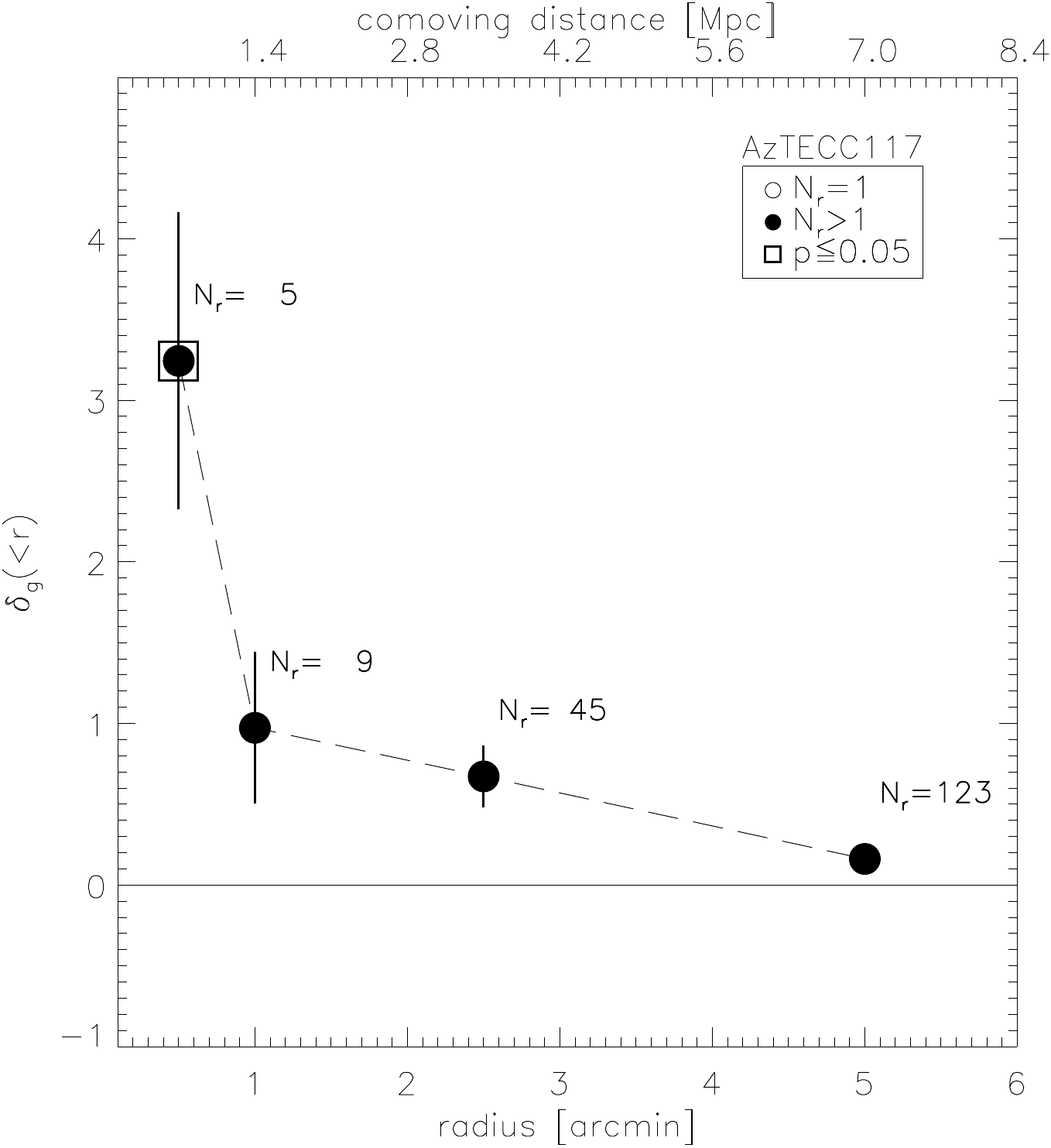}
\includegraphics[width=0.23\textwidth]{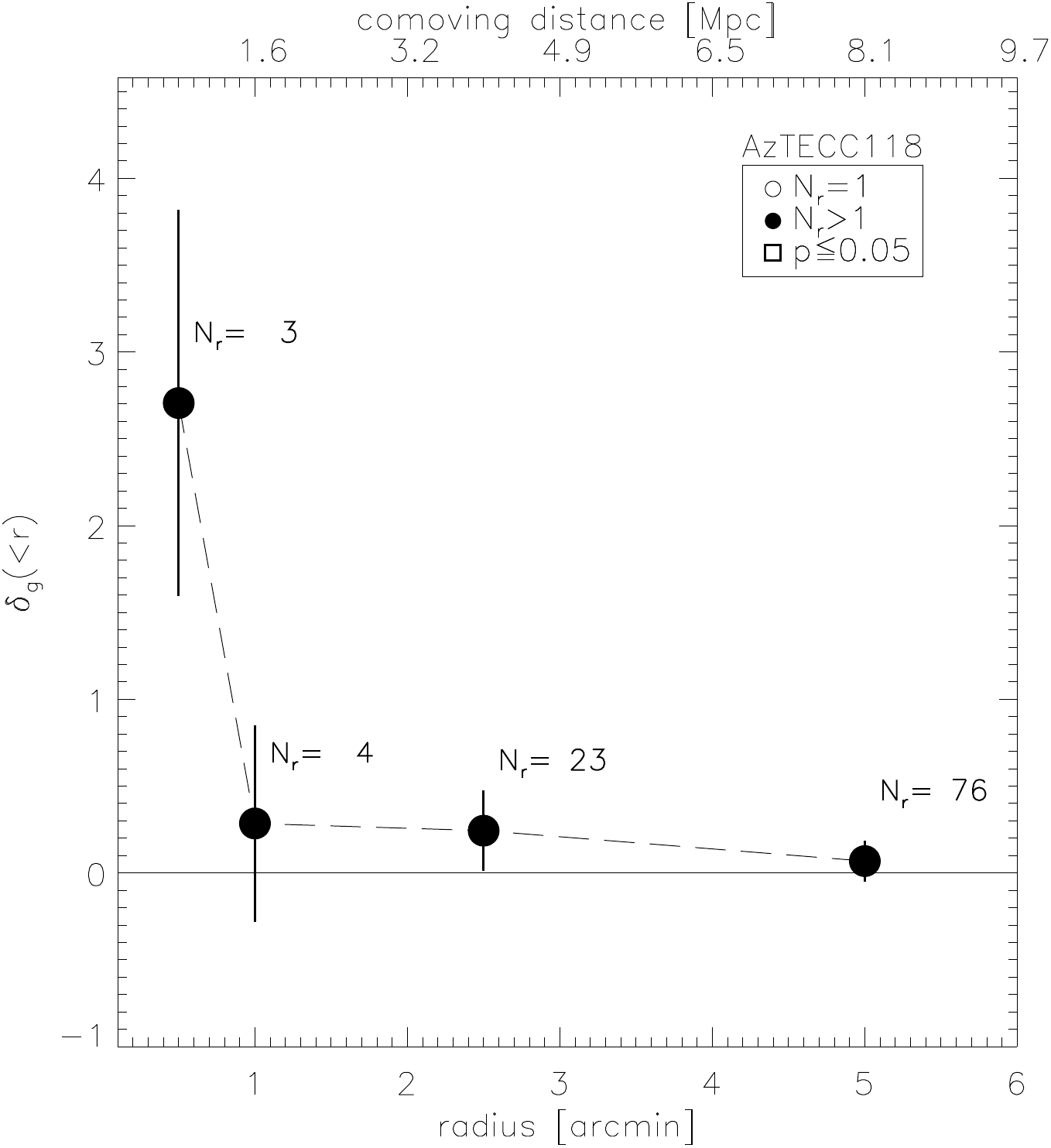}
\includegraphics[width=0.23\textwidth]{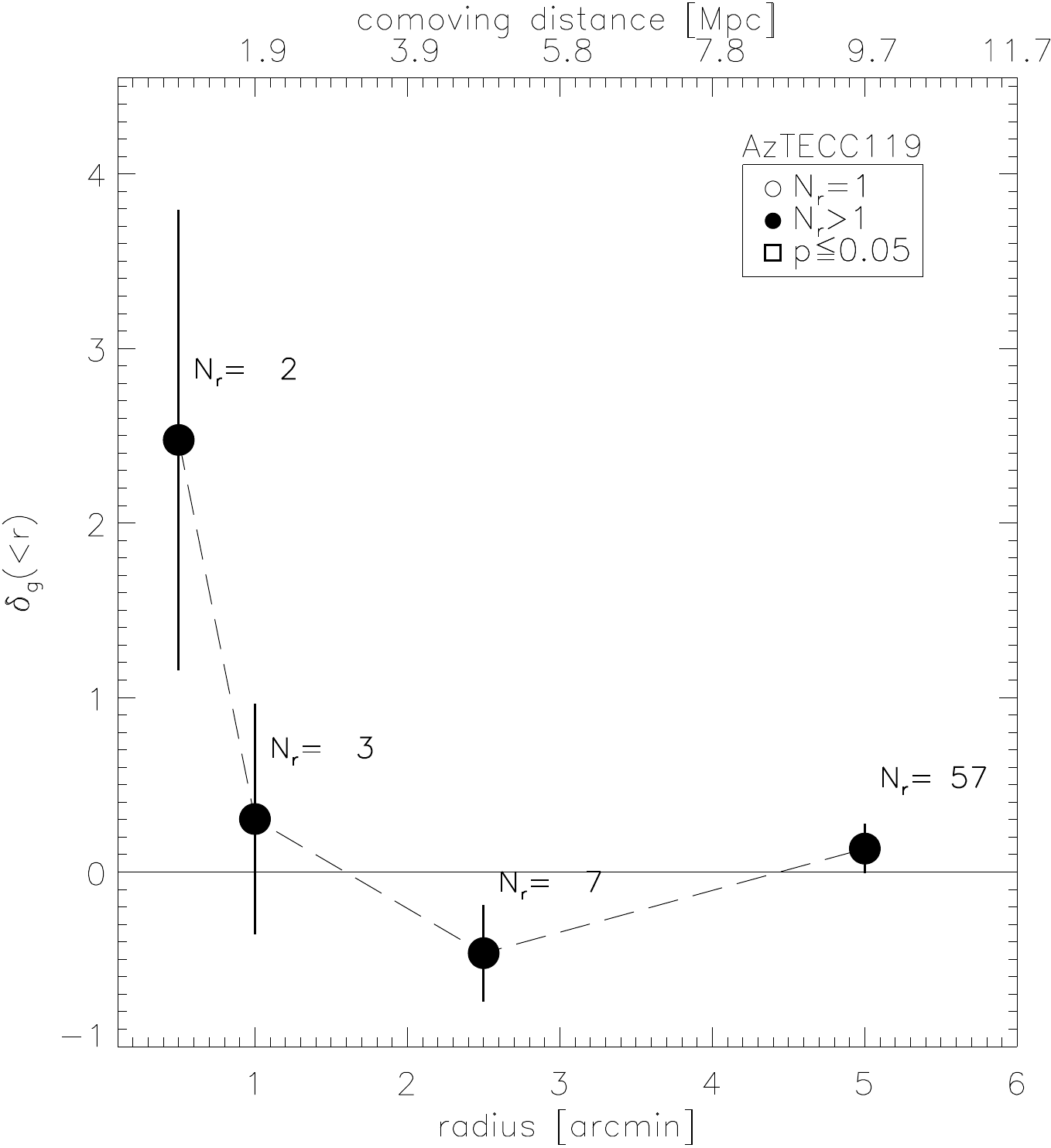}
\includegraphics[width=0.23\textwidth]{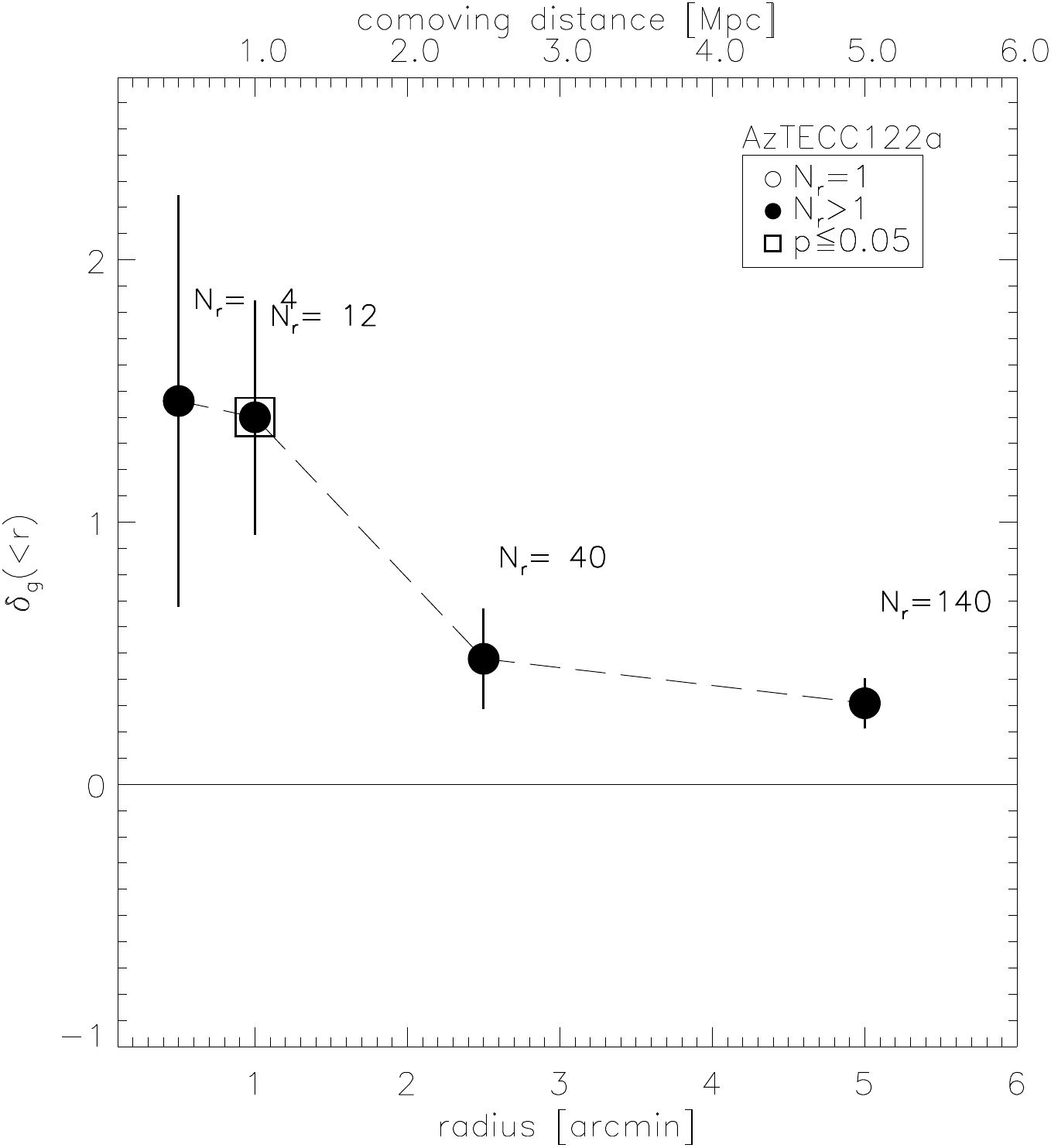}
\includegraphics[width=0.23\textwidth]{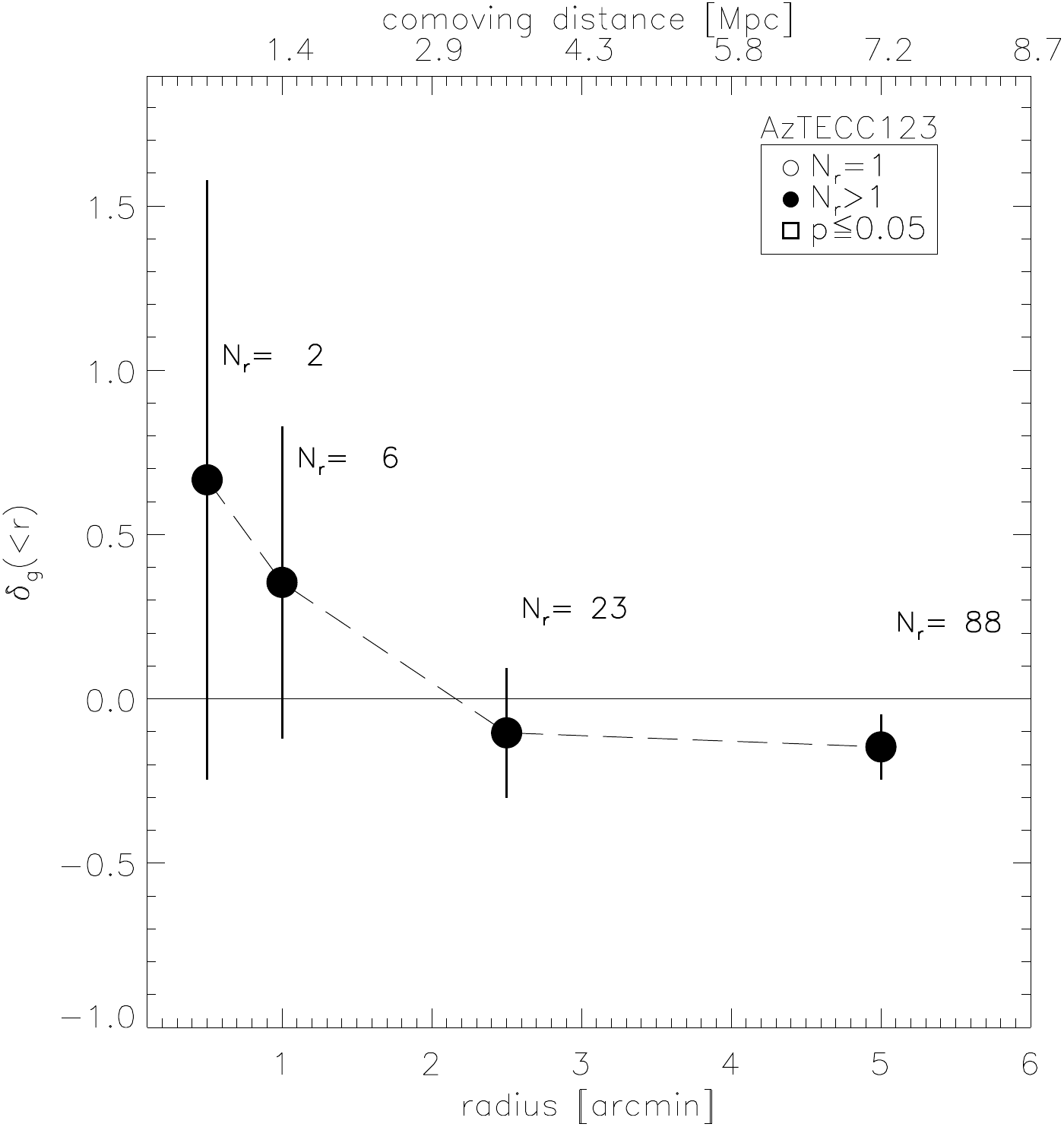}
\includegraphics[width=0.23\textwidth]{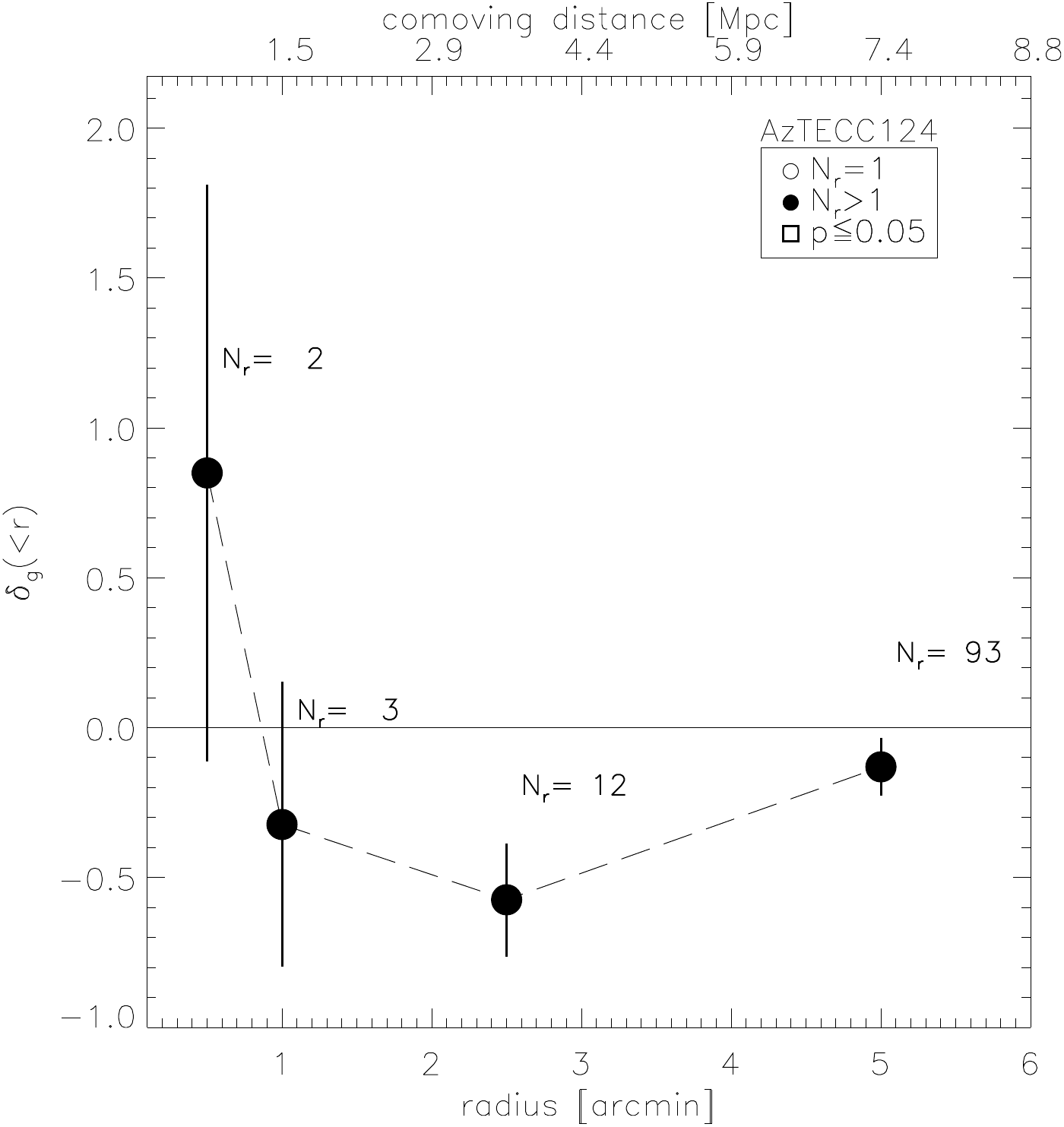}
\includegraphics[width=0.23\textwidth]{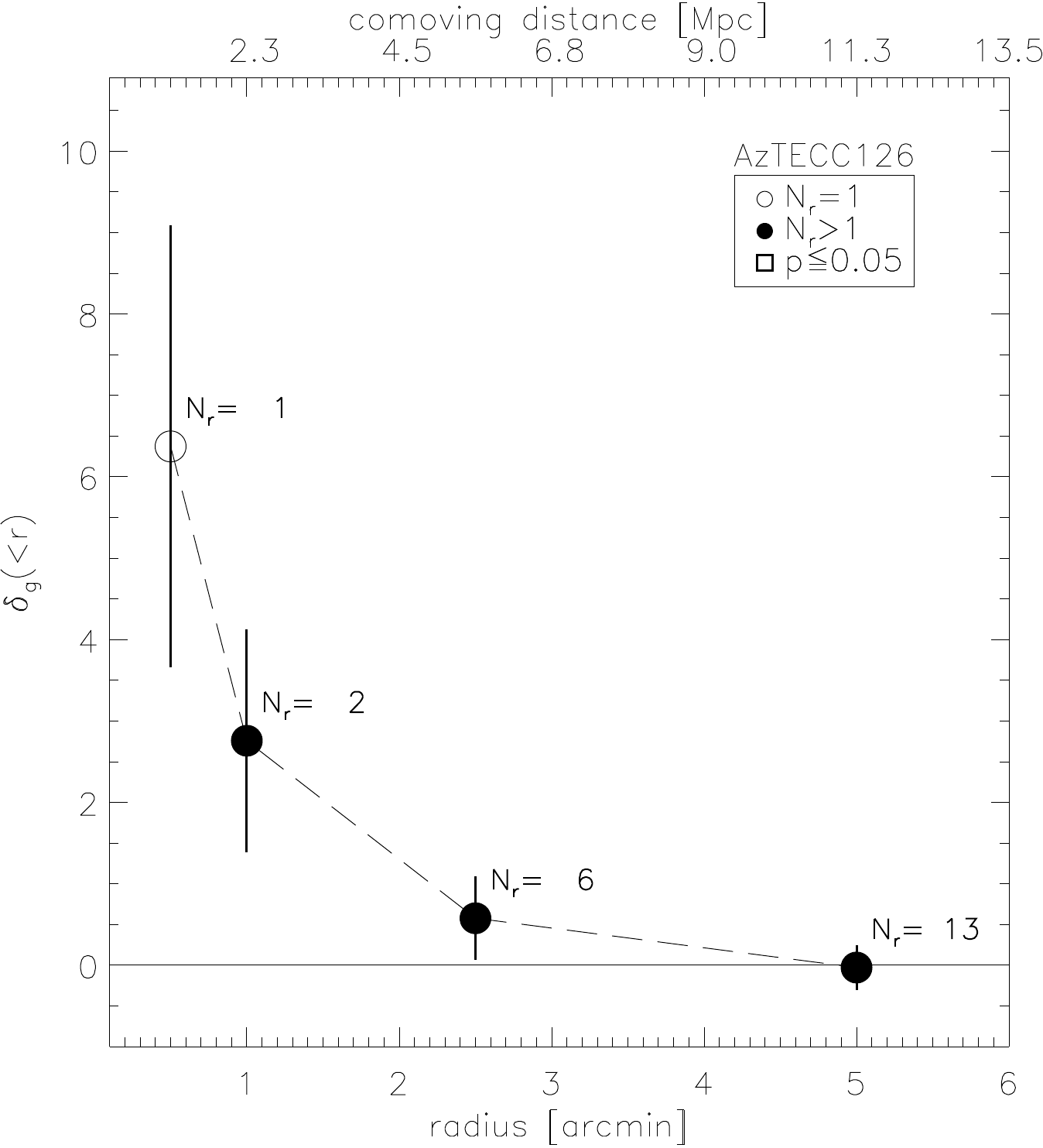}
\includegraphics[width=0.23\textwidth]{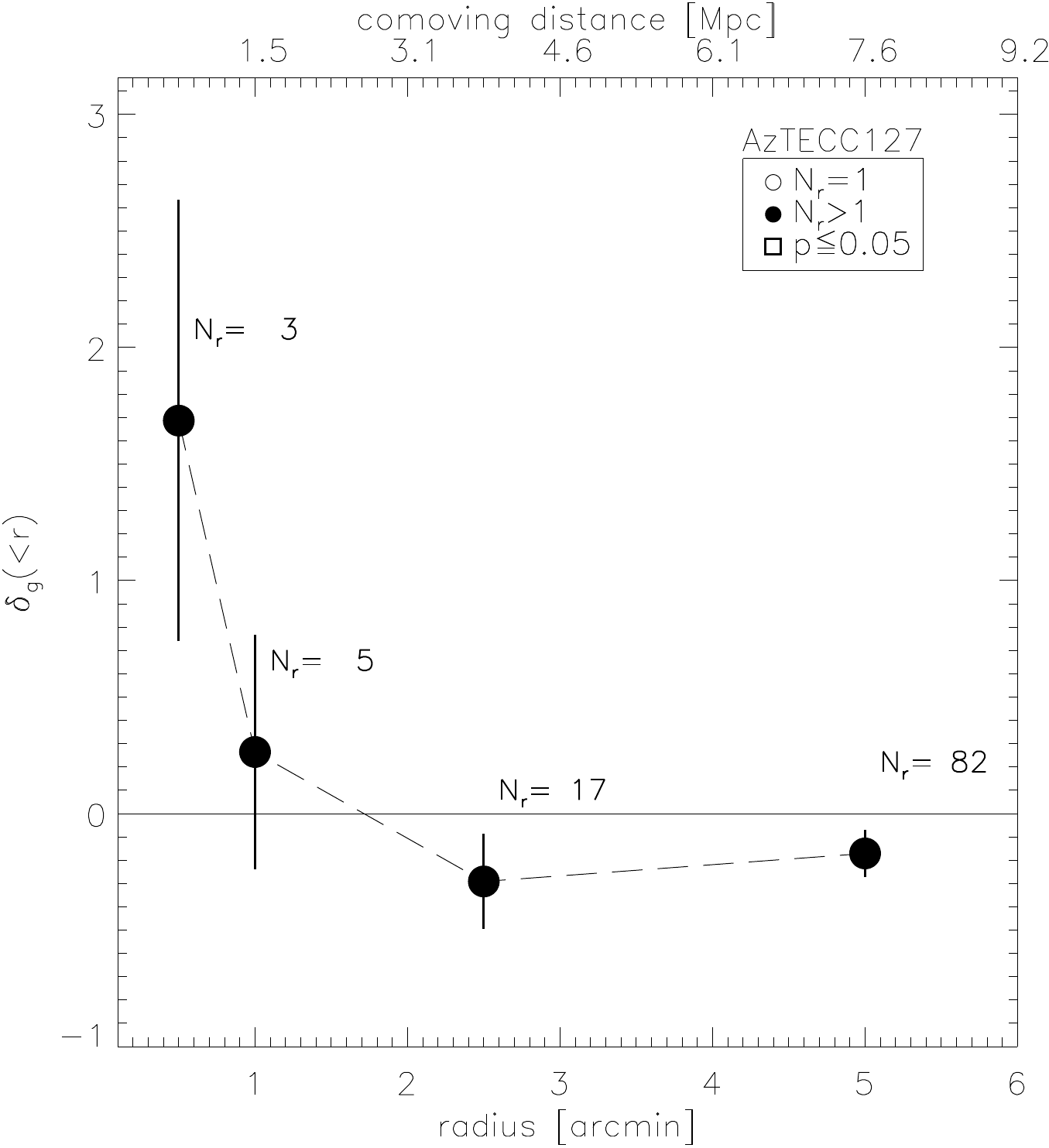}
\includegraphics[width=0.23\textwidth]{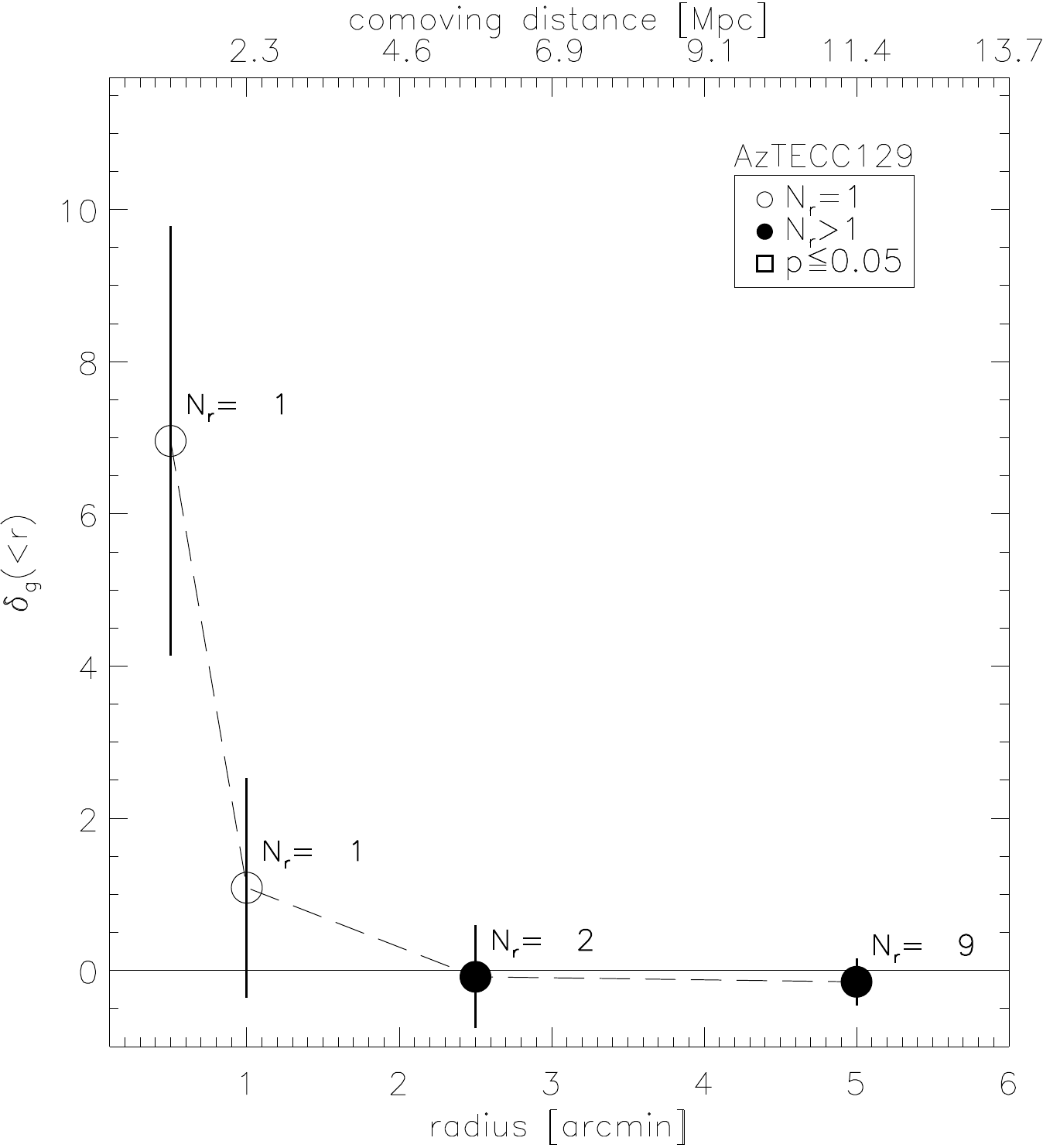}
\caption{continued.}
\end{center}
\end{figure*}

\end{document}